\documentclass[12pt]{report}
\usepackage{graphicx}
\usepackage{amsfonts}
\usepackage{amssymb,amsmath}
\usepackage{latexsym}
\usepackage{color}
\usepackage{cite}
\usepackage{bm}
\usepackage{ulem}

\input{colordvi.tex}

\renewcommand{\thefootnote}{\#\arabic{footnote}}

\begin{document}

\renewcommand{\thepage}{\arabic{page}}
\setcounter{page}{1}
\renewcommand{\thefootnote}{\#\arabic{footnote}}

\begin{titlepage}

\begin{center}

\vspace*{130pt}

{\LARGE
Constraints on the neutrino parameters by future cosmological 21cm line 
and precise CMB polarization observations
}

\vspace{40pt}

{\Large Yoshihiko~Oyama}

\vspace{100pt}

{\Large DOCTOR OF PHILOSOPHY}

\vspace{20pt}

{\Large Department of Particle and Nuclear Physics \\
School of High Energy Accelerator Science \\
The Graduate University for Advanced Studies (SOKENDAI)}

\vspace{20pt}

{\Large 2014 }

\end{center}

\end{titlepage}


\begin{abstract}
Observations of the 21 cm line radiation coming from the epoch of reionization
have a great capacity to study the cosmological growth of the Universe.
Also, CMB polarization produced by gravitational lensing
has a large amount of information about 
the growth of matter fluctuations at late time.
In this thesis, we investigate 
%
their sensitivities to the impact of neutrino property 
on the growth of density fluctuations,
such as the total neutrino mass, the neutrino mass hierarchy, 
the effective number of neutrino species (extra radiation), 
and the lepton asymmetry of our Universe. 

We will show that by combining the precise CMB polarization observations with 
Square Kilometer Array (SKA) 
we can measure the impact of non-zero neutrino mass
on the growth of density fluctuation,
%
and determine the neutrino mass hierarchy at 2~$\sigma$ level
if the total neutrino mass is smaller than 0.1~eV. 

Additionally, we will show that by using these combinations  we 
can constrain the lepton asymmetry better than big-bang nucleosynthesis (BBN).
Besides we discuss constraints on that in the presence of some extra radiation, 
and show that the 21 cm line observations can substantially improve 
the constraints obtained by CMB alone, and allow us to distinguish 
the effects of the lepton asymmetry from those of extra radiation.
\end{abstract}

\tableofcontents

\chapter{Introduction}
\label{ch:intro}

\section{Observations of 21 cm line radiation}

\begin{figure}[tb]
\begin{center}
\includegraphics[bb= 19 494 792 829, width=0.8\linewidth]{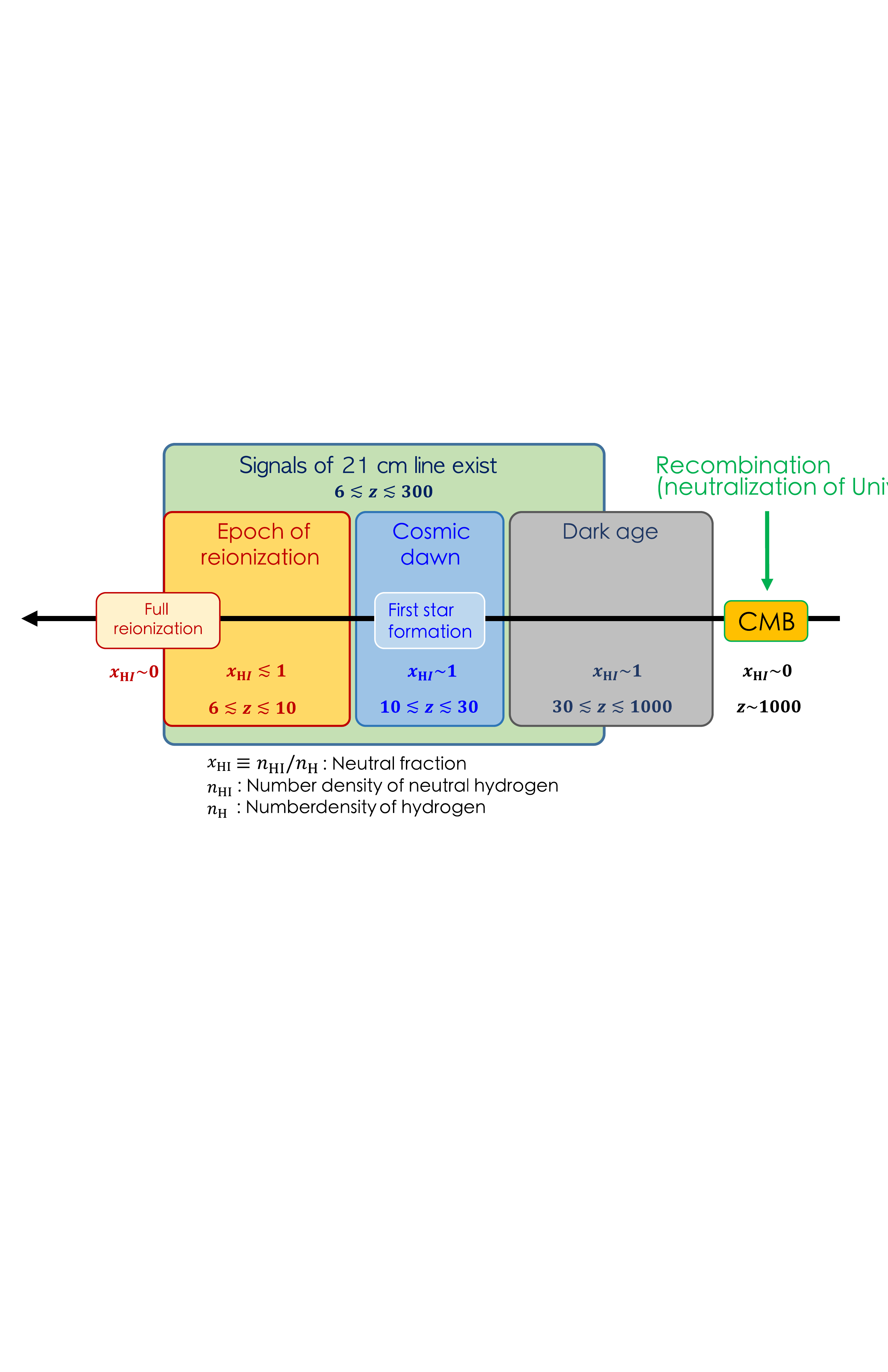}
\caption{Epochs in which  cosmological 21 cm line was emitted.}
\label{fig:darkage}
\end{center}
\end{figure}

Observations of the high redshift Universe 
($6 \hspace{0.3em} \raisebox{0.4ex}{$<$}\hspace{-0.75em}
\raisebox{-.7ex}{$\sim $}\hspace{0.3em} z$) 
with the 21 cm
line of neutral hydrogen attracts attention because it opens a new
window to the early phases of the cosmological structure formations.
After recombination ($z\sim1100$), because the
Universe is neutral, and there had not existed any luminous objects
yet, this era is called ``the cosmic dark age''.
After the dark age, first luminous objects formed at around $z \sim
30$, and this epoch is called ``the cosmic dawn'' or just ``the late
time of the dark age''. Finally, X-rays are emitted from the remnants
of the luminous objects, which ionizes inter-galactic medium
(IGM). This epoch is called ``the epoch of reionization (EOR)''.
So far, it has been a big challenge to observe such past epochs.
However, 
there are a large amount of neutral hydrogen gas in these epochs, 
%
%
and we can
observe the high redshift Universe by using the 21 cm line which are emitted from the gas
(Fig.\ref{fig:darkage}).
This is the reason why the observation of 
the 21 cm line attracts attention quite recently.

%
Using the observation of the 21 cm line, 
we can not only study how the Universe was ionized
but also obtain 
information about the 
%
%
density fluctuations of matter 
because the distribution of neutral hydrogen
traces that of cold dark matter (CDM).
Therefore, we can use the observation of the 21 cm line like those of CMB
or Galaxy surveys, and constrain cosmological parameters such as the
density parameter for the energy density of CDM $\Omega_{c}$ or of dark
energy $\Omega_{\Lambda}$.
Besides, the observation of the 21 cm line has some advantages.  First,
the observation enables us to survey very past eras and wide redshift
ranges (21 cm tomography). 
Secondly, in such high redshift eras, 
the non-linear growth of the fluctuation is smaller than that in later
epochs. Therefore, theoretical uncertainties of the predictions for the
21cm line observation are much smaller than those for galaxy surveys.

\section{21 cm line}
\begin{figure}[tb]
\begin{center}
\includegraphics[bb= 0 20 595 252, width=0.7\linewidth]{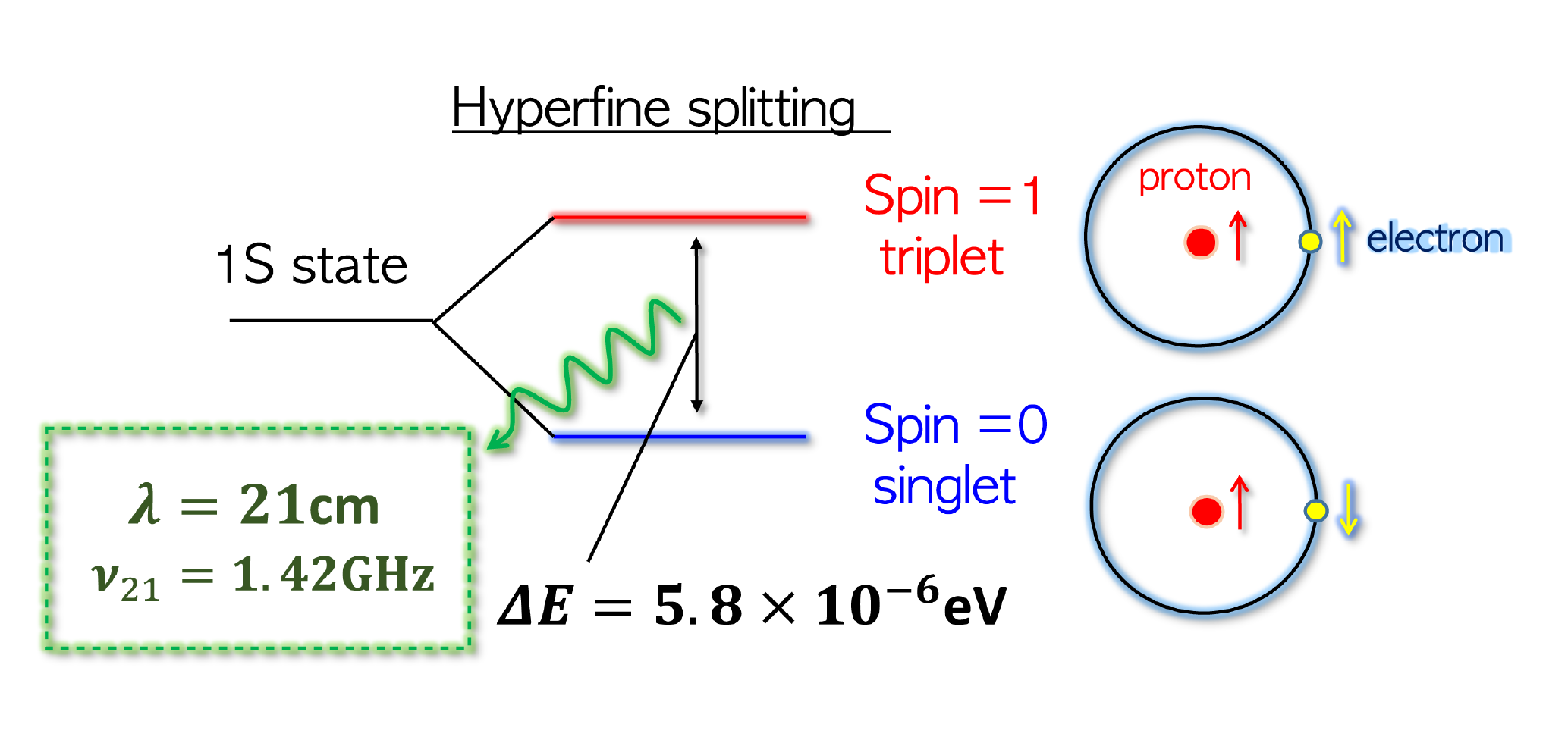} 
\caption{Hyperfine structure of neutral hydrogen atom.\label{fig:21cmline}}
\end{center}
\end{figure}

The 21 cm line of neutral hydrogen atom is emitted by transition
between the hyperfine levels of the 1S ground state,
and the hyperfine structure is induced by an interaction of magnetic moments 
between proton and electron (see Appendix~\ref{ap:bisai}).
%
The energy difference of the hyperfine structure is $\Delta E \sim 5.8\times 10^{-6}$eV,
and this energy corresponds to the frequency $\nu_{21} \simeq 1.4$GHz (the wave length
is $\lambda \simeq 21$cm). Therefore this spectral line is called the 21 cm line
(see Fig.\ref{fig:21cmline}).

\section{Neutrino mass and its properties}

Due to the discovery of non-zero neutrino masses by Super-Kamiokande 
through neutrino oscillation experiments in 1998, 
the standard model of particle physics was forced to 
be modified so as to theoretically include the neutrino masses.

So far only the mass-squared differences of neutrino species
have been measured by neutrino oscillation experiments, 
which are reported  to be $\Delta m^2_{21}\equiv m_{2}^2 -m_{1}^2
=7.59^{+0.19}_{-0.21}\times 10^{-5} {\rm eV}^2$~\cite{Aharmim:2008kc}
and $\Delta m^2_{32}\equiv m_{3}^2 -m_{2}^2
=2.43^{+0.13}_{-0.13}\times 10^{-3} {\rm eV}^2$~\cite{Adamson:2008zt}.
However, absolute values and their hierarchical structure 
(normal or inverted) have not been obtained yet
although information on them is indispensable to build new particle physics models.

In particle physics, some new ideas and new future experiments
have been proposed to measure the absolute values
and/or determine the hierarchy of neutrino masses, 
e.g., through tritium beta decay in KATRIN experiment~\cite{KATRIN}, 
neutrinoless double-beta decay~\cite{GomezCadenas:2010gs}, 
atmospheric neutrinos in the proposed iron calorimeter at INO~\cite{INO,Blennow:2012gj}
and the upgrade of the IceCube detector (PINGU) \cite{Akhmedov:2012ah},
and long-baseline oscillation experiments, 
e.g., NO$\nu$A~\cite{Ayres:2004js},
J-PARC to Korea (T2KK)~\cite{Ishitsuka:2005qi,Hagiwara:2005pe} 
or Oki island (T2KO)~\cite{Badertscher:2008bp},
and CERN to Super-Kamiokande
with high energy (5~GeV) neutrino beam~\cite{Agarwalla:2012zu}.

On the other hand, 
such non-zero neutrino masses affect cosmology significantly
through suppression of growth of density fluctuation 
%
%
because relativistic neutrinos have large large thermal velocity
and erase their own density fluctuations up to horizon scales 
due to their free streaming behavior.
%
%
By measuring power spectra of density fluctuations,
we can constrain the total neutrino mass
$\Sigma~m_{\nu}$ \cite{Komatsu:2008hk,Komatsu:2010fb,Reid:2009xm,
Hannestad:2010yi,Elgaroy:2003yh,Reid:2009nq,Crotty:2004gm,
Goobar:2006xz,Seljak:2004xh,Ichikawa:2004zi,Seljak:2006bg,Fukugita:2006rm,
Ichiki:2008ye,Thomas:2009ae,RiemerSorensen:2011fe,
Hamann:2010pw,Saito:2010pw,Ade:2013zuv}
and the effective number of neutrino species $N_{\nu}$
\cite{Komatsu:2008hk,Komatsu:2010fb,Reid:2009xm,Reid:2009nq,Crotty:2004gm,
Pierpaoli:2003kw,Crotty:2003th,Seljak:2006bg,Hamann:2010pw,Ade:2013zuv}
through observations of cosmic microwave background (CMB) 
anisotropies and large-scale structure (LSS).
%
%
The robust upper bound on $\Sigma~m_{\nu}$ obtained so far is
$\Sigma m_{\nu} < 0.23$~eV (95 $\%$ C.L.) 
by the CMB observation by Planck (see Ref.~\cite{Ade:2013zuv}).
For forecasts for future CMB observations, see also
Refs.~\cite{Lesgourgues:2005yv,dePutter:2009kn,Abazajian:2013oma,Wu:2014hta}.

Moreover, by observing the power spectrum of 
cosmological 21 cm line radiation fluctuation, 
we will be able to obtain useful information on the neutrino
masses~\cite{McQuinn:2005hk,Loeb:2008hg,Pritchard:2008wy,Pritchard:2009zz,Abazajian:2011dt,Oyama:2012tq}. 
That is because the 21 cm line radiation is emitted 
(1) long after the recombination (at a redshift $z \ll 10^3$ ) and 
(2) before an onset of the LSS formation. 
The former condition (1) gives us information on
smaller neutrino mass ( $\lesssim 0.1 $~eV),
and the latter condition (2) means we
can treat only a linear regime of the matter perturbation, 
which can be analytically calculated unlike the LSS case.

%
In actual analyses, it is essential that we combine 
data of the 21 cm line with those of CMB
because the constrained cosmological parameter space 
is complementary to each other. 
%
For example, 
the former is quite sensitive to the dark energy density, 
but the latter is relatively insensitive to it. 
On the other hand, the former has only a mild sensitivity to
the normalization of matter perturbation, 
but the latter has an obvious sensitivity to it by definition. 
In pioneering work by~\cite{Pritchard:2008wy}, 
the authors tried to make a forecast for 
constraint on the neutrino mass hierarchy 
by combining Planck with future 21 cm line observations 
in case of relatively degenerate neutrino masses 
$\Sigma m_{\nu}\sim 0.3 $~eV.

Additionally, 
there is another issue related to the neutrino properties,
it is the lepton asymmetry of the Universe.
The issue of the asymmetry of matter and antimatter 
in the Universe is one of the important subject 
in cosmology and particle physics. 
The baryon asymmetry is now accurately determined 
by using the combination of cosmological observations such as 
cosmic microwave background (CMB), 
big bang nucleosynthesis (BBN), large scale structure, 
type Ia supernovae and so on. 
It is represented in term of the baryon-photon ratio
%
$\eta =  (n_b - n_{\bar{b}})/n_\gamma \simeq 6 \times 10^{-10}$,
where $n_b, n_{\bar{b}}$ and $n_\gamma$ are
the number densities of baryon, anti-baryon and photon, respectively.
On the other hand, the asymmetry in the leptonic sector
is not well determined and
only a weak constraint on the neutrino degeneracy parameter 
$\xi_\nu = \mu_\nu / T_\nu$ is obtained~\footnote{
  So far constraints on $\xi_\nu$ have been obtained by BBN (e.g., see
  ~\cite{Kohri:1996ke,Sato:1998nf} and Fig. \ref{fig:etaXi2014} in
  Appendix~\ref{sec:BBNrelation}), which is sometimes combined with
  CMB and/or some other observations (e.g., see
  Refs.~\cite{Popa:2008tb,Shiraishi:2009fu,Caramete:2013bua}).
}.
%
Although the lepton asymmetry is expected to be the same order 
as the baryon asymmetry due to the spharelon effect, in some models, 
it can be much larger than the baryonic asymmetry
\cite{Casas:1997gx,MarchRussell:1999ig,McDonald:1999in,Kawasaki:2002hq,Takahashi:2003db}.
Furthermore, if  the lepton asymmetry is large, it may significantly 
affect some aspects of the evolution of the Universe:
QCD phase transition \cite{Schwarz:2009ii}, 
large-scale cosmological magnetic field \cite{Semikoz:2009ye}, 
density fluctuations if primordial fluctuation is generated via 
the curvaton mechanism 
\cite{Gordon:2003hw,Lyth:2002my,DiValentino:2011sv} 
and so on. 

Thus, it would be worth investigating to what extent 
the lepton asymmetry can be probed beyond the accuracy 
of current cosmological observations. 
%
%
Since the signals from the 21 cm line can cover a wide redshift range, 
they can be complementary to other observations such as CMB. 
In addition, the effects of the lepton asymmetry
 mainly appear on small scales, which can be well
measured by 21 cm line observations. 
Thus such a survey would provide useful information.

\section{Purposes and organization of this thesis}

\begin{figure}[tb]
\begin{center}
\includegraphics[bb= 92 353 512 589, width=0.5\linewidth]{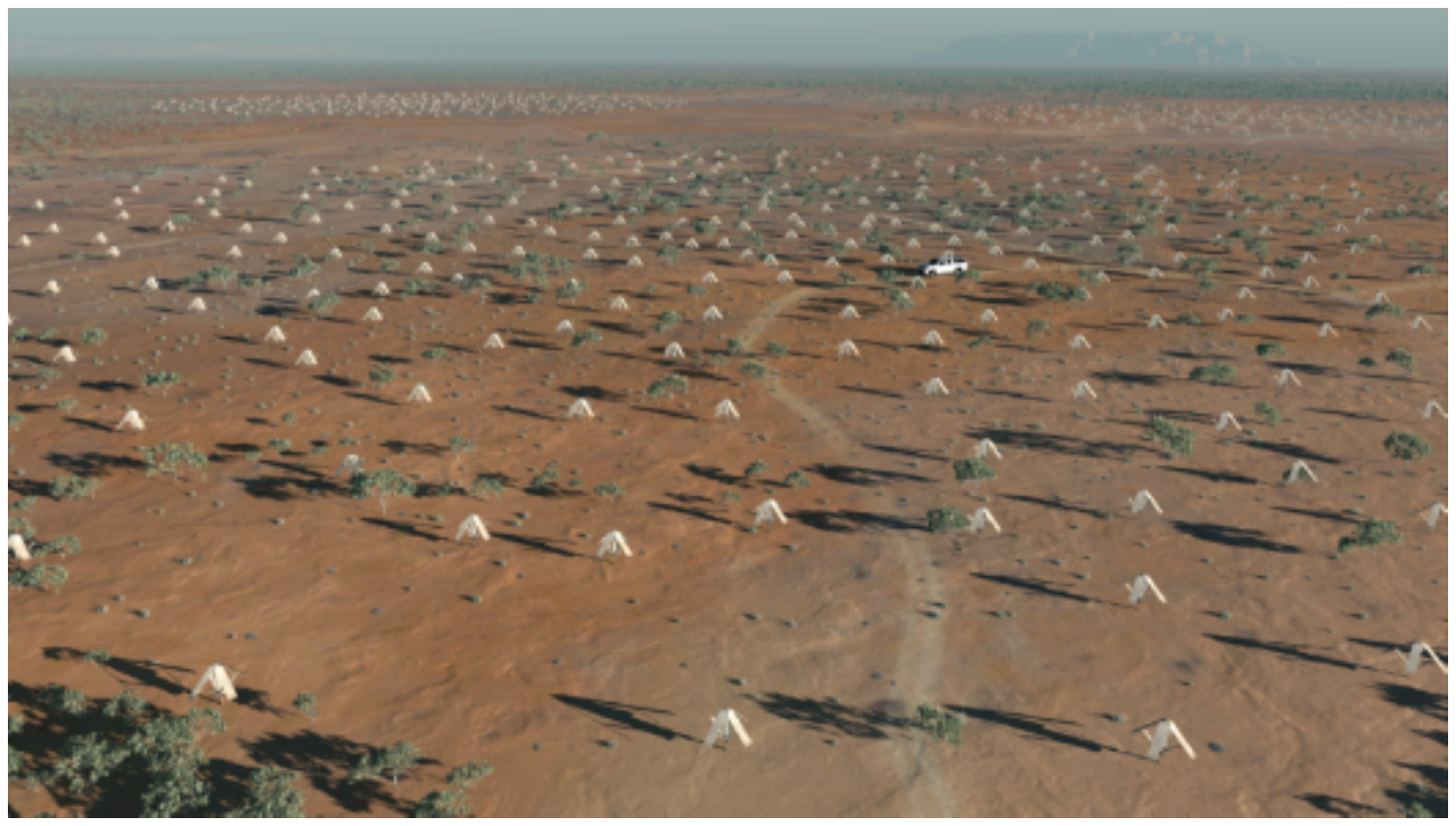} 
\caption{An image of the Square Kilometer Array (SKA)
\label{fig:SKAgazou}}{http://www.skatelescope.org/}
\end{center}
\end{figure}

In this thesis, we focus on future observations of both the 21 cm
line radiation coming from the epoch of the reionization 
($7\leq z\leq 10$) and the CMB polarization produced by a gravitational lensing, 
in order to study their sensitivities to the neutrino properties such as
the total neutrino mass, the neutrino mass hierarchy, the effective
number of neutrino species (extra radiation component), and the lepton
asymmetry of our Universe.
As 21 cm line observation, we particularly focus on future
experiments such as the Square Kilometer Array (SKA)
(Fig.\ref{fig:SKAgazou}) \cite{SKA} and Omniscope
\cite{Tegmark:2009kv,Tegmark:2008au}.

This thesis is organized as follows.
In Chapter \ref{chap:21cm}, \ref{chap:spin} and \ref{chap:powerspectrum},
we review the brightness temperature of the 21 cm radiation,
the spin temperature (excitation temperature of the hyperfine splitting),
and the power spectrum of the 21 cm radiation, respectively.
In Chapter \ref{chap:neutrino}, we briefly explain the growth of the
density fluctuations and some effects due to neutrino properties.
In Chapter \ref{chap:Fisher}, \ref{chap:21cmFisher},
\ref{chap:CMBFisher} and \ref{chap:BAOFisher}, we summarize our
analytical methods (Fisher information analysis), and review Fisher
matrices of each experiment (21 cm line, CMB and BAO
(baryon acoustic  oscillation), respectively).
In Chapter \ref{Chap:result_mass}, and \ref{chap:result_lepton}, 
we present our results as forecasts for specific observations, 
paying particular attention to 
how the 21 cm observations will help to measure neutrino parameters.
%

\section{Basic variables and constants}

Here, we present a summary table of basic symbols in this thesis.
\\

\begin{tabular}{cl} 
\hline 
Symbol & Definition \hspace{10cm} \\ 
\hline  
$a$ & Scale factor. \\
$A_{ul}$ & Einstein A coefficient (spontaneous decay rate). \\
$A_{21}$ & Einstein A coefficient of the hyperfine splitting. \\
$B_{ul}$ & Einstein B coefficient (stimulated emission). \\
$B_{lu}$ & Einstein B coefficient (absorption). \\
$c$ & Light speed. \\
$f_{\nu}$ & Energy fraction of neutrino to matter,
$f_{\nu} \equiv \rho_{\nu}/(\rho_{b}+\rho_{c}+\rho_{\nu})$, \\
& where $\rho_{\nu}$ includes both of neutrino and anti-neutrino. \\
$g_{u}$ & Degree of freedom of the upper state. \\
$g_{l}$ & Degree of freedom of the lower state. \\
$g_{0}$ & Degree of freedom of the spin singlet state. \\
$g_{1}$ & Degree of freedom of the spin triplet state. \\
$g_{\mu\nu}$& Metric tensor. \\
$H$ & Hubble parameter, $H \equiv (da/dt)a^{-1}$, where $t$ is the cosmic time. \\
$\mathcal{H}$& Conformal Hubble parameter, $\mathcal{H}=aH$. \\
$H_{0}$ & Present value of the Hubble parameter. \\
$h$ & Dimensionless Hubble parameter, $h\equiv H_{0}/(100 {\rm km \ s^{-1} \ Mpc^{-1}})$. \\
$h_{P}$ & Planck constant. \\
$\hbar$ & Reduced Planck constant, $\hbar\equiv h_{P}/2\pi$. \\
$I_{\nu}$ & Specific intensity. \\
$I_{\nu}^{BB}$ & Specific intensity of black body. \\
$\mbox{\boldmath $k$}$& Wave number vector (Fourier dual of the comoving coordinate). \\
$k$ & Absolute value of $\mbox{\boldmath $k$}$, $k=|\mbox{\boldmath $k$}|$. \\
$k_{B}$ & Boltzmann constant. \\
$ m_{e}$ & Electron mass. \\
$ m_{H}$ & Hydrogen mass. \\
\hline
\end{tabular}

\begin{tabular}{cl} 
\hline 
Symbol & Definition \hspace{10cm} \\ 
\hline
$ m_{\nu}$ & Neutrino mass. \\
$\Sigma m_{\nu}$ & Total neutrino mass, $\Sigma m_{\nu} \equiv \sum_{i=1}^{3} m_{i}$, \\
& where $m_{i}$ is each mass eigenstate. \\
$N_{\nu}$ & Effective number of neutrino species. \\
$\Delta N_{\nu}$ & Difference between the effective number of neutrino species. \\
&and the standard value, $\Delta N_{\nu} \equiv N_{\nu} - 3.046$. \\
$p$ & Pressure. \\
$P_{21} $ & Power spectrum of $\delta_{21}$ (21 cm line power spectrum). \\ 
$P_{\delta \delta} $ & Power spectrum of matter density fluctuation. \\ 
$T$ & Temperature. \\
$T_{b}$ & Brightness temperature. \\
$t_{g}$ & Proper time of a radiation source. \\
$T_{S}$ & Spin temperature. \\ 
$T_{\gamma}$& CMB temperature. \\
$w$ & $w \equiv p/\rho$. \\
$Y_{p}$ & Primordial $^4$He mass fraction. \\ 
$x_{HI}$ & Neutral fraction \\ 
& (the ratio of neutral hydrogen atoms and total protons). \\
$x_{i}$ & Ionization fraction, $x_{i} = 1-x_{HI}$. \\
$z$& Redshift, $z=a^{-1}-1$. \\
$\alpha_{EM}$ & Fine structure constant. \\
$\delta$& Density fluctuation. \\
$\delta_{21}$& Fluctuation of $\Delta T_{b}^{obs}$.  \\
$\delta_{b}$& Density fluctuation of baryons. \\
$\delta_{c}$& Density fluctuation of cold dark matter. \\
$\delta^{D}(x) $ & Dirac delta function. \\ 
$\delta_{H}$ & Density fluctuation of hydrogen. \\
$\delta_{HI}$& Density fluctuation of neutral fraction. \\
$\delta_{\nu}$& Density fluctuation of neutrinos. \\
$\Delta T_{b}$ & Difference between the 21 cm line brightness temperature \\
&  and CMB temperature, $T_{b}-T_{\gamma}$.  \\
$\Delta T_{b}^{obs}$ & observed $\Delta T_{b}$. \\
$\eta$ & Conformal time. \\
$\lambda$ & Wave length. \\
$\lambda_{21}$ & Wave length of 21 cm line. \\
$\mu$ & Cosine of the angle of a wave vector between a line of sight direction. \\
$\mu_{\nu_{i}}$ & Chemical potential of each neutrino flavor $\nu_{i}$. \\
$\nu$ & Frequency. \\
$\nu_{21}$ & Frequency of 21 cm line at a rest frame. \\
$\nu_{ul}$ & Transition frequency between the upper and lower state. \\
$\xi $& Degeneracy parameter. \\
\hline
\end{tabular}

\begin{tabular}{cl} 
\hline 
Symbol & Definition \hspace{10cm} \\
\hline
$\xi_{\nu_{i}} $& Degeneracy parameter of each neutrino flavor, $\xi_{\nu_{i}} \equiv \mu_{\nu_{i}}/T_{\nu}$. \\
$\rho$& Energy density. \\
$\rho_{b}$ & Energy density of baryons. \\
$\rho_{c}$ & Energy density of cold dark matter. \\
$\rho_{m}$ & Energy density of matter, $\rho_{m}\equiv \rho_{c} + \rho_{b}+\rho_{\nu}$. \\
$\rho_{\nu}$ & Energy density of neutrinos. \\
& In the section \ref{sec:lepton_asym} and the appendix \ref{sec:app_lept}, \\ 
& $\rho_{\nu}$ includes only the energy density of neutrino. \\
& In the other chapters and sections, it includes both of neutrino \\
& and anti-neutrino. \\
$\tau_{\nu}$& Optical depth. \\
$\phi(\eta,\mbox{\boldmath $x$})$&  Perturbation of gravitational potential, \\
& where $\mbox{\boldmath $x$}$ is the comoving coordinate. \\
$\psi(\eta,\mbox{\boldmath $x$})$&  Perturbation of spatial curvature, \\
& where $\mbox{\boldmath $x$}$ is the comoving coordinate. \\
$\Omega_{b}$& Density parameter of baryons at present. \\
$\Omega_{m}$& Density parameter of matter at present. \\ 
$\Omega_{\Lambda }$& Density parameter of dark energy at present. \\ 
$\Omega_{\nu}$& Density parameter of neutrino at present. \\
\hline
\end{tabular}

\chapter[Brightness temperature of 21 cm radiation]
{Brightness temperature of 21 cm radiation\label{chap:21cm}
\normalsize{\cite{Furlanetto:2006jb,Arashiba:2009,Pritchard:2011xb}}}

%
In this chapter, we review basic physical quantities
about the cosmological 21 cm line observation.
For further details, we refer
the readers to Refs.~\cite{Furlanetto:2006jb,Pritchard:2011xb}.

\section[Brightness temperature and transfer equation]
{Brightness temperature and transfer equation \label{sec:brightness}
\normalsize{\cite{text:rybicki,text:Nakai}}}

\subsection{Brightness temperature}

Brightness temperature $T_{b}$ means intensity of radiation,
and it is defined by specific intensity of black body 
in the Rayleigh-Jeans approximation ($k_{B}T>>h_{P} \nu$).
In the approximation, the specific intensity of 
black body $I^{BB}_{\nu}$ is given by
\begin{eqnarray}
I^{BB}_{\nu}(T)=\frac{2\nu^{2}}{c^{2}} k_{B}T.
\end{eqnarray}
By using the specific intensity of black body $I^{BB}_{\nu}$, 
the brightness temperature $T_{b}$ is defined as
\begin{eqnarray}
I_{\nu}& =&\frac{2\nu^{2}}{c^{2}} k_{B}T_{b} \nonumber\\
\longrightarrow  \ \ \ T_{b} &\equiv& \frac{c^{2}}{2\nu^{2}k_{B}} I_{\nu},
\end{eqnarray}
where $I_{\nu}$ is intensity of radiation
(specific intensity : emitted energy per unit area, unit time, 
unit frequency and unit solid angle).

When the Rayleigh-Jeans approximation is not applicable,
the specific intensity of black body $I^{BB}_{\nu}$ is given by
\begin{eqnarray}
I^{BB}_{\nu}(T)=\frac{2h_{P} \nu^{3}}{c^{2}} \frac{1}{\exp\left( \frac{h_{P} \nu }{k_{B} T} \right)-1}. 
\label{eq:Blackbody}
\end{eqnarray}
Therefore, we can define the equivalent brightness temperature $J_{\nu}(T)$ as
\begin{eqnarray}
J_{\nu}(T) &\equiv& \frac{h \nu}{k_{B}} \frac{1}{\exp\left( \frac{h_{P} \nu }{k_{B} T} \right)-1} \\
\longrightarrow 
I_{\nu} &=& \frac{2\nu^{2}}{c^{2}} k_{B}J_{\nu}(T).
\end{eqnarray}
From now on, we do not distinguish $J_{\nu}(T)$ from $T_{b}$,
and express them as just $T_{b}$.

\subsection{Transfer equation}

The flux intensity obeys the transfer equation,
\begin{eqnarray}
\frac{1 }{c}\frac{dI_{\nu}(\mbox{\boldmath $r$},t,\mbox{\boldmath $n$})}{dt}
= \eta_{\nu}(\mbox{\boldmath $r$},t,\mbox{\boldmath $n$}) 
- \alpha_{\nu} (\mbox{\boldmath $r$},t,\mbox{\boldmath $n$}) 
I_{\nu}(\mbox{\boldmath $r$},t,\mbox{\boldmath $n$}), \label{eq:transfer}
\end{eqnarray}
where $t$ is the cosmic time , $\eta_{\nu}$ is the emission coefficient, 
which represents the contribution of spontaneous emission, 
$\alpha_{\nu}$ is the absorption coefficient, which represents the contribution 
of absorption and stimulated emission (it is interpreted as negative absorption),
%
$\mbox{\boldmath $r$}$ is the comoving coordinate, 
and $\mbox{\boldmath $n$}$ is the unit vector which points to the direction of radiation.

Here, we define the optical depth $\tau_{\nu}$ as,
\begin{eqnarray}
d\tau_{\nu} \equiv \alpha_{\nu}c dt = \alpha_{\nu}ds 
\longleftrightarrow \tau_{\nu}(s)\equiv \int_{0}^{s} \alpha_{\nu}(\mbox{\boldmath $r$}(s'),s',
\mbox{\boldmath $n$}(s')) ds', \label{eq:optical}
\end{eqnarray}
where $s$ is the physical length.
This quantity $\tau_{\nu}$ represents the degree of diffusion of radiation.
By using a transformation of $t \longrightarrow \tau_{\nu}$,
the transfer equation becomes
\begin{eqnarray}
\frac{dI_{\nu}(\mbox{\boldmath $r$},\tau_{\nu},\mbox{\boldmath $n$})}{d\tau_{\nu}}
= - I_{\nu}(\mbox{\boldmath $r$},\tau_{\nu},\mbox{\boldmath $n$})
+\frac{
\eta_{\nu}(\mbox{\boldmath $r$},\tau_{\nu},\mbox{\boldmath $n$})
}
{
 \alpha_{\nu} (\mbox{\boldmath $r$},\tau_{\nu},\mbox{\boldmath $n$})
}, \label{eq:transfer2}
\end{eqnarray}
%
%
%
and we can rewrite this equation as 
the following equation relevant to the brightness temperature $T_{b}$,
\begin{eqnarray}
\frac{dT_{b}(\nu,\mbox{\boldmath $r$},\tau_{\nu},\mbox{\boldmath $n$})}{d\tau_{\nu}}
= - T_{b}(\nu,\mbox{\boldmath $r$},\tau_{\nu},\mbox{\boldmath $n$})
+\frac{c^{2}}{2k_{B}\nu^{2}}
\frac{
\eta_{\nu}(\mbox{\boldmath $r$},\tau_{\nu},\mbox{\boldmath $n$})
}
{
 \alpha_{\nu} (\mbox{\boldmath $r$},\tau_{\nu},\mbox{\boldmath $n$})
}.\label{eq:transfer3}
\end{eqnarray}
%
By solving this equation,
we can get the solution of brightness temperature $T_{b}$.

\subsection{Emission and absorption coefficients 
in a two level system}

%
In this subsection, we discuss a two-level system 
because the hyperfine structure is described by such a system.
In the two-level system, the emission and absorption coefficients 
are expressed by the Einstein A and B coefficients,
which represent the probability of a transition between two energy levels.
The A coefficient corresponds to the spontaneous emission,
and the B coefficient corresponds to the absorption and 
the stimulated emission, respectively.

\subsubsection{A coefficient : $A_{ul}$}
%
The Einstein coefficient $A_{ul}$ represents 
the probability of a transition between two energy levels 
per unit time. The unit is inverse of time.

\subsubsection{B coefficients : $B_{lu}$, $B_{ul}$}
%
The probability of absorption and stimulated emission 
are proportional to the intensity of incoming radiation.
Therefore, we introduce the average intensity $\bar{J}$,
\begin{eqnarray}
\bar{J} \equiv \frac{1}{4\pi} \int_{4\pi}d\Omega \int_{0}^{\infty}I_{\nu}\phi(\nu)d\nu ,
\end{eqnarray}
where $\phi(\nu)$ is a line profile,
and by using  $\bar{J}$, we define the Einstein B coefficients as 
\begin{eqnarray}
B_{lu}\bar{J} &:& {\rm The \ probability \ of \ absorption, } \nonumber  \\
B_{ul}\bar{J} &:& {\rm The \ probability \ of \ stimulated \ emission.}  \nonumber  
\end{eqnarray}

\vspace{-20pt}
\subsubsection{}
These coefficients are related to the variation of intensity,
and the relations are expressed as
\begin{subequations}
\begin{eqnarray}
&& {\rm Spontaneous \ emission} \ : \ 
dI_{\nu} = \frac{h_{P}\nu_{ul}}{4\pi} n_{u}\phi_{e}(\nu)A_{ul}cdt, \\
&& {\rm Absorption} \hspace{59pt} : \ 
dI_{\nu} = \frac{h_{P}\nu_{ul}}{4\pi} n_{l}\phi_{a}(\nu)B_{lu}I_{\nu}cdt,  \\
&& {\rm Stimulated \ emission} \hspace{13pt} : \
dI_{\nu} = \frac{h_{P}\nu_{ul}}{4\pi} n_{u}\phi_{e}(\nu)B_{ul}I_{\nu}cdt.
\end{eqnarray}
\end{subequations}
Here, 
$n_{l}(\mbox{\boldmath $r$},t)$ and $n_{u}(\mbox{\boldmath $r$},t)$
are the number density of atom in the lower state and the upper state, respectively,
$\phi_{e}(\nu)$ and $\phi_{a}(\nu)$ are the line profiles 
of emission and absorption, respectively. 
Therefore, the equation of intensity $I_{\nu}$ can be written as
\begin{eqnarray}
&&dI_{\nu} = dI_{\nu}\mid _{{\rm Spontaneous}}+ dI_{\nu}\mid _{{\rm Absorption}} 
+ dI_{\nu}\mid _{{\rm Stimulated}} \nonumber\\
\longrightarrow
&&\frac{1}{c}\frac{dI_{\nu}}{dt}
=  h_{P}\nu_{ul}n_{u}\phi_{e}\frac{A_{ul}}{4\pi} 
 - h_{P}\nu_{ul}
   \left\{  n_{l}\frac{B_{lu}}{4\pi}\phi_{a} 
       - n_{u}\phi_{e}\frac{B_{ul}}{4\pi}  
   \right\} I_{\nu}, \label{eq:Ein}
\end{eqnarray}
%
where $\nu_{ul}$ is the transition frequency between the upper and lower states.
In comparison between (\ref{eq:transfer}) and (\ref{eq:Ein}),
we can express the emission and absorption coefficients as 
\begin{subequations}
\begin{eqnarray}
&&\eta_{\nu}(\mbox{\boldmath $r$},t)
=h_{P}\nu_{ul}n_{u}(\mbox{\boldmath $r$},t)\phi_{e}(\nu)\frac{A_{ul}}{4\pi}\\
&&\alpha_{\nu}(\mbox{\boldmath $r$},t) 
= h_{P}\nu_{ul}\left\{ n_{l}(\mbox{\boldmath $r$},t)\frac{B_{lu}}{4\pi} \phi_{a}(\nu)
- n_{u}(\mbox{\boldmath $r$},t)\frac{B_{ul}}{4\pi} \phi_{e}(\nu) \right\}. \label{kyusyu}
\end{eqnarray}
\end{subequations}
In this situation, $\alpha_{\nu}$ and $\eta_{\nu}$ do not depend 
on the direction of radiation $\mbox{\boldmath $n$}$.
From now on, we assume that
the line profiles of emission and absorption are same function
($\phi(\nu) \equiv \phi_{a}(\nu)=\phi_{e}(\nu)$) for simplicity.

\subsection{Spin temperature\label{sub:spinkyusyu}}

We introduce the spin temperature $T_{S}$, 
which is the excitation temperature of hyperfine structure
(the detail is shown in the Chapter \ref{chap:spin}),
\begin{eqnarray}
\frac{n_{u}}{n_{l}} \equiv 
\frac{g_{u}}{g_{l}}\exp \left( -\frac{h_{P} \nu_{ul}}{k_{B}T_{S}} \right) \label{eq:spinondo},
\end{eqnarray}
where $g_{u}$ and $g_{l}$ are the degree of freedom of
the upper and lower states, respectively.
According to the following relation
(the detail is shown in the Appendix \ref{ap:Ein})
\begin{subequations}
\begin{eqnarray}
A_{ul} &=& \frac{2h_{P}\nu_{ul}^{3}}{c^{2}}B_{ul},  \label{eq:Einsteinrelation1}\\
g_{u}B_{ul} &=& g_{l}B_{lu}, \label{eq:Einsteinrelation2}
\end{eqnarray}
\end{subequations}
we can express $\alpha_{\nu}$ as 
\begin{eqnarray}
\alpha_{\nu} &=& \frac{h_{P}\nu_{ul}}{4\pi}\phi(\nu) 
n_{l}B_{lu}\left\{ 1- \frac{n_{u}}{n_{l}}\frac{B_{ul}}{B_{lu}} 
\right\}\nonumber\\
&=& \frac{c^{2}}{8\pi\nu_{ul}^{2}}\frac{g_{u}}{g_{l}}
A_{ul}n_{l}\phi(\nu)
\left\{ 1- \exp\left( -\frac{h_{P}\nu_{ul}}{k_{B}T_{S}}\right) \right\}. 
\label{eq;kyusyu}
\end{eqnarray}
Therefore, the source term of Eq.(\ref{eq:transfer3})  
(the second term on the right hand side)
can be written as
\begin{eqnarray}
\frac{\eta_{\nu}}{\alpha_{\nu}} 
&=&  
h_{P}\nu_{ul}n_{u}\phi_{e}(\nu)\frac{A_{ul}}{4\pi}
\left[ 
\frac{c^{2}}{8\pi\nu_{ul}^{2}}\frac{g_{u}}{g_{l}}
A_{ul}n_{l}\phi(\nu)\left\{ 1- \exp\left( -\frac{h_{P}\nu_{ul}}{k_{B}T_{S}}\right)\right\}
\right]^{-1} \nonumber\\
&=&
\frac{2h_{P}\nu_{ul}^{3}}{c^{2}}\frac{g_{l}}{g_{u}}\frac{n_{u}}{n_{l}}
\left\{ 1- \exp\left( -\frac{h_{P}\nu_{ul}}{k_{B}T_{S}}\right) \right\}^{-1}\nonumber\\
&=&
\frac{2h_{P}\nu_{ul}^{3}}{c^{2}}
\frac{1}{ \exp \left( \frac{h_{P}\nu_{ul}}{k_{B}T_{S}} \right) -1}\nonumber\\
&\approx &
\frac{2\nu_{ul}^{2}}{c^{2}}k_{B}T_{S}. \label{eq:kyusyuhousya}
\end{eqnarray}
In the last line, we use the approximation of $h_{P}\nu_{ul}<<k_{B}T_{S}$.
This approximation is valid in observations of 21 cm line
because $h_{P}\nu_{lu}/k_{B} \simeq 0.068{\rm K}$ and generally $0.068{\rm K} << T_{S}$ 
(the detail is shown in the Chapter \ref{chap:spin}).

\subsection{The solution of the transfer equations\label{sub:solution_transfer}}

By using Eq.(\ref{eq:kyusyuhousya}), we can rewrite Eq.(\ref{eq:transfer3}) as 
\begin{eqnarray}
\frac{dT_{b}(\nu,\mbox{\boldmath $r$}(\tau_{\nu}),\tau_{\nu})}{d\tau_{\nu}}
= - T_{b}(\nu,\mbox{\boldmath $r$}(\tau_{\nu}),\tau_{\nu})
+\left(\frac{\nu_{ul}}{\nu}\right)^{2}
T_{S}(\mbox{\boldmath $r$}(\tau_{\nu}),\tau_{\nu}).
\label{eq:transfer4}
\end{eqnarray}
The solution of Eq.(\ref{eq:transfer4}) is given by
\begin{eqnarray}
T_{b}(\nu,\mbox{\boldmath $r$}(\tau_{\nu}),\tau_{\nu})
=e^{-\tau_{\nu}}T_{b}(\nu,\mbox{\boldmath $r$}(0),0) 
+ \left(\frac{\nu_{ul}}{\nu}\right)^{2}\int^{\tau_{\nu}}_{0}
e^{\tau'_{\nu}-\tau_{\nu}}T_{S}(\mbox{\boldmath $r$}(\tau'_{\nu}),\tau'_{\nu}) d\tau'_{\nu}. \label{eq:Tbkai1}
\end{eqnarray}
By using approximation of $T_{S}(\mbox{\boldmath $r$}(\tau'_{\nu}),\tau'_{\nu})
\approx T_{S}(\mbox{\boldmath $r$}(0),0)$, 
we can rewrite Eq.(\ref{eq:Tbkai1}) as
\begin{eqnarray}
T_{b}(\nu,\mbox{\boldmath $r$}(\tau_{\nu}),\tau_{\nu})
\approx
e^{-\tau_{\nu}}T_{b}(\nu,\mbox{\boldmath $r$}(0),0) 
+ \left(\frac{\nu_{ul}}{\nu}\right)^{2}
T_{S}(\mbox{\boldmath $r$}(0),0)
\left[
1-e^{-\tau_{\nu}}
\right]. \label{eq:Tbkai2}
\end{eqnarray}
%
Therefore, 
the difference between the brightness temperature $T_{b}(\nu,\mbox{\boldmath $r$}(\tau_{\nu}),\tau_{\nu})$
and the incoming radiation $T_{b}(\nu,\mbox{\boldmath $r$}(0),0)$
can be written as
\begin{eqnarray}
\Delta T_{b} (\nu,\mbox{\boldmath $r$}(0),0)
&\equiv &
T_{b}(\nu,\mbox{\boldmath $r$}(\tau_{\nu}),\tau_{\nu})
 - 
T_{b}(\nu,\mbox{\boldmath $r$}(0),0) \nonumber \\
&\approx&
\left(
1-e^{-\tau_{\nu}}
\right)
\left[
\left(\frac{\nu_{ul}}{\nu}\right)^{2}
T_{S}(\mbox{\boldmath $r$}(0),0)
-T_{b}(\nu,\mbox{\boldmath $r$}(0),0) 
\right].
\end{eqnarray}
Here, the brightness temperature of incoming radiation 
$T_{b}(\nu,\mbox{\boldmath $r$}(0),0)$ is 
that of CMB radiation $T_{\gamma}$.
%
%
By the comoving coordinate \mbox{\boldmath $r$}(0)
at the incident point 
and the conformal time $\eta(z)$ at the point,
the temperature of incoming radiation
$T_{b}(\nu,\mbox{\boldmath $r$}(0),0)$ can be expressed as
\begin{eqnarray}
T_{b}(\nu,\mbox{\boldmath $r$}(0),0) 
= T_{\gamma}(\mbox{\boldmath $r$}(0),\eta(z)).
\end{eqnarray}
%
%
From now on, 
by the comoving coordinate \mbox{\boldmath $r$}(0)
and the conformal time $\eta(z)$, 
we express the spin temperature and $\Delta T_{b}$
to be $T_{S}(\mbox{\boldmath $r$}(0),\eta(z))$ and
$\Delta T_{b}(\nu,\mbox{\boldmath $r$}(0),\eta(z))$,
respectively.

Because the observed brightness temperature is
redshifted (the frequency and the temperature of CMB
rise $1/(1+z)$-fold),
the difference of observed brightness temperature 
$\Delta T_{b}^{obs}$ is given by
\begin{eqnarray}
\Delta T_{b}^{obs} \left( \frac{\nu}{1+z},\mbox{\boldmath $r$}(0),\eta(z)\right) 
&=&  \frac {\Delta T_{b} \left( \nu ,\mbox{\boldmath $r$}(0),\eta(z) \right)} {1+z} 
\nonumber \\
&=& 
\frac
{
\left(\frac{\nu_{ul}}{\nu}\right)^{2}
T_{S}(\mbox{\boldmath $r$}(0),\eta(z))
-T_{\gamma}(\mbox{\boldmath $r$}(0),\eta(z)) 
}
{1+z}
\left(
1-e^{-\tau_{\nu}}
\right).
\end{eqnarray}
In the case of $\nu=\nu_{ul}$, $\Delta T_{b}^{obs}$ is expressed as
\begin{eqnarray}
\Delta T_{b}^{obs} \left( \frac{\nu_{ul}}{1+z},\mbox{\boldmath $r$}(0),\eta(z) \right) 
&=& 
\frac
{
T_{S}(\mbox{\boldmath $r$}(0),\eta(z))
-T_{\gamma}(\mbox{\boldmath $r$}(0),\eta(z)) 
}
{1+z}
\left(
1-e^{-\tau_{\nu_{ul}}}
\right). \label{eq:kidoondo}
\end{eqnarray}
From Eq.(\ref{eq:kidoondo}),
when the spin temperature $T_{S}$ is higher than the CMB temperature $T_{\gamma}$,
the observed radiation becomes an emission line ($ 0 < \Delta T_{b}^{obs}$).
By contrast, when $T_{S}$ is lower than $T_{\gamma}$,
the observed radiation becomes absorption line ($\Delta T_{b}^{obs} < 0$).
From now on, the ``difference of
the brightness temperature $\Delta T_{b}^{obs}$''
is called just the ``brightness temperature''.

\section{Optical depth and line profile}

\subsection{Optical depth}

In this section, we estimate the optical depth $\tau_{\nu_{ul}}$,
which appears in Eq.(\ref{eq:kidoondo}).
The optical depth is defined by Eq.(\ref{eq:optical}) to be,
\begin{eqnarray}
\tau_{\nu}(\mbox{\boldmath $r$}(s),\mbox{\boldmath $r$}(0),\eta(z)) 
= \int_{0}^{s} \alpha_{\nu}(\mbox{\boldmath $r$}(s'),s',\eta(z)) ds'. \label{eq:optical2}
\end{eqnarray}
By using Eqs.(\ref{eq;kyusyu}), (\ref{eq:optical2}) and 
the approximation of $h\nu_{ul} << k_{B}T_{S}$,
the absorption coefficient $\alpha_{\nu}$ is expressed as
\begin{eqnarray}
\alpha_{\nu} (\mbox{\boldmath $r$},\eta) &=& 
\frac{c^{2}}{8\pi\nu_{ul}^{2}}\frac{g_{u}}{g_{l}}
A_{ul}n_{l}(\mbox{\boldmath $r$},\eta) 
\phi(\nu)\left\{ 1- 
\exp\left( -\frac{h_{P}\nu_{ul}}{k_{B}T_{S}(\mbox{\boldmath $r$},\eta) }\right) \right\} \nonumber\\
&\approx &
\frac{c^{2}}{8\pi\nu_{ul}^{2}}\frac{g_{u}}{g_{l}}
A_{ul}
n_{l}(\mbox{\boldmath $r$},\eta) 
\phi(\nu)
\frac{
h_{P}\nu_{ul}
}{
k_{B}T_{S}(\mbox{\boldmath $r$},\eta)
}.
\end{eqnarray}
Therefore, the optical depth is given by
\begin{eqnarray}
\tau_{\nu}(\mbox{\boldmath $r$}(s),\mbox{\boldmath $r$}(0),\eta(z)) 
=
\frac{c^{2}h_{P}A_{ul}}{8\pi k_{B}\nu_{ul}}
\frac{g_{u}}{g_{l}}
\phi(\nu)
 \int_{0}^{s} 
n_{l}(\mbox{\boldmath $r$}(s'),\eta(z)) \frac{1}{T_{S}(\mbox{\boldmath $r$}(s'),\eta(z)) }ds'. 
\label{eq:optical3}
\end{eqnarray}

\subsection{Line profile}

\begin{figure}[tb]
\begin{center}
\includegraphics[bb= 12 27 136 112, height=3cm,width=5cm]{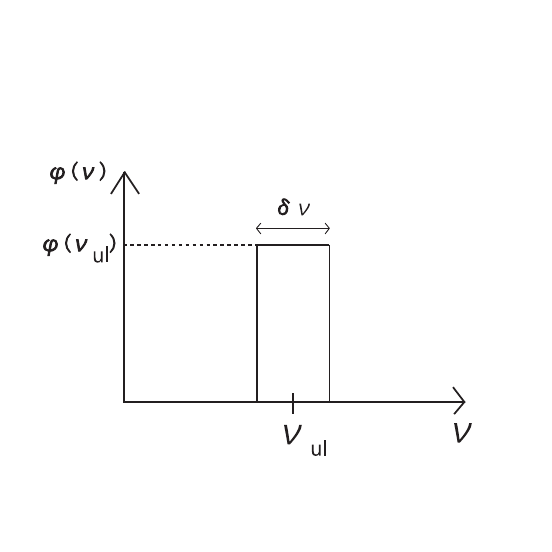} 
\hspace{10pt}\caption{Line profile} \label{fig:lineprofile}
\end{center}
\end{figure}

In this subsection,
we estimate the line profile $\phi(\nu)$ in Eq.(\ref{eq:optical3}).
The line profile is normalized as follows,
\begin{eqnarray}
\int_{0}^{\infty}\phi(\nu)d\nu=1.
\end{eqnarray}
We can consider that
$\phi(\nu_{ul})$ has non-zero value between only $\delta \nu$ 
which is the frequency region near the transition frequency $\nu_{ul}$
(Fig.\ref{fig:lineprofile}),
and the line profile can be written as
\begin{eqnarray}
\phi(\nu_{ul})\delta \nu \approx 1
\longrightarrow
\phi(\nu_{ul})
\approx 
\frac{1}{\delta \nu}. \label{eq:lineprofile2}
\end{eqnarray}
The frequency width $\delta \nu$ is caused by 
the Doppler effect of radiation.
Therefore, by using the velocity width of a hydrogen gas region  (radiation source)
along line of sight (LOS) $\Delta v_{\|}$,
$\delta\nu$ is written as
\begin{eqnarray}
\delta \nu = \frac{\Delta v_{\|}}{c} \nu_{ul}  \label{eq:sindousuu},
\end{eqnarray}
and $\Delta v_{\|}$ can be expressed as
\begin{eqnarray}
\Delta v_{\|} (\mbox{\boldmath $r$}(0),s,\eta(z))
\approx \frac{dv_{\|}(\mbox{\boldmath $r$}(0),\eta(z))}{dr_{\|}} \frac{s}{a(z)},
\end{eqnarray}
where $s$ is the physical size of the gas region,
%
%
$dv_{\|}/dr_{\|}$ is the derivative of $v_{\|}$
with respect to the direction of LOS $r_{\|}$, 
and $a(z)$ is the scale factor at the time
when the background radiation enters the gas region.
The gradient of velocity $dv_{\|}/dr_{\|}$ 
has main two contributions:
one comes from the expansion of the Universe,
the other comes from the peculiar motion of the gas region.

\subsubsection{Expansion of the Universe}
The velocity width due to the expansion of the Universe is given by
\begin{eqnarray}
\left.
\Delta v (\mbox{\boldmath $r$}(0),s,\eta(z)) \right|_{{\rm Hubble}}
&=&
\left. \frac{dv_{\|}(\mbox{\boldmath $r$}(0),\eta(z))}{dr_{\|}} \right|_{{\rm Hubble}}
\frac{s}{a(z)}\nonumber \\
&=&
s\left. \frac{dv_{\|}(\mbox{\boldmath $r$}(0),\eta(z))}{a(z)dr_{\|}} \right|_{{\rm Hubble}}.
\label{eq:hubblekouka}
\end{eqnarray}
Because $d v_{\|}/d(a r_{\|})$ is the derivative of velocity 
with respect to the physical distance $ar_{\|}$,
it represents the Hubble parameter.
Therefore, Eq.(\ref{eq:hubblekouka}) can be rewritten as
\begin{eqnarray}
\left.
\Delta v (\mbox{\boldmath $r$}(0),s,\eta(z)) \right|_{{\rm Hubble}}
= H(z)s,
\end{eqnarray}
where $H(z)$ is the Hubble parameter at the redshift z.

\subsubsection{Peculiar motion}
%
The velocity width due to the peculiar motion $v_{p\|}$ 
of a radiation source along LOS  is expressed as 
\begin{eqnarray}
\left.
\Delta v (\mbox{\boldmath $r$}(0),s,\eta(z)) \right|_{{\rm Peculiar}}
= \frac{dv_{p\|}(\mbox{\boldmath $r$}(0),\eta(z))}{dr_{\|}}\frac{s}{a(z)}.
\end{eqnarray}

\subsubsection{}

The net velocity width $\Delta v_{\|}$
is given by the sum of the above two contributions,
and it is written as
\begin{eqnarray}
\Delta v _{\|} (\mbox{\boldmath $r$}(0),s,\eta(z))
&=&
\left.
\Delta v (\mbox{\boldmath $r$}(0),s,\eta(z)) \right|_{{\rm Hubble}}
+
\left.
\Delta v (\mbox{\boldmath $r$}(0),s,\eta(z)) \right|_{{\rm Peculiar}} \nonumber \\
&=&
H(z)s
\left[
1+ \frac{1}{a(z)H(z)}
\frac{dv_{p\|}(\mbox{\boldmath $r$}(0),\eta(z))}{dr_{\|}}
\right]. \label{eq:sokudo}
\end{eqnarray}
Therefore, by using
Eqs.(\ref{eq:lineprofile2}), (\ref{eq:sindousuu}) and (\ref{eq:sokudo}),
the value of the line profile at $\nu = \nu_{ul}$ is given by
\begin{eqnarray}
\phi(\nu_{ul})
&\approx & 
\frac{c}{\Delta v_{\|}\nu_{ul}}\nonumber \\
&=&
\frac{c}{sH(z)\nu_{ul}}
\left[
1+\frac{1}{a(z)H(z)}
\frac{dv_{p\|}(\mbox{\boldmath $r$}(0),\eta(z))}{dr_{\|}}
\right]^{-1} \nonumber\\
&\approx &
\frac{c}{sH(z)\nu_{ul}}
\left[
1- \frac{1+z}{H(z)}
\frac{dv_{p\|}(\mbox{\boldmath $r$}(0),\eta(z))}{dr_{\|}}
\right]. \label{eq:lineprofilefinal}
\end{eqnarray}
In the last line of Eq.(\ref{eq:lineprofilefinal}), 
we use $a(z)=(1+z)^{-1}$ and 
consider that the velocity of the peculiar motion is much smaller than 
that of the cosmological expansion.
By using this estimated line profile,
we can obtain the optical depth at $\nu=\nu_{ul}$ as
\begin{eqnarray}
\tau_{\nu_{ul}}(\mbox{\boldmath $r$}(s),\mbox{\boldmath $r$}(0),\eta(z)) 
&=&
\frac{c^{2}h_{P}A_{ul}}{8\pi k_{B}\nu_{ul}}
\frac{g_{u}}{g_{l}}
\phi(\nu_{ul})
\int_{0}^{s} 
n_{l}(\mbox{\boldmath $r$}(s'),\eta(z)) \frac{1}{T_{S}(\mbox{\boldmath $r$}(s'),\eta(z)) }
ds\nonumber'\\
\nonumber \\
&=&
\frac{c^{3}h_{P}A_{ul}}{8\pi k_{B} \nu_{ul}^{2}}
\frac{g_{u}}{g_{l}}
\frac{1}{sH(z)}
\left[
1- \frac{1+z}{H(z)}
\frac{dv_{p\|}(\mbox{\boldmath $r$}(0),\eta(z))}{dr_{\|}}
\right] \nonumber\\
\nonumber \\
&& 
\times\int_{0}^{s} 
n_{l}(\mbox{\boldmath $r$}(s'),\eta(z)) \frac{1}{T_{S}(\mbox{\boldmath $r$}(s'),\eta(z)) }
ds' \label{eq:optical4}.
\end{eqnarray}

We consider that the size of the gas region is much smaller than 
Mpc scale, and average variations of physical quantities up to the gas size.
%
In this case, the number density (of lower state) and the spin temperature 
can be expressed as
\begin{eqnarray}
n_{l}(\mbox{\boldmath $r$}(s'),\eta(z)) 
& \approx & n_{l}(\mbox{\boldmath $r$}(0),\eta(z)), \nonumber \\
T_{S}(\mbox{\boldmath $r$}(s'),\eta(z)) 
&\approx & T_{S}(\mbox{\boldmath $r$}(0),\eta(z)). \nonumber 
\end{eqnarray}
Therefore, the third line of Eq.(\ref{eq:optical4}) can be written as
\begin{eqnarray}
\int_{0}^{s} 
n_{l}(\mbox{\boldmath $r$}(s'),\eta(z)) \frac{1}{T_{S}(\mbox{\boldmath $r$}(s'),\eta(z)) }
ds'
\approx
\frac{
n_{l}(\mbox{\boldmath $r$}(0),\eta(z)) }
{ T_{S}(\mbox{\boldmath $r$}(0),\eta(z))}s.
\label{eq:approx_Ts}
\end{eqnarray}
By using Eq.(\ref{eq:approx_Ts}), we can estimate the optical depth of 21 cm line 
for diffuse inter galactic medium (IGM) at
\begin{eqnarray}
\hspace{-10pt}
\tau_{\nu_{ul}}(\mbox{\boldmath $r$}(0),\eta(z)) 
&=&
\frac{c^{3}h_{P}A_{ul}}{8\pi k_{B}\nu_{ul}^{2}}
\frac{g_{u}}{g_{l}}
\frac
{n_{l}(\mbox{\boldmath $r$}(0),\eta(z)) }
{T_{S}(\mbox{\boldmath $r$}(0),\eta(z))}
\frac{1}{H(z)}
\left[
1- \frac{1+z}{H(z)}
\frac{dv_{p\|}(\mbox{\boldmath $r$}(0),\eta(z))}{dr_{\|}}
\right].
\label{eq:optical5}
\end{eqnarray}
%

\section{Observed brightness temperature}

In this section, we estimate the brightness temperature of 
the 21 cm line observation by using the optical depth
which is estimated in the previous section.
From now on, we use \mbox{\boldmath $r$} as the location of a gas region,
instead of \mbox{\boldmath $r$}(0).
By assuming that the optical depth is sufficiently small,
i.e. $1-\exp(-\tau_{\nu}) \approx \tau_{ul}$
\footnote{
This assumption is valid at almost all eras related to the 21 cm line observation
($\mathcal{O}(10)<z<\mathcal{O}(100)$).
We can estimate the optical depth at
$\tau_{\nu_{ul}} \sim \mathcal{O}(1)\times 10^{-1} 
\times \left(\frac{1+z}{10}\right)^{3/2} \left( \frac{{\rm K}}{T_{S}}\right)$.
From Figs.\ref{fig:spin1}-\ref{fig:spinhosi2} and in the next chapter,
we find that $\tau_{\nu_{ul}}<<1$ is valid in the redshift range.
},
%
and substituting Eq.(\ref{eq:optical5}) into Eq.(\ref{eq:kidoondo}),
the brightness temperature can be rewritten as
\begin{eqnarray}
\Delta T_{b}^{obs} \left( \frac{\nu_{ul}}{1+z},\mbox{\boldmath $r$},\eta(z) \right) 
&\approx & 
\frac
{T_{S}(\mbox{\boldmath $r$},\eta(z))-T_{\gamma}(\mbox{\boldmath $r$},\eta(z)) }
{1+z}\tau_{\nu_{ul}}(\mbox{\boldmath $r$},\eta(z)) \nonumber\\
&\approx&
\frac{c^{3}h_{P}A_{ul}}{8\pi k_{B}\nu_{ul}^{2}}
\frac{g_{u}}{g_{l}}
\frac
{n_{l}(\mbox{\boldmath $r$},\eta(z)) }
{(1+z)H(z)}
\left[
1-\frac{T_{\gamma}(\mbox{\boldmath $r$},\eta(z)) }{T_{S}(\mbox{\boldmath $r$},\eta(z))}
\right]
\nonumber\\
&&\times
\left[
1- \frac{1+z}{H(z)}
\frac{dv_{p\|}(\mbox{\boldmath $r$},\eta(z))}{dr_{\|}}
\right]. \label{eq:obsbrightness}
\end{eqnarray}

Next, we rewrite the number density of the lower state $n_{l}$
in the Eq.(\ref{eq:obsbrightness})
by using the number density of protons $n_{H}$.
The ground state of neutral hydrogen
splits into the upper (spin triplet $1 _1 S_{1/2}$ : $g_{u}=3$) 
and the lower (spin singlet $1 _0 S_{1/2}$ : $g_{l}=1$) state.
Therefore, by using the approximation 
in which neutral hydrogen atoms of the lower and upper state
exist in $1:3$, respectively, 
the number density of the lower state $n_{l}$ is expressed as
\begin{eqnarray}
n_{l} \approx
\frac{g_{l}}{g_{u}+g_{l}} n_{HI}=\frac{1}{4}n_{HI},
\end{eqnarray}
where $n_{HI}$ is the number density of neutral hydrogen atoms.
Here, we introduce the neutral fraction $x_{HI}$,
which means the ratio of neutral hydrogen atoms and total protons, 
and express the number density of neutral hydrogen
as $n_{HI}=x_{HI}n_{H}$, where $n_{H}$ is the number density 
of total protons. 
By using this relation, $n_{l}$ can be rewritten as
\begin{eqnarray}
n_{l} (\mbox{\boldmath $r$},\eta(z))
\approx
\frac{1}{4}x_{HI}(\mbox{\boldmath $r$},\eta(z)) n_{H}(\mbox{\boldmath $r$},\eta(z)).
\label{eq:n_l_x_HI}
\end{eqnarray}
By using Eq.(\ref{eq:n_l_x_HI}), the observed brightness temperature (\ref{eq:obsbrightness})
is given by
\begin{eqnarray}
\Delta T_{b}^{obs} \left( \frac{\nu_{21}}{1+z},\mbox{\boldmath $r$},\eta(z) \right) 
&\approx&
\frac{3c^{3}h_{P}A_{21}}{32\pi k_{B}\nu_{21}^{2}}
\frac
{x_{HI}(\mbox{\boldmath $r$},\eta(z))n_{H}(\mbox{\boldmath $r$},\eta(z)) }
{(1+z)H(z)}
\left[
1-\frac{T_{\gamma}(\mbox{\boldmath $r$},\eta(z)) }{T_{S}(\mbox{\boldmath $r$},\eta(z))}
\right]
\nonumber\\
&&\times
\left[
1-\frac{1+z}{H(z)}
\frac{dv_{p\|}(\mbox{\boldmath $r$},\eta(z))}{dr_{\|}}
\right], \label{eq:obsbrightness2}
\end{eqnarray}
where, instead of the index $ul$, 
we use $21$ (i.e. $\nu_{ul}$, $A_{ul}$ 
$\longrightarrow$ $\nu_{21}$, $A_{21}$)
to emphasis that those quantities are related to the 21 cm line.
Additionally, the spatial average of 
the brightness temperature $\Delta \bar{T}_{b}^{obs}(z)$ 
at the redshift $z$ is expressed as 
\begin{eqnarray}
\Delta \bar{T}_{b}^{obs} \left( \frac{\nu_{21}}{1+z}\right) 
&=&
\frac{3c^{3}h_{P}A_{21}}{32\pi k_{B}\nu_{21}^{2}}
\frac
{
\bar{x}_{HI}(z)
\bar{n}_{H}(z) 
}
{(1+z)H(z)}
\left[
1-\frac{\bar{T}_{\gamma}(z) }{\bar{T}_{S}(z)}
\right],
\label{eq:obsbrightness3}
\end{eqnarray}
where $\bar{x}_{HI},\bar{n}_{H},\bar{T}_{S}$ and $\bar{T}_{\gamma}$
mean the spatial averaged quantities.

The brightness temperature of 21 cm line $\Delta \bar{T}_{b}^{obs}(z)$ 
can be estimated at
\begin{eqnarray}
\Delta \bar{T}_{b}^{obs} \left( \frac{\nu_{21}}{1+z}\right) 
&\approx&
26.8\bar{x}_{HI}(z)  \left( \frac{1-Y_{p}}{1-0.25} \right)
\left(
\frac{\Omega_{b}h^{2}}{0.023}
\right)
\left(
\frac{0.15}{\Omega_{m}h^{2}}
\frac{1+z}{10}
\right)^{1/2} 
\left[
1-\frac{\bar{T}_{\gamma}(z) }{\bar{T}_{S}(z)}
\right]{\rm mK}, \nonumber \\
\label{eq:obsbrightness4}
\end{eqnarray}
From this equation, 
we find that
the brightness temperature of 21 cm line 
is about several mK at $z\sim10$.
When we estimate $\Delta \bar{T}_{b}^{obs}(z)$, 
we use the following relations and quantities:
%
%
The Friedmann equation in the matter dominated era,
%
\begin{eqnarray}
H^{2} = \frac{H_{0}^{2}\Omega_{m}}{a^{3}};
\end{eqnarray}
the relation between mass abundance of hydrogens and baryons,
%
\begin{eqnarray}
m_{H}c^2\bar{n}_{H} = \bar{\rho}_{H} 
\simeq(1-Y_{p})\bar{\rho}_{b},
\end{eqnarray}
where 
$\rho_{b}$ and $\rho_{H}$ is the energy density of baryons
and hydrogens respectively,
$Y_{p}$ is the helium mass fraction,
and $m_{H}$ is the mass of hydrogen;
the transition frequency of the hyperfine splitting,
\begin{eqnarray}
\nu_{21} =1.420405751786    \ \ {\rm GHz};
\end{eqnarray}
the Einstein coefficient of the splitting,
\begin{eqnarray}
A_{21} =\frac{2\pi \alpha_{EM} \nu^{3}_{21}h_{P}^{2}}{3c^{4}m_{e}^{2}} 
= 2.86888 \times 10^{-15} {\rm s^{-1}}.
\end{eqnarray}
%

\chapter[Spin temperature]{Spin temperature
\normalsize{\cite{Furlanetto:2006jb,Arashiba:2009}}}
\label{chap:spin}

In this chapter, we review 
the evolution of spin temperature 
at the dark age, the cosmic dawn and the epoch of reionization.

\section{Time evolution of the spin temperature}

\subsection{The evolution equation of spin temperature}

By Eq.(\ref{eq:spinondo}) in the Chapter \ref{ch:intro},
the spin temperature $T_{S}$ is defined as 
\begin{eqnarray}
\frac{n_{1}}{n_{0}}
&\equiv & \frac{g_{1}}{g_{0}}\exp \left( -\frac{h_{P} \nu_{21}}{k_{B}T_{S}} \right)\\
&=&
3\exp \left( -\frac{T_{\star}}{T_{S}} \right), \label{eq:spinondo2}
\end{eqnarray}
where $T_{\star} \equiv h_{P}\nu_{21}/k_{B}=0.068$K,
and subscripts $0$ and $1$ mean the quantities related to 
the spin 0 (lower) and spin 1 (upper) states, respectively.
This spin temperature is the excitation temperature of 
the hyperfine splitting of neutral hydrogen.
The excitation temperature is defined by
viewing the distribution of the upper and lower states
to be the Boltzmann distribution.

By differentiating Eq.(\ref{eq:spinondo2}) with respect to the time, 
we can obtain the following evolution equation of spin temperature,
\begin{eqnarray}
\frac{n_{1}}{n_{0}}	=
3\exp \left( -\frac{T_{\star}}{T_{S}} \right) 
\ \ & \longrightarrow  &\ \
\frac{T_{\star}}{T_{S}} = \ln3 - \ln n_{1}+\ln n_{0} \nonumber, \\
& \longrightarrow &
\frac{\partial}{\partial t_{g}} \left( \frac{T_{\star}}{T_{S}} \right)
= \frac{1}{n_{0}} \frac{\partial n_{0}}{\partial t_{g}}
-\frac{1}{n_{1} }\frac{\partial n_{1}}{\partial t_{g}},
\label{eq:spinondo3}  
\end{eqnarray}
where $t_{g}$ is the proper time of a radiation source. 
%
%
From this equation, we find that
the spin temperature depends on 
the time evolution of number densities $n_{1}$ and $n_{0}$.
Therefore, 
transition processes between the upper and lower states
influences the evolution of spin temperature,
and %
such processes are the following:

\begin{enumerate}
\item Collisions (Spin flip due to 
 hydrogen-hydrogen (H-H), electron-hydrogen (e-H)
and proton-hydrogen (p-H) collisions)
\item Transition due to absorption and emission of background photons (CMB)
\item Transition due to Lyman $\alpha$ photons through other energy states
\item Time variation of neutral fraction $x_{HI}$
\end{enumerate}

\subsubsection{\underline{\large{1. Collisions (H-H, e-H and p-H)}}}

The time evolutions of the number densities due to collisions obey
the following equations,
\begin{eqnarray}
\left.
\frac{\partial n_{1}}{\partial t_{g}} 
\right|_{\rm{collision}}& \equiv & C_{01}n_{0} - C_{10}n_{1}, \\
\left.
\frac{\partial n_{0}}{\partial t_{g}} 
\right|_{\rm{collision}} &\equiv & -C_{01}n_{0} + C_{10}n_{1},
\end{eqnarray}
where $C_{01}$ and $C_{10}$ are the reaction ratios of 
excitation and deexcitation due to the collisions
(H-H, e-H and p-H), respectively.

\subsubsection {\underline{\large{2.
Absorption and emission of background photons (CMB)}}}

By the definition of the Einstein coefficients,
the time variations of the number densities 
due to absorption and emission of CMB photons of $\nu=\nu_{21}$
can be written as
\begin{eqnarray}
&&\left.
\frac{\partial n_{1}}{\partial t_{g}} 
\right|_{\rm{BGphoton}} = B_{01}I^{CMB}_{\nu_{21}}n_{0}
 - \left(A_{10} + B_{01}I^{CMB}_{\nu_{21}} \right)n_1, \\
&&\left.
\frac{\partial n_{0}}{\partial t_{g}} 
\right|_{\rm{BGphoton}} = -B_{01}I^{CMB}_{\nu_{21}}n_{0}
 + \left(A_{10} + B_{01}I^{CMB}_{\nu_{21}} \right)n_1,
\end{eqnarray}
where $I_{\nu_{21}}^{CMB}$ is the specific intensity of CMB at $\nu=\nu_{21}$.
Here, we only consider the CMB photons as background radiation.
In practice, photons due to the transition of the hyperfine splitting
also contributes to the background radiation \cite{Lewis:2007kz}.
However, that effect is smaller than that of the CMB photons.
Therefore, we neglect such contribution in the background radiation.

\subsubsection{\underline{\large{
3. Transition due to Lyman $\alpha$ photons through other energy states}}}

The time evolutions of the number densities due to Ly$\alpha$ photons
obey the following equations,
\begin{eqnarray}
\left.
\frac{\partial n_{1}}{\partial t_{g}} 
\right|_{\rm{Ly\alpha}}& \equiv & P_{01}n_{0} - P_{10}n_{1}, \\
\left.
\frac{\partial n_{0}}{\partial t_{g}} 
\right|_{\rm{Ly\alpha}} &\equiv & -P_{01}n_{0} + P_{10}n_{1},
\end{eqnarray}
where $P_{01}$ and $P_{10}$ are the reaction ratios of 
excitation and deexcitation due to absorption or emission 
of Ly$\alpha$ photons, respectively.
This transition occurs through the 2P state of neutral hydrogen
(the spin state changes when the 1S ground state is excited to the 2P state
 and subsequently deexcited to the 1S state).
This coupling of spin temperature and Ly$\alpha$ radiation
is called the Wouthuysen-Field effect (or coupling)
\footnote{
``Wouthuysen'' is pronounced as roughly 
``Vowt-how-sen''~\cite{Furlanetto:2006jb}.
}\cite{Wouthuysen:1952,Field:1958}.

\subsubsection{\underline{\large{
4. Time variation of neutral fraction $x_{HI}$}} \cite{Lewis:2007kz}}

This effect 
comes from the own time variation of neutral hydrogen number density
($n_{HI}=x_{HI}n_{H}$).
%
The time evolution of $n_{HI}$ due to 
the variation of neutral fraction $x_{HI}$ obeys the following equation,
\begin{eqnarray}
\left.
\frac{\partial n_{HI}}
{\partial t_{g}} 
\right|_{NF}
=\frac{\partial x_{HI}}{\partial t_{g}} n_{H}. \label{eq:neutral}
\end{eqnarray}
According to the degree of statistical freedom ($g_{0}=1, g_{1}=3$), 
we can consider that the variation 
affects 
$n_{0}$ and $n_{1}$ in the proportion of 1:3.
Therefore, the contributions due to the variation can be written as
\begin{eqnarray}
\left.
\frac{\partial n_{1}}{\partial t_{g}} 
\right|_{\rm{NF}} & \approx & 
 \frac{g_{1}}{g_{1}+g_{0}} 
\frac{\partial x_{HI}}{\partial t_{g}}
n_{H}
=\frac{3}{4}\frac{1}{x_{HI}}\frac{\partial x_{HI}}{\partial t_{g}} (n_{0}+n_{1}),
\\
\left.
\frac{\partial n_{0}}{\partial t_{g}} 
\right|_{\rm{NF}} & \approx & 
 \frac{g_{0}}{g_{1}+g_{0}} 
\frac{\partial x_{HI}}{\partial t_{g}}
n_{H}
=\frac{1}{4}\frac{1}{x_{HI}}\frac{\partial x_{HI}}{\partial t_{g}} (n_{0}+n_{1}),
\end{eqnarray}
where we consider that the states of all neutral hydrogen atoms 
are the ground states (i.e. $n_{HI}=n_{0}+n_{1}$).

\subsubsection*{}

According to the above contributions, the derivatives of the number densities
$\partial n_{1}/\partial t_{g}$ and $\partial n_{0}/\partial t_{g}$
are given by
\begin{eqnarray}
\frac{\partial n_{1}}{\partial t_{g}}
&= &
\left.
\frac{\partial n_{1}}{\partial t_{g}} 
\right|_{\rm{collision}}
+
\left.
\frac{\partial n_{1}}{\partial t_{g}} 
\right|_{\rm{BGphotons}}
+
\left.
\frac{\partial n_{1}}{\partial t_{g}} 
\right|_{\rm{Ly\alpha}}
+
\left.
\frac{\partial n_{1}}{\partial t_{g}} 
\right|_{\rm{NF}} \nonumber \\
\nonumber \\
&=&
\left(
C_{01} + B_{01}I_{\nu_{21}}^{CMB} + P_{01} 
\right) n_{0}
-
\left(
C_{10} + A_{10} + B_{10}I_{\nu_{21}}^{CMB} + P_{10} 
\right) n_{1} \nonumber \\
\nonumber \\
&&+
\frac{3}{4}\frac{1}{x_{HI}}\frac{\partial x_{HI}}{\partial t_{g}} (n_{0}+n_{1}),
\label{eq:n1henka}
\end{eqnarray}
\begin{eqnarray}
\frac{\partial n_{0}}{\partial t_{g}}
&= &
\left.
\frac{\partial n_{0}}{\partial t_{g}} 
\right|_{\rm{collision}}
+
\left.
\frac{\partial n_{0}}{\partial t_{g}} 
\right|_{\rm{BGphotons}}
+
\left.
\frac{\partial n_{0}}{\partial t_{g}} 
\right|_{\rm{Ly\alpha}}
+
\left.
\frac{\partial n_{0}}{\partial t_{g}} 
\right|_{\rm{NF}} \nonumber \\
\nonumber \\
&=&
-\left(
C_{01} + B_{01}I_{\nu_{21}}^{CMB} + P_{01} 
\right) n_{0}
+
\left(
C_{10} + A_{10} + B_{10}I_{\nu_{21}}^{CMB} + P_{10} 
\right) n_{1} \nonumber \\
\nonumber \\
&&+
\frac{1}{4}\frac{1}{x_{HI}}\frac{\partial x_{HI}}{\partial t_{g}} (n_{0}+n_{1}).
\label{eq:n0henka}
\end{eqnarray}
By using Eqs.(\ref{eq:n1henka}) and (\ref{eq:n0henka}),
we can rewrite the evolution equation of spin temperature 
Eq.(\ref{eq:spinondo3}) as
\begin{eqnarray}
\frac{\partial}{\partial t_{g}} \left( \frac{T_{\star}}{T_{S}} \right)
&=&
\frac{1}{n_{0}} \frac{\partial n_{0}}{\partial t_{g}}
-\frac{1}{n_{1} }\frac{\partial n_{1}}{\partial t_{g}} \nonumber\\
&=&
-\left(1+\frac{n_{0}}{n_{1}} \right) C_{01}
+\left(1+\frac{n_{1}}{n_{0}} \right) C_{10} \nonumber \\
&&-\left(1+\frac{n_{0}}{n_{1}} \right) P_{01}
+\left(1+\frac{n_{1}}{n_{0}} \right) P_{10} \nonumber \\
&&-\left(1+\frac{n_{0}}{n_{1}} \right) B_{01}I_{\nu_{21}}^{CMB}
+\left(1+\frac{n_{1}}{n_{0}} \right) \left( A_{10} + B_{10}I_{\nu_{21}}^{CMB} \right) \nonumber \\
&&-\frac{1}{4}\frac{1}{x_{HI}}\frac{\partial x_{HI}}{\partial t_{g}} 
\left(
3\frac{n_{0}}{n_{1}} - \frac{n_{1}}{n_{0}} +2 
\right). \label{eq:spinhatten}
\end{eqnarray}
%
%
%
Here, we introduce the following temperatures related to 
the reaction ratios of collision ($T_{g}$:gas temperature) and  
Ly$\alpha$ ($T_{\alpha}$:color temperature),
\begin{subequations}
\begin{eqnarray}
\frac{C_{01}}{C_{10}}
&\equiv &
3\exp \left(-\frac{T_{\star}}{T_{g}} \right), \\ 
\frac{P_{01}}{P_{10}}
&\equiv &
3\exp \left(-\frac{T_{\star}}{T_{\alpha}}\right).
\end{eqnarray}
\end{subequations}
%
$T_{g}$ and $T_{a}$ correspond to
temperatures in thermal equilibrium between the upper and lower state 
through 
only collisions or Ly$\alpha$ process, respectively,
\begin{subequations}
\begin{eqnarray}
n_{1}C_{10} = n_{0}C_{01} \ : \ 
&& {\rm only \ collisions,} \\
n_{1}P_{10} = n_{0}P_{01} \ : \ 
&& {\rm only \ Ly}\alpha \ {\rm process.}
\end{eqnarray}
\end{subequations}

By using $T_{g}$ and the definition  of 
spin temperature Eq.(\ref{eq:spinondo2}),
we can rewrite the first line of Eq.(\ref{eq:spinhatten}) as
\begin{eqnarray}
-\left(1+\frac{n_{0}}{n_{1}} \right) C_{01}
+\left(1+\frac{n_{1}}{n_{0}} \right) C_{10}
&=& C_{10}
\left[
1 + \frac{n_{1}}{n_{0}} - \left( 1 + \frac{n_{0}}{n_{1}} \right) \frac{C_{01}}{C_{10}}
\right] \nonumber \\
&=&
C_{10}
\left[
1 + 3\exp \left( -\frac{T_{\star}}{T_{S}}\right) - 3\exp\left( -\frac{T_{\star}}{T_{g}} \right)
\right.\nonumber\\
&& \ \ \ \ \ \ \ \ \ \ \ \ \ \ \ \ \ \ \ \ \ \ \ \ \ \ \
\left.
 -\exp \left( \frac{T_{\star}}{T_{S}} - \frac{T_{\star}}{T_{g}} \right)
\right] \nonumber\\
&\approx&
C_{10}
\left[
1 + 3\left(1 -\frac{T_{\star}}{T_{S}}\right) - 3\left(1- \frac{T_{\star}}{T_{g}} \right)
\right.\nonumber\\
&& \ \ \ \ \ \ \ \ \ \ \ \ \ \ \ \ \ \ \ \ \ \ \ \ \ \ \
\left.
 - \left(1+ \frac{T_{\star}}{T_{S}} - \frac{T_{\star}}{T_{g}} \right)
\right]\nonumber\\
&=&
4C_{10}\left[
\frac{T_{\star}}{T_{g}} - \frac{T_{\star}}{T_{S}}
\right], \label{eq:dai1gyou}
\end{eqnarray}
%
where, in the third line, we assume that $T_{S}>>T_{\star}$, $T_{g}>>T_{\star}$
and expand $\exp(\cdot)$ up to the first order~\footnote{
$T_{g},T_{\gamma},T_{\alpha}$
are generally higher than the present CMB temperature
($T_{\gamma 0}\approx 2.7 \rm{K}$),
and $T_{\gamma0}>>T_{\star}=0.068$K.
Therefore, we can assume that $T_{g}>>T_{\star},T_{\alpha}>>T_{\star},
T_{\gamma}>>T_{\star}$. 
Additionally, $T_{S}>>T_{\star}$ is also valid
because $T_{S}$ takes values close to any of $T_{g}$, $T_{\alpha}$ or $T_{\gamma}$. 
\label{foot:T}
}.
In the same way, 
by using the color temperature $T_{\alpha}$,
the second line of Eq.(\ref{eq:spinhatten})
can be rewritten as
\begin{eqnarray}
-\left(1+\frac{n_{0}}{n_{1}} \right) P_{01}
+\left(1+\frac{n_{1}}{n_{0}} \right) P_{10}
&=&
4P_{10}\left[
\frac{T_{\star}}{T_{\alpha}} - \frac{T_{\star}}{T_{S}}
\right], \label{eq:dai2gyou}
\end{eqnarray}
where we also use the similar assumption 
which is $T_{S}>>T_{\star}$, $T_{\alpha}>>T_{\star}$.
Next, the third line of Eq. (\ref{eq:spinhatten})
is expressed as
\begin{eqnarray}
&&-\left(1+\frac{n_{0}}{n_{1}} \right) B_{01}I_{\nu_{21}}^{CMB}
+\left(1+\frac{n_{1}}{n_{0}} \right) \left( A_{10} + B_{10}I_{\nu_{21}}^{CMB} \right) \nonumber \\
&&=
\left( A_{10} + B_{10}I_{\nu_{21}}^{CMB} \right) 
\left[
1+\frac{n_{1}}{n_{0}}
-\left(1+\frac{n_{1}}{n_{0}} \right) 
\frac{B_{01}I_{\nu_{21}}^{CMB}}{A_{10} + B_{10}I_{\nu_{21}}^{CMB} }
\right]. \label{eq:CMB1}
\end{eqnarray}
Here, by using the following relations of the Einstein coefficients
Eqs.(\ref{eq:Einsteinrelation1}) and (\ref{eq:Einsteinrelation2}),
\begin{eqnarray}
A_{10}& =& \frac{2h_{P}\nu_{ul}^{3}}{c^{2}}B_{10}, \nonumber\\
g_{1}B_{10} &=& g_{0}B_{01}, \nonumber
\end{eqnarray}
and the specific intensity of black body Eq.(\ref{eq:Blackbody}),
\begin{eqnarray}
I^{CMB}_{\nu_{21}}=\frac{2h_{P} \nu_{21}^{3}}{c^{2}} 
\frac{1}{\exp\left( \frac{h_{P} \nu_{21} }{k_{B} T_{\gamma}} \right)-1},
 \nonumber 
\end{eqnarray}
we can rewrite the coefficient of Eq.(\ref{eq:CMB1}) as
\begin{eqnarray}
A_{10} + B_{10}I_{\nu_{21}}^{CMB}
&=&
 A_{10} +  \frac{c^{2}}{2h_{P}\nu_{ul}^{3}}A_{10} 
 \frac{2h_{P}\nu_{21}^{3}}{c^{2}}\frac{1}{\exp\left(\frac{T_{\star}}{T_{\gamma}}\right)-1} \nonumber \\
&=&
\frac{A_{10}}{1-\exp\left(-\frac{T_{\star}}{T_{\gamma}}\right)}.
\label{eq:coeff_CMB_spin1}
\end{eqnarray}
Additionally, by using the following relation,
\begin{eqnarray}
 \frac{B_{01}I_{\nu_{21}}^{CMB}}{A_{10} + B_{10}I_{\nu_{21}}^{CMB}}
&=&
 3\frac{c^{2}}{2h_{P}\nu_{ul}^{3}}A_{10} 
 \frac{2h_{P}\nu_{21}^{3}}{c^{2}}\frac{1}{\exp\left(\frac{T_{\star}}{T_{\gamma}}\right)-1}
\left[
\frac{A_{10}}{1-\exp\left(-\frac{T_{\star}}{T_{\gamma}}\right)}
\right]^{-1} \nonumber \\
&=&3\exp\left(- \frac{T_{\star}}{T_{\gamma}} \right),
\end{eqnarray}
and Eqs.(\ref{eq:coeff_CMB_spin1}), Eq.(\ref{eq:CMB1}) can be rewritten as
\begin{eqnarray}
&&-\left(1+\frac{n_{0}}{n_{1}} \right) B_{01}I_{\nu_{21}}^{CMB}
+\left(1+\frac{n_{1}}{n_{0}} \right) 
\left( A_{10} + B_{10}I_{\nu_{21}}^{CMB} \right) \nonumber \\
&&=
\frac{A_{10}}{1-\exp\left(-\frac{T_{\star}}{T_{\gamma}}\right)}
\left[
1+\frac{n_{1}}{n_{0}}
-\left(1+\frac{n_{1}}{n_{0}} \right) 
3\exp\left(- \frac{T_{\star}}{T_{\gamma}} \right)
\right] \nonumber\\
&&=
\frac{A_{10}}{1-\exp\left(-\frac{T_{\star}}{T_{\gamma}}\right)}
\left[
1
+3\exp \left( -\frac{T_{\star}}{T_{S}} \right)
-3\exp \left(-\frac{T_{\star}}{T_{\gamma}} \right) 
-\exp\left(\frac{T_{\star}}{T_{S}} - \frac{T_{\star}}{T_{\gamma}} \right)
\right] \nonumber\\
&&\approx
\frac{A_{10}}{1-\left(1-\frac{T_{\star}}{T_{\gamma}}\right)}
\left[
1
+3\left(1 -\frac{T_{\star}}{T_{S}} \right)
-3\left(-\frac{T_{\star}}{T_{\gamma}} \right) 
-\left(1+\frac{T_{\star}}{T_{S}} - \frac{T_{\star}}{T_{\gamma}} \right)
\right] \nonumber\\
&&=
4 A_{10} \frac{T_{\gamma}}{T_{\star}}
\left[
\frac{T_{\star}}{T_{\gamma}} - \frac{T_{\star}}{T_{S}}
\right] \label{eq:dai3gyou}, \\
\nonumber 
\end{eqnarray}
where, in the fourth line,
we also use the similar assumption which
is $T_{S}>>T_{\star}$, $T_{\gamma}>>T_{\star}$,
and expand $\exp(\cdot)$ up to the first order.
Finally, by using $T_{S}>>T_{\star}$
and expand $\exp(\cdot)$ up to the first order,
the fourth line of Eq.(\ref{eq:spinhatten}) is expressed as
\begin{eqnarray}
-\frac{1}{4}\frac{1}{x_{HI}}\frac{\partial x_{HI}}{\partial t_{g}} 
\left(
3\frac{n_{0}}{n_{1}} - \frac{n_{1}}{n_{0}} +2 
\right)\nonumber
&=&
-\frac{1}{4}\frac{1}{x_{HI}}\frac{\partial x_{HI}}{\partial t_{g}} 
\left[
\exp \left( \frac{T_{\star}}{T_{S}} \right)
-3\exp \left( -\frac{T_{\star}}{T_{S}} \right) +2
\right] \nonumber\\
&\approx&
-\frac{1}{4}\frac{1}{x_{HI}}\frac{\partial x_{HI}}{\partial t_{g}} 
\left[
\left(1+ \frac{T_{\star}}{T_{S}} \right)
-3\left( 1-\frac{T_{\star}}{T_{S}} \right) +2
\right] \nonumber\\
&=&
-\frac{1}{x_{HI}}
\frac{\partial x_{HI}}
{\partial t_{g}} \frac{T_{\star}}{T_{S}}.
\label{eq:dai4gyou}
\end{eqnarray}
Therefore, by using
Eqs.(\ref{eq:dai1gyou}), (\ref{eq:dai2gyou}), (\ref{eq:dai3gyou})
and (\ref{eq:dai4gyou}),
we can rewrite the evolution equation 
of spin temperature Eq.(\ref{eq:spinhatten}) as
\begin{eqnarray}
&&\frac{\partial}{\partial t_{g}} \left( \frac{T_{\star}}{T_{S}} \right)
=
4C_{10}\left[
\frac{T_{\star}}{T_{g}} - \frac{T_{\star}}{T_{S}}
\right]
+4P_{10}\left[
\frac{T_{\star}}{T_{\alpha}} - \frac{T_{\star}}{T_{S}}
\right]
+4 A_{10} \frac{T_{\gamma}}{T_{\star}}
\left[
\frac{T_{\star}}{T_{\gamma}} - \frac{T_{\star}}{T_{S}}
\right] \nonumber\\
&& \hspace{260pt}
-\frac{1}{x_{HI}}\frac{\partial x_{HI}}{\partial t_{g}} \frac{T_{\star}}{T_{S}}, \nonumber \\
\nonumber\\
&&\longrightarrow \ \ \ \
\frac{\partial}{\partial t_{g}} \left( \frac{1}{T_{S}} \right)
+\frac{1}{x_{HI}}\frac{\partial x_{HI}}{\partial t_{g}} \frac{1}{T_{S}}
=
4\left[ C_{10}
\left(
\frac{1}{T_{g}} - \frac{1}{T_{S}}
\right)
+P_{10}
\left(
\frac{1}{T_{\alpha}} - \frac{1}{T_{S}}
\right),
\right.\nonumber\\
&&  \hspace{260pt}
\left.
+A_{10}
 \frac{T_{\gamma}}{T_{\star}}
\left(
\frac{1}{T_{\gamma}} - \frac{1}{T_{S}}
\right)
\right] \label{eq:spinequation}\\
&&\longrightarrow \ \ \ \
\frac{\partial T_{S}}{\partial t_{g}}
-\frac{T_{S}}{x_{HI}}\frac{\partial x_{HI}}{\partial t_{g}}
=
-4T_{S}^{2}\left[ C_{10}
\left(
\frac{1}{T_{g}} - \frac{1}{T_{S}}
\right)
+P_{10}
\left(
\frac{1}{T_{\alpha}} - \frac{1}{T_{S}}
\right)
\right.\nonumber\\
&& \hspace{260pt}
\left.
+A_{10}
 \frac{T_{\gamma}}{T_{\star}}
\left(
\frac{1}{T_{\gamma}} - \frac{1}{T_{S}}
\right)
\right].
\end{eqnarray}

\subsection[Spin temperature in thermal equilibrium]
{Spin temperature in  thermal equilibrium
\normalsize{\cite{Field:1958}}}

The time variation term of neutral fraction
can be neglected through the most of the epochs
related to the 21 cm line observation.
In addition, when each process is in thermal equilibrium
(time scales of each interaction are sufficiently shorter than 
that of the Hubble expansion),
the time derivative of spin temperature also can be neglected.
In this situation, by Eq.(\ref{eq:spinequation}),
the spin temperature $T_{S}$ is given by
\begin{eqnarray}
&&0
=
4\left[ C_{10}
\left(
\frac{1}{T_{g}} - \frac{1}{T_{S}}
\right)
+P_{10}
\left(
\frac{1}{T_{\alpha}} - \frac{1}{T_{S}}
\right)
+A_{10}
 \frac{T_{\gamma}}{T_{\star}}
\left(
\frac{1}{T_{\gamma}} - \frac{1}{T_{S}}
\right)
\right], \nonumber\\
\nonumber\\
&&\longrightarrow \ \
T_{S}
=
\left(
T_{\gamma} 
+ \frac{C_{10}}{A_{10}}\frac{T_{\star}}{T_{g}} T_{g} 
+ \frac{P_{10}}{A_{10}} \frac{T_{\star}}{T_{\alpha}} T_{\alpha}
\right)
\left(
1
+ \frac{C_{10}}{A_{10}}\frac{T_{\star}}{T_{g}} 
+ \frac{P_{10}}{A_{10}} \frac{T_{\star}}{T_{\alpha}} 
\right)^{-1}.
\label{eq:spinequiliblium0}
\end{eqnarray}
Here, we define the coupling coefficients,
\begin{subequations}
\begin{eqnarray}
y_{c}
\equiv \frac{C_{10}}{A_{10}}\frac{T_{\star}}{T_{g}}, \\
y_{\alpha}
\equiv
\frac{P_{10}}{A_{10}} \frac{T_{\star}}{T_{\alpha}}.
\end{eqnarray}
\end{subequations}
By using these coupling coefficients,  
the spin temperature $T_{S}$ can be expressed as
\begin{eqnarray}
T_{S}
=
\frac{
T_{\gamma} +  y_{c}T_{g} + y_{\alpha} T_{\alpha}
}
{
1 + y_{c} + y_{\alpha}
}. \label{eq:spinequiliblium2}\\[-10pt]
\nonumber 
\end{eqnarray}
%
In this way,
the spin temperature in thermal equilibrium is determined by
the temperatures of CMB $T_{\gamma}$,
gas $T_{g}$ and Ly$\alpha$ $T_{\alpha}$
(in other words, $T_{S}$ is a weighted average of them,
and the weights are $1:y_{c}:y_{\alpha}$).
According to the Eq.(\ref{eq:spinequiliblium2}),
we find that the spin temperature takes values
close to the temperature related to the transition process 
which has the strongest coupling.

\section[Global history of the spin temperature]{Global history of spin temperature
\normalsize{\cite{Furlanetto:2006jb}}
 \label{section:spinhistory}}

In this section, we briefly explain the global history of spin temperature.

\subsection{Before the star formation 
$30 \lesssim z \lesssim 300$ : \\ 
The dark age}

\begin{figure}[t]
\begin{center}
\includegraphics[bb=0 0 809 556, width=1\linewidth]{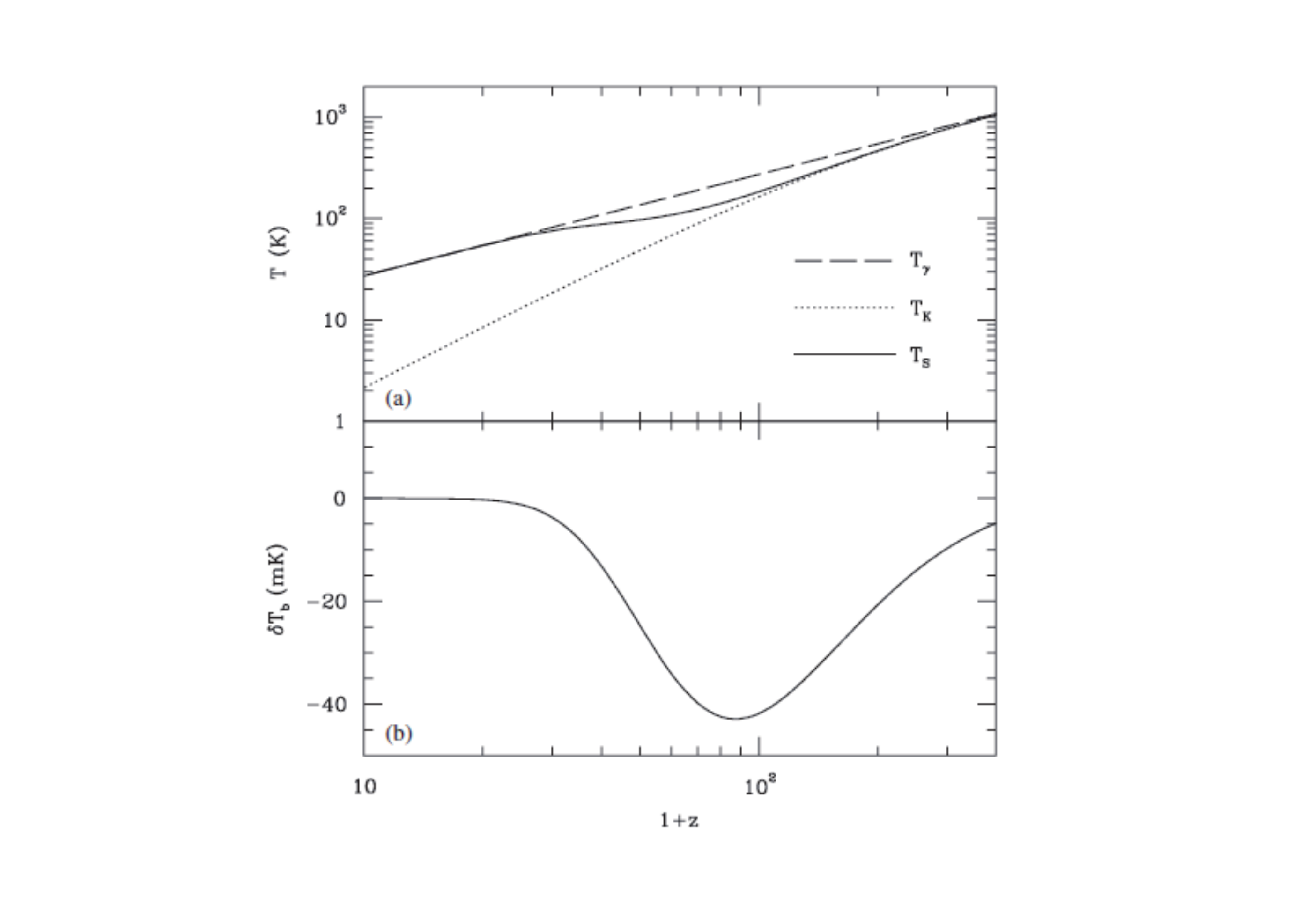} 
\hspace{10pt}\caption{
(a)The global history of spin temperature before the star formation
($30 \lesssim z  \lesssim 300$).
$T_{\gamma}$ is the CMB temperature,
$T_{g}$ is the gas temperature 
and $T_{S}$ is the spin temperature.
(B)The evolution of 21 cm line brightness temperature 
(in this figure, $\delta T_{b}$ represents the brightness temperature)
\cite{Furlanetto:2006jb}.
\label{fig:spin1}}
\end{center}
\end{figure}

In the very high redshifted era 
$300 \lesssim z$,
baryons and CMB photons are combined through the Compton scattering~\footnote{
This Compton scattering is due to free electrons which 
do not form atoms at the recombination era.}.
Therefore, the CMB $T_{\gamma}$ and the gas temperature $T_{g}$
are $T_{\gamma} \sim T_{g}$ (Fig.\ref{fig:spin1}),
and the spin temperature is also 
$T_{S}\sim T_{g} \sim T_{\gamma}$.
By the reasons stated above,
signals of 21 cm line do not exist in this era~\footnote{
However, in a recent work \cite{Fialkov:2013kna},
the authors indicate the possible presence of the signals in this era.}.

%
After that and before the star formation
($30 \lesssim z  \lesssim 300$),
the Compton scattering becomes not effective,
and it can be neglected at about $ z \sim 150 $.
Therefore, the gas temperature $T_{g}$
decreases adiabatically,
and the spin temperature becomes $T_{S} \sim T_{g}$ 
through the collisions.
Next, around $z \sim 70$,
the value of spin temperature starts to approach 
that of CMB temperature $T_{\gamma}$
because the density of baryons becomes smaller
and the transition process due to collisions becomes
not effective in comparison with transition due to CMB photons.
In this era, signals of 21 cm line can be detected.
Finally, around $z \sim  30$,
the spin temperature becomes $T_{S} \sim T_{\gamma}$,
and the signals disappear around the redshift.

Although signals of 21 cm line exist in theses eras,  
the signals ($70 \lesssim z$) 
can not pass through the ionosphere of the Earth.
Furthermore, in the frequency range 
($30 \lesssim z$), 
the galactic foregrounds are very strong.
Therefore, detection of the high redshifted signals is significantly difficult
(we need to use a lunar or space-based observatory, 
e.g. the lunar radio array, which is a planned array at the Moon in the future
\cite{Jester:2009dw}).

\subsection{After star formation 
$ z \lesssim 30
$ :
The cosmic dawn and the epoch of reionization}

\begin{figure}[t]
\begin{center}
\includegraphics[bb= 0 0 809 426, width=1\linewidth]{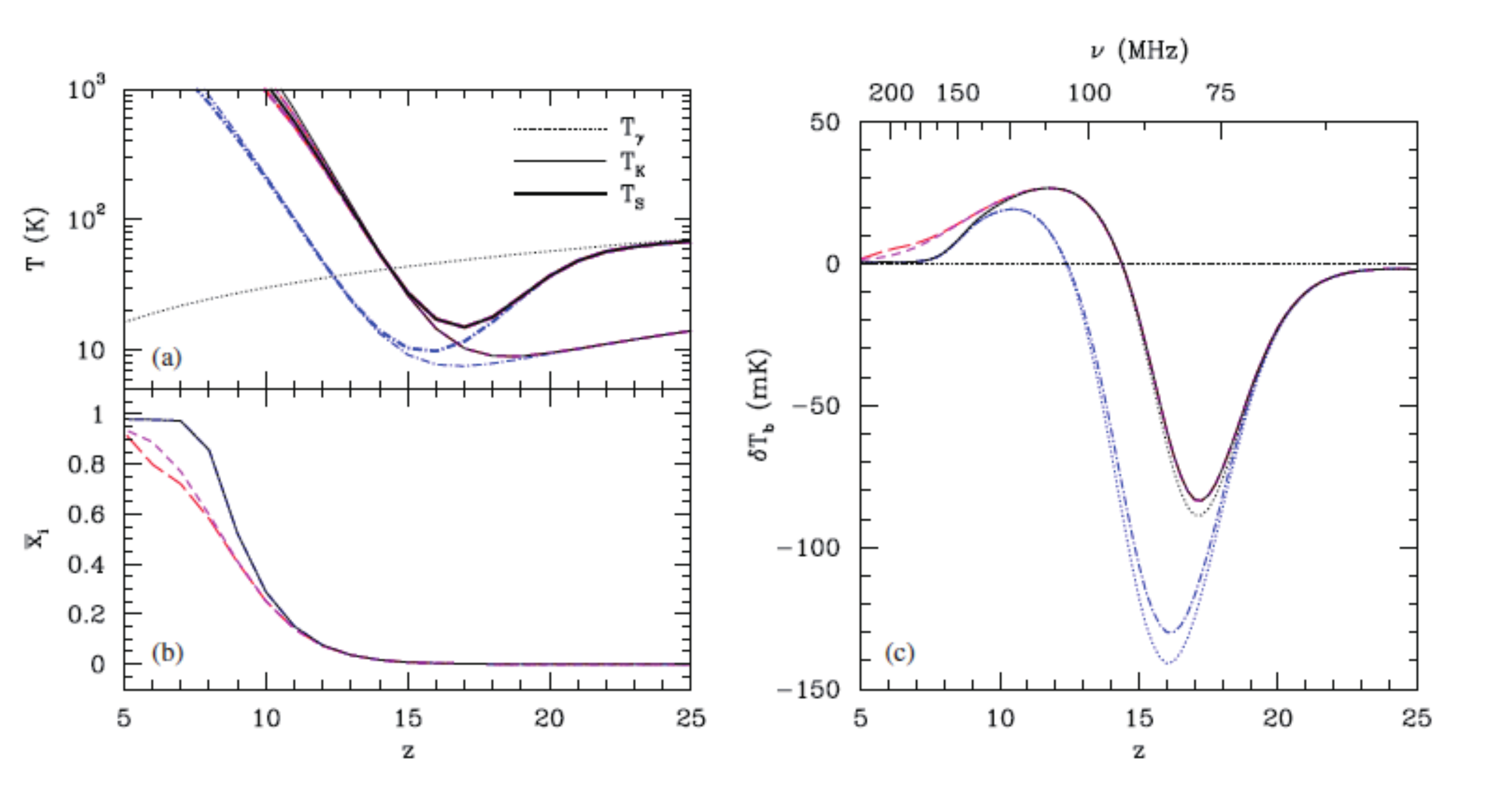} 
\hspace{10pt}\caption{
The global history of IGM (Inter-Galactic Medium)
when PoP II dominates in the Cosmic dawn and the epoch of reionization
\cite{Furlanetto:2006jb,Furlanetto:2006tf}:
(a) temperatures (b) ionization fraction $x_{i}=1-x_{HI}$
(c) brightness temperature of 21 cm line 
(in this figure, $\delta T_{b}$ represents the brightness temperature).
Each line corresponds to different models of the star formation:
the black ($f_{X}=1$),
the blue dot-dashed ($f_{X}=0.2$)
and the red dashed lines (strong photoheating feedback),
respectively.
$f_{X}$ is a renormalization factor,
which is necessary when the relation between star formation rate and 
X-ray luminosity is extrapolated to the high redshift.
The photoheating feedback means
suppression of the star formation due to photoionization of stars.
\label{fig:spinhosi1}}
\end{center}
\end{figure}

\begin{figure}[tb]
\begin{center}
\includegraphics[bb=  0 0 809 438, width=1\linewidth]{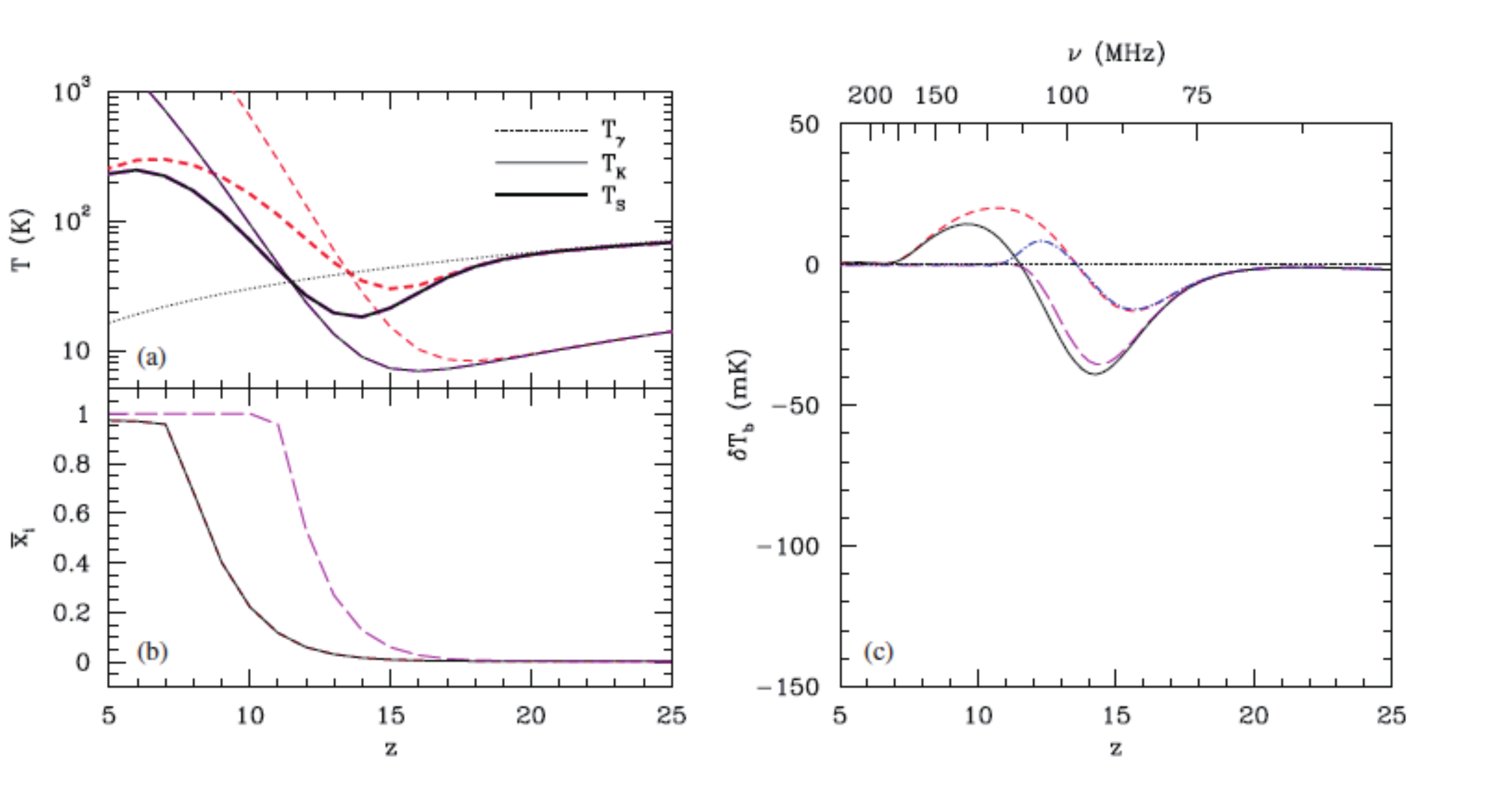} 
\hspace{10pt}\caption{
Same as Fig.\ref{fig:spinhosi1}, but PoP III dominates
in the Cosmic dawn and the epoch of reionization
\cite{Furlanetto:2006tf,Furlanetto:2006jb}:
Each line corresponds to the different models 
of the star formation:
the black ($f_{esc}=0.1,f_{X}=1$),
the red short-dashed ($f_{esc}=0.1,f_{X}=5$),
the pink long-dashed ($f_{esc}=1,f_{X}=1$),
the blue dot-dashed lines 
($f_{esc}=1, f_{X} = 5$, only shown in (c)),
respectively.
Here, $f_{esc}$ is the escape fraction 
(the ratio of ionization photons which escape from host galaxies).
\label{fig:spinhosi2}}
\end{center}
\end{figure}

After the star formation,
the transition due to Ly$\alpha$ photons which come from stars becomes effective.
Therefore, the spin temperature $T_{S}$ takes values close to $T_{\alpha}$
because of the Wouthuysen-Field effect
\cite{Furlanetto:2006jb,Wouthuysen:1952}.
%
Although the gas and Ly$\alpha$ temperatures
are generally different,
there are a lot of situations in which 
these temperature take same values.
%
When a large amount of neutral hydrogen gas exists 
and the optical depth is large,
Ly$\alpha$ photons are scattered with the gas many times.
In that case, the distribution of the photons
is close to that of black body with $T_{g}$ around the frequency of Ly$\alpha$.
%
Therefore,
the Ly$\alpha$ temperature becomes $T_{\alpha}\sim T_{g}$,
and as a consequence $T_{S}\sim T_{\alpha} \sim T_{g}$ in the Cosmic dawn.
(Figs.\ref{fig:spinhosi1} and \ref{fig:spinhosi2}).

After that, since the X-ray which comes form remnants of luminous objects 
heats the IGM in the epoch of reionization, 
the gas temperature becomes
$T_{\gamma}<T_{g}$ and $T_{\gamma}<T_{S} \sim T_{g}$
around $z \sim 10$.
In this situation, 
the power spectrum of 21 cm line becomes a relatively simple form
(the detail is shown in the Chapter \ref{chap:powerspectrum}).
Therefore, this era has the much advantage to determine the cosmological parameters.
However, the evolution of spin temperature 
strongly depends on the detail of the star formation,
and there are some uncertainties due to astrophysics.

\chapter[Fluctuation and power spectrum of the 21 cm radiation]
{Fluctuation and power spectrum of the 21 cm radiation
\label{chap:powerspectrum}
\normalsize{\cite{Furlanetto:2006jb,Arashiba:2009,McQuinn:2005hk}}}

In this chapter, we introduce the fluctuation of
21 cm line brightness temperature
$\delta_{21}\equiv
(\Delta T_{b}^{obs} - \Delta \bar{T}_{b}^{obs})
/\Delta \bar{T}_{b}^{obs})$
and its power spectrum.

\section{Fluctuation of brightness temperature}

\subsection{Fluctuation of brightness temperature}

The brightness temperature of 21 cm line $\delta _{21}$ is defined as
\begin{eqnarray}
\delta_{21}\left( \mbox{\boldmath $x$}, \eta ; z\right) 
\equiv 
\frac{
\Delta T_{b}^{obs}\left( \mbox{\boldmath $x$}, \eta ; z\right) 
-\Delta \bar{T}_{b}^{obs}\left( \mbox{\boldmath $x$}, \eta ; z\right) 
}
{
\Delta \bar{T}_{b}^{obs}\left( \mbox{\boldmath $x$}, \eta ; z\right) 
}
=
\frac{
\Delta T_{b}^{obs}\left( \mbox{\boldmath $x$}, \eta ; z\right) 
}
{
\Delta \bar{T}_{b}^{obs}\left( \mbox{\boldmath $x$}, \eta ; z\right) 
}
-1\label{eq:delta21},
\end{eqnarray}
where $\Delta \bar{T}_{b}^{obs}\left( \mbox{\boldmath $x$}, \eta ; z\right)$
is the spatial average of the brightness temperature.
By Eqs.(\ref{eq:obsbrightness3}) and (\ref{eq:obsbrightness4}), 
the brightness temperature is expressed as
\begin{eqnarray}
\Delta \bar{T}_{b}^{obs} \left( \frac{\nu_{21}}{1+z}\right) 
&=&
\frac{3c^{3}h_{P}A_{21}}{32\pi k_{B}\nu_{21}^{2}}
\frac
{
\bar{x}_{HI}(z)
\bar{n}_{H}(z) 
}
{(1+z)H(z)}
\left[
1-\frac{\bar{T}_{\gamma}(z) }{\bar{T}_{S}(z)}
\right] \nonumber \\
& \approx &
27\bar{x}_{HI}(z)
\left(
\frac{\Omega_{b}h^{2}}{0.023}
\right)
\left(
\frac{0.15}{\Omega_{m}h^{2}}
\frac{1+z}{10}
\right)^{1/2} 
\left[
1-\frac{\bar{T}_{\gamma}(z) }{\bar{T}_{S}(z)}
\right] {\rm mK}. \label{eq:21bright}
\end{eqnarray}
Here, we introduce the following 
fluctuations of each physical quantity,
\begin{eqnarray}
\delta_{X} \equiv \frac{X-\bar{X}}{\bar{X}},
\end{eqnarray}
where $X$ means each quantity and $\bar{X}$ is its spatial average.
Additionally, we treat the contribution due to the peculiar velocity 
in Eq.(\ref{eq:obsbrightness2}) as a perturbation,
\begin{eqnarray}
\delta_{\partial v}
\equiv \frac{1+z}{H(z)}\frac{d v_{p\|}}{dr_{\|}}.
\end{eqnarray}

By using the fluctuations of each quantity,
the brightness temperature Eq.(\ref{eq:obsbrightness2})
is expressed as
\begin{eqnarray}
\Delta T_{b}^{obs}
&=&
\frac{3c^{3}h_{P}A_{21}}{32\pi k_{B}\nu_{21}^{2}}
\frac
{
\bar{x}_{HI}(1+\delta_{x_{HI}})
\bar{n}_{H}(1+\delta_{{H}})
}
{(1+z)H(z)}
\left[
1-\frac{\bar{T}_{\gamma}
(1+\delta_{T_{\gamma}})}{\bar{T}_{S}(1+\delta_{T_{S}})}
\right]
(1- \delta_{\partial v} ) \nonumber \\
\nonumber \\
&=&
\frac{3c^{3}h_{P}A_{21}}{32\pi k_{B}\nu_{21}^{2}}
\frac
{\bar{x}_{HI}\bar{n}_{H}}{(1+z)H(z)}
\left(
1 - \frac{\bar{T}_{\gamma}}{\bar{T}_{S}}
\right)
(1+\delta_{x_{HI}})(1+\delta_{H})(1- \delta_{\partial v} ) \nonumber\\
&& \ \ \ \ \ \ \ \ \ \ \ \ \ \ \ \ \
\times 
\frac{1}{1+\delta_{T_{S}}}
\left(
1 - \frac{\bar{T}_{\gamma}}{\bar{T}_{S}}
\right)^{-1}
\left[
1+\delta_{T_{S}}-\frac{\bar{T}_{\gamma}+\bar{T}_{\gamma}
\delta_{T_{\gamma}}}{\bar{T}_{S}}
\right] \nonumber\\
\nonumber \\
&=&
\Delta \bar{T}_{b}^{obs}
(1+\delta_{x_{HI}})(1+\delta_{H})
(1- \delta_{\partial v} )\frac{1}{1+\delta_{T_{S}}}
\left[
1+\frac{\bar{T}_{S}\delta_{T_{S}}-\bar{T}_{\gamma}
\delta_{T_{\gamma}}}{\bar{T}_{S} - \bar{T}_{\gamma}}
\right]. \label{eq:delta212}
\end{eqnarray}
From now on, we treat the fluctuations of 
the number density of hydrogen ($\sim$ proton) $\delta_{H}$,
the CMB temperature $\delta_{T_{\gamma}}$
and the contribution of peculiar velocity $\delta_{\partial v}$
as small perturbations,
and neglect more than the second order terms of them.
However, there are several situations
in which the fluctuation of neutral fraction
$\delta_{x_{HI}}$ becomes ${\cal O}(1)$,
e.g. at the late stage of the epoch of reionization.
Additionally, the spin temperature depends on the neutral fraction 
through the collisional transition process 
because the process depends on the amount of neutral hydrogen.
Therefore, when $\delta_{x_{HI}}$ becomes ${\cal O}(1)$,
the fluctuation of spin temperature $\delta_{T_{S}}$
has the potential to become ${\cal O}(1)$.
According to these reasons, 
Eq.(\ref{eq:delta212}) is expressed as
\begin{eqnarray}
\Delta T_{b}^{obs}
&=&
\Delta \bar{T}_{b}^{obs}
(1+\delta_{x_{HI}})(1+\delta_{H})
(1- \delta_{\partial v} )\frac{1}{1+\delta_{T_{S}}}
\left[
1+\frac{\bar{T}_{S}\delta_{T_{S}}-\bar{T}_{\gamma}
\delta_{T_{\gamma}}}{\bar{T}_{S} - \bar{T}_{\gamma}}
\right] \nonumber \\
&\approx&
\Delta \bar{T}_{b}^{obs}\frac{1}{1+\delta_{T_{S}}}
\left[
1+\delta_{H}+\delta_{x_{HI}}- \delta_{\partial v}
+\frac{\bar{T}_{S}}{\bar{T}_{S} - \bar{T}_{\gamma}}\delta_{T_{S}}
-\frac{\bar{T}_{\gamma}}{\bar{T}_{S} - \bar{T}_{\gamma}}
\delta_{T_{\gamma}}
\right. \nonumber \\
&&+\left.
\left\{
\delta_{H}\delta_{x_{HI}}- \delta_{\partial v}\delta_{x_{HI}}
+\frac{\bar{T}_{S}}{\bar{T}_{S} 
- \bar{T}_{\gamma}}\delta_{T_{S}}\delta_{x_{HI}}
-\frac{\bar{T}_{\gamma}}{\bar{T}_{S} 
- \bar{T}_{\gamma}}\delta_{T_{\gamma}}\delta_{x_{HI}}
\right. \right. \nonumber \\
&& \ \ \ \ \ \ 
\left. \left.
+\frac{\bar{T}_{S}}{\bar{T}_{S} - \bar{T}_{\gamma}}
\delta_{T_{S}}\delta_{H}
-\frac{\bar{T}_{S}}{\bar{T}_{S} 
- \bar{T}_{\gamma}}\delta_{T_{S}}\delta_{\partial v}
\right\}
\right],
\label{eq:delta213}
\end{eqnarray}
where the terms in $\{\cdot\}$ are 
second order contributions as a consequence of considering that 
$\delta_{x_{HI}}$ and $\delta_{T_{S}}$ may become ${\cal O}(1)$.
By using Eqs.(\ref{eq:delta21}) and (\ref{eq:delta213}),
$\delta_{21}$ is given by
\begin{eqnarray}
\delta_{21}
&=&
\frac{1}{1+\delta_{T_{S}}}
\left[
1+\delta_{H}+\delta_{x_{HI}}- \delta_{\partial v}
+\frac{\bar{T}_{S}}{\bar{T}_{S} - \bar{T}_{\gamma}}\delta_{T_{S}}
-\frac{\bar{T}_{\gamma}}{\bar{T}_{S} 
- \bar{T}_{\gamma}}\delta_{T_{\gamma}}
\right. \nonumber \\
&&\ \ \ \ \ \ \ \ \ \ \ \ \ \
+\left.
 \left\{
\delta_{H}\delta_{x_{HI}}- \delta_{\partial v}\delta_{x_{HI}}
+\frac{\bar{T}_{S}}{\bar{T}_{S} 
- \bar{T}_{\gamma}}\delta_{T_{S}}\delta_{x_{HI}}
-\frac{\bar{T}_{\gamma}}{\bar{T}_{S} 
- \bar{T}_{\gamma}}\delta_{T_{\gamma}}\delta_{x_{HI}}
\right. \right. \nonumber \\
&& \ \ \ \ \ \ \ \ \ \ \ \ \ \ \ \ \ \ \
\left. \left.
+\frac{\bar{T}_{S}}{\bar{T}_{S} 
- \bar{T}_{\gamma}}\delta_{T_{S}}\delta_{H}
-\frac{\bar{T}_{S}}{\bar{T}_{S} 
- \bar{T}_{\gamma}}\delta_{T_{S}}\delta_{\partial v}
\right\}
\right] - 1.
\label{eq:delta214}
\end{eqnarray}

\subsection{In the case of $T_{\gamma}<<T_{S}$}
 \label{subsection:Ts>Tgamma}

If the spin temperature is higher than the CMB one,
the fluctuation of 21 cm line brightness temperature 
becomes a simpler form
%
because we can 
neglect the dependences of $T_{S}$ and $T_{\gamma}$ in it.
As we saw in the the Chapter \ref{chap:spin},
after the star formation, around $z \sim 10$,
the condition
$T_{\gamma} << T_{S}$ is valid.
%
Because the dependences on $T_{S}$ and $T_{\gamma}$ 
do not exist in the brightness temperature under this condition,
we can also neglect their fluctuations
$\delta _{T_{S}}$ and $\delta_{T_{\gamma}}$.
In this case, Eq.(\ref{eq:delta214}) reduces to
\begin{eqnarray}
\delta_{21}
& \approx &
\left[\
1+\delta_{H}+\delta_{x_{HI}}- \delta_{\partial v}
+\{
\delta_{H}\delta_{x_{HI}}- \delta_{\partial v}\delta_{x_{HI}}
\}
\right] - 1 \nonumber \\
&=&
\delta_{H}+\delta_{x_{HI}}- \delta_{\partial v}
+\{
\delta_{H}\delta_{x_{HI}}- \delta_{\partial v}\delta_{x_{HI}}
\}.
\label{eq:delta215}
\end{eqnarray}
%
Since the fluctuation of the neutral fraction 
is also sufficiently small $\delta_{x_{HI}}<<1$
except for the end of the epoch of reionization, 
we can also neglect the terms in $\{\cdot\}$ in the Eq.(\ref{eq:delta215}).
Therefore, the brightness temperature becomes 
the following simple form,
\begin{eqnarray}
\delta_{21}
& \approx &
\delta_{H}+\delta_{x_{HI}}- \delta_{\partial v}.
\label{eq:delta216}
\end{eqnarray}
%
Since we can use the above simple form as 
the 21 cm line fluctuations,
%
we mainly focus on the epoch of reionization (around $z \sim  10$) in this thesis.

\section{Power spectrum of 21 cm line radiation}

\subsection{Power spectrum of 21 cm line radiation}

In this section, we introduce the power spectrum of
21 cm line radiation. 
By using the fluctuation of 21 cm line brightness temperature,
the power spectrum of 21 cm line $P_{21}$ is defined as 
\begin{eqnarray}
\langle \tilde{\delta}_{21}(\mbox{\boldmath $k$})
 \tilde{\delta}_{21} (\mbox{\boldmath $k$}')\rangle
= (2\pi)^{3} \delta^{D}(\mbox{\boldmath $k$} + \mbox{\boldmath $k$}')
P_{21}({\mbox{\boldmath $k$}}),
\label{eq:21powerspectrum}
\end{eqnarray}
where $\left\langle \cdot \right\rangle$
means ensemble average, 
$\delta^{D}(\cdot)$ is the Dirac delta function
and $(\tilde{\cdot})$ means the Fourier component of the quantity.
Here, we define the Fourier and the inverse Fourier transformation as 
\begin{subequations}
\begin{eqnarray}
\tilde {A}(\mbox{\boldmath $k$}) &=& 
\int_{-\infty}^{\infty} \int_{-\infty}^{\infty} \int_{-\infty}^{\infty}dx^{3} 
e^{-i \mbox{\boldmath $k$} \cdot \mbox{\boldmath $x$}} A(\mbox{\boldmath $x$}), \label{eq:fourier1}\\
A(\mbox{\boldmath $k$}) &=& 
\int_{-\infty}^{\infty} \int_{-\infty}^{\infty} \int_{-\infty}^{\infty}
\frac{dk^{3}}{(2\pi)^{3}} 
e^{i \mbox{\boldmath $k$}\cdot \mbox{\boldmath $x$}} \tilde{A}(\mbox{\boldmath $k$}).
\end{eqnarray}
\end{subequations}

The power spectrum of 21 cm line has dependences on 
not only the absolute value of a wave vector $k=|\mbox{\boldmath $k$}|$
but also the angle between $\mbox{\boldmath $k$}$ and a LOS direction.
The dependence comes from the peculiar motion of 
a radiation source along the LOS.
Therefore, we use $\mbox{\boldmath $k$}$ as the argument of $P_{21}$.
The power spectrum of 21 cm line generally has a very complicated form.
However, under the conditions of  
$T_{\gamma}<<T_{S}$ and $\delta_{x_{HI}}<<1$,
which are considered in the section \ref{subsection:Ts>Tgamma},
the spectrum reduces to a relatively simple form.
In particular, under the condition of $\delta_{x_{HI}}<<1$,
we can neglect the second order terms of the fluctuations.
In this case, the fluctuation of 21 cm line can be expressed as
\begin{eqnarray}
\tilde{\delta}_{21}
& = &
\tilde{\delta}_{H}+\tilde{\delta}_{x_{HI}} 
- \tilde{\delta}_{\partial v}.
\label{eq:delta217}
\end{eqnarray}
By substituting
Eq.(\ref{eq:delta217}) into Eq.(\ref{eq:21powerspectrum}),
the power spectrum of 21 cm line $P_{21}$ can be given by
\begin{eqnarray}
\langle \tilde{\delta}_{21}(\mbox{\boldmath $k$})
 \tilde{\delta}_{21} (\mbox{\boldmath $k$}') \rangle
& = &
\left\langle 
\left[
\tilde{\delta}_{H}(\mbox{\boldmath $k$}) 
+\tilde{\delta}_{x_{HI}}(\mbox{\boldmath $k$}) 
 - \tilde{\delta}_{\partial v}(\mbox{\boldmath $k$}) 
\right]
\left[
\tilde{\delta}_{H}(\mbox{\boldmath $k$}') 
+\tilde{\delta}_{x_{HI}} (\mbox{\boldmath $k$}') 
- \tilde{\delta}_{\partial v}(\mbox{\boldmath $k$}') 
\right]
\right\rangle \nonumber \\ 
&=&
\langle 
\tilde{\delta}_{H}
(\mbox{\boldmath $k$})\tilde{\delta}_{H}(\mbox{\boldmath $k$}') 
\rangle 
+
\langle 
\tilde{\delta}_{x_{HI}}(\mbox{\boldmath $k$})
 \tilde{\delta}_{x_{HI}} (\mbox{\boldmath $k$}') 
\rangle 
+
\langle 
\tilde{\delta}_{\partial v}
(\mbox{\boldmath $k$})\tilde{\delta}_{\partial v}(\mbox{\boldmath $k$}') 
\rangle \nonumber \\
&&
+
\langle 
\tilde{\delta}_{H}(\mbox{\boldmath $k$})
\tilde{\delta}_{x_{HI}} (\mbox{\boldmath $k$}') 
\rangle 
+
\langle 
\tilde{\delta}_{x_{HI}} 
(\mbox{\boldmath $k$})\tilde{\delta}_{H}(\mbox{\boldmath $k$}') 
\rangle 
\nonumber \\
&& -
\langle 
\tilde{\delta}_{H}(\mbox{\boldmath $k$})
\tilde{\delta}_{\partial v} (\mbox{\boldmath $k$}') 
\rangle 
-
\langle 
\tilde{\delta}_{\partial v} 
(\mbox{\boldmath $k$})\tilde{\delta}_{H}(\mbox{\boldmath $k$}') 
\rangle \nonumber \\
&& -
\langle 
\tilde{\delta}_{x_{HI}} 
(\mbox{\boldmath $k$})\tilde{\delta}_{\partial v} (\mbox{\boldmath $k$}') 
\rangle 
-
\langle 
\tilde{\delta}_{\partial v}
 (\mbox{\boldmath $k$})\tilde{\delta}_{x_{HI}} (\mbox{\boldmath $k$}')
\rangle \nonumber \\
&=&
(2\pi)\delta^{D}(\mbox{\boldmath $k$} + \mbox{\boldmath $k$}')
\left[
P_{\delta_{H}\delta_{H}}(\mbox{\boldmath $k$})+
P_{\delta_{x_{HI}}\delta_{x_{HI}}}(\mbox{\boldmath $k$})+
P_{\delta_{\partial v}\delta_{\partial v}}(\mbox{\boldmath $k$})
\right.\nonumber \\
&&\ \ \ \ \ \ \ \ \ \ \ \ \ \ \ \ \ \ \ \ \ \ \
\left.
+\{
P_{\delta_{H}\delta_{x_{HI}}}
(\mbox{\boldmath $k$})+P_{\delta_{x_{HI}}\delta_{H}}(\mbox{\boldmath $k$})
\}
\right.\nonumber \\
&&\ \ \ \ \ \ \ \ \ \ \ \ \ \ \ \ \ \ \ \ \ \ \
\left.
-\{
P_{\delta_{H}\delta_{\partial v}}(\mbox{\boldmath $k$})
+P_{\delta_{\partial v}\delta_{H}}(\mbox{\boldmath $k$})
\}
\right.\nonumber \\
&&\ \ \ \ \ \ \ \ \ \ \ \ \ \ \ \ \ \ \ \ \ \ \
\left.
-\{
P_{\delta_{x_{HI}}\delta_{\partial v}}(\mbox{\boldmath $k$})
+P_{\delta_{\partial v}\delta_{x_{HI}}}(\mbox{\boldmath $k$})
\}
\right],
\label{eq:21powerspectrum3}
\end{eqnarray}
\begin{eqnarray}
\longrightarrow  \ \ \ \ \ \ \ \ 
P_{21}(\mbox{\boldmath $k$}) &=&
P_{\delta_{H}\delta_{H}}(\mbox{\boldmath $k$})+
P_{\delta_{x_{HI}}\delta_{x_{HI}}}(\mbox{\boldmath $k$})+
P_{\delta_{\partial v}\delta_{\partial v}}(\mbox{\boldmath $k$})
\nonumber \\
&&
\left.
+\{
P_{\delta_{H}\delta_{x_{HI}}}
(\mbox{\boldmath $k$})+P_{\delta_{x_{HI}}\delta_{H}}(\mbox{\boldmath $k$})
\}
\right.\nonumber \\
&&
\left.
-\{
P_{\delta_{H}\delta_{\partial v}}(\mbox{\boldmath $k$})
+P_{\delta_{\partial v}\delta_{H}}(\mbox{\boldmath $k$})
\}
\right.\nonumber \\
&&
-\{
P_{\delta_{x_{HI}}\delta_{\partial v}}(\mbox{\boldmath $k$})
+P_{\delta_{\partial v}\delta_{x_{HI}}}(\mbox{\boldmath $k$})
\},
\label{eq:power214}
\end{eqnarray}
where we define the power spectra $P_{\delta_{A}\delta_{B}}$ of 
fluctuations $\delta_{A},\delta_{B}$ as
\begin{eqnarray}
\langle \tilde{\delta}_{A}(\mbox{\boldmath $k$})
 \tilde{\delta}_{B} (\mbox{\boldmath $k$}') \rangle
= (2\pi)^{3} \delta^{D}(\mbox{\boldmath $k$} + \mbox{\boldmath $k$}')
P_{\delta_{A}\delta_{B}}({\mbox{\boldmath $k$}}).
\label{eq:define_power_spectra}
\end{eqnarray}
By introducing the following notation,
\begin{eqnarray}
P_{\{\delta_{A}\delta_{B}\}} ({\mbox{\boldmath $k$}})
= \frac{1}{2} \{ P_{\delta_{A}\delta_{B}} ({\mbox{\boldmath $k$}})
+ P_{\delta_{B}\delta_{A}}({\mbox{\boldmath $k$}}) \},
\end{eqnarray}
we can rewrite Eq.(\ref{eq:power214}) as
\begin{eqnarray}
P_{21}(\mbox{\boldmath $k$}) &=&
P_{\delta_{H}\delta_{H}}(k)+
P_{\delta_{x_{HI}}\delta_{x_{HI}}}(k)+
P_{\delta_{\partial v}\delta_{\partial v}}(\mbox{\boldmath $k$})
\nonumber \\
&&
+
2P_{\delta_{H}\delta_{x_{HI}}}(k)
-
2P_{\{\delta_{H}\delta_{\partial v}\}}(\mbox{\boldmath $k$})
-
2P_{\{\delta_{x_{HI}}\delta_{\partial v}\}}(\mbox{\boldmath $k$}),
\label{eq:power215}
\end{eqnarray}
where we use that $P_{\delta_{H}\delta_{H}}(\mbox{\boldmath $k$})$, 
$P_{\delta_{HI}\delta_{HI}}(\mbox{\boldmath $k$})$ and
$P_{\{\delta_{H}\delta_{x_{HI}}\}}(\mbox{\boldmath $k$})$
have only the dependence on the absolute value of $\mbox{\boldmath $k$}$.
%

Next, we introduce the cosine of 
the angle between a LOS direction and a wave vector,
\begin{eqnarray}
&&\mu \equiv \frac{k_{\|}}{|\mbox{\boldmath $k$}|}, \\
\nonumber \\ 
&&k_{\|}:{\rm the \ component \ 
of \ the \ wave \ vector \ \mbox{\boldmath $k$} \ along \ a \ LOS}, \nonumber
\end{eqnarray}
and the following growth factor,
which is defined by the growing mode $D^{+}$ of a density fluctuation~\footnote{
The growth factor $f$ is generally $\cal{O}$(1),
e.g. in the matter dominated Universe,
$f=1$ because of $D^{+}\propto a$.},
%
\begin{eqnarray}
f\equiv \frac{a}{D^{+}}\frac{D^{+}}{da} 
= \frac{d \ln D^{+} }{d\ln a}.
\end{eqnarray}
By using $\mu$, $f$ and density fluctuation of baryons $\tilde{\delta}_{H}$,
we can rewrite the contribution of peculiar velocity
$\delta_{\partial v}$ as
\begin{eqnarray}
\tilde{\delta}_{\partial v} \approx
-\mu^{2} f\tilde{\delta}_{H}. \label{eq:velocity}
\end{eqnarray}
The detail of this relation is shown
in the Chapter \ref{chap:neutrino}.
By using this relation,
the power spectrum of 21 cm line is expressed as
\begin{eqnarray}
P_{21}(k,\mu) &=&
P_{\delta_{H}\delta_{H}}(k)+
P_{\delta_{x_{HI}}\delta_{x_{HI}}}(k)
+2P_{\delta_{H}\delta_{x_{HI}}}(k)
\nonumber \\
&&
+
2\mu^{2}fP_{\delta_{H}\delta_{H}}(k)
+
2\mu^{2}fP_{\delta_{x_{HI}}\delta_{H}}(k)
+\mu^{4}f^{2}P_{\delta_{H}\delta_{H}}(k),
\label{eq:power216}
\end{eqnarray}
In the Eq.(\ref{eq:power216}),
it is important that the term of $\mu^{4}$
only depends on the power spectrum of baryons,
which almost traces the matter power spectrum in large scales.
%
Therefore, in principle, we can get the information of 
matter fluctuations from the 21 cm line power spectrum
even if we can not understand the behavior of 
fluctuations of neutral fraction $\delta_{x_{HI}}$.

\subsection{Power spectrum of ionization fraction}

Since the epoch of reionization is the matter dominated era, 
the growth factor becomes $f\approx1$.
%
Furthermore, in large scales, we can assume that
fluctuations of hydrogen ($\sim$ baryons) trace those of matters
($P_{\delta_{H}\delta_{H}}(k)=P_{\delta \delta}(k)$
and
$P_{\delta_{x_{HI}}\delta_{H}}(k)=P_{\delta_{x_{HI}}\delta}(k)$).
Therefore, the power spectrum can be rewritten as
\begin{equation}
P_{T_{b}} (k,\mu)
\equiv
(\Delta \bar{T}_{b}^{obs})^2 P_{21} (k,\mu) 
=
P_{\mu^0} (k) + \mu^2 P_{\mu^2} (k)  + \mu^4 P_{\mu^4} (k), 
\end{equation}
where we use the following notations,
\begin{subequations}
\begin{eqnarray}
P_{\mu^0} (k) & \equiv & \mathcal{P}_{\delta \delta}(k) 
- 2 \mathcal{P}_{x\delta}(k) + \mathcal{P}_{xx}(k),\\
P_{\mu^2}(k)  & \equiv & 
2 \left( \mathcal{P}_{\delta \delta}(k) - \mathcal{P}_{x\delta}(k) \right), \\
P_{\mu^4}(k)  & \equiv &   \mathcal{P}_{\delta \delta}(k),
\end{eqnarray}
\end{subequations}
\begin{subequations}
\begin{eqnarray}
\mathcal{P}_{\delta \delta}(k) & \equiv & 
(\Delta \bar{T}_{b}^{obs})^2 P_{\delta \delta}(k), \\
\mathcal{P}_{x \delta}(k) & \equiv &
(\Delta \bar{T}_{b}^{obs})^2 
\frac{\bar{x}_i}{\bar{x}_{HI}} 
P_{x\delta}(k), \\
\mathcal{P}_{x x}(k) & \equiv &
(\Delta \bar{T}_{b}^{obs})^2 
\left( \frac{\bar{x}_i}{\bar{x}_{HI}}
\right)^2 P_{ x x}(k).
\end{eqnarray}
\end{subequations}
Here, instead of the power spectra of neutral fraction $x_{HI}$ ($=1-x_{i}$),
we use those of ionization fraction $x_{i}$, $P_{x \delta}$ and $ P_{ x x} $.
The ionization fraction spectra are defined in the same manner as
Eq.~(\ref{eq:define_power_spectra}) for 
its fluctuation $\delta_{xi}$.

$P_{x\delta}$ and $P_{xx}$ can be neglected 
as long as we consider eras when the IGM is completely neutral. 
However, after the reionization starts,
these two spectra significantly contribute 
to the 21 cm line power spectrum.  
Although a rigorous evaluation of these power spectra
may need some numerical simulations, 
we adopt the treatment given in Ref.~\cite{Mao:2008ug}, 
where it is assumed that 
$\mathcal{P}_{x\delta}$ and $\mathcal{P}_{xx}$ have specific forms
which match simulations of radiative transfer 
in Refs.~\cite{McQuinn:2006et,McQuinn:2007dy}.  
The explicit forms of the power spectra are parametrized to be 
\begin{subequations}
\begin{eqnarray}
\label{eq:Pxx}
\mathcal{P}_{x x}  (k) 
& = & 
b_{xx}^2 \left[ 1 + \alpha_{xx} (k R_{xx}) + (k R_{xx})^2 \right]^{-\gamma_{xx} / 2} \mathcal{P}_{\delta\delta} (k), \\
\label{eq:Pxdelta}
\mathcal{P}_{x \delta} (k) 
& = &  
b_{x\delta}^2 ~e^{ - \alpha_{x\delta} (k R_{x\delta}) - (k R_{x\delta})^2} \mathcal{P}_{\delta\delta} (k),
\end{eqnarray}
\end{subequations}
where $b_{xx}$, $b_{x\delta}$, $\alpha_{xx}$, $\gamma_{xx}$ and $\alpha_{x\delta}$
are parameters which characterize the amplitudes and the shapes of the spectra.
$R_{xx}$ and $R_{x\delta}$ represent the effective size of ionized bubbles.  
%
%
In our analysis,
we adopt the values listed in Table~\ref{tab:Pxx_xdelta}
as the fiducial values of these parameters.

\begin{table}[t]
  \centering 
  \begin{tabular}{ccccccccc}
\hline \hline
~~$z$~~ & ~~$\bar{x}_H$~~
& ~~$b_{xx}^2$~~ & ~~$R_{xx}$~~  & ~~$\alpha_{xx}$~~ & ~~$\gamma_{xx}$~~ 
& ~~$b_{x\delta}^2$~~ & ~~$R_{x\delta}$~~  & ~~$\alpha_{x\delta}$~~ \\ 
&  &  & $[{\rm Mpc}]$& & & &$[{\rm Mpc}]$ \\
\hline
$9.2$  &  $0.9$ & $0.208$ & $1.24$ & $-1.63$ & $0.38$ & $0.45$ & $0.56$   & $-0.4$ \\ 
$8.0$  &  $0.7$ & $2.12$   & $1.63$ & $-0.1$   & $1.35$ & $1.47$ & $0.62$   & $0.46$ \\ 
$7.5$  &  $0.5$ & $9.9$     & $1.3$   & $1.6$    & $2.3$   & $3.1$   & $0.58$   & $2.0$ \\ 
$7.0$  &  $0.3$ & $77.0$   & $3.0$   & $4.5$    & $2.05$ & $8.2$   & $0.143$ & $28.0$ \\
\hline \hline
\end{tabular}
\caption{Fiducial values for the parameters in $\mathcal{P}_{xx}(k)$
  and $\mathcal{P}_{x\delta}(k)$ (See Eqs.~\eqref{eq:Pxx} and
  \eqref{eq:Pxdelta}) \cite{Mao:2008ug}.}
  \label{tab:Pxx_xdelta}
\end{table}

\chapter
[Density fluctuations and neutrino properties]
{Density fluctuations and neutrino properties
\normalsize{\cite{Wong:2011ip,Lesgourgues:2006nd}}}
\label{chap:neutrino}

In this chapter, we start discussing influences of neutrinos 
on the growth of density fluctuations. 
Note that we use Planck units 
($c=1$, $\hbar =1$, $k_{B}=1$),
and metric signature is $(- + + +)$,
in this chapter.

\section[Density fluctuations]{Density fluctuations \normalsize 
{\cite{Lesgourgues:2006nd,weinberg}} }

\subsection{Equations of density fluctuations}

In this section, we review
the treatment of density fluctuations in first order.
Here, we assume the homogeneous and isotropic Universe,
and neglect the spatial curvature.
Besides, we only consider the scalar component.

In the conformal Newtonian gauge,
the perturbed Friedmann-Lema$\hat{i}$tre-Robertson-Walker metric is given by
\begin{eqnarray}
ds^{2}=
       -a^{2}(\eta)
       \left[
            \left\{ 1 + 2\psi(\eta,\mbox{\boldmath $x$})\right\}d\eta^{2}
            -\left\{ 1- 2\phi(\eta,\mbox{\boldmath $x$})\right\} \delta_{ij}dx^{i}dx^{j}
       \right] \label{eq:metric},
\end{eqnarray}
where \mbox{\boldmath $x$} is a comoving coordinate 
and we use conformal time $dt\equiv ad\eta$ ($t$ is the cosmic time).
%
The energy momentum tensor is given by
\begin{eqnarray}
T^{\mu \nu} = pg^{\mu \nu} + (\rho + p)u^{\mu}u^{\nu} + \Sigma ^{i}_{\ j},
\end{eqnarray}
where $\Sigma ^{i}_{\ j}$ is the traceless ($\Sigma^{i}_{\ i}=0$)
anisotropic stress tensor,
which is treated as a perturbation,
$p$ is the pressure,
$\rho$ is the energy density,
$u^{\mu}$ is the four velocity,
$g^{\mu\nu}$ is the metric tensor,
the Greece induces are  $\mu = 0,1,2,3$,
and roman induces are $i=1,2,3$.
Here, the velocity of fluid is expressed as
\begin{eqnarray}
v^{i} \equiv 
      \frac{u^{i}}{u^{0}}=\frac{dx^{i}}{d\eta},
\end{eqnarray}
and we also treat this $v^{i}$ as a perturbation.
The four velocity is written as
\begin{eqnarray}
u^{\mu} = \frac{dx^{\mu}}
{\sqrt[]{\mathstrut -ds^{2}}} = \frac{d\eta}{\sqrt[]{\mathstrut -ds^{2}}}(1,v^{1},v^{2},v^{3}) \label{eq:4gen1}.
\end{eqnarray}
By using the condition of $u^{\mu}u_{\nu}=1$,
we find the following relation,
\begin{eqnarray}
&&g_{\mu\nu}u^{\mu}u^{\nu}= g_{00}u^{0}u^{0} + g_{0i}u^{0}u^{i} + g_{i0}u^{i}u^{0} + g_{ij}u^{i}u^{j} = 1, \nonumber\\
   &&\ \ \longrightarrow \ \ 
   g_{00}\left(
   \frac{d\eta}
   {\sqrt[]{\mathstrut -ds^{2}}} \right)^{2}+{\cal O}({\rm Second \ order})=1, \nonumber\\
   && \ \ \longrightarrow \ \ 
   \frac{d\eta}
   {\sqrt[]{\mathstrut -ds^{2}}} = (-g_{00})^{-1/2} \approx \frac{1}{a}(1-\psi).
\end{eqnarray}
In addition, the four velocity Eq.(\ref{eq:4gen1})
can be rewritten as
\begin{eqnarray}
u^{\mu} = \frac{1}{a}(1 - \psi ,v^{1},v^{2},v^{3}). \label{eq:4gen2}
\end{eqnarray}
Therefore, the components of the energy momentum tensor 
in first order are given as
\begin{subequations}
\begin{eqnarray}
T^{0}_{\ 0} &=& T^{0\alpha}g_{\alpha 0} \approx -\rho, \\
T^{0}_{\ i} &=& T^{0\alpha}g_{\alpha i} \approx (\rho +p)v_{i}, \\
T^{i}_{\ 0} &=& T^{i\alpha}g_{\alpha 0} \approx -(\rho +p)v^{i}, \\
T^{i}_{\ j} &=& T^{i\alpha}g_{\alpha j} \approx p\delta^{i}_{\ j} + \Sigma^{i}_{\ j},
\end{eqnarray}
\end{subequations}
By writing the perturbations of $\rho$, $p$
to be $\delta \rho, \delta p$
and its spatial averages to be $\bar{\rho}$, $\bar{p}$, 
the total energy momentum tensor is given by
\begin{subequations}
\begin{eqnarray}
T^{0}_{\ 0}  &\approx &  -\bar{\rho} + \delta \rho = -\bar{\rho}(1+ \delta),\\
T^{0}_{\ i}  &\approx & (\bar{\rho} + \bar{p})v_{i}, \\
T^{i}_{\ 0}  &\approx & -(\bar{\rho} + \bar{p})v^{i}, \\
T^{i}_{\ j}  &\approx & \bar{p}\delta^{i}_{\ j}
 + \delta p\delta^{i}_{\ j} + \Sigma^{i}_{\ j},
\label{eq:energymome}
\end{eqnarray}
\end{subequations}
where we define the following density fluctuation,
\begin{eqnarray}
\delta \equiv \frac{\delta \rho}{\bar{\rho}},
\end{eqnarray}
From now on, we use this density fluctuation
as the perturbation of energy density.

By using Eqs.(\ref{eq:metric}), (\ref{eq:energymome})
and the Einstein equation,
we can get the following equations of 
these perturbations, 
\begin{subequations}
\begin{eqnarray}
-k^{2}\tilde{\phi}-3\mathcal{H}(\tilde{\phi}'+\mathcal{H}\tilde{\psi}) 
      &=& 4\pi Ga^{2}\sum^{}_{a}\bar{\rho}_{a}\tilde{\delta}_{a} 
      \label{eq:einA}\\
k^{2} ( \tilde{\phi}' + \mathcal{H}\tilde{\psi} ) 
      &=& 4\pi Ga^{2}\sum^{}_{a}(\bar{\rho}_{a}+\bar{p}_{a})\tilde{\theta}_{a} 
      \label{eq:einB}\\
\tilde{\phi}''+ \mathcal{H}(\tilde{\psi}'+2\tilde{\phi}')+(2\mathcal{H}'+\mathcal{H}^{2})\tilde{\phi}
+\frac{k^{2}}{3}(\tilde{\phi}-\tilde{\psi}) 
      &=& 4\pi Ga^{2}\sum^{}_{a}\tilde{\delta p}_{a}\\
k^{2}(\tilde{\phi}-\tilde{\psi}) 
&=& 12\pi Ga^{2}\sum^{}_{a}(\bar{\rho_{a}}+\bar{p}_{a})\tilde{\sigma}_{a}, \label{eq:einD}
\end{eqnarray}
\end{subequations}
where index $a$ means each fluid component,
and $(\tilde{\cdot})$ means the Fourier component
(the Fourier transformation is given by Eq.(\ref{eq:fourier1})),
$(')$ means the derivative with respect to
the conformal time $(')\equiv \partial/\partial \eta$
and $\mathcal{H}$ is the comoving Hubble parameter
$\mathcal{H}\equiv \frac{a'}{a}=aH$
($H$ is the physical Hubble parameter $H=\frac{1}{a}\frac{da}{dt}$).
Additionally, we define the following quantities,
\begin{subequations}
\begin{eqnarray}
\tilde{\theta}_{a} &\equiv& ik^{i}\tilde{v}_{ai}, \\
(\bar{\rho}_{a}+\bar{p}_{a})\tilde{\sigma}_{a} &\equiv &
            -\frac{1}{k^{2}}\left( k_{i}k^{j} -\frac{1}{3}\delta_{i}^{j}\right) 
            \tilde{\Sigma}^{\ i}_{a \ j}, 
\label{eq:sigmateigi}
\end{eqnarray}
\end{subequations}
%
 %
By using Eq.(\ref{eq:einA}) and (\ref{eq:einB}),
we can obtain the equation about $\phi$,
\begin{eqnarray}
k^{2}\tilde{\phi} 
     = -4\pi Ga^{2}\sum^{}_{a}
     \left( \bar{\rho}_{a}\tilde{\delta}_{a}
     + \frac{3\mathcal{H}}{k^{2}}(\bar{\rho}_{a}+\bar{p}_{a})\tilde{\theta}_{a} \right).
\label{eq:pissoneq}
\end{eqnarray}

According to the conservation of energy and momentum
$T^{\mu\nu}_{\ \ ;\mu}=0$ (where $; \mu$ means 
a conformal derivative with respect to $\mu$),
we get the following equations,
\begin{subequations}
\begin{eqnarray}
{\rm time \ component :}\ \nu& =&0 \nonumber \\
\tilde{\delta}'&=&-(1+w)(\tilde{\theta}-3\tilde{\phi}')
               -3\mathcal{H}\left( \frac{\tilde{\delta p}}{\tilde{\delta \rho}} -w \right)\tilde{\delta}, \\
{\rm spatial \ component :}\ \nu&=&i \nonumber \\
\tilde{\theta}' &=& -\mathcal{H}(1-3w)\tilde{\theta} - \frac{w'}{1+w}\tilde{\theta}
+\frac{1}{1+w}\frac{\tilde{\delta p}}{\tilde{\delta \rho}}k^{2} \tilde{\delta}
-k^{2}\tilde{\delta} - k^{2}\tilde{\sigma} + k^{2}\tilde{\psi}, \nonumber \\
\label{eq:energymom2}
\end{eqnarray}
\end{subequations}
where $\theta\equiv \partial^{i}v_{i}$, 
and we use the equation of state $p = w\rho$.
Note that these equations are not independent of the Einstein equation.

\subsection{Equations of matter fluctuations}

Below we consider matter fluctuations.
We include cold dark matter (c),
baryons (b), non-relativistic neutrino ($\nu$)
as the matter
(we show the behavior of neutrinos in the next section).

\subsubsection*{Cold Dark Matter (CDM)}

Cold dark matter (CDM) is a component of 
non-relativistic particles
and does not interact (or weakly) with
the other particles except for the gravity,
and we can neglect the pressure.
Therefore, we can treat it as 
a perfect fluid.
By Eq.(\ref{eq:energymom2}) and
the conservation law of energy and momentum about CDM,
the evolution equations of CDM fluctuations $\delta_{c}$ are given by
\begin{subequations}
\begin{eqnarray}
\tilde{\delta}_{c}'&=&-ik\tilde{v}_{c}+3\tilde{\phi}', \\
\tilde{v_{c}}' &=& -\mathcal{H}\tilde{v}_{c}-ik\tilde{\psi},
\label{eq:energymom3}
\end{eqnarray}
\end{subequations}
where we set $\theta_{c}=ik^{i}v_{c i}=ikv_{c}$.

\subsubsection*{Baryons}

Since baryons are strongly combined with electrons
through the Coulomb interaction,
we treat them as mixed fluid here.
%
After electrons and positrons annihilate each other,
baryons behave as a non-relativistic fluid. 
Therefore, we can neglect their pressure and anisotropic stress.
According to the transportation of the energy and the momentum 
through the scattering, 
we can get the following equations of fluctuations of baryons,
\begin{subequations}
\begin{eqnarray}
\tilde{\delta}_{b}'&=&-ik\tilde{v}_{b}+3\tilde{\phi}', \\
\tilde{v_{b}}' &=& -\mathcal{H}\tilde{v}_{b}-ik\tilde{\psi} 
                - \frac{4\bar{\rho}_{\gamma}}
                  {3\bar{\rho}_{b}}a(\eta)n_{e}\sigma_{T}
                  (\tilde{v}_{b}-\tilde{v}_{\gamma}),
                  \label{eq:baryon2}
\label{eq:energymom4}
\end{eqnarray}
\end{subequations}
where 
the third term of the left hand side of Eq.(\ref{eq:baryon2})
means the interaction between baryons and photons,
$\gamma$ means quantities of photons, 
$a(\eta)$ is the scale factor,
$n_{e}$ is the number density of electrons,
and $\sigma_{T}$ is the cross section of the Thomson scattering.

\subsubsection*{}

Below, we consider epochs related to the observation of 21 cm line
$10
\hspace{0.3em}\raisebox{0.4ex}{$<$}
\hspace{-0.75em}\raisebox{-.7ex}{$\sim $}\hspace{0.3em}$
$z$
$\hspace{0.3em}\raisebox{0.4ex}{$<$}
\hspace{-0.75em}\raisebox{-.7ex}{$\sim $}\hspace{0.3em}
300$.
Since the interaction between baryons and photons is decoupled
in these epochs,
we neglect the term related to the interaction in Eq.(\ref{eq:baryon2}).
Therefore, the equations of fluctuations of baryons reduce to those of CDM.
%
Furthermore, we can neglect the anisotropic stress of non-relativistic matters,
and get the relation $\tilde{\phi}=\tilde{\psi}$ from Eq.(\ref{eq:einD}).
%
Because the Universe is dominated by the matter in these epochs,
the energy density of radiation components $\bar{\rho}_{\gamma}$
in Eq.(\ref{eq:pissoneq}) can be neglected 
in comparison with those of matter components
$\bar{\rho}_{m}\equiv \bar{\rho}_{c}+\bar{\rho}_{b}+\bar{\rho}_{\nu}$,
where $\bar{\rho_{\nu}}$ includes the energy density of neutrino and anti-neutrino.
Besides, we consider only sub-horizon scale ($aH=\mathcal{H}<<k$) here,
and can neglect the terms related to $\tilde{\theta}_{a}$
in the right hand side of Eq.(\ref{eq:pissoneq}).
According to these conditions, we obtain the following 
equations of the fluctuations,
\begin{subequations}
\begin{eqnarray}
\tilde{\delta}_{c}'&=&-ik\tilde{v}_{c}-3\tilde{\phi}', \\
\tilde{v_{c}}' &=& -\mathcal{H}\tilde{v}_{c}-ik\tilde{\phi}, \\
\nonumber \\ 
\tilde{\delta}_{b}'&=&-ik\tilde{v}_{b}-3\tilde{\phi}', \\
\tilde{v_{b}}' &=& -\mathcal{H}\tilde{v}_{b}-ik\tilde{\phi}, \\
\nonumber\\ 
k^{2}\tilde{\phi} 
    & =& -4\pi Ga^{2}
   ( \bar{\rho}_{b}\tilde{\delta}_{b} +\bar{\rho}_{c}\tilde{\delta}_{c} 
     +\bar{\rho}_{\nu}\tilde{\delta}_{\nu} ).
     \label{eq:poissonB}
\end{eqnarray}
\end{subequations}

\section
[Free-streaming behavior of neutrinos ]
{Free-streaming behavior of neutrinos 
\normalsize{ \cite{Lesgourgues:2006nd,Strumia:2006db}}}

\subsection{Free-streaming length}

Neutrinos are very light collisionless particles.
Therefore, they have free motion (the free-streaming)
due to their large thermal velocity $v_{th}$,
and a typical scale of the motion is 
about $\sim v_{th} /H$. 
%
%
The scale is called the free-streaming length $\lambda_{FS}$.
Here, we can define it as
\begin{subequations}
\begin{eqnarray}
k_{FS}(t) &\equiv & 
\left( \frac{4\pi G\bar{\rho}(t)a^{2}(t)}{ v_{th}^{2} (t)} \right)^{1/2}, \\
\lambda_{FS}(t)&=&
            2\pi\frac{a(t)}{k_{FS}(t)}=2\pi \ \sqrt[]{\mathstrut \frac{2}{3}}\frac{v_{th}(t)}{H(t)}.
\label{eq:freestream}
\end{eqnarray}
\end{subequations}
This definition is similar to the Jeans-length.

When neutrinos are relativistic particles
($m_{\nu}<<T_{\nu}$, where $m_{\nu}$ is the neutrino mass,
$T_{\nu}$ is the temperature of the neutrinos.),
their velocity is almost light speed.
Therefore, the free-streaming scale is same as the Hubble horizon scale.
However, after the neutrinos become non-relativistic particles
($m_{\nu}>>T_{\nu}$),
the thermal velocity becomes smaller and we can estimate it at
\begin{eqnarray}
v_{th}\simeq \frac{a_{nr}}{a}
      =  \frac{T_{\nu 0}}{T_{nr}}\frac{a_{0}}{a} 
      \simeq T_{\nu 0}\frac{3}{m_{\nu}}  \frac{a_{0}}{a}   
      \simeq 150(1+z)\left(\frac{1 {\rm eV}}{m_{\nu}}\right) {\rm km \ s^{-1}},
\label{eq:freestream2}
\end{eqnarray}
where we use the relation between 
the present temperature of the neutrinos $T_{\nu 0}$ 
and the CMB temperature $T_{\gamma 0}$,
$T_{\nu 0}=(4/11)^{1/3}T_{\gamma 0}$.
When the Universe is dominated by matters and dark energy,
the free-free steaming length 
and the corresponding wave number are expressed as
\begin{subequations}
\begin{eqnarray}
\lambda_{FS}(t) &=&
        7.7\frac{1+z}
           {\sqrt[]{\mathstrut \Omega_{\Lambda}+\Omega_{m}(1+z)^{3}}}
           \left(\frac{1{\rm eV}}{m_{\nu}}\right) 
           h^{-1}{\rm Mpc}, \\ \label{eq:freestreamingwave}
k_{FS}(t)&=&
0.82\frac{\sqrt[]{\mathstrut \Omega_{\Lambda}+\Omega_{m}(1+z)^{3}}}
            {(1+z)^{2}}
            \left(\frac{m_{\nu}}{1{\rm eV}}\right) 
            h{\rm Mpc^{-1}},
\label{eq:freestream3}
\end{eqnarray}
\end{subequations}
where $\Omega_{m}$ is the present density parameter of the matters,
$\Omega_{\Lambda}$ is that of the cosmological constant,
and $h$ is the dimensionless Hubble parameter 
$H_{0}=100 h {\rm km \ s^{-1} \ Mpc^{-1}}$.

When we consider the matter dominated Universe,
the free-streaming length behaves like
$\lambda_{FS}\propto (aH)^{-1}\propto t^{1/3}$.
However, the ``comoving'' free-streaming length
behaves like $\lambda_{FS}/a \propto (a^{2}H)^{-1}\propto t^{-1/3}$
because the time dependence of the scale factor is $a\propto t^{2/3}$.
%
According to the transition temperature 
$3T_{nr}=m_{\nu}$, we obtain the following relation,
\begin{eqnarray}
\frac{a_{nr}}{a_{0}}&=&\frac{T_{\nu 0}}{T_{nr}}, \nonumber \\
\longrightarrow \ \ (1+z)&=&\frac{T_{nr}}{T_{\nu 0}}
                       =\frac{m_{\nu}}{3T_{\nu 0}}
\simeq2.0\times 10^{3}\left(\frac{m_{\nu}}{{\rm eV}}\right).
\end{eqnarray}
Therefore, by using Eq.(\ref{eq:freestreamingwave}),
the minimum wave number $k_{nr}$ (the maximum free streaming length)
is given by
\begin{eqnarray}
k_{nr} \simeq
        0.018\Omega_{m}^{1/2}\left(\frac{m_{\nu}}{1{\rm eV}}\right)^{1/2}
        h{\rm Mpc^{-1}}.
\label{eq:freestream4}
\end{eqnarray}
Because of the free-streaming behavior,
the fluctuation of neutrinos is erased
in scales which are smaller than the free-streaming length.
Therefore, the energy density of neutrinos
does not contribute to the gravitational growth 
of the other matter fluctuations in such scales.
In contrast, the free-streaming behavior is neglected 
in scales which are larger than the free-streaming length.
In particular, in the scales of $k<k_{nr}$,
the fluctuation of neutrinos have never been affected 
by the free-streaming behavior.
Therefore, the neutrinos contributes to 
the growth of the density fluctuations like CDM in such a large scale.

\subsection{Large scale behavior of neutrinos}

First, we consider scales which are larger than 
the free-streaming length of neutrinos ($k<k_{FS}$).
Since we can treat the neutrinos same as CDM in this case,
the equations of CDM, baryons and neutrinos are same ones. 
Therefore, we can express these fluctuations as
$\delta_{c}\sim \delta_{b} \sim \delta_{\nu} $
and $v_{c}\sim v_{b} \sim v_{\nu}$.
By writing these fluctuations as $\delta_{m}$,
the equations of these fluctuations are given by
\begin{subequations}
\begin{eqnarray}
\tilde{\delta}_{m}'&=&-ik\tilde{v}_{m}+3\tilde{\phi}' \label{eq:yuragi1}, \\
\tilde{v_{m}}' &=& -\mathcal{H}\tilde{v}_{m}-ik\tilde{\phi}\label{eq:yuragi2},\\
k^{2}\tilde{\phi} 
    & =& -4\pi Ga^{2}
   ( \bar{\rho}_{b} +\bar{\rho}_{c}+\bar{\rho}_{\nu} )\tilde{\delta}_{m}
    = -4\pi Ga^{2}\bar{\rho}_{m}\tilde{\delta}_{m}. 
    \label{eq:yuragi3}
\end{eqnarray}
\end{subequations}
By using Eqs.(\ref{eq:yuragi1}) and (\ref{eq:yuragi2}),
we can obtain the following second order derivative equation,
\begin{eqnarray}
\tilde{\delta}_{m}''+\mathcal{H}\tilde{\delta}_{m}'
       =-k^{2}\tilde{\phi} + 3 \left( \tilde{\phi}''
       +\mathcal{H}\tilde{\phi}' \right). \label{eq:matteryuragi}
\end{eqnarray}
In the sub-horizon scale, the main contribution of 
the source term in the left hand side
is $-k^{2}\tilde{\phi}$.
%
By substituting Eq.(\ref{eq:yuragi3}) into Eq.(\ref{eq:matteryuragi}) 
and using the Friedmann equation,
\begin{eqnarray}
\mathcal{H}^{2}= \frac{8\pi G}{3}\bar{\rho}a^{2},
\label{eq:freidmanconformal}
\end{eqnarray}
we can obtain
\begin{eqnarray}
\left(
     1+\frac{9}{2}\frac{\mathcal{H}^{2}}{k^{2}}
\right)\tilde{\delta}_{m}''
+\mathcal{H}
\left(
    1 - \frac{9}{2}\frac{\mathcal{H}^{2}}{k^{2}}
\right)\tilde{\delta}_{m}'
      =
\frac{3}{2}\mathcal{H}^{2}
\left(
   1 - \frac{3}{2}\frac{\mathcal{H}^{2}}{k^{2}}
\right)\tilde{\delta}_{m},
\label{eq:matteryuragi-2}
\end{eqnarray}
where we assume the matter dominated Universe $\bar{\rho} \propto a^{-3}$.
When we consider sub-horizon scale
$\mathcal{H}/k << 1$,
Eq.(\ref{eq:matteryuragi-2}) is approximated as
\begin{eqnarray}
\tilde{\delta}_{m}''+\mathcal{H}\tilde{\delta}_{m}' 
=\frac{3}{2}\mathcal{H}^{2}\tilde{\delta}_{m}.
\label{eq:matteryuragi-3}
\end{eqnarray}
We can get the same equation if  we neglect the terms
$\tilde{\phi}'$ and $\tilde{\phi}''$ in Eq.(\ref{eq:matteryuragi}).
By Eq.(\ref{eq:matteryuragi-3}), the equation of $\tilde{\delta}_{m}$ is given by
\begin{eqnarray}
\tilde{\delta}_{m}''+\mathcal{H}\tilde{\delta}_{m}' 
= 4\pi Ga^{2}\bar{\rho}_{m}\tilde{\delta}_{m}. 
\label{eq:matteryuragi2}
\end{eqnarray}

By the translation of variables from $\eta$ to the scale factor $a$,
the differential of $\eta$ is written as
\begin{eqnarray}
\frac{\partial}{\partial \eta} 
      =\frac{da}{\partial \eta}\frac{\partial}{\partial a}
      =a\mathcal{H}\frac{\partial}{\partial a}
      =a^{2}H\frac{\partial}{\partial a}.
\end{eqnarray}
Therefore, the left hand side of Eq.(\ref{eq:matteryuragi2}) becomes
\begin{eqnarray}
\tilde{\delta}_{m}''+\mathcal{H}\tilde{\delta}_{m}' 
 &=&a^{2}H\frac{\partial}{\partial a}\left( a^{2}H\frac{\partial}{\partial a} \tilde{\delta}_{m}  \right) 
  +aHa^{2}H\frac{\partial}{\partial a}\tilde{\delta}_{m} \nonumber \\
  &=& a^{4}H^{2}\frac{\partial^{2}\tilde{\delta}_{m}}{\partial a^{2}}
  + a^{3}H^{2}\left( 3+ \frac{d \ln H}{d\ln a}\right)\frac{\partial \tilde{\delta}_{m}}{\partial a}.
  \label{eq:matteryuragi3}
\end{eqnarray}
By differentiating the following Friedmann equation
with respect to the scale factor $a$,
%
\begin{eqnarray}
H^{2}&=& \frac{8\pi G}{3}\bar{\rho} \nonumber \\ 
&=& H^{2}_{0}
\left(
\frac{\Omega_{m}}{a^{3}}+\Omega_{\Lambda}
\right),
\label{eq:friedman}
\end{eqnarray}
%
we can obtain the following relation,
\begin{eqnarray}
2H\frac{dH}{da} = -3H_{0}^{2}\Omega_{m}\frac{1}{a^{4}}. 
\label{eq:yuragiuhen1}
\end{eqnarray}
Furthermore, by differentiating Eq.(\ref{eq:yuragiuhen1})$\times a^3$,
\begin{eqnarray}
\frac{d}{da} \left( a^{3}H\frac{dH}{da} \right) 
    &=& \frac{d}{da} \left(-\frac{3}{2}a^{3}H^{2}_{0}\Omega_{m}\frac{1}{a^{4}} \right) \nonumber \\
    &=& -\frac{3}{2}H^{2}_{0}\frac{\Omega_{m}}{a^{3}} a \nonumber \\
    &=& 4\pi G\bar{\rho}_{m}a, \nonumber \\
\longrightarrow \ \
   4\pi Ga^{2}\bar{\rho}_{m} 
   &=& a\frac{d}{da}\left( a^{3}H\frac{dH}{da} \right),
\label{eq:yuragiuhen2}
\end{eqnarray}
we can write the right hand side of  Eq.(\ref{eq:matteryuragi2}) as
\begin{eqnarray}
4\pi Ga^{2}\bar{\rho}_{m} \tilde{\delta}_{m}
   = a\frac{d}{da}\left( a^{3}H\frac{dH}{da} \right) \tilde{\delta}_{m}.
\label{eq:RHS_matteryuragi2}
\end{eqnarray}
By using Eqs.(\ref{eq:matteryuragi3}) and (\ref{eq:RHS_matteryuragi2}),
Eq (\ref{eq:matteryuragi2}) is expressed to be
\begin{eqnarray}
a^{4}H^{2}\frac{\partial^{2}\tilde{\delta}_{m}}{\partial a^{2}}
  + a^{3}H^{2}\left( 3+ \frac{d \ln H}{d\ln a}\right)\frac{\partial \tilde{\delta}_{m}}{\partial a}
  - a\frac{d}{da}\left( a^{3}H\frac{dH}{da} \right)
  \tilde{\delta}_{m} = 0. \label{eq:yuragieq4}
\end{eqnarray}
Additionally, the factor of the second term of Eq.(\ref{eq:yuragieq4})
is rewritten as
\begin{eqnarray}
a^{3}H^{2}\left( 3+ \frac{d \ln H}{d\ln a}\right)
 &=& a^{3}H^{2}\left(\frac{d \ln (a^{3}H^{2}/H)}{d\ln a}\right) \nonumber \\
 &=& a^{4}H^{2}\left\{ \frac{d \ln (a^{3}H^{2})}{da} -\frac{d \ln H}{da} \right\} \nonumber \\
 &=& a \left\{  \frac{d(a^{3}H^{2})}{da} - a^{3}H\frac{d H}{da} \right\}. \label{eq:yuragidai2kou}
\end{eqnarray}
Therefore, Eq. (\ref{eq:yuragieq4}) reduces to
\begin{eqnarray}
&&a\left[\left\{
   a^{3}H^{2}\frac{\partial^{2}\tilde{\delta}_{m}}{\partial a^{2}}
   +  \frac{d(a^{3}H^{2})}{da} \frac{\partial \tilde{\delta}_{m}}{\partial a}\right\} 
   -\left\{
    a^{3}H\frac{d H}{da}  \frac{\partial \tilde{\delta}_{m}}{\partial a}
   +\frac{d}{da}\left( a^{3}H\frac{dH}{da} \right)\tilde{\delta}_{m} 
   \right\}
   \right] = 0 \nonumber \\ 
 && \ \ \longrightarrow  \ \
   \left[
   \frac{\partial}{\partial a}
   \left(
   a^{3}H^{2}\frac{\partial \tilde{\delta}_{m}}{\partial a}
   \right)
   -
   \frac{\partial}{\partial a}
   \left(
   a^{3}H\frac{dH}{da}\tilde{\delta}_{m}
   \right)
   \right] = 0 \nonumber\\
&& \ \ \longrightarrow  \ \
    \frac{\partial}{\partial a}
   \left[
   a^{3}H^{2}\frac{\partial }{\partial a}   
   \left( \frac{\tilde{\delta}_{m}}{H}  \right)
   \right] = 0. \label{eq:yuragieq6}
\end{eqnarray}
The solution of this equation can be
derived analytically and it is given by
\begin{eqnarray}
&&\frac{\partial}{\partial a}
   \left[
   a^{3}H^{2}\frac{\partial }{\partial a}   
   \left( \frac{\tilde{\delta}_{m}}{H}  \right)
   \right] = 0 
   \ \ \longrightarrow \ \ 
   a^{3}H^{2}\frac{\partial }{\partial a}   
   \left( \frac{\tilde{\delta}_{m}}{H}  \right)
    = A(\mbox{\boldmath $k$}), \nonumber \\
    \nonumber \\
&& \ \longrightarrow  \ \ 
   \tilde{\delta}_{m}(\mbox{\boldmath $k$},a)=
   A(\mbox{\boldmath $k$})H(a)\int^{a}\frac{da'}{a'^{3}H(a')^{3}} + B(\mbox{\boldmath $k$})H(a),
\label{eq:yuragieq7}
\end{eqnarray}
where  $A(\mbox{\boldmath $k$})$ and $B(\mbox{\boldmath $k$})$
are the arbitrary function with respect to $\eta$ (or $a$).
The first and second terms mean the growing and 
decaying solutions, respectively.
When the Universe is dominated by matter,
by using Eq.(\ref{eq:friedman}), this solution reduces to
\begin{eqnarray}
   \tilde{\delta}_{m}(\mbox{\boldmath $k$},a)
   &=&
   A(\mbox{\boldmath $k$})H_{0}\left(\frac{\Omega_{m}}{a^{3}}\right)^{\frac{1}{2}}
   \int^{a}\frac{da'}{a'^{3}H_{0}^{3}
   \left( \Omega_{m}/a'^{3}\right)^{3/2}} 
   + B(\mbox{\boldmath $k$})H_{0}\left(\frac{\Omega_{m}}{a^{3}}\right)^{\frac{1}{2}} \nonumber \\
   \nonumber \\
   &=&
   A(\mbox{\boldmath $k$})\frac{2}{5H^{2}_{0}\Omega_{m}}a
    + B(\mbox{\boldmath $k$})H_{0} \ \sqrt[]{\Omega_{m}}a^{-\frac{3}{2}}.
\label{eq:yuragieq8}
\end{eqnarray}
From now on, we express the growing and decaying solutions 
as $D^{+}(a)$ and $D^{-}(a)$, respectively.
In Eq.(\ref{eq:yuragieq8}),
we find that 
the growing solution (first term) behaves like $D^{+}(a) \propto a$.
In contrast the decaying solution (second term)
behaves like $D^{-}(a) \propto a^{-3/2}$.

\subsection{Small scale behavior of neutrinos}
%
Here, we consider scales which are smaller than 
the free-streaming length ($k_{FS}<k$).
In this case, the fluctuation of neutrinos does not grow,
but those of CDM and baryons can do.
According to this, these fluctuations become
$\tilde{\delta}_{\nu}<<\tilde{\delta}_{b},\tilde{\delta}_{c}$.
%
Since the energy density of neutrinos is 
$\bar{\rho}_{\nu}<<\bar{\rho}_{b},\bar{\rho}_{c}$,
the product of the energy density and the density fluctuation
also becomes
$\bar{\rho}_{\nu}\tilde{\delta}_{\nu}
<<\bar{\rho}_{b}\tilde{\delta}_{b},\bar{\rho}_{c}\tilde{\delta}_{c}$.
Therefore,
the source term of the left hand side of Eq.(\ref{eq:poissonB})
reduces to
\begin{eqnarray}
   -4\pi Ga^{2}
   ( \bar{\rho}_{b}\tilde{\delta}_{b} +\bar{\rho}_{c}\tilde{\delta}_{c}+\bar{\rho}_{\nu}\tilde{\delta}_{\nu})
   \approx -4\pi Ga^{2}( \bar{\rho}_{b}\tilde{\delta}_{b} +\bar{\rho}_{c}\tilde{\delta}_{c}). \label{eq:gravsorce}
\end{eqnarray}
Because baryons and CDM obey the same equation,
these are $\tilde{\delta}_{b}\sim\tilde{\delta}_{c}$
and we can express them as $\tilde{\delta}_{m}$ below.

Here, we define the ratio of 
the energy density of non-relativistic neutrino 
to that of matter as
\begin{eqnarray}
f_{\nu} \equiv \frac{\bar{\rho_{\nu}}}{\bar{\rho}_{m}}
  =\frac{\Omega_{\nu }}{\Omega_{m}}
  =\frac{\Omega_{\nu }}{\Omega_{c}+\Omega_{b}+\Omega_{\nu }}.
\end{eqnarray}
By using this, we can rewrite Eq.(\ref{eq:gravsorce}) as
\begin{eqnarray}
-4\pi Ga^{2}( \bar{\rho}_{b}\tilde{\delta}_{b} +\bar{\rho}_{c}\tilde{\delta}_{c}) 
  &=& -4\pi Ga^{2}( \bar{\rho}_{b} + \bar{\rho}_{c}) \tilde{\delta}_{m}\nonumber\\
  &=& -4\pi Ga^{2}( \bar{\rho}_{m} - \bar{\rho}_{\nu}) \tilde{\delta}_{m} \nonumber \\
  &=& -4\pi Ga^{2}\bar{\rho}_{m}(1 - f_{\nu}) \tilde{\delta}_{m}.
  \label{eq:gravsorce2}
\end{eqnarray}
According to them,
we can obtain the following equations of the density fluctuation,
\begin{subequations}
\begin{eqnarray}
\tilde{\delta}_{m}'&=&-ik\tilde{v}_{m}+3\tilde{\phi}' \label{eq:yuraginu1},\\
\tilde{v_{m}}' &=& -\mathcal{H}\tilde{v}_{m}
-ik\tilde{\phi}\label{eq:yuraginu2}, \\
k^{2}\tilde{\phi} 
    & =& -4\pi Ga^{2}\bar{\rho}_{m}(1 - f_{\nu}) \tilde{\delta}_{m}.
\end{eqnarray}
\end{subequations}
By using these equations,
we can get the following second order differential equation,
\begin{eqnarray}
\tilde{\delta}_{m}''+\mathcal{H}\tilde{\delta}_{m}'
       -4\pi Ga^{2}\bar{\rho}_{m}(1 - f_{\nu}) \tilde{\delta}_{m}=0, \label{eq:freeyuragi}
\end{eqnarray}
where the terms of $\phi'$ and $\phi''$ can be neglected
in the same way as the case of $k_{FS}>k$.
Furthermore, by using Eq.(\ref{eq:matteryuragi3})
and translating the variable from $\eta$ to $a$,
Eq.(\ref{eq:freeyuragi}) can be rewritten as
\begin{eqnarray}
a^{4}H^{2}\frac{\partial^{2}\tilde{\delta}_{m}}{\partial a^{2}}
  + a^{3}H^{2}\left( 3+ \frac{d \ln H}{d\ln a}\right)\frac{\partial \tilde{\delta}_{m}}{\partial a}
  - \frac{3}{2}a^{2}H^{2}(1-f_{\nu})\tilde{\delta}_{m}=0.
  \label{eq:yuraginu3}
\end{eqnarray}
When the Universe are dominated by the matter
$H\propto a^{-\frac{3}{2}}$,
the second term of Eq.(\ref{eq:yuraginu3}) reduces to
\begin{eqnarray}
 a^{3}H^{2}\left( 3+ \frac{d \ln H}{d\ln a}\right)\frac{\partial \tilde{\delta}_{m}}{\partial a}
  =  a^{3}H^{2}\left( 3 - \frac{3}{2}\right)\frac{\partial \tilde{\delta}_{m}}{\partial a}
  = a^{3}H^{2}\frac{3}{2}\frac{\partial \tilde{\delta}_{m}}{\partial a}
\end{eqnarray}
Therefore, we can obtain the following equation,
\begin{eqnarray}
\frac{\partial^{2}\tilde{\delta}_{m}}{\partial a^{2}}
  + \frac{3}{2a}\frac{\partial \tilde{\delta}_{m}}{\partial a}
  - \frac{3}{2a^{2}}(1-f_{\nu})\tilde{\delta}_{m}=0. 
  \label{eq:yuraginu4}
\end{eqnarray}

To solve Eq.(\ref{eq:yuraginu4}) we substitute 
$\tilde{\delta}_{m}\propto a^{y}$ into Eq.(\ref{eq:yuraginu4}),
and obtain the following equation,
\begin{eqnarray}
&&y(y-1)+\frac{3}{2}y - \frac{3}{2}(1-f_{\nu})=0, \nonumber\\
&&\longrightarrow \ \
y^{2}+\frac{1}{2}y-\frac{3}{2}(1-f_{\nu})=0,
\end{eqnarray}
The solution of this equation is 
\begin{eqnarray}
y=\frac{-1 \pm 5 \ \sqrt[]{1-\frac{24}{25}f_{\nu}}}{4}.
\end{eqnarray}
By using a approximation of $f_{\nu}<<1$, 
this solution reduce to
%
\begin{subequations}
\begin{eqnarray}
y_{+}\equiv 
\frac{-1 + 5 \ \sqrt[]{1-\frac{24}{25}f_{\nu}}}{4} 
&\approx & 1-\frac{3}{5}f_{\nu}, \\
y_{-} \equiv 
\frac{-1 - 5 \ \sqrt[]{1-\frac{24}{25}f_{\nu}}}{4}
&\approx & -\frac{3}{2}+\frac{3}{5}f_{\nu}.
\end{eqnarray}
\end{subequations}
Finally, 
the growing $D^{+}(a)$ and decaying modes $D^{-}(a)$
are given by
\begin{subequations}
\begin{eqnarray}
D^{+} & \propto & a^{y_{+}} \approx a^{1-\frac{3}{5}f_{\nu}}, \\
D^{-} & \propto & a^{y_{-}} \approx a^{-\frac{3}{2}+\frac{3}{5}f_{\nu}}.
\end{eqnarray}
\end{subequations}
%
%
From these solutions,  we find that 
the growth of fluctuations including the free-streaming effect
is suppressed in comparison with 
the growth not including the effect (Eq.(\ref{eq:yuragieq8})).
This suppression is an influence due to massive neutrinos.

\section[Peculiar velocity]
{Peculiar velocity
\normalsize{\cite{Bharadwaj:2004nr}}}

When we estimate the power spectrum of
21 cm line (Eq.(\ref{eq:power216})),
we use the following relation~\footnote{
Instead of $\tilde{\delta}_{H}$,
we use  $\tilde{\delta}_{b}$ as the density fluctuation of baryons.}.
%
\begin{eqnarray}
\tilde{\delta}_{\partial v} &\approx& -f(a)\mu^{2}\tilde{\delta}_{b}, \\
\nonumber \\
\delta_{\partial v} & \equiv& \frac{1}{aH(a)}\frac{dv_{p\|}}{dr_{\|}}.
\end{eqnarray}
In this section,we derive this relation.

By Eq.(\ref{eq:yuraginu1}),
the peculiar velocity of the matter $v_{m}$ is given by
\begin{eqnarray}
\tilde{\delta}_{m}'&=&-ik\tilde{v}_{m}', \label{eq:pvelo}
\end{eqnarray}
where we assume $v_{m}\sim v_{b} \sim v_{c}$
and neglect the terms of $\tilde{\phi}'$ and $\tilde{\phi}''$
in the same manner as the previous sections.
First of all, we rewrite this equation as
\begin{eqnarray}
\tilde{v}_{m}&=&\frac{i}{k}\tilde{\delta}'_{m} \nonumber \\
             &=&\frac{i}{k}a^{2}H\frac{\partial \tilde{\delta}_{m}}{\partial a} \nonumber \\
             &=&\frac{i}{k}a^{2}H\frac{\partial}{\partial a} 
            \left( D^{+}\frac{\tilde{\delta}_{m}}{D^{+}}\right). \label{eq:pecuvhenkei1}
\end{eqnarray}
By using an approximation of $\tilde{\delta}_{m} \propto D^{+}$
(neglecting the decaying mode), the growing mode is expressed as
$D^{+}/\tilde{\delta}_{m} = {\rm constant}$.
Therefore, Eq.(\ref{eq:pecuvhenkei1}) is rewritten as
\begin{eqnarray}
\tilde{v}_{m}
             &=&\frac{iaH}{k}
            \frac{a}{D^{+}}
            \frac{\partial D^{+}}{\partial a} 
            \tilde{\delta}_{m} 
             =\frac{iaH}{k}
            f(a)
            \tilde{\delta}_{m},
\end{eqnarray}
where we define the growth factor $f(a)$ as
\begin{eqnarray}
f(a) \equiv
     \frac{a}{D^{+}}
     \frac{\partial D^{+}}{\partial a} 
     = \frac{d \ln D^{+}}{d \ln a}.
\end{eqnarray}
This quantity is generally $\mathcal{O}(1)$.
For example, when the Universe is dominated by matter,
$D^{+}\propto a$ and $f(a)=1$.
According to the above discussion,
the Fourier component of the three velocity
$\mbox{\boldmath $\tilde{v}$}_{m}$
can be written as
\begin{eqnarray}
\mbox{\boldmath $\tilde{v}$}_{m} 
      = \frac{ \mbox{\boldmath $k$}}{k}\tilde{v}_{m} 
      \approx
      \frac{iaH\mbox{\boldmath $k$}}{k^{2}}f(a)\tilde{\delta}_{m}.
\end{eqnarray}
By using inverse Fourier transformation,
the quantity becomes
\begin{eqnarray}
\mbox{\boldmath $v$}_{m} 
      (\mbox{\boldmath $x$},a) 
      =  \int \frac{d^{3}k}{(2\pi)^{3}} e^{i \mbox{\boldmath $k$} \cdot \mbox{\boldmath $x$}}
       \mbox{\boldmath $\tilde{v}$}_{m} (\mbox{\boldmath $k$},a) 
      \approx
      iaH(a)f(a)
      \int \frac{d^{3}k}{(2\pi)^{3}} e^{i \mbox{\boldmath $k$} \cdot \mbox{\boldmath $x$}}
      \frac{\mbox{\boldmath $k$}}{k^{2}}\tilde{\delta}_{m}
      (\mbox{\boldmath $k$},a). \label{eq:mattervelocity}
\end{eqnarray}
%
This three velocity of matter represents 
the peculiar velocity $v_{p}$ of neutral hydrogen gas
because we assume  $v_{m}\sim v_{b}$ here.

Next, we estimate the line of sight component $v_{m\|}$ of $v_{m}$.
The inner product between 
Eq.(\ref{eq:mattervelocity}) and a unit vector $\hat{\mbox{\boldmath $x$}}_{\|}$
which points to the LOS direction is given by
\begin{eqnarray}
v_{m \|} (\mbox{\boldmath $x$},a) 
      =      
      \mbox{\boldmath $v$}_{m}  (\mbox{\boldmath $x$},a) \cdot \hat{\mbox{\boldmath $x$}}_{\|} 
      &\approx &
      iaH(a)f(a)
      \int \frac{d^{3}k}{(2\pi)^{3}} e^{i \mbox{\boldmath $k$} \cdot \mbox{\boldmath $x$}}
      \frac{\mbox{\boldmath $k$}\cdot \hat{\mbox{\boldmath $x$}}_{\|}}{k^{2}}
      \tilde{\delta}_{m}(\mbox{\boldmath $k$},a) \nonumber\\
      \nonumber \\
      & = &
      iaH(a)f(a)
      \int \frac{d^{3}k}{(2\pi)^{3}} e^{i \mbox{\boldmath $k$} \cdot \mbox{\boldmath $x$}}
      \frac{\mu}{k}
      \tilde{\delta}_{m}(\mbox{\boldmath $k$},a),
      \label{eq:mattervelocity2}
\end{eqnarray}
where we define $\mu$ as
\begin{eqnarray}
\mu \equiv \frac{ \mbox{\boldmath $k$}\cdot \hat{\mbox{\boldmath $x$}}_{\|}}{k}.
\end{eqnarray}
This quantity is the cosine between
a wave number vector $\mbox{\boldmath $k$}$
and the direction of LOS $\hat{\mbox{\boldmath $x$}}_{\|}$.
By differentiating Eq.(\ref{eq:mattervelocity2})
with respect to the LOS direction $r_{\|}=| \mbox{\boldmath $x$}_{\|}|$,
where $\mbox{\boldmath $x$}_{\|}$ is the vector of 
the LOS component of $\mbox{\boldmath $x$}$,
we can obtain
\begin{eqnarray}
\frac{d v_{m \|} (\mbox{\boldmath $x$},a)}{dr_{\|}} 
      & \approx &
      iaH(a)f(a)
      \int \frac{d^{3}k}{(2\pi)^{3}} 
      \left(
      \frac{d}{dr_{\|}}
      e^{i \mbox{\boldmath $k$} \cdot \mbox{\boldmath $x$}}
      \right)
      \frac{\mu}{k}
      \tilde{\delta}_{m}(\mbox{\boldmath $k$},a).
      \label{eq:mattervelocity3}
\end{eqnarray}
%
Here, we assume that 
the focusing region, whose position is $\mbox{\boldmath $x$}$,
is not apart form the position of $\mbox{\boldmath $x$}_{\|}$.
In this case, the argument of the exponential part 
of Eq.(\ref{eq:mattervelocity3}) can be approximated as
\begin{eqnarray}
\mbox{\boldmath $k$} \cdot \mbox{\boldmath $x$}
     = \mbox{\boldmath $k$} \cdot \mbox{\boldmath $x$}_{\|} +
      \mbox{\boldmath $k$} \cdot (\mbox{\boldmath $x$}-\mbox{\boldmath $x$}_{\|})
     \approx kr_{\|}\mu.
\end{eqnarray}
Therefore, Eq.(\ref{eq:mattervelocity3})
can be expressed as
\begin{eqnarray}
\frac{d v_{m \|} (\mbox{\boldmath $x$},a)}{dr_{\|}} 
      & \approx &
      iaH(a)f(a)
      \int \frac{d^{3}k}{(2\pi)^{3}} 
      \left(
      \frac{d}{dr_{\|}}
      e^{i kr_{\|}\mu}
      \right)
      \frac{\mu}{k}
      \tilde{\delta}_{m}(\mbox{\boldmath $k$},a) \nonumber \\
      & \approx & 
      -aH(a)f(a)
      \int \frac{d^{3}k}{(2\pi)^{3}} 
      e^{i kr_{\|}\mu}
      \mu^{2}
      \tilde{\delta}_{m}(\mbox{\boldmath $k$},a), \nonumber \\
\longrightarrow \ \
\frac{1}{aH(a)}\frac{d v_{m \|} (\mbox{\boldmath $x$},a)}{dr_{\|}}   
      &=&  
      \int \frac{d^{3}k}{(2\pi)^{3}} 
      e^{i kr_{\|}\mu}
      \left(
      -f(a)\mu^{2}
      \tilde{\delta}_{m}(\mbox{\boldmath $k$},a) 
      \right).
\label{eq:mattervelocity4}
\end{eqnarray}
Since $\tilde{\delta}_{\partial_{v}}$
is the Fourier transformed quantity of  
Eq.(\ref{eq:mattervelocity4}), we can obtain
\begin{eqnarray}
\tilde{\delta}_{\partial_{v}}(\mbox{\boldmath $k$},a) 
      = 
      -f(a)\mu^{2}
      \tilde{\delta}_{m}(\mbox{\boldmath $k$},a).
\label{eq:mattervelocity5}
\end{eqnarray}
Thus, the derivative of the peculiar velocity
is proportional to the density fluctuation 
in the Fourier space.
The fluctuation $\tilde{\delta}_{\partial_{v}}$
depends on the direction through $\mu$,
and induces an anisotropic distortion in the density fluctuation.
This effect is called the ``redshift space distortion''.

\section{Other effects due to neutrino properties}


In comparison with the standard $\Lambda$CDM models 
where three massless active neutrinos are assumed, 
we can introduce two more freedoms.
A first additional freedom is 
the effective number of neutrino species $N_{\nu}$, 
and it represents generations of relativistic neutrinos
before the matter-radiation equality epoch.
$N_{\nu}$ can include other relativistic components,
and may not be equal to three. 
A second additional freedom is the neutrino mass hierarchy. 
The difference of the neutrino mass hierarchy affects 
both the free-streaming scales and 
the expansion rate of the Universe~\cite{Lesgourgues:2004ps}.  
%
In terms of 21 cm line observation, 
the minimum cutoff of wave number is given by
$k_{{\rm min}}=2\pi/(yB)\sim 6\times10^{-2} h{\rm Mpc}^{-1}$
(see the Section \ref{sec:21cm_spec} in the Chapter \ref{chap:21cmFisher}).
However, the wave number corresponding to the
neutrino free-streaming scale is 
$k_{{\rm free}}\lesssim 10^{-2} h{\rm Mpc}^{-1}$.
Therefore, the main feature due to the difference of the mass hierarchy  
comes from the impact on the cosmic expansion rate
when we focus on the 21 cm line observations. 
%
In this thesis, we separately study the following two cases:

%
\subsubsection{(A) Effective number of neutrino species}

In this analysis, 
we add the effective number of neutrino species $N_{\nu}$ 
to the fiducial parameter set,
and the fiducial value of this parameter is $N_{\nu}=3.046$.
%
This parameter represents three species of massive neutrinos 
and an extra relativistic component.

\subsubsection{(B) Neutrino mass hierarchy}

The normal and inverted mass hierarchies mean  
$m_{1}<m_{2} \ll m_{3}$ 
and $m_{3} \ll m_{1}<m_{2}$, respectively. 
In a cosmological context, many different parameterizations of the
mass hierarchy have been proposed
\cite{Takada:2005si,Slosar:2006xb,DeBernardis:2009di,Jimenez:2010ev}.
In our analysis, we adopt $r_{\nu} \equiv 
(m_{3} - m_{1})/\Sigma m_{\nu}$~\cite{Jimenez:2010ev} 
as an additional parameter to 
discriminate the true neutrino mass hierarchy from the other. 
%
$r_{\nu}$ becomes positive for the normal hierarchy, 
and negative for the inverted hierarchy.
Besides, the difference between $r_{\nu}$ of  these two hierarchies
becomes larger as the total mass $\Sigma m_{\nu}$ becomes smaller.
Therefore, $r_{\nu}$ is particularly useful 
for distinguishing the mass hierarchy.  
In Fig \ref{fig:hie_ellipse} (in the Chapter \ref{Chap:result_mass}), 
we plot behaviors of $r_{\nu}$ as 
a function of $\Sigma m_{\nu}$. 

Note that there is a lowest value of
$\Sigma m_{\nu}$ which depends on the mass hierarchy
by the neutrino oscillation experiments. 
The lowest value is
$\Sigma m_{\nu}\sim$0.1 eV for the inverted hierarchy
or $\Sigma m_{\nu}\sim$0.06 eV for the normal hierarchy. 
Therefore, if we obtain a clear constraint like 
$\Sigma m_{\nu} \ll $~0.10~eV, 
we can determine
that the mass hierarchy is obviously 
normal without any ambiguities. 
However, we can discriminate the
mass hierarchy even when the mass hierarchy is 
larger than $0.01$~eV
if we use $r_{\nu}$, as will be shown later.

\section{Influence of lepton asymmetry of the Universe}
\label{sec:lepton_asym}


In this section, we briefly explain 
influence of the non-zero lepton number asymmetry in the Universe, 
or non-zero chemical potential for neutrinos.
%
%
When there are non-zero chemical potentials for neutrinos, 
they affect their energy density and pressure,
which influence the background evolution. 
The existence of non-zero chemical potential also modifies 
the perturbation equation of neutrinos. 
Below we describe the changes of the background and perturbation parts. 

\subsection{Background}

The distribution function for neutrino species $\nu_i$ and 
its anti-particle $\bar{\nu}_i$ with $i = e, \mu, \tau$ are given by
\begin{equation}
f_{\nu_i} (p_i) =  \frac{1}{e^{p_i/T_\nu + \xi_{\nu i}} +1}, 
\qquad
f_{\bar{\nu}_i} (p_i)  =  \frac{1}{e^{p_i/T_\nu - \xi_{\nu i}} +1},
\end{equation}
where $p_i$ is momentum of $\nu_i$,
$T_\nu$ is the temperature of neutrinos
and related to that at the present epoch 
$T_{\nu_0}$ as $T_\nu = T_{\nu 0}/a$ with $a$ being the scale factor.
$\xi_{\nu i}$ is the degeneracy parameter 
which is defined as 
$\xi_{\nu i} \equiv \mu_{\nu i} / T_{\nu}$,
where $\mu_{\nu i}$ is the chemical potential for $\nu_i$,
In the following, 
we omit the subscript $i$ for simplicity and give the formulas
for one neutrino species including its mass $m$.

The effects of the lepton asymmetry on the background evolution 
appear as the changes in its energy density and pressure. 
The energy density and pressure of a neutrino species are given by
\begin{eqnarray}
\label{eq:rho_nu}
\rho_\nu + \rho_{\bar{\nu}} 
&=& \frac{1}{2 \pi^2} \int_0^{\infty} p^2 dp \sqrt{p^2 + m^2} 
\left( f_\nu + f_{\bar{\nu}} \right), \\
\label{eq:rho_p}
p_\nu + p_{\bar{\nu}} 
&=& \frac{1}{2 \pi^2} \int_0^{\infty} p^2 dp \frac{p^2}{3\sqrt{p^2 + m^2}}
\left( f_\nu + f_{\bar{\nu}} \right).
\end{eqnarray}
where $\rho_{\nu}$ and $\rho_{\bar{\nu}}$ are the energy densities of 
neutrino and anti-neutrino respectively,
and $p_{\nu}$ and $p_{\bar{\nu}}$ are the pressure of neutrino and anti-neutrino respectively.
By using the comoving momentum $q \equiv pa $,  the above integral can be rewritten as 
\begin{eqnarray}
\label{eq:rho2}
\rho_\nu + \rho_{\bar{\nu}} 
&=& 
\frac{T_{\nu }^4}{2\pi^2 } \int_0^{\infty} y^3 dy \sqrt{1 + \left( \frac{a \tilde{m}}{y} \right)^2 }
\left( \frac{1}{e^{y + \xi} +1}  +  \frac{1}{e^{y - \xi} +1} \right),  \\
\label{eq:p2}
p_\nu + p_{\bar{\nu}} 
&=&
\frac{T_{\nu }^4}{6\pi^2 } \int_0^{\infty} y^3 dy \frac{1}{\sqrt{1 + \left( a \tilde{m} / y \right)^2 }}
\left( \frac{1}{e^{y + \xi} +1}  +  \frac{1}{e^{y - \xi} +1} \right),  
\end{eqnarray}
where we have defined $y$ and $\tilde{m}$ as 
\begin{equation}
y \equiv \frac{q}{T_{\nu 0}}, 
\qquad\qquad
\tilde{m} \equiv \frac{m}{T_{\nu 0 }}.
\end{equation}
Although generally the above integrals should be performed numerically, 
some useful approximation can be adopted, in relativistic and non-relativistic limits,
in particular, when $ | \xi | \ll \mathcal{O}(1)$.  
Below we give explicit formulas for each case.

\subsubsection[Relativistic limit]
{\underline{Relativistic limit} \cite{Kohri:2014hea}}

When 
$\frac{a \tilde{m}}{y} \ (= \frac{m}{p}) \ll 1$,
by expanding the integrand in Eqs.~\eqref{eq:rho_nu} and \eqref{eq:rho_p} 
up to the second order in 
$(a \tilde{m})/y$, 
the energy density and pressure can be written as 
\begin{eqnarray}
\label{eq:rho_r}
\rho_\nu + \rho_{\bar{\nu}} 
&\simeq& 
\frac{T_{\nu }^4}{2\pi^2 } \int_0^{\infty} y^3 dy \left( 1 + \frac12  \left( \frac{a \tilde{m}}{y} \right)^2 \right)
\left( \frac{1}{e^{y + \xi} +1}  +  \frac{1}{e^{y - \xi} +1} \right),  \\
p_\nu + p_{\bar{\nu}} 
&\simeq&
\frac{T_{\nu }^4}{6\pi^2 } \int_0^{\infty} y^3 dy \left( 1 - \frac12  \left( \frac{a \tilde{m}}{y} \right)^2 \right)
\left( \frac{1}{e^{y + \xi} +1}  +  \frac{1}{e^{y - \xi} +1} \right).
\end{eqnarray}
These integrals can be performed exactly and we obtain 
\begin{eqnarray}
\label{eq:rho_r2}
\hspace{-30pt}
\rho_\nu + \rho_{\bar{\nu}} 
&\simeq& 
 \frac{7 \pi^2 }{120} T_\nu^4 \left[ 
\left\{ 1+ \frac{30}{7} \left( \frac{\xi}{\pi} \right)^2 + \frac{15}{7} \left( \frac{\xi}{\pi} \right)^4  \right\}
+ \frac{5}{7 \pi^2} (a \tilde{m})^2 \left\{ 1+ 3 \left( \frac{\xi}{\pi} \right)^2 \right\} 
\right], 
\\
\hspace{-30pt}
p_\nu + p_{\bar{\nu}} 
&\simeq&
\frac13 \frac{7 \pi^2 }{120} T_\nu^4 \left[ 
\left\{ 1+ \frac{30}{7} \left( \frac{\xi}{\pi} \right)^2 + \frac{15}{7} \left( \frac{\xi}{\pi} \right)^4  \right\}
- \frac{5}{7 \pi^2} (a \tilde{m})^2 \left\{ 1+ 3 \left( \frac{\xi}{\pi} \right)^2 \right\} 
\right].
\end{eqnarray}

\subsubsection[Non-relativistic limit]
{\underline{Non-relativistic limit} \cite{Kohri:2014hea}}

When $\frac{a \tilde{m} }{ y } \ (= \frac{m}{p}) \gg 1$,  
we can expand Eq.~\eqref{eq:rho_nu}  around $y /(a\tilde{m}) = 0$ 
and $\xi = 0$~\footnote{
In a non-relativistic limit for any $\xi$ values, 
the exact solutions of $\rho_{\nu}+\rho_{\bar{\nu}}$ and $p_{\nu}+p_{\bar{\nu}}$
are  expressed by using polylogarithm. 
The formulas are given in Appendix~\ref{sec:app_lept2}. 
}
as
\begin{eqnarray}
\rho_\nu + \rho_{\bar{\nu}} 
&=&
\frac{T_{\nu }^4}{2\pi^2 } \int_0^{\infty} y^3 dy  
 \frac{a \tilde{m}}{y}
\sqrt{\left( \frac{y}{a \tilde{m}} \right)^2  + 1}
\left( \frac{1}{e^{y + \xi} +1}  +  \frac{1}{e^{y - \xi} +1} \right) \notag \\
&\simeq  &
\frac{T_{\nu }^4 a \tilde{m}}{2\pi^2 } \int_0^{\infty} y^2 dy  
\left[ 1 + \frac12 \left( \frac{y}{a \tilde{m}} \right)^2  \right]
\left( \frac{1}{e^{y + \xi} +1}  +  \frac{1}{e^{y - \xi} +1} \right) \notag \\
&\simeq &
\frac{T_{\nu }^4 a \tilde{m}}{2\pi^2 } \int_0^{\infty} y^2 dy  
\left[ 1 + \frac12 \left( \frac{y}{a \tilde{m}} \right)^2  \right]
\sum_{i} C_i (y) \xi^i ,
\end{eqnarray} 
where $C_i (y)$ are the coefficients for the expansion of $ \left( (e^{y+\xi} +1)^{-1} + (e^{y-\xi} +1)^{-1} \right)$ around 
$\xi =  0$. 
We note that the terms with odd power for $\xi$ do not appear.
Explicit formulas for $C_i (y)$ are given in Appendix~\ref{sec:app_lept1}. 
Having the expressions for  $C_i (y)$, we can analytically perform the integral of the form:
\begin{equation}
\int_0^\infty C_i(y) y^2 dy, 
\qquad
{\rm and}
\qquad
\int_0^\infty C_i(y) y^4 dy.
\end{equation}
By taking into account the terms up to the 10th order in $\xi$, we obtain
\begin{eqnarray}
\label{eq:rho_nr}
\rho_\nu + \rho_{\bar{\nu}}  
&\simeq&
 \frac{T_{\nu }^4}{2\pi^2 } ( a \tilde{m}) 
\left[  3 \zeta (3) + (\log 4) \xi^2  + \frac{1}{24} \xi^4  -\frac{1}{1440}  \xi^6  + \frac{1}{40320}\xi^8  - \frac{17}{14515200} \xi^{10} \right] \notag \\
&& 
+  \frac{T_{\nu }^4}{4\pi^2 }\frac{1}{ a \tilde{m}}
\left[ 45 \zeta(5)   + 18 \zeta (3) \xi^2 + (\log 4) \xi^4 + \frac{1}{60} \xi^6  - \frac{1}{6720} \xi^8 + \frac{1}{302400}\xi^{10} \right],
\notag \\
\end{eqnarray}
where $\zeta(x)$ means the Riemann zeta function.
Similar calculations also hold for the pressure, and we have, up to the 10th order in $\xi$, 
\begin{eqnarray}
\label{eq:p_nr}
p_\nu + p_{\bar{\nu}} 
&\simeq&
\frac{T_{\nu }^4}{6\pi^2 } 
 \frac{1}{a \tilde{m}}
\left[ 45 \zeta(5)   + 18 \zeta (3) \xi^2 + (\log 4) \xi^4 + \frac{1}{60} \xi^6  
- \frac{1}{6720} \xi^8 + \frac{1}{302400}\xi^{10} \right]
\notag \\
&& \hspace{-30pt}
- \frac{T_{\nu }^4 }{12\pi^2 } 
\left( \frac{1}{a \tilde{m}} \right)^3
\left[  \frac{2835 \zeta (7)}{2}  + 675 \zeta(5) \xi^2 + 45 \zeta (3) \xi^4  
+ (\log 4) \xi^6  + \frac{1}{112} \xi^8  -\frac{1}{20160} \xi^{10}\right]. 
\nonumber \\
&& \hspace{-30pt}
\end{eqnarray} 
We have checked that above formulas are accurate as $10^{-7}$ for $|\xi | < 1$ to obtain $\rho_\nu$ and $p_\nu$ 
with non-zero $\xi$. 

\subsection{Perturbation equation}

Here, we discuss the perturbation equation for massive neutrinos including the chemical potential.
By perturbing the phase-space distribution function $f_\nu $ as \cite{Ma:1995ey}
\begin{equation}
\delta f_\nu (\eta, \mbox{\boldmath $x$}, \mbox{\boldmath $p$}) 
+ \delta f_{\bar{\nu}} (\eta, \mbox{\boldmath $x$}, \mbox{\boldmath $p$}) 
= \left(\bar{f}_\nu (p)  
+  \bar{f}_{\bar{\nu}} (p) \right)\Psi_\nu (\eta, \mbox{\boldmath $x$}, \mbox{\boldmath $p$}),
\end{equation}
where $\eta$ is the conformal time, 
$\bar{f}_\nu$ and $\bar{f}_{\bar{\nu}}$ are the background distribution functions,
and $\Psi_\nu$ represents its perturbation.
%
The perturbed Boltzmann equation for $\Psi_\nu$ for the Fourier mode $\mbox{\boldmath $k$}$ 
in the synchronous gauge is given by
\begin{equation}
\tilde{\Psi}'_\nu+ i \frac{y}{\sqrt{y^2 + a^2 \tilde{m}^2}} 
(\mbox{\boldmath $k$}\cdot \hat{\mbox{\boldmath $n$}} ) \tilde{\Psi}_\nu 
+ \frac{d \ln (\bar{f}_\nu + \bar{f}_{\bar{\nu}} ) }{d \ln y} \left[  
\tilde{\eta}'_{{\rm T}} - \frac12 \left( \tilde{h}'_{{\rm L}} 
+ 6 \tilde{\eta}'_{{\rm T}} \right) (\mbox{\boldmath $k$}\cdot \hat{\mbox{\boldmath $n$}} )^2
\right] =0,
\end{equation}
where 
%
$(\cdot)'$ represents the derivative with respect to the conformal time,
$\tilde{(\cdot)}$ represents the Fourier component of the quantity
(except for $\tilde{m}\equiv m/T_{\nu 0}$),
$h_{{\rm L}}$ and  $\eta_{{\rm T}}$ are metric perturbations \cite{Ma:1995ey},
%
and $\hat{\mbox{\boldmath $n$}}$ is the direction of the momentum $\mbox{\boldmath $p$}$.

We expand $\tilde{\Psi}_\nu$ with the Legendre polynomial as 
\begin{equation}
\tilde{\Psi}_\nu (\eta, \mbox{\boldmath $k$}, \mbox{\boldmath $p$}) 
= \sum_{l=0}^{\infty} (-i)^l (2l+1)\tilde{\Psi}_{\nu l} (\eta, \mbox{\boldmath $k$}, p) P_l 
(\hat{\mbox{\boldmath $k$}}\cdot \hat{\mbox{\boldmath $n$}}),
\end{equation}
where $\hat{\mbox{\boldmath $k$}}$ is the direction of $\mbox{\boldmath $k$}$. 
The evolution equations for each multiple moment in the synchronous gauge take 
the following forms:
\begin{eqnarray}
\tilde{\Psi}'_{\nu 0} & = &  
- \frac{y k}{\sqrt{y^2 + a^2 \tilde{m}^2}} \tilde{\Psi}_{\nu 1} 
+ \frac16 \tilde{h}'_{{\rm L}} \frac{ d \ln (\bar{f}_\nu + \bar{f}_{\bar\nu})}{d \ln y}, \\ \notag \\
\tilde{\Psi}'_{\nu 1} & = &   
\frac{y k}{3 \sqrt{y^2 + a^2 \tilde{m}^2}} \left( \tilde{\Psi}_{\nu 0} 
- 2 \tilde{\Psi}_{\nu 2} \right), \\ \notag \\
\tilde{\Psi}'_{\nu 2} & = &   
\frac{y k}{5 \sqrt{y^2 + a^2 \tilde{m}^2}} \left( 2 \tilde{\Psi}_{\nu 1} - 3 \tilde{\Psi}_{\nu 3} \right)  
-\left(  \frac{1}{15} \tilde{h}'_{{\rm L}}  
+ \frac25 \tilde{\eta}'_{{\rm T}} \right) \frac{ d \ln (\bar{f}_\nu + \bar{f}_{\bar\nu})}{d \ln y}, \\ \notag \\
\tilde{\Psi}'_{\nu l} & = &   
\frac{y k}{(2l+1) \sqrt{y^2 + a^2 \tilde{m}^2}} \left( l \tilde{\Psi}_{\nu (l-1)} 
- (l+1) \tilde{\Psi}_{\nu (l+1)} \right), ~~ ({\rm for} ~l \ge 3).
\end{eqnarray}
The dependence on the chemical potential appears in the factor
 $d \ln (\bar{f}_\nu + \bar{f}_\nu)/d \ln y$, which 
can be written as \cite{Lesgourgues:1999wu}
\begin{equation}
\frac{ d \ln (\bar{f}_\nu + \bar{f}_{\bar\nu})}{d \ln y} 
= - \frac{y \left( 1+ \cosh \xi \cosh y \right)}{(\cosh \xi + \exp(-y) ) (\cosh \xi + \cosh y)}.
\label{eq:dlnfdlny}
\end{equation} 

By making the modifications given above as well as those for the background quantities to 
{\tt CAMB} \cite{Lewis:1999bs,CAMB}, 
we calculate power spectra of 21cm and CMB fluctuations and make a Fisher information analysis 
(the details of the analysis are shown in the next chapter).

\chapter
[Fisher information matrix ]
{Fisher information matrix \normalsize{\cite{Tegmark:1996bz,Arashiba:2009}}}
\label{chap:Fisher}

In this chapter, we introduce the Fisher information matrix,
which is a analysis method used to estimate 
sensitivities of experiments 
to constraints on theoretical parameters.
In this thesis, we use the analysis in order to 
estimate sensitivities of 21 cm line, CMB and BAO observations.

\section{Fisher information analysis}

\subsection{Definition of statistical quantities}

First of all, we define statistical quantities used in this chapter.
We express a data vector which is obtained by an observation as
\begin{eqnarray}
\mbox{\boldmath $x$} = (x_{1},x_{2},\cdot \cdot \cdot, x_{n}).
\end{eqnarray}
We regard the vector as stochastic variables,
and they obey the following probability density function
\begin{eqnarray}
f(\mbox{\boldmath $x$}|\mbox{\boldmath $\theta$}),
\end{eqnarray}
where
\begin{eqnarray}
\mbox{\boldmath $ \theta $} 
= (\theta_{1},\theta_{2},\cdot \cdot \cdot, \theta_{m}),
\end{eqnarray}
is a vector consisting of theoretical parameters in a model
($m$ represents the number of the parameters).
The probability density function
$f(\mbox{\boldmath $x$}|\mbox{\boldmath $ \theta $} $)
is a normalized as
\begin{eqnarray}
\int d^{n}x f(\mbox{\boldmath $x$}|\mbox{\boldmath $\theta$} )=1. \label{eq:kikakuka}
\end{eqnarray}
%
By the probability density function, 
we express the average value of a quantity as
\begin{eqnarray}
\left\langle \cdot \right\rangle
=
\int d^{n}x  ( \cdot) 
f(\mbox{\boldmath $x$}|\mbox{\boldmath $\theta$} ).
\end{eqnarray}

\subsection{Variance-covariance matrix}

We introduce the following covariance matrix 
${\rm Cov}(\mbox{\boldmath $X$},\mbox{\boldmath $Y$})$
for $\mbox{\boldmath $X$}(\mbox{\boldmath $x$}) 
=\hspace{-2pt}( X_{1}( \mbox{\boldmath $x$}),
X_{2}(\mbox{\boldmath $x$}),
\cdot \cdot \cdot
X_{p}( \mbox{\boldmath $x$}))^{T}$
and
$\mbox{\boldmath $Y$}(\mbox{\boldmath $x$}) 
= \hspace{-2pt}( Y_{1}( \mbox{\boldmath $x$} ),
Y_{2}(\mbox{\boldmath $x$}),$
$\cdot \cdot \cdot 
Y_{q}( \mbox{\boldmath $x$}))^{T}$~\footnote{
$\mbox{\boldmath $X$}$ and $\mbox{\boldmath $Y$}$ are
row vectors,
and
$(\cdot )^{T}$ means a transposed vector or a matrix.
},
%
which are $p$ or $q$ dimension stochastic variables,
respectively,
%
%
%
%
\begin{subequations}
\begin{eqnarray}
{\rm Cov}(\mbox{\boldmath $X$},\mbox{\boldmath $Y$}) 
&\equiv &
\left\langle
[\mbox{\boldmath $X$}-\left\langle \mbox{\boldmath $X$} 
\right\rangle]
[ \mbox{\boldmath $Y$}-\left\langle \mbox{\boldmath $Y$} \right\rangle ]^{T}
\right\rangle, \\
{\rm Cov}(\mbox{\boldmath $X$},\mbox{\boldmath $Y$})_{ij} 
&\equiv&
\left\langle
[X_{i}-\left\langle X_{i} \right\rangle]
[Y_{j}-\left\langle Y_{j} \right\rangle]
\right\rangle,
\end{eqnarray}
\end{subequations}
where this covariance matrix is a $p$-by-$q$ matrix.
If $\mbox{\boldmath $Y$}=\mbox{\boldmath $X$}$,
%
the matrix reduces to the following
variance-covariance matrix
\begin{subequations}
\begin{eqnarray}
V(\mbox{\boldmath $X$}) &\equiv &
\left\langle
[\mbox{\boldmath $X$}-\left\langle \mbox{\boldmath $X$} 
\right\rangle]
[ \mbox{\boldmath $X$}-\left\langle \mbox{\boldmath $X$} \right\rangle ]^{T}
\right\rangle, \\
V(\mbox{\boldmath $X$})_{ij} 
&\equiv&
\left\langle
[X_{i}-\left\langle X_{i} \right\rangle]
[X_{j}-\left\langle X_{j} \right\rangle]
\right\rangle. 
\label{eq:kyoubunsan} \\[-12pt]  \nonumber 
\end{eqnarray}
\end{subequations}
By definition, the matrix is a $p \times p$ symmetric matrix
$V(\mbox{\boldmath $X$})= (V(\mbox{\boldmath $X$}))^{T}$,
and it reduces to the variance of $\mbox{\boldmath $X$}$ when $p=1$.
Therefore, the matrix means the extension of variance
to a multi-dimensional space.
Because the matrix is a semi-positive definite matrix,
the following relation is valid 
for any vector $\mbox{\boldmath $u$}
= (u_{1},\cdot \cdot \cdot,u_{p})^{T} \in {\rm R}^{p}$,
\begin{eqnarray}
\mbox{\boldmath $u$}^{T}V(\mbox{\boldmath $X$})\mbox{\boldmath $u$}
=\sum_{i,j=1}^{p} u_{i}V(\mbox{\boldmath $X$})_{ij}u_{j} \geq 0.
\end{eqnarray}
%

\subsection{Unbiased estimator}

When an expectation value 
$\langle\hat{\theta}_{k} \rangle$
of an estimator 
$\hat{\theta}_{k}(\mbox{\boldmath $x$})$
estimated from a sample satisfies the following relation,
\begin{eqnarray}
\langle \hat{\theta}_{k} \rangle
\equiv 
\int d^{n}x \hat{\theta}_{k}
(\mbox{\boldmath $x$})
f(\mbox{\boldmath $x$}|\mbox{\boldmath $ \theta $})
=\theta_{k},
\label{eq:fuhensuiteiryou}
\end{eqnarray}
$\hat{\theta}_{k}$ is called an unbiased estimator of $\theta_{k}$.
In the Fisher information analysis,
we estimate the minimum variance limit of 
theoretical parameters by using the unbiased estimator~\footnote{
Generally, values of parameters estimated from a sample
are different from values of population's parameters.
For example, we consider that we take
samples $X_{1},\cdot \cdot \cdot X_{n}$
which obey a population.
The population has an average value $\mu$ and a variance $\sigma^{2}$.
For $X_{i}$, we calculate the expectation value of 
the following estimators.
\begin{itemize}
\item[1.] \underline{Average of the population}
\begin{eqnarray*}
\bar{X}=\frac{1}{n}\sum^{n}_{i=1} X_{i}.
\end{eqnarray*}
Since the expectation value of this quantity is
$\langle \bar{X} \rangle = \mu$,
this estimator is an unbiased estimator.
\item[2.] \underline{Variance}
\begin{eqnarray*}
\bar{\sigma}^{2}=\frac{1}{n}\sum^{n}_{i=1} (X_{i}-\bar{X})^{2}.
\end{eqnarray*}
The expectation value of this estimator is
$\left\langle \bar{\sigma}^{2} \right\rangle 
= \frac{n-1}{n}\sigma^{2} $,
this value is different from 
the variance of the population $\sigma^{2}$.
Therefore, it is not an unbiased estimator.
An unbiased estimator of the variance is
the following ``unbiased variance''
\begin{eqnarray*}
\bar{\sigma}^{2}=\frac{1}{n-1}\sum^{n}_{i=1} (X_{i}-\bar{X})^{2}.
\end{eqnarray*}
\end{itemize}}.

\subsection{Fisher information matrix}

Thus far, we regard
$f(\mbox{\boldmath $x$}|\mbox{\boldmath $ \theta $})$
as a probability density of an observed data $\mbox{\boldmath $x$}$,
and the theoretical parameters are fixed.
However, we can also regard 
$f(\mbox{\boldmath $x$}|\mbox{\boldmath $ \theta $})$
as a function of $\mbox{\boldmath $\theta $}$
when we fix the data $\mbox{\boldmath $x$}$.
%
In this case,
$f(\mbox{\boldmath $x$}|\mbox{\boldmath $ \theta $})$
means a quantity which has some information of the theoretical parameters.
%
%
From this standpoint, we rewrite
$f(\mbox{\boldmath $x$}|\mbox{\boldmath $ \theta $})$
as
$L(\mbox{\boldmath $ \theta $} | \mbox{\boldmath $x$})$,
and it is called the likelihood function.
By the likelihood function,
we can define the Fisher information matrix (or just Fisher matrix) as
\begin{subequations}
\begin{eqnarray}
F\equiv
 -\left\langle
  \frac{\partial^{2} 
  \ln L(\mbox{\boldmath $ \theta $} | \mbox{\boldmath $x$})}
       {\partial\mbox{\boldmath $ \theta $} 
       \partial \mbox{\boldmath $ \theta $}^{T}}
 \right\rangle, \\
 \nonumber \\
F_{ij} \equiv
 -\left\langle
  \frac{\partial^{2} 
  \ln L(\mbox{\boldmath $ \theta $} | \mbox{\boldmath $x$})}
       {\partial \theta_{i} \partial \theta _{j}}
 \right\rangle. \label{eq:fisherdef0} 
\end{eqnarray}
\end{subequations}
For this Fisher information matrix,
we can derive the Cram$\acute{{\rm e}}$r-Rao bound,
which makes a connection between 
the minimum variances of theoretical parameters
and the Fisher matrix.

\subsection{Cram$\acute{{\rm e}}$r-Rao bound}

Between the variance-covariance matrix of unbiased estimators 
$V(\mbox{\boldmath $\hat{\theta}$})$ and the Fisher matrix $F$,
the following inequality holds~\footnote{
Here, $(\cdot)_{ii}$ means a diagonal component of the matrix.
},
%
\begin{eqnarray}
V_{ii}(\mbox{\boldmath $\hat{\theta}$}) \geq (F^{-1})_{ii}.
\end{eqnarray}
As we see below,
this inequality is derived by the Cram$\acute{{\rm e}}$r-Rao bound.

First of all, by differentiating 
the definition of unbiased estimator Eq.(\ref{eq:fuhensuiteiryou})
with respect to a theoretical parameter $\theta_{l}$,
we obtain the following relation
\begin{eqnarray}
\int d^{n}x \hat{\theta}_{k}(\mbox{\boldmath $x$})
\frac{ \partial f(\mbox{\boldmath $x$}|\mbox{\boldmath $ \theta $})}
{\partial \theta_{l}}
=\delta_{kl}  
\ \ \longrightarrow \ \
\int d^{n}x \hat{\theta}_{k}(\mbox{\boldmath $x$})
\frac{ \partial 
\ln f(\mbox{\boldmath $x$}|\mbox{\boldmath $ \theta $})}
{\partial \theta_{l}}
(\mbox{\boldmath $x$}|\mbox{\boldmath $ \theta $})
=\delta_{kl}.
\label{eq:fuhensuiteiryou2}
\end{eqnarray}
Additionally, by differentiating the normalization condition of 
$f(\mbox{\boldmath $x$}|\mbox{\boldmath $ \theta $})$
(Eq.(\ref{eq:kikakuka}))
with respect to $\theta_{l}$,
we obtain
\begin{eqnarray}
\int d^{n}x 
\frac{ \partial f(\mbox{\boldmath $x$}|\mbox{\boldmath $ \theta $})}
{\partial \theta_{l}}
= 0
\ \ \longrightarrow \ \
\int d^{n}x 
\frac{ \partial 
\ln f(\mbox{\boldmath $x$}|\mbox{\boldmath $ \theta $})}
{\partial \theta_{l}}
f(\mbox{\boldmath $x$}|\mbox{\boldmath $ \theta $})
=0.
\label{eq:fuhensuiteiryou3}
\end{eqnarray}
Taking difference between
Eq.(\ref{eq:fuhensuiteiryou2}) and 
$ \theta_{k}\times$Eq.(\ref{eq:fuhensuiteiryou3}),
we can obtain the following relation
\begin{eqnarray}
\int d^{n}x
    \left[  
    \hat{\theta}_{k}(\mbox{\boldmath $x$}) 
    - \theta_{k}
    \right]
    \frac{ \partial 
    \ln f(\mbox{\boldmath $x$}|\mbox{\boldmath $ \theta $})}
         {\partial \theta_{l}}
    f(\mbox{\boldmath $x$}|\mbox{\boldmath $ \theta $})
    &=& \delta_{kl}, \nonumber \\
    \nonumber \\
\longrightarrow \ \
    \left\langle  
    \left[
    \hat{\theta}_{k}(\mbox{\boldmath $x$}) - \theta_{k}
    \right]
    \frac{ \partial
    \ln f(\mbox{\boldmath $x$}|\mbox{\boldmath $ \theta $})}
         {\partial \theta_{l}}    
    \right\rangle
    &=& \delta_{kl}.
    \label{eq:fuhensuiteiryou4}
\end{eqnarray}
Furthermore, Eq.(\ref{eq:fuhensuiteiryou3}) means
\begin{eqnarray}
\int d^{n}x 
\frac{ \partial 
\ln f(\mbox{\boldmath $x$}|\mbox{\boldmath $ \theta $})}
{\partial \theta_{l}}
f(\mbox{\boldmath $x$}|\mbox{\boldmath $ \theta $})
=0
\ \ \ \longleftrightarrow \ \ \
\left\langle 
\frac{ \partial 
\ln f(\mbox{\boldmath $x$}|\mbox{\boldmath $ \theta $})}
{\partial \theta_{l}} \right\rangle = 0.
\label{eq:fuhensuiteiryou5}
\end{eqnarray}
Therefore, the expectation value of
$\frac{\partial 
\ln f(\mbox{\boldmath $x$}|\mbox{\boldmath $ \theta $})}
{\partial \mbox{\boldmath $\theta $}}$ is zero.
By using Eq.(\ref{eq:fuhensuiteiryou5})
and the definition of unbiased estimator
$\theta_{k}=\langle \hat{\theta}_{k} \rangle$,
the Eq.(\ref{eq:fuhensuiteiryou4}) is rewritten as
the following covariance matrix between
$\hat{\mbox{\boldmath $\theta $}}$
and
$\frac{\partial \ln f(\mbox{\boldmath $x$}|\mbox{\boldmath $ \theta $})}
{\partial \mbox{\boldmath $\theta $}}$
\begin{eqnarray}
\hspace{-10pt}
{\rm Cov}\left(\hat{\theta}_{k},\frac{ \partial \ln f(\mbox{\boldmath $x$}|\mbox{\boldmath $ \theta $})}
{\partial \theta_{l}}    \right)_{kl}
=
\left\langle  
    \left[
    \hat{\theta}_{k} 
    - \langle \hat{\theta}_{k} \rangle
    \right]
    \left[
    \frac{ \partial 
    \ln f(\mbox{\boldmath $x$}|\mbox{\boldmath $ \theta $})}
         {\partial \theta_{l}}   
    - \left\langle 
\frac{ \partial 
\ln f(\mbox{\boldmath $x$}|\mbox{\boldmath $ \theta $})}
    {\partial \theta_{l}} \right\rangle
    \right]
    \right\rangle
    = \delta_{kl},
    \label{eq:fuhensuiteiryou6}
\end{eqnarray}
where this covariance matrix is a $m\times m$ matrix
($m$ means the number of the theoretical parameters).

Next, we introduce the following vector
consisting of $\mbox{\boldmath $\theta $}$ and
$\frac{\partial 
\ln f(\mbox{\boldmath $x$}|\mbox{\boldmath $ \theta $})}
{\partial \mbox{\boldmath $\theta $}}$,
\begin{eqnarray}
\mbox{\boldmath $A$}(\mbox{\boldmath $x$})
=\left( 
\begin{array}{c}
\mbox{\boldmath $\theta $} \\
\frac{\partial \ln f(\mbox{\boldmath $x$}|\mbox{\boldmath $ \theta $})}
{\partial \mbox{\boldmath $\theta $}} \\
\end{array} 
\right),
\end{eqnarray}
and the variance-covariance matrix of
the vector $\mbox{\boldmath $A$}(\mbox{\boldmath $x$})$
is written as
\begin{eqnarray}
V(\mbox{\boldmath $A$})
&=&\left\langle 
[\mbox{\boldmath $A$}-\left\langle \mbox{\boldmath $A$}\right\rangle]
[\mbox{\boldmath $A$} - \left\langle \mbox{\boldmath $A$} \right\rangle]^{T} \nonumber 
    \right\rangle\\
    \nonumber \\[-5pt]
    &=&\left(
\begin{array}{cc}
    \\ [-5pt]
    V(\mbox{\boldmath $\hat{\theta}$}) 
    & {\rm Cov} \left(\mbox{\boldmath $\hat{\theta}$} ,
          \frac{\partial 
          \ln f(\mbox{\boldmath $x$}|\mbox{\boldmath $ \theta $})}
          {\partial \mbox{\boldmath $\theta $}}
          \right)\\
          \\
    {\rm Cov} \left(
          \frac{\partial
           \ln f(\mbox{\boldmath $x$}|\mbox{\boldmath $ \theta $})}
          {\partial \mbox{\boldmath $\theta $}} ,
          \mbox{\boldmath $\hat{\theta}$} \right)
     & V\left(
          \frac{\partial
           \ln f(\mbox{\boldmath $x$}|\mbox{\boldmath $ \theta $})}
          {\partial \mbox{\boldmath $\theta $}} \right) \\
          \\
\end{array} 
\right), \label{eq:Anobunsan}
\end{eqnarray}
where each block is $m\times m$ matrix.
By using Eq.(\ref{eq:fuhensuiteiryou6})
and the definition of the Fisher matrix Eq.(\ref{eq:fisherdef0}),
we can express the variance-covariance matrix 
$V(\mbox{\boldmath $A$})$ as
%
\begin{eqnarray}
V(\mbox{\boldmath $A$})
     &=&\left(
    \begin{array}{cc}
    V(\mbox{\boldmath $\hat{\theta}$}) 
    & {\rm 1}_{m}\\
    {\rm 1}_{m}
    & F \\
    \end{array} 
    \right).
\end{eqnarray}
where ${\rm 1}_{m}$ is a $m \times m$ identity matrix.
Because the $V(\mbox{\boldmath $A$})$ is 
a semi-positive definite matrix,
the following inequality holds 
for any vector $\mbox{\boldmath $U$}$,
\begin{eqnarray}
0 \leq 
\mbox{\boldmath $U$}^{T}
V(\mbox{\boldmath $A$})\mbox{\boldmath $U$}.
\label{eq:seiteichi}
\end{eqnarray}
Here, we introduce any two vectors
$\mbox{\boldmath $u$},
\mbox{\boldmath $v$} \in {\rm R}^{m}$,
and express the vector
$\mbox{\boldmath $U$}$ as
\begin{eqnarray}
\mbox{\boldmath $U$}
=\left( 
\begin{array}{c}
\mbox{\boldmath $u$} \\
\mbox{\boldmath $v$} \\
\end{array} 
\right).
\end{eqnarray}
From this expression,
the inequality Eq.(\ref{eq:seiteichi}) can be expressed as
\begin{eqnarray}
0\leq
\mbox{\boldmath $u$}^{T}
V(\mbox{\boldmath $\hat{\theta}$})\mbox{\boldmath $u$} 
+ \mbox{\boldmath $u$}^{T}\mbox{\boldmath $v$} 
+ \mbox{\boldmath $v$}^{T}\mbox{\boldmath $u$} 
+ \mbox{\boldmath $v$}^{T}F\mbox{\boldmath $v$}. 
\label{eq:seiteichi2}
\end{eqnarray}
Because the Fisher matrix is symmetric matrix,
the inequality Eq.(\ref{eq:seiteichi2}) can be rewritten as
\begin{eqnarray}
0 \leq
\hspace{-1pt}\mbox{\boldmath $u$}^{T}
\left(
V(\mbox{\boldmath $\hat{\theta}$})
-F^{-1}
\right)
\mbox{\boldmath $u$} 
+
\left(
\mbox{\boldmath $v$} +F^{-1}\mbox{\boldmath $u$} 
\right)^{T}
F
\left(
\mbox{\boldmath $v$} +F^{-1}\mbox{\boldmath $u$} 
\right),
\label{eq:kuramerurao}
\end{eqnarray}
where we use that the Fisher matrix has 
an inverse matrix~\footnote{
$F$ reduces to the following variance-covariance matrix of 
$\frac{ \partial \ln f(\mbox{\boldmath $x$}|\mbox{\boldmath $ \theta $})}
{\partial \mbox{\boldmath $\theta $}} $,
\begin{eqnarray}
F&=&\left\langle
  \frac{ \partial \ln f(\mbox{\boldmath $x$}|\mbox{\boldmath $ \theta $})}
  {\partial \mbox{\boldmath $\theta$}} 
  \frac{ \partial \ln f(\mbox{\boldmath $x$}|\mbox{\boldmath $ \theta $})}
  {\partial \mbox{\boldmath $\theta$}^{T}} 
  \right\rangle \nonumber \\
  &=&
  \left\langle 
      \left[
      \frac{ \partial 
      \ln f(\mbox{\boldmath $x$}|\mbox{\boldmath $ \theta $})}
      {\partial \mbox{\boldmath $\theta$}} 
      -
      \left\langle 
        \frac{ \partial 
        \ln f(\mbox{\boldmath $x$}|\mbox{\boldmath $ \theta $})}
        {\partial \mbox{\boldmath $\theta$}} \right\rangle
      \right]
           \left[
      \frac{ \partial 
      \ln f(\mbox{\boldmath $x$}|\mbox{\boldmath $ \theta $})}
      {\partial \mbox{\boldmath $\theta$}^{T}} 
      -
      \left\langle 
        \frac{ \partial 
        \ln f(\mbox{\boldmath $x$}|\mbox{\boldmath $ \theta $})}
        {\partial \mbox{\boldmath $\theta$}^{T}} \right\rangle
      \right] \right\rangle  \nonumber \\
      &=&
      V\left(
       \frac{ \partial 
       \ln f(\mbox{\boldmath $x$}|\mbox{\boldmath $ \theta $})}
       {\partial \mbox{\boldmath $\theta$}} 
       \right). \nonumber 
\end{eqnarray}
Therefore, $F$ is a semi-positive-definite matrix 
and has an inverse matrix.
}.
%

Since $F$ is a semi-positive definite matrix,
the left hand side of (\ref{eq:kuramerurao})
take the minimum value when
\begin{eqnarray}
\mbox{\boldmath $v$} +F^{-1}\mbox{\boldmath $u$} 
= 0.
\end{eqnarray}
Therefore, we can obtain the following inequality,
\begin{eqnarray}
0 \leq
\mbox{\boldmath $u$}^{T}
\left(
V(\mbox{\boldmath $\hat{\theta}$})
-F^{-1}
\right)
\mbox{\boldmath $u$},
\label{eq:kuramerurao2}
\end{eqnarray}
this inequality is called 
the Cram$\acute{{\rm e}}$r-Rao bound (or inequality).
%
When we choose $\mbox{\boldmath $u$}=\mbox{\boldmath $e$}_{i}$,
where $\mbox{\boldmath $e$}_{i}$ is an unit vector
pointing in the $i$th direction,
the Cram$\acute{{\rm e}}$r-Rao bound reduces to
\begin{eqnarray}
(F^{-1})_{ii}
\leq
V(\mbox{\boldmath $\hat{\theta}$})_{ii}
\ \ \ \ \ (1 \leq i \leq m), 
\label{eq:variancefisher}
\end{eqnarray}
where $V(\mbox{\boldmath $\hat{\theta}$})_{ii}$ 
represents the variance of the estimated theoretical parameter 
$\hat{\theta}_{i}(\mbox{\boldmath $x$})$.
This inequality  means
the limit of the minimum variance of $\hat{\theta}_{i}$.
When the equality holds, the unbiased estimator is called 
an unbiased efficient estimator.
As stated above, 
we can estimate the minimum variance
of theoretical parameters
by calculating the inverse of the Fisher information matrix.

\section{Fisher information matrix for Gaussian likelihood}

When we estimate the sensitivity of observations
of 21 cm line, CMB and BAO,
we assume that the likelihood function is 
the multi-dimension Gaussian distribution function,
\begin{subequations}
\begin{eqnarray}
L(\mbox{\boldmath $\mu$},C|\mbox{\boldmath $x$})
&=& \frac{1}
{(2\pi)^{\frac{n}{2}}\sqrt{\det C}}
\exp\left[
- \frac{1}{2} (\mbox{\boldmath $x$}-\mbox{\boldmath $\mu$})^{T} 
C^{-1}(\mbox{\boldmath $x$}-\mbox{\boldmath $\mu$})
\right], \\
\mbox{\boldmath $\mu$} 
&\equiv & \left\langle \mbox{\boldmath $x$} \right\rangle,
\label{eq:xkitaiti}\\
C & \equiv &   
\left\langle( \mbox{\boldmath $x$}-\mbox{\boldmath $\mu$} )
( \mbox{\boldmath $x$}-\mbox{\boldmath $\mu$} )^{T}
\right\rangle.
\label{eq:likelihoodC}
\end{eqnarray}
\end{subequations}
In this section, we calculate the Fisher matrix
of the Gaussian likelihood.
When we choose the Gaussian likelihood,
the theoretical parameters
$\mbox{\boldmath $\theta $}$
are related to the likelihood 
through the expectation value $\mbox{\boldmath $\mu $}$
and the variance-covariance matrix $C$.
%

First of all, we can express
its log likelihood $\mathcal{L}\equiv -\ln L$ as
\begin{eqnarray}
\mathcal{L}
 = -\ln L = 
 {\frac{n}{2}}\ln 2\pi  
 +\frac{1}{2}\ln \det C
 +\frac{1}{2} (\mbox{\boldmath $x$}-\mbox{\boldmath $\mu$})^{T}
 C^{-1}(\mbox{\boldmath $x$}-\mbox{\boldmath $\mu$}). \label{eq:loglikelihood}
\end{eqnarray}
By using $\ln \det C = {\rm Tr}[\ln C]$ and
the following cyclic behavior of trace ${\rm Tr}[\ \cdot \ ]$
\begin{eqnarray}
\hspace{-10pt}
(\mbox{\boldmath $x$}-\mbox{\boldmath $\mu$})^{T}
C^{-1}(\mbox{\boldmath $x$}-\mbox{\boldmath $\mu$})
= {\rm Tr}\left[ 
(\mbox{\boldmath $x$}-\mbox{\boldmath $\mu$})^{T}
C^{-1}(\mbox{\boldmath $x$}-\mbox{\boldmath $\mu$})
\right]
=
{\rm Tr}\left[
 C^{-1} (\mbox{\boldmath $x$}-\mbox{\boldmath $\mu$})
(\mbox{\boldmath $x$}-\mbox{\boldmath $\mu$})^{T}
\right],
\end{eqnarray}
the log likelihood Eq.(\ref{eq:loglikelihood}) can be written as
\begin{eqnarray}
2\mathcal{L} &=& n\ln 2\pi + {\rm Tr} [\ln C + 
C^{-1}D]  \label{eq:loglikelihood1}, \\
D &\equiv & (\mbox{\boldmath $x$}-\mbox{\boldmath $\mu$})
(\mbox{\boldmath $x$}-\mbox{\boldmath $\mu$})^{T}.
\label{eq:loglikelihood2}
\end{eqnarray}

Next, we differentiate Eq.(\ref{eq:loglikelihood1})
with respect to the parameter $\theta_{i}$~\footnote{
Here, derivatives with respect to
the parameter $\theta_{i}$ are written as
$\frac{\partial}{\partial \theta_{i}} = _{ \ ,i}$.
}.
%
By using $(\ln C)_{,i} = C^{-1}C$
and $(C^{-1})_{,i} = -C^{-1}C_{,i}C^{-1}$,
the derivative is given by
\begin{eqnarray}
2\mathcal{L}_{,i} &=& 
{\rm Tr} [ \ln C + C^{-1}D ]_{,i}\nonumber \\
&=&
{\rm Tr} [ \ln C_{,i} + (C^{-1})_{,i}D + C^{-1}D_{,i} ] \nonumber \\
&=&
{\rm Tr} [ C^{-1} C_{,i} - C^{-1}C_{,i}C^{-1}D + C^{-1}D_{,i} ]. \label{eq:likelihoodbibun}
\end{eqnarray}
Furthermore, by differentiating Eq.(\ref{eq:likelihoodbibun})
with respect to $\theta_{j}$, we obtain
\begin{eqnarray}
2\mathcal{L}_{,ij} &=& 
{\rm Tr} [C^{-1} C_{,i} 
- C^{-1}C_{,i}C^{-1}D + C^{-1}D_{,i} \ ] _{,j} \nonumber \\
&=&
{\rm Tr} \left[
(C^{-1})_{j} C_{,i} + C^{-1} C_{,ij} 
\right. \nonumber\\
&& \ \ \ \ \left.
- (C^{-1})_{,j}C_{,i}C^{-1}D 
- C^{-1}C_{,ij}C^{-1}D - C^{-1}C_{,i}(C^{-1})_{,j}D
- C^{-1}C_{,i}C^{-1}D_{,j} 
\right. \nonumber \\
&& \ \ \ \ \left.
+ (C^{-1})_{,j}D_{,i} \ 
+ C^{-1}D_{,ij}  
\right] \nonumber \\
&=&
{\rm Tr} \left[
-C^{-1}C_{,j}C^{-1} C_{,i} + C^{-1} C_{,ij} 
\right. \nonumber\\
&& \ \ \ \ \left.
+ C^{-1}C_{,j}C^{-1} C_{,i}C^{-1}D 
- C^{-1}C_{,ij}C^{-1}D 
\right. \nonumber \\
&& \ \ \ \ \left.
+ C^{-1}C_{,i}C^{-1}C_{,j}C^{-1}D
- C^{-1}C_{,i}C^{-1}D_{,j} 
\right. \nonumber \\
&& \ \ \ \ \left.
- C^{-1}C_{,j}C^{-1}D_{,i} \ 
+ C^{-1}D_{,ij}  
\right]. \label{eq:logLbibunn1}
\end{eqnarray}
The Fisher matrix is given by
the expectation value of Eq.(\ref{eq:logLbibunn1}).

Since $C$ has already been expectation value,
it is only necessary to calculate 
the expectation values related to $D$.
The expectation values of $D$,$D_{,i}$ and $D_{,ij}$
are given by
\begin{eqnarray}
\left\langle
D
\right\rangle 
&=&
\left\langle
(\mbox{\boldmath $x$}-\mbox{\boldmath $\mu$})
(\mbox{\boldmath $x$}-\mbox{\boldmath $\mu$})^{T}
\right\rangle = C, \\
\nonumber 
\left\langle
D_{,i}
\right\rangle 
&=&
\left\langle
[ (\mbox{\boldmath $x$}-\mbox{\boldmath $\mu$})
(\mbox{\boldmath $x$}-\mbox{\boldmath $\mu$})^{T}]_{,i}
\right\rangle  \nonumber \\
&=&
-\left\langle
\mbox{\boldmath $\mu$}_{,i} (\mbox{\boldmath $x$}-\mbox{\boldmath $\mu$})^{T} 
\right\rangle 
- \left\langle
(\mbox{\boldmath $x$}-\mbox{\boldmath $\mu$})\mbox{\boldmath $\mu$}_{,i}^{T}  
\right\rangle    \nonumber \\
&=&
-\mbox{\boldmath $\mu$}_{,i}
\left\langle
(\mbox{\boldmath $x$}-\mbox{\boldmath $\mu$})^{T}
\right\rangle 
- \left\langle
(\mbox{\boldmath $x$}-\mbox{\boldmath $\mu$})
\right\rangle
\mbox{\boldmath $\mu$}_{,i}^{T} = 0,  \label{eq:Dkitaiti2} \\
\nonumber 
\left\langle
D_{,ij}
\right\rangle 
&=&
\left\langle
[ (\mbox{\boldmath $x$}-\mbox{\boldmath $\mu$})
(\mbox{\boldmath $x$}-\mbox{\boldmath $\mu$})^{T} ]_{,ij}
\right\rangle    \nonumber \\
&=&
-\left\langle
\mbox{\boldmath $\mu$}_{,ij}(\mbox{\boldmath $x$}-\mbox{\boldmath $\mu$})^{T}
\right\rangle 
+\left\langle
\mbox{\boldmath $\mu$}_{,i}\mbox{\boldmath $\mu$}_{,j}^{T}
\right\rangle 
\nonumber \\
&&
+ \left\langle
\mbox{\boldmath $\mu$}_{,j}\mbox{\boldmath $\mu$}_{,i}^{T}  
\right\rangle    
- \left\langle
(\mbox{\boldmath $x$}-\mbox{\boldmath $\mu$})\mbox{\boldmath $\mu$}_{,ij}^{T}  
\right\rangle \nonumber \\
&=&
-\mbox{\boldmath $\mu$}_{,ij}
\left\langle
(\mbox{\boldmath $x$}-\mbox{\boldmath $\mu$})^{T}
\right\rangle 
+\mbox{\boldmath $\mu$}_{,i}\mbox{\boldmath $\mu$}_{,j}^{T}
\nonumber \\
&&
+ \mbox{\boldmath $\mu$}_{,j}\mbox{\boldmath $\mu$}_{,i}^{T}     
- \left\langle   
(\mbox{\boldmath $x$}-\mbox{\boldmath $\mu$})
\right\rangle     
\mbox{\boldmath $\mu$}_{,ij}^{T}
\nonumber \\
&=&
\mbox{\boldmath $\mu$}_{,i}\mbox{\boldmath $\mu$}_{,j}^{T}
+
\mbox{\boldmath $\mu$}_{,j}\mbox{\boldmath $\mu$}_{,i}^{T}, 
\label{eq:Dkitaiti3}
\end{eqnarray}
where we use 
$\left\langle (\mbox{\boldmath $x$}-\mbox{\boldmath $\mu $} ) 
\right\rangle =0$ in Eqs.(\ref{eq:Dkitaiti2}) 
and (\ref{eq:Dkitaiti3})
because 
$\mbox{\boldmath $\mu $}$ has already been 
the expectation value of $\mbox{\boldmath $x $}$.
By using the expectation values of
$D$,$D_{,i}$ and $D_{,ij}$,
we obtain the following 
expectation value of Eq.(\ref{eq:logLbibunn1}),
\begin{eqnarray}
\left\langle
2\mathcal{L}_{,ij} 
\right\rangle 
&=&
{\rm Tr} \left[
-C^{-1}C_{,j}C^{-1} C_{,i} + C^{-1} C_{,ij} 
\right. \nonumber\\
&& \ \ \ \ \left.
+ C^{-1}C_{,j}C^{-1} C_{,i}C^{-1}
\left\langle D  \right\rangle
- C^{-1}C_{,ij}C^{-1}
\left\langle D \right\rangle
\right. \nonumber \\
&& \ \ \ \ \left.
+ C^{-1}C_{,i}C^{-1}C_{,j}C^{-1}
\left\langle D \right\rangle
- C^{-1}C_{,i}C^{-1}
  \left\langle D_{,j} \right\rangle 
\right. \nonumber \\
&& \ \ \ \ \left.
- C^{-1}C_{,j}C^{-1}                             
\left\langle D_{,j} \right\rangle 
+ C^{-1}
\left\langle D_{,ij} \right\rangle  
\right] \nonumber \\
&=&
{\rm Tr} \left[
-C^{-1}C_{,j}C^{-1} C_{,i} + C^{-1} C_{,ij} 
\right. \nonumber\\
&& \ \ \ \ \left.
+ C^{-1}C_{,j}C^{-1} C_{,i}C^{-1}C 
- C^{-1}C_{,ij}C^{-1}C
\right. \nonumber \\
&& \ \ \ \ \left.
+ C^{-1}C_{,i}C^{-1}C_{,j}C^{-1}C
\right. \nonumber \\
&& \ \ \ \ \left.
+ C^{-1}
(\mbox{\boldmath $\mu$}_{,i} \mbox{\boldmath $\mu$}_{,j}^{T}
+
\mbox{\boldmath $\mu$}_{,j} \mbox{\boldmath $\mu$}_{,i}^{T}  )
\right] \nonumber \\
&=&
{\rm Tr} \left[ C^{-1}C_{,i}C^{-1}C_{,j}
+ C^{-1}
(\mbox{\boldmath $\mu$}_{,i}\mbox{\boldmath $\mu$}_{,j}^{T}
+
\mbox{\boldmath $\mu$}_{,j} \mbox{\boldmath $\mu$}_{,i}^{T}  )
\right]. \label{eq:likelihoodtrace}
\end{eqnarray}
According to the properties of trace
${\rm Tr}[AB]={\rm Tr}[BA],
\ {\rm Tr}[A]={\rm Tr}[A^{T}]$
and the symmetric behavior of 
the variance-covariance matrix $C=C^{T}$,
the second term of Eq.(\ref{eq:likelihoodtrace})
reduces to
\begin{eqnarray}
{\rm Tr} \left[
C^{-1}
(\mbox{\boldmath $\mu$}_{,i} \mbox{\boldmath $\mu$}_{,j}^{T}
+
\mbox{\boldmath $\mu$}_{,j} \mbox{\boldmath $\mu$}_{,i}^{T}  )
\right]
&=&
{\rm Tr} \left[
( C^{-1}\mbox{\boldmath $\mu$}_{,i}\mbox{\boldmath $\mu$}_{,j}^{T})^{T}
+
C^{-1}\mbox{\boldmath $\mu$}_{,j}\mbox{\boldmath $\mu$}_{,i}^{T}  
\right] \nonumber \\
&=&
{\rm Tr} \left[
\mbox{\boldmath $\mu$}_{,j}\mbox{\boldmath $\mu$}_{,i}^{T} (C^{-1})^{T}
+
C^{-1}\mbox{\boldmath $\mu$}_{,j}\mbox{\boldmath $\mu$}_{,i}^{T}
\right] \nonumber \\
&=&
{\rm Tr} \left[    
\mbox{\boldmath $\mu$}_{,i}^{T} C^{-1}\mbox{\boldmath $\mu$}_{,j}
+
\mbox{\boldmath $\mu$}_{,i}^{T}  C^{-1}\mbox{\boldmath $\mu$}_{,j} 
\right] \nonumber \\
&=&
2\mbox{\boldmath $\mu$}_{,i}^{T} C^{-1}\mbox{\boldmath $\mu$}_{,j}.
\end{eqnarray}
Therefore, Eq.(\ref{eq:likelihoodtrace})
can be rewritten as
\begin{eqnarray}
2\left\langle
\mathcal{L}_{,ij} 
\right\rangle  
&=&
{\rm Tr} \left[C^{-1}C_{,i}C^{-1}C_{,j}\right]
+2\mbox{\boldmath $\mu$}_{,i}^{T}
C^{-1}\mbox{\boldmath $\mu$}_{,j}.
\label{eq:likelihoodtrace2}
\end{eqnarray}
Since the definition of the Fisher matrix is 
$F \equiv \left\langle \mathcal{L}_{,ij} \right\rangle$,
we can obtain the following 
Fisher matrix of the Gaussian likelihood,
\begin{eqnarray}
\\[-10pt]\nonumber 
F= &=&
\frac{1}{2}{\rm Tr} \left[C^{-1}C_{,i}C^{-1}C_{,j}\right]
+\mbox{\boldmath $\mu$}_{,i}^{T} C^{-1}\mbox{\boldmath $\mu$}_{,j}.
\label{eq:likelihoodtrace3}\\[-10pt]\nonumber 
\end{eqnarray}
In the analysis of this thesis,
we use this Fisher matrix formula.

\chapter{Fisher information matrix of 21 cm line observation}
\label{chap:21cmFisher}

In this chapter, we calculate the Fisher matrix of 21 cm line observations.
At first, we introduce visibility,
which is the observed quantity of 21 cm line observations.
Next, we treat the visibility as observed data
and calculate the formula of the Fisher matrix. 

\section[Visibility]
{Visibility\normalsize{\cite{Furlanetto:2006jb,text:Nakai,Arashiba:2009}}}

\subsection[Definition of visibility]{Definition of visibility}

In 21 cm line observations,
signals are observed by a radio interferometer.
When radio waves arrive at antennae of an interferometer,
voltage is generated and pairs of the antennae output 
cross-correlations of the voltages.
The cross-correlations are called visibility.

\begin{figure}[ht]
\begin{center}
\includegraphics[bb=  0 50 809 427, width=1\linewidth]{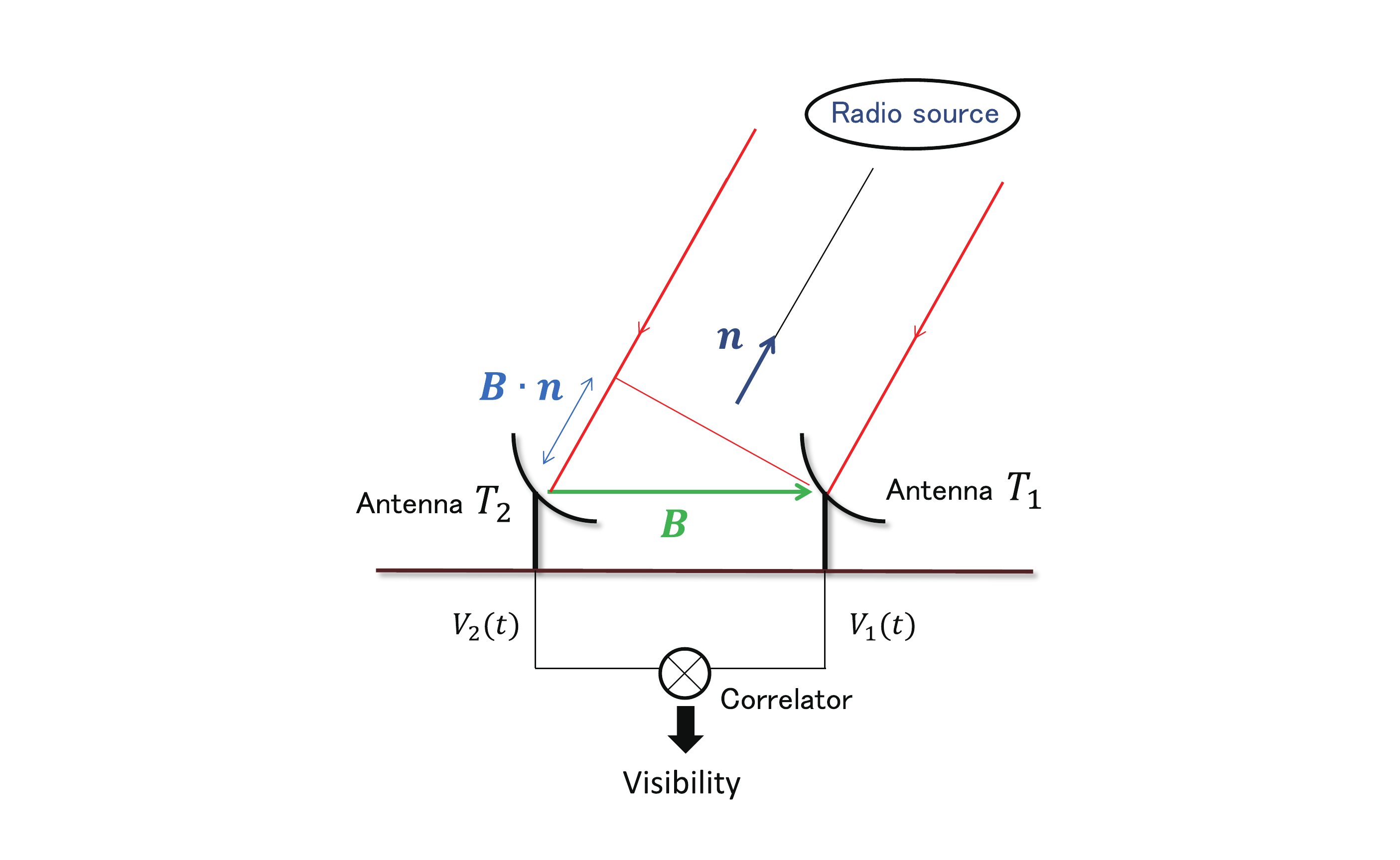} 
\hspace{10pt}\caption{An antenna pair of an interferometer} 
\label{fig:denpa}
\end{center}
\end{figure}

Here, we consider a pair of antennae $T_{1}$ and $T_{2}$ 
(Fig.\ref{fig:denpa}), and express generated voltages of each antenna as
$V_{1}(t)$ and $V_{2}(t)$, respectively.
The generated voltage $V_{1}(t)$ of antenna $T_{1}$ is given by
\begin{eqnarray}
V_{1}(t) = V_{0}e^{-2\pi i\nu t},
\end{eqnarray}
where $\nu$ is the frequency of the radio wave, 
$t$ is the time at the Earth, $V_{0}$ is as amplitude of the voltage.
On the other hand, 
the radio wave which has the same phase
arrives at the antenna $T_{2}$ late.
Therefore, the (geometric) time delay $\tau_{g}$ is given by
\begin{eqnarray}
\tau_{g} = 
\frac{
\mbox{\boldmath $B$} \cdot
\mbox{\boldmath $n$}
}{c},
\end{eqnarray}
where
$\mbox{\boldmath $B$}$ is the baseline vector
between the pair of the two antennae,
and $\mbox{\boldmath $n$}$ is a unit vector 
pointing to the direction of the radio source.
By the time delay $\tau_{g}$,
the generated voltage of $T_{2}$ is expressed as
\begin{eqnarray}
V_{2}(t) = V_{0}e^{-2\pi i\nu (t-\tau_{g})}.
\end{eqnarray}
The pair of the antennae outputs a cross-correlation function
$C_{12}(\tau)$ between $V_{1}$ and $V_{2}$.
Generally, the cross-correlation function
between some quantities $A$ and $B$ is  defined as
\begin{eqnarray}
C_{AB}(\tau)\equiv \lim_{T\rightarrow \infty}\frac{1}{2T}\int^{T}_{-T}A(t)B^{*}(t-\tau)dt.
\end{eqnarray}
By this definition, the cross correlation $C_{12}(\tau)$ of 
$V_{1}$ and $V_{2}$ is given by
\begin{eqnarray}
C_{12}(\tau) & \equiv& \lim_{T\rightarrow \infty}\frac{1}{2T}\int^{T}_{-T}V_{1}(t)V_{2}^{*}(t-\tau)dt 
      \nonumber \\
      &=& \lim_{T\rightarrow \infty}\frac{1}{2T}
          \int^{T}_{-T}V_{0}e^{-2\pi i\nu t}V_{0}e^{2\pi i \nu (t-\tau_{g}-\tau)}dt 
      \nonumber \\
      &=& \lim_{T\rightarrow \infty}\frac{1}{2T}
          \int^{T}_{-T}V_{0}^{2}e^{-2\pi i \nu (\tau_{g}+\tau)}dt 
      \nonumber \\
      &=& 
      V_{0}^{2}e^{-2\pi i \nu (\tau_{g}+\tau)}, \label{eq:sougosoukan}
\end{eqnarray}
where the coefficient $V_{0}^{2}$ 
is proportional to the power of the radio wave
$\epsilon(\mbox{\boldmath $n$})$ ($=$ the energy per unit time).
The power of the radio wave $\epsilon(\mbox{\boldmath $n$})$
can be written as
\begin{eqnarray}
\epsilon(\mbox{\boldmath $n$}) 
    = A_{\nu}(\mbox{\boldmath $n$})I_{\nu}(\mbox{\boldmath $n$})
      d\nu    \Omega_{\mbox{\boldmath $n$}},                  
\end{eqnarray}
where $I_{\nu}(\mbox{\boldmath $n$})$ is 
the specific intensity of the radio source,
$d\Omega_{\mbox{\boldmath $n$}}$ is the solid angle
and $A_{\nu}(\mbox{\boldmath $n$})$ is 
the effective area of the antenna.
Since $V^{2}_{0}\propto \epsilon(\mbox{\boldmath $n$}) $,
we introduce the following quantity,
\begin{eqnarray}
dR(\tau;\mbox{\boldmath $n$} ,\mbox{\boldmath $B$}, \nu) \equiv
      A_{\nu}(\mbox{\boldmath $n$})I_{\nu}(\mbox{\boldmath $n$})
      d\nu    d\Omega_{\mbox{\boldmath $n$}}   
      e^{-2\pi i\nu (\tau_{g}+\tau)}.
\end{eqnarray}
We integrate this quantity $dR$ which comes from various directions, 
and obtain
\begin{eqnarray}
R(\tau;\mbox{\boldmath $B$}, \nu) &=&
      \int_{\Omega_{{\rm source}}}
      A_{\nu}(\mbox{\boldmath $n$})I_{\nu}(\mbox{\boldmath $n$})
      d\nu      e^{-2\pi i\nu (\tau_{g}+\tau)} d\Omega_{\mbox{\boldmath $n$}}  \nonumber \\    
      &=& 
      e^{-2\pi i\nu \tau} d\nu   
      \int_{\Omega_{{\rm source}}}
      A_{\nu}(\mbox{\boldmath $n$})I_{\nu}(\mbox{\boldmath $n$})
      e^{-2\pi i\nu\frac{
\mbox{\boldmath $B$} \cdot
\mbox{\boldmath $n$}
}
{c}   }d\Omega_{\mbox{\boldmath $n$}}, \label{eq:sougosoukan3}
\end{eqnarray}
where $\Omega_{{\rm  source}}$ is the total solid angle of the radio source.
By the integral of the second line of Eq.(\ref{eq:sougosoukan3}),
we define the visibility $V(\mbox{\boldmath $B$}, \nu)$ as
\begin{eqnarray}
V(\mbox{\boldmath $B$}, \nu)      
      \equiv
      \int_{\Omega_{{\rm source}}}
      A_{\nu}(\mbox{\boldmath $n$})I_{\nu}(\mbox{\boldmath $n$})
      e^{-2\pi i \nu\frac{
      \mbox{\boldmath $B$} \cdot
\mbox{\boldmath $n$}
}{c}} d\Omega_{\mbox{\boldmath $n$}}. 
\label{eq:visibility}
\end{eqnarray}
In an observation by an interferometer,
we can get the original specific intensity 
$I_{\nu}(\mbox{\boldmath $n$})$
by measuring visibilities of various base lines $\mbox{\boldmath $B$}$.
From now on,
instead of the baseline vector $\mbox{\boldmath $B$}$, 
we introduce the following vector 
$\mbox{\boldmath $u$}_{B}$, which is defined as
\begin{eqnarray}
\mbox{\boldmath $u$}_{B} = (u,v,w)
   \equiv
   \frac{\nu}{c} \mbox{\boldmath $B$}.
\end{eqnarray}
%
and we can rewrite the visibility as
\begin{eqnarray}
V(\mbox{\boldmath $u$}_{B}}, \nu)      
      =
      \int_{\Omega_{{\rm source}}}
      A_{\nu}(\mbox{\boldmath $n$})I_{\nu}(\mbox{\boldmath $n$})
       e^{-2\pi i\mbox{\boldmath $u$}_{B} \cdot 
      \mbox{\boldmath $n$}}
      d\Omega_{\mbox{\boldmath $n$}.  \label{eq:visibility2}
\end{eqnarray}
%

\subsection{
Visibility for a narrow radio source 
}

By polar coordinate
$\mbox{\boldmath $n$}$
$=(\sin\theta \cos\phi,\sin\theta\sin\phi,\cos\theta)$
$\longrightarrow d\Omega_{\mbox{\boldmath $n$}}=\sin \theta d\theta d\phi$,
the visibility can be expressed as
\begin{eqnarray}
V(u,v,w, \nu )      
      &=&
      \int_{\Omega_{{\rm source}}}
      \sin \theta d\theta d\phi
      A_{\nu}(\theta,\phi)I_{\nu}(\theta,\phi)
       \nonumber \\    
       &&\ \ \ \ \ \ \ \ \  \ \ \ 
       \times 
       \exp [-2\pi i( u\sin\theta \cos\phi + v\sin\theta \sin\phi + w\cos\theta)].
       \label{eq:visibility4}
\end{eqnarray}
Additionally, by doing the following transformation of the variables,
\begin{eqnarray}
&&\left\{
\begin{array}{c}
\xi = \sin\theta \cos\phi, \\
\eta = \sin\theta \sin\phi, \\
\end{array}
\right.
\\
&&\longrightarrow \ \
d\theta d\phi = \frac{d\xi d\eta}{\sqrt[]{(\xi^{2}+\eta^{2})(1-\xi^{2}-\eta^{2})}},
\end{eqnarray}
the visibility is rewritten as
\begin{eqnarray}
V(u,v,w, \nu )      
      &=&
      \int_{\Omega_{{\rm source}}}
       \frac{d\xi d\eta}{\sqrt[]{1-\xi^{2}-\eta^{2}}}
      A_{\nu}(\xi,\eta)I_{\nu}(\xi,\eta)
       \nonumber \\    
       &&\ \ \ \ \ \ \ \ \  \ \ \ 
       \times 
       \exp [-2\pi i ( u\xi + v\eta + w \ \sqrt[]{1-\xi^{2}-\eta^{2}})].
       \label{eq:visibility5}
\end{eqnarray}
When the region of a radio source is sufficiently narrow 
and the source exists near only 
the direction $\mbox{\boldmath $n$}=(0,0,1)$,
we can use an approximations of
$|\theta|<<1 \  \longrightarrow \ |\xi|$
and
$|\eta|<<1 \ \longrightarrow 
\ \sqrt[]{1-\xi^{2}-\eta^{2}} \approx 1$.
%
In this case, the visibility becomes 
\begin{eqnarray}
V(u,v,w, \nu )      
      &\approx&
      \int_{\Omega_{{\rm source}}}
      d\xi d\eta
      A_{\nu}(\xi,\eta)I_{\nu}(\xi,\eta)
       \exp [-2\pi i( u\xi + v\eta + w )] \nonumber \\
      &=&
      e^{-2\pi i  w }
      \int_{-\infty}^{\infty}
      d\xi 
      \int_{-\infty}^{\infty} 
       d\eta
      A_{\nu}(\xi,\eta)I_{\nu}(\xi,\eta)
       \exp [-2\pi i ( u\xi + v\eta )],
       \label{eq:visibility6}
\end{eqnarray}
where we can take the integration range as
$-\infty <\xi< \infty$ and $-\infty <\eta< \infty$
because the effective area $A_{\nu}(\xi,\eta)$ 
of an antenna is zero 
outside of the region where the radio source exits.
Hence, $A_{\nu}(\xi,\eta)$  means a window function.
Since $|\theta|<<1$, we can use the following approximations,
\begin{subequations}
\begin{eqnarray}
\xi &=& \sin\theta\cos\phi \approx \theta\cos\phi\equiv \theta^{1},\\
\eta &=& \sin\theta\sin\phi \approx \theta\sin\phi \equiv \theta^{2}.
\end{eqnarray}
\end{subequations}
By using these approximations, we can write the visibility as 
\begin{eqnarray}
\hspace{-20pt}
V(u,v,w, \nu )      
      &\approx&
       e^{-2\pi i  w }
      \int_{-\infty}^{\infty}
      d\theta^{1}
      \int_{-\infty}^{\infty} 
       d\theta^{2}
      A_{\nu}(\theta^{1},\theta^{2}) I_{\nu}(\theta^{1},\theta^{2})
       \exp [-2\pi i ( u\theta^{1} + v\theta^{2} )],
       \label{eq:visibility7}
\end{eqnarray}
where $\theta^{1}$ and $\theta^{2}$
means visual angle of the radio source,
and their directions are perpendicular to the LOS. 
From Eq.(\ref{eq:visibility7}),
we find that the visibility $V(u,v,w,\nu)$ is 
the Fourier transformation of the specific intensity 
$I_{\nu}(\theta^{1},\theta^{2})$ of the radio source 
multiplied by the window function 
$A_{\nu}(\theta^{1},\theta^{2})$.
%
%
Therefore, by using the inverse Fourier transformation,
we can get the original specific intensity from the visibility.
Form now on, we set $w=0$ and omit $e^{-2\pi i  w }$
from Eq.(\ref{eq:visibility7}).

By substituting Eq.(\ref{eq:visibility7})
into Eq.(\ref{eq:sougosoukan3})
and integrating it with respect to the frequency,
we can define the following quantity,
\begin{eqnarray}
S(u,v,\tau)      
      &\equiv&
      \int e^{-2\pi i \nu\tau }
       d\nu
      V(u,v,w=0,\nu)
       \nonumber\\
       &=&
       \int_{-\infty}^{\infty} e^{-2\pi i \nu\tau }
       d\nu F_{\mbox{\boldmath $ \theta $}}(\nu)
      \int_{-\infty}^{\infty}
      d\theta^{1}
      \int_{-\infty}^{\infty} 
       d\theta^{2}
      A_{\nu}(\theta^{1},\theta^{2}) I_{\nu}(\theta^{1},\theta^{2})
       \exp [-2\pi i ( u\theta^{1} + v\theta^{2} ) ] \nonumber\\
 && \hspace{-30pt}  =    
       \int_{-\infty}^{\infty} 
       d\nu 
      \int_{-\infty}^{\infty}
      d\theta^{1}
      \int_{-\infty}^{\infty} 
       d\theta^{2}
      W(\theta^{1},\theta^{2},\nu)
      I_{\nu}(\theta^{1},\theta^{2})
       \exp [-2\pi i ( u\theta^{1} + v\theta^{2} +\nu\tau) ],
       \label{eq:visibility8}
\end{eqnarray}
where 
we take the integration range as $-\infty < \nu < \infty$
by introducing a window function 
$ F_{\mbox{\boldmath $ \theta $}}(\nu)$,
and we use $W(\theta^{1},\theta^{2},\nu) \equiv  
A_{\nu}(\theta^{1},\theta^{2}) F_{\mbox{\boldmath $ \theta $}}(\nu)$
as the window function of the specific intensity 
$I_{\nu}(\theta^{1},\theta^{2})$.

\subsection{Visibility of 21cm line observations}

\begin{figure}[thb]
\begin{center}
\includegraphics[bb= 0 50 809 409, width=1\linewidth]{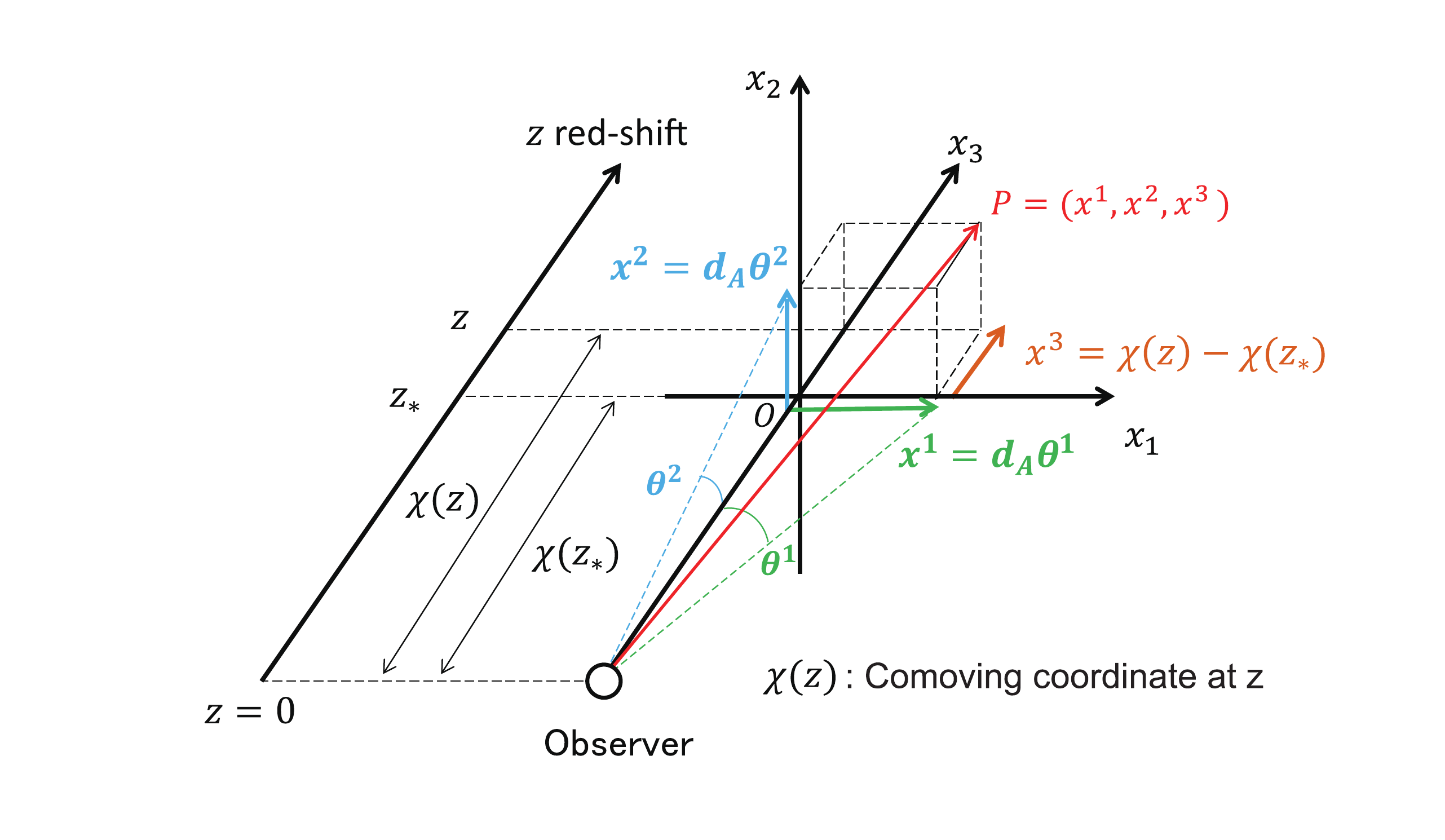} 
\hspace{10pt}\caption{Comoving coordinate 
($x^{1},x^{2},x^{3}$).} \label{fig:zahyou}
\end{center}
\end{figure}

Here, we calculate the visibility of 21 cm line observations.
We consider a radio wave which comes from a region
near redshift  $z_{*}$,
which is a reference redshift of the radio source. 
In this case, observed wave length and frequency are 
close to
$\lambda_{*}\equiv\lambda_{21}(1+z_{*})$
and 
$\nu_{*}\equiv\frac{c}{\lambda_{*}}=\frac{\nu_{21}}{1+z_{*}}$,
respectively.
By using the comoving angular diameter distance $d_{A}$
and the Hubble parameter, 
we can make a connection with 
the coordinate of Eq.(\ref{eq:visibility8})
$(\theta^{1},\theta^{2},\nu)$ and
the comoving coordinate 
$\mbox{\boldmath $x$}=(x^{1},x^{2},x^{3})$,
where $x^{1}$ and $x^{2}$ are the components of 
the comoving coordinate perpendicular to the LOS,
$x^{3}$ is the component along the LOS,
and their origin is 
the reference redshift position $z=z_{*}$ (see Fig.\ref{fig:zahyou}).
The comoving angular diameter distance $d_{A}$ is given by
\begin{eqnarray}
d_{A}(z) \equiv \frac{l}{\theta},
\end{eqnarray}
where $l$ is a comoving size and $\theta$ is a visual angle.
By using $d_{A}$, $\theta^{1}$ and $\theta^{2}$,
the components of the comoving coordinate 
$x^{1}$ and $x^{2}$ are expressed as
\begin{eqnarray}
x^{1} = d_{A}(z_{*})\theta^{1}, \ , \ \ \
x^{2} = d_{A}(z_{*})\theta^{2}.
\end{eqnarray}
From the Fig.\ref{fig:zahyou},
the $x^{3}$ is given by the difference between
the comoving coordinate $\chi(z)$ and 
the central redshift $\chi(z_{\star})$.
Therefore, $x^{3}$ is written as
\begin{eqnarray}
x^{3} &=& 
\int^{z}_{0}\frac{cdz'}{H(z')}
-
\int^{z_{*}}_{0}\frac{cdz'}{H(z')}\nonumber\\
&=&
\int^{z}_{z_{*}}\frac{cdz'}{H(z')} \nonumber \\
&\approx&
\frac{c(z-z_{*})}{H(z_{*})},
\end{eqnarray}
where we use that
$z$ is approximately close to the reference redshift $z_{*}$
(the region of the radio source is sufficiently narrow).
%
The following is a summary of the above,
\begin{eqnarray}
\mbox{\boldmath $x$} = (x^{1},x^{2},x^{3})
=\left(
   d_{A}(z_{*})\theta^{1}, d_{A}(z_{*})\theta^{2}, \frac{c(z-z_{*})}{H(z_{*})}
   \right). \label{eq:kyoudouzahyou}
\end{eqnarray}
Additionally, by using the relation $1+z=\nu_{21}/\nu$,
$x^{3}$ is rewritten as
\begin{eqnarray}
x^{3}=\frac{c(z-z_{*})}{H(z_{*})}
     &=&\frac{c}{H(z_{*})} 
      \left(
           \frac{\nu_{21}}{\nu} - \frac{\nu_{21}}{\nu_{*}}
      \right) \nonumber \\
     &\approx&
      \frac{c\nu_{21}}{H(z_{*})} 
      \left(
           -\frac{1}{\nu_{*}^{2}}(\nu - \nu_{*}) 
      \right) \nonumber \\
     &=& -\frac{c\nu_{21}}{H(z_{*})\nu_{*}^{2}} 
      \left(
\nu - \nu_{*} \right),
\label{eq:x3}
\end{eqnarray}
where we use that the difference 
between $\nu$ and $\nu_{*}$ are sufficiently small.
Since the minus sign in the above equation
can be eliminated by the Fourier transformation,
we omit it from now on.
Moreover, we define the conversion factor from $\nu$ to $x^{3}$ as
\begin{eqnarray}
y(z_{*}) \equiv
      \frac{c\nu_{21}}{H(z_{*})\nu_{*}^{2}} =
      \frac{c(1+z_{*})^{2}}{\nu_{21}H(z_{*})}.
      \label{eq:y}
\end{eqnarray}
Instead of $\nu$, by using $\Delta \nu \equiv \nu-\nu_{*}$ and $y(z_{*})$,
Eq.(\ref{eq:kyoudouzahyou}) is rewritten as
\begin{eqnarray}
\mbox{\boldmath $x$} = (x^{1},x^{2},x^{3})
=\left( \ 
   d_{A}(z_{*})\theta^{1}, d_{A}(z_{*})\theta^{2},y(z_{*})\Delta \nu \ 
   \right). \label{eq:kyoudouzahyou2}
\end{eqnarray}
This relation is the connection with the $(\theta^{1}, \theta^{2},\nu)$
and the comoving coordinate $\mbox{\boldmath $x$}$.

Next, we rewrite the variable of Eq.(\ref{eq:visibility8}) by $\Delta \nu$, 
and we can omit the extra factor $e^{-2\pi i\nu_{*}\tau}$
because the factor does not affect the following discussion.
In this case, $S(u,v,\tau)$ is expressed as
\begin{eqnarray}
S(u,v,\tau)
      &=&    
       \int_{-\infty}^{\infty} 
       d(\Delta\nu )
      \int_{-\infty}^{\infty}
      d\theta^{1}
      \int_{-\infty}^{\infty} 
       d\theta^{2}
      W(\theta^{1},\theta^{2},\Delta\nu)
      I_{\nu}(\theta^{1},\theta^{2}) \nonumber \\
       && \ \ \ \ \ \ \ \ \ \ \ \ \ \ \ \  \ \ \
        \ \ \ \ \ \ \ \ \ \ \ \ \ \ \ \ \
       \times
       \exp [-2\pi i ( u\theta^{1} + v\theta^{2} +\Delta\nu\tau) ] .
\label{eq:visibility9}
\end{eqnarray}
Since the formulae related to the 21 cm line
is written by using the brightness temperature,
we translate the intensity  $I_{\nu}$
into $T_{b}(\nu) = \frac{c^{2}}{2\nu^{2}k_{B}}I_{\nu}$.
Furthermore, 
the difference between the brightness temperature and 
CMB temperature $\Delta T_{b} \equiv T_{b}- T_{\gamma}$
is generally used in observations of 21 cm line. 
Therefore, we use this quantity from no on.
By using the difference of observed brightness temperature 
$\Delta T_{b}^{obs}$, 
the visibility and its integration $S(u,v,\tau)$
with respect to $\nu$ are given by
\begin{eqnarray}
V_{T_{b}}(u,v, \Delta\nu )
      &=&
      \int_{-\infty}^{\infty}
      d\theta^{1}
      \int_{-\infty}^{\infty} 
       d\theta^{2}
      A_{\nu}(\theta^{1},\theta^{2}) \Delta T^{obs}_{b}
       (\theta^{1},\theta^{2}, \Delta\nu)
       \exp [-2\pi i ( u\theta^{1} + v\theta^{2} )] \nonumber \\
       &=&
       \int_{-\infty}^{\infty}
       d\theta^{1}
       \int_{-\infty}^{\infty} 
       d\theta^{2}
       A_{\nu}(\theta^{1},\theta^{2}) 
       \Delta \bar{T}^{obs}_{b}
       (1 + \delta_{21} 
       (\theta^{1},\theta^{2}, \Delta\nu)
       ) \nonumber \\ 
       && \hspace{80pt}
       \times \exp [-2\pi i ( u\theta^{1} + v\theta^{2} )],
       \label{eq:visibility10} \\
       \nonumber \\
S_{T_{b}}(u,v,\tau)
      &=&
       \int_{-\infty}^{\infty} 
       d(\Delta\nu )
       F_{\mbox{\boldmath $\theta$}}(\Delta \nu)
       V_{T_{b}}(u,v, \Delta\nu )   e^{-2\pi i\Delta\nu \tau}
      \nonumber \\
      &=&    
       \int_{-\infty}^{\infty} 
       d(\Delta\nu )
      \int_{-\infty}^{\infty}
      d\theta^{1}
      \int_{-\infty}^{\infty} 
       d\theta^{2}
       W(\theta^{1},\theta^{2},\Delta\nu)
       \Delta \bar{T}^{obs}_{b}
       (1 + \delta_{21} 
       (\theta^{1},\theta^{2}, \Delta\nu)
       ) 
       \nonumber \\
       && \hspace{140pt}
       \times
       \exp [-2\pi i ( u\theta^{1} + v\theta^{2} +\Delta\nu\tau) ].
       \label{eq:visibility11}
\end{eqnarray}
For the brightness temperature
$  \Delta  T_{b}^{obs} (\theta^{1},\theta^{2}, \Delta\nu) 
= \Delta \bar{T}^{obs}_{b}
(1 + \delta_{21} 
(\theta^{1},\theta^{2}, \Delta\nu))$
in Eqs.(\ref{eq:visibility10}) and (\ref{eq:visibility11}),
the averaged component term $\Delta \bar{T}^{obs}_{b}$ 
does not depend on the location.
Therefore, we only need to consider the fluctuation term
$ \Delta \bar{T}^{obs}_{b}\delta_{21} 
(\theta^{1},\theta^{2}, \Delta\nu)$
in  Eqs.(\ref{eq:visibility10}) and (\ref{eq:visibility11}).
From now on, 
we define and use the following vectors,
\begin{subequations}
\begin{eqnarray}
\mbox{\boldmath $\Theta$} & \equiv & (\theta^{1},\theta^{2},\Delta \nu),
\label{eq:defTheta}\\
\mbox{\boldmath $u$}_{\perp} & \equiv & (u,v), \\
\mbox{\boldmath $u$} & \equiv & 
(u,v,\tau) = (\mbox{\boldmath $u$}_{\perp},\tau), \label{eq:defu}
\end{eqnarray}
\end{subequations}
and we call the coordinate space of \mbox{\boldmath $u$} u-space.

\section{Fisher information matrix of 21 cm line observations}

In this section, 
we calculate the Fisher matrix of 21 cm line.observations.
We treat visibilities as observed data,
and only consider their fluctuation component.
From now on, 
$V_{T_{b}}(\mbox{\boldmath $u$}_{\perp},\Delta \nu)$
and $S_{T_{b}}(\mbox{\boldmath $u$}_{\perp},\tau)$
represent only their fluctuation components.
Therefore, we can assume that 
expected values of the visibilities are given by 
$\left\langle 
V_{T_{b}i} \right\rangle
= $ $\left\langle 
V_{T_{b}}(\mbox{\boldmath $u$}_{\perp i},\Delta \nu) 
\right\rangle = 0 $,
where $i$ is index of baseline vectors,
and the visibilities
obey the following multi-dimensional Gaussian likelihood,
%
\begin{eqnarray}
P(V_{T_{b}1}, V_{T_{b}2}, \cdot \cdot \cdot , V_{T_{b}N})
  &=& \frac{1}
           {(2\pi)^{\frac{n}{2}}\sqrt{\det C}}
           \exp\left[
               - \frac{1}{2} 
                 \sum^{N}_{i,j=1}
                 V_{T_{b}i}^{*}
                 \left( C_{V_{T_{b}}}^{-1} \right)_{ij} V_{T_{b}j}
           \right], \\ \label{eq:21cmgauss}
   \left( C_{V_{T_{b}}}\right)_{ij} &=&
          \left\langle
             V_{T_{b}i}V_{T_{b}j}^{*}
          \right\rangle, \label{eq:21cmvariance}
\end{eqnarray}
where the number of the data is determined 
by the experimental resolution of $u$, $v$ and $\Delta\nu$ 
($N=$ $N_{u}\times N_{v} \times N_{\Delta \nu}$).
Below we calculate the Fisher matrix 
of the Fourier transformation of the visibility,
i.e. $S_{T_{b}}$. 
The Fisher matrix is calculated through Eq.(\ref{eq:likelihoodtrace3}),
and we need to estimate the 
variance-covariance matrix $C_{S_{T_{b}}}$ of $S_{T_{b}}$
The variance-covariance matrix has the contributions
of sample variance
$C^{SV} _{S_{T_{b}}}\equiv 
\left\langle
S_{T_{b}i}S_{T_{b}j}^{*}
\right\rangle $
and detector noise $C^{N}_{S_{T_{b}}}$.
Below, we first calculate the former contribution.

\subsection[Sample variance]
{Sample Variance \normalsize{\cite{Arashiba:2009,McQuinn:2005hk}}}

Here, we calculate the variance-covariance matrix 
$C^{SV} _{S_{T_{b}}}\equiv          
\left\langle S_{T_{b}i}S_{T_{b}j}^{*} \right\rangle $
of 21 cm line signals, 
and it is called the sample variance.
By the fluctuation component of Eq.(\ref{eq:visibility11}), 
the matrix is expressed as
\begin{eqnarray}
\left( C^{SV} _{S_{T_{b}}}\right)_{ij}
       &\equiv &
          \left\langle
             S_{T_{b}}( \mbox{\boldmath $u$}_{i} ) S_{T_{b}}^{*}( \mbox{\boldmath $u$}_{j})
          \right\rangle \nonumber \\
       &=&
       \left\langle
        \left(
       \int_{-\infty}^{\infty} 
       \int_{-\infty}^{\infty}
       \int_{-\infty}^{\infty} 
       d^{3}\Theta
       W(\mbox{\boldmath $\Theta$})
       \Delta \bar{T}^{obs}_{b}
       \delta_{21} 
       (\mbox{\boldmath $\Theta$})
       e^{-2\pi i \mbox{\boldmath $u$}_{i} \cdot \mbox{\boldmath $\Theta$} }
       \right)
       \right.
       \nonumber \\
      && \times
       \left.
       \left(
       \int_{-\infty}^{\infty} 
       \int_{-\infty}^{\infty}
       \int_{-\infty}^{\infty} 
       d^{3}\Theta'
       W(\mbox{\boldmath $\Theta$}')
       \Delta \bar{T}^{obs}_{b}
       \delta_{21} 
       (\mbox{\boldmath $\Theta$}')
       e^{-2\pi i \mbox{\boldmath $u$}_{j} \cdot \mbox{\boldmath $\Theta$}' }
       \right)^{*}
       \right\rangle. \label{eq:samplevariance}
\end{eqnarray}
By defining the following Fourier transformation,
\begin{subequations}
\begin{eqnarray}
\tilde{A}(\mbox{\boldmath $u$}) 
      &\equiv & \int^{\infty}_{-\infty}\int^{\infty}_{-\infty}\int^{\infty}_{-\infty}
      d^{3}\Theta A(\mbox{\boldmath $\Theta$}) 
      e^{-2\pi i \mbox{\boldmath $u$} \cdot \mbox{\boldmath $\Theta$}},  \\
A(\mbox{\boldmath $\Theta$}) 
      &\equiv& \int^{\infty}_{-\infty}\int^{\infty}_{-\infty}\int^{\infty}_{-\infty}
      d^{3}u \tilde{A}(\mbox{\boldmath $u$}) 
      e^{2\pi i \mbox{\boldmath $u$} \cdot \mbox{\boldmath $\Theta$}}, \nonumber \\
\end{eqnarray}\label{eq:fourierdef}
\end{subequations}
and using their property
\begin{eqnarray}
\int^{\infty}_{-\infty}\int^{\infty}_{-\infty}\int^{\infty}_{-\infty}
     d^{3}\Theta A(\mbox{\boldmath $\Theta$}) B(\mbox{\boldmath $\Theta$}) 
    e^{-2\pi i \mbox{\boldmath $u$} \cdot \mbox{\boldmath $\Theta$}}  
    =
    \int^{\infty}_{-\infty}\int^{\infty}_{-\infty}\int^{\infty}_{-\infty}
    d^{3}u' \tilde{A}(\mbox{\boldmath $u$}-\mbox{\boldmath $u$}')
     \tilde{B}(\mbox{\boldmath $u$}'),
\end{eqnarray}
we can estimate Eq.(\ref{eq:samplevariance}) as follows,
\begin{eqnarray}
\left( C^{SV} _{S_{T_{b}}}\right)_{ij}
       &=&
       \left\langle
        \left(
       \int_{-\infty}^{\infty} 
       \int_{-\infty}^{\infty}
       \int_{-\infty}^{\infty} 
       d^{3}u'
       \tilde{W}(\mbox{\boldmath $u$}_{i}-\mbox{\boldmath $u$}')
       \Delta \bar{T}^{obs}_{b}
       \tilde{\delta}_{21} 
       (\mbox{\boldmath $u$}')
       \right)
       \right.
       \nonumber \\
      && \times
       \left.
       \left(
       \int_{-\infty}^{\infty} 
       \int_{-\infty}^{\infty}
       \int_{-\infty}^{\infty} 
       d^{3}u''
       \tilde{W}(\mbox{\boldmath $u$}_{j}-\mbox{\boldmath $u$}'')
       \Delta \bar{T}^{obs}_{b}
       \tilde{\delta}_{21} 
       (\mbox{\boldmath $u$}'')
       \right)^{*}  
       \right\rangle \nonumber \\
       &=&
       \int_{-\infty}^{\infty} 
       \int_{-\infty}^{\infty}
       \int_{-\infty}^{\infty} 
       d^{3}u'
       \int_{-\infty}^{\infty} 
       \int_{-\infty}^{\infty}
       \int_{-\infty}^{\infty} 
       d^{3}u''
       \tilde{W}(\mbox{\boldmath $u$}_{i}-\mbox{\boldmath $u$}')
       \tilde{W}^{*}(\mbox{\boldmath $u$}_{j}-\mbox{\boldmath $u$}'') \nonumber \\
       && \  \ \ \ \ \ \ \ \ \ \ \   \ \ \ \ \ \ \ 
       \  \ \ \ \ \ \ \ \ \ \ \  \ \ \ \ \ \ \ \ \ \ \ \ \ \ \ \ \
       \times \left(\Delta \bar{T}^{obs}_{b}\right)^{2}
       \langle
       \tilde{\delta}_{21} 
       (\mbox{\boldmath $u$}')
       \tilde{\delta}_{21} ^{*}
       (\mbox{\boldmath $u$}'')
       \rangle \nonumber \\
       &=&
       \int_{-\infty}^{\infty} 
       \int_{-\infty}^{\infty}
       \int_{-\infty}^{\infty} 
       d^{3}u'
       \int_{-\infty}^{\infty} 
       \int_{-\infty}^{\infty}
       \int_{-\infty}^{\infty} 
       d^{3}u''
       \tilde{W}(\mbox{\boldmath $u$}_{i}-\mbox{\boldmath $u$}')
       \tilde{W}^{*}(\mbox{\boldmath $u$}_{j}-\mbox{\boldmath $u$}'') \nonumber \\
       && \  \ \ \ \ \ \ \ \ \ \ \   \ \ \ \ \ \ \ 
       \  \ \ \ \ \ \ \ \ \ \ \  \ \ \ \ \ \ \ \ \ \ \ \ \ \ \ \ \
       \times \left(\Delta \bar{T}^{obs}_{b}\right)^{2} 
       P_{21}(\mbox{\boldmath $u$}')
       \delta^{D}(\mbox{\boldmath $u$}'-\mbox{\boldmath $u$}'') \nonumber \\
       &=&
       \int_{-\infty}^{\infty} 
       \int_{-\infty}^{\infty}
       \int_{-\infty}^{\infty} 
       d^{3}u'
       \tilde{W}(\mbox{\boldmath $u$}_{i}-\mbox{\boldmath $u$}')
       \tilde{W}^{*}(\mbox{\boldmath $u$}_{j}-\mbox{\boldmath $u$}')
      \left(\Delta \bar{T}^{obs}_{b}\right)^{2} 
       P_{21}(\mbox{\boldmath $u$}') \nonumber \\
        &=&
       \int_{-\infty}^{\infty} 
       \int_{-\infty}^{\infty}
       \int_{-\infty}^{\infty} 
       d^{3}u'
       \left|
       \tilde{W}(\mbox{\boldmath $u$}_{i}-\mbox{\boldmath $u$}')
       \right|^{2} \delta_{ij}
       \left(\Delta \bar{T}^{obs}_{b}\right)^{2}
       P_{21}(\mbox{\boldmath $u$}'),
       \label{eq:samplevariance2}
\end{eqnarray}
where we define the following power spectrum of 21 cm line,
\begin{eqnarray}
       \langle
       \tilde{\delta}_{21} 
       (\mbox{\boldmath $u$})
       \tilde{\delta}_{21} ^{*}
       (\mbox{\boldmath $u$}')
       \rangle
       \equiv 
       P_{21}(\mbox{\boldmath $u$})
       \delta^{D}(\mbox{\boldmath $u$}-\mbox{\boldmath $u$}'), \label{eq:21poweru}
\end{eqnarray}
and in the final line of Eq.(\ref{eq:samplevariance2}),
we use the diagonal property of 
the window function $\tilde{W}(\mbox{\boldmath $u$})$.

Next, we set the following normalization condition of the
window function,
\begin{eqnarray}
\int_{-\infty}^{\infty} 
       \int_{-\infty}^{\infty}
       \int_{-\infty}^{\infty} 
       d^{3}u
       \tilde{W}(\mbox{\boldmath $u$})=1.
\end{eqnarray}
Since this window function has non-zero value 
in a small region near $\mbox{\boldmath $u$}=0$
with its volume $\delta u \delta v \delta \tau$,
we can express it as
\begin{eqnarray}
\delta u \delta v \delta \tau
       \tilde{W}(\mbox{\boldmath $u$}) \approx 1 
       \ \ \longrightarrow  \ \ \
       \tilde{W}(\mbox{\boldmath $u$}) \approx \frac{1}{\delta u \delta v \delta \tau},
\end{eqnarray}
By using this window function,
Eq.(\ref{eq:samplevariance2})
reduces to
\begin{eqnarray}
\left(C^{SV} _{S_{T_{b}}}\right)_{ij}
       &\approx&
       \frac{1}{\delta u \delta v \delta \tau}
       \delta_{ij}
       \left(\Delta \bar{T}^{obs}_{b}\right)^{2} 
       P_{21}(\mbox{\boldmath $u$}_{i}) \nonumber \\
       &=&
       \frac{\delta_{ij}}{\delta u \delta v \delta \tau}
       P_{T_{b}}(\mbox{\boldmath $u$}_{i}) \label{eq:samplevariance3}, \\
       \nonumber \\
 P_{T_{b}}(\mbox{\boldmath $u$}_{i})
      &\equiv &
\left(\Delta \bar{T}^{obs}_{b}\right)^{2} 
       P_{21}(\mbox{\boldmath $u$}_{i}).
\end{eqnarray}
Because the resolutions of $u$, $v$ and $\tau$ are 
$\delta u \delta v \approx \frac{A_{e}}{\lambda^{2}}$
and $\delta\tau \approx \frac{1}{B}$,
the sample variance can be given by \cite{McQuinn:2005hk}
\begin{eqnarray}
\left(C^{SV} _{S_{T_{b}}}\right)_{ij}
       &\approx&
       \frac{\lambda^{2}B}{A_{e}}
       P_{T_{b}}(\mbox{\boldmath $u$}_{i}) \delta_{ij}.
       \label{eq:samplevariance4}
\end{eqnarray}
%

\subsection[Detector Noise]
{Detector Noise \normalsize{\cite{Arashiba:2009,McQuinn:2005hk}}}

The detector noise of visibility 
per a pair of two antennae is give by
\cite{McQuinn:2005hk,denpatext2}
\begin{eqnarray}
\Delta V^{N}_{T_{b}}
       = \frac{\lambda^{2}T_{{\rm sys}}}{A_{e}\ \sqrt[]{\delta(\Delta\nu)t_{0}}},
\end{eqnarray}
where $t_{0}$ is the observation time of a frequency channel,
$\delta(\Delta\nu)$ is the frequency resolution,
$A_{e}$ is the effective area of the antenna,
$ \lambda=\lambda_{21} (1+z)$ is the observed  wave length
and $T_{{\rm sys}}$ is the system temperature.

If a number of pairs of antennae correspond to 
one baseline vector $\mbox{\boldmath $u$}_{\perp}$,
the detector noise reduces to
\begin{eqnarray}
\Delta V^{N}_{T_{b}}(\mbox{\boldmath $u$}_{\perp},\Delta\nu)
       = \frac{\lambda^{2}T_{{\rm sys}}}
         {A_{e}\ \sqrt[]{\delta(\Delta\nu)N_{b}(\mbox{\boldmath $u$}_{\perp})t_{0}}},
\end{eqnarray}
where $N_{b}(\mbox{\boldmath $u$}_{\perp})$ is 
the number of the antenna pairs.
From this formula, the noise of $S_{T_{b}}$ is given by
\begin{eqnarray}
\Delta S^{N}_{T_{b}}(\mbox{\boldmath $u$}_{\perp},\tau)
      & = &
     \int^{\infty}_{-\infty} d(\Delta\nu)F_{\mbox{\boldmath $\theta$}}(\Delta\nu)  
      \Delta V^{N}_{T_{b}}(\mbox{\boldmath $u$}_{\perp},\Delta\nu)
      e^{-2\pi i \Delta\nu \tau} \nonumber \\
      &=&
      \sum_{j=1}^{B/\delta(\Delta\nu)}\delta(\Delta\nu)
      \Delta V^{N}_{T_{b}}(\mbox{\boldmath $u$}_{\perp},\Delta\nu_{j})
      e^{-2\pi i \Delta\nu_{j} \tau},
      \label{eq:noise_of_S}
\end{eqnarray}
where in the second line we use that the window function 
$F_{\mbox{\boldmath $ \theta $}}(\Delta \nu)$
has non-zero values in a narrow frequency range $B$,
and the number of the data is $B/\delta(\Delta\nu)$.
The frequency band $B$ is called the bandwidth.
By using Eq.(\ref{eq:noise_of_S}),
the variance-covariance matrix of $\Delta S^{N}_{T_{b}}$
is calculated as follows,
%
\begin{eqnarray}
\left( C_{S_{T_{b}}}^{N}\right)_{ij} 
&\equiv& \left\langle 
      \Delta S^{N}_{T_{b}}(\mbox{\boldmath $u$}_{i})
      \Delta S^{N*}_{T_{b}}(\mbox{\boldmath $u$}_{j})
      \right\rangle \nonumber \\
      &=&
      \sum_{m,l=1}^{B/\delta(\Delta\nu)}[\delta(\Delta\nu)]^{2}
      e^{-2\pi i ( \Delta\nu_{m} \tau_{i}-\Delta\nu_{l} \tau_{j})}
      \left\langle 
      \Delta V^{N}_{T_{b}}(\mbox{\boldmath $u$}_{\perp i},\Delta\nu_{m})   
      \Delta V^{N*}_{T_{b}}(\mbox{\boldmath $u$}_{\perp j},\Delta\nu_{l})
      \right\rangle \nonumber \\
      &=&
      \sum_{m,l=1}^{B/\delta(\Delta\nu)}[\delta(\Delta\nu)]^{2}
      e^{-2\pi i ( \Delta\nu_{m} \tau_{i}-\Delta\nu_{l} \tau_{j})}
      \left[ 
      \Delta V^{N}_{T_{b}}(\mbox{\boldmath $u$}_{\perp i},\Delta\nu_{m})
      \right]^{2} \delta_{ij}\delta_{ml} \nonumber \\
      &=&
      \sum_{m=1}^{B/\delta(\Delta\nu)}[\delta(\Delta\nu)]^{2}
      \left[ 
      \Delta V^{N}_{T_{b}}(\mbox{\boldmath $u$}_{\perp i},\Delta\nu_{m})
      \right]^{2} \delta_{ij} \nonumber \\
      &=&
      \frac{B}{\delta(\Delta\nu)}
      [\delta(\Delta\nu)]^{2}
      \left(
      \frac{\lambda^{2}T_{{\rm sys}}}
         {A_{e}\ \sqrt[]{\delta(\Delta\nu)N_{b}(\mbox{\boldmath $u$}_{\perp i})t_{0}}}
      \right)^{2}\delta_{ij} \nonumber \\
      &=&
      B\left(
      \frac{\lambda^{2}T_{{\rm sys}}}
         {A_{e}}
      \right)^{2}
      \frac{\delta_{ij}}{N_{b}(\mbox{\boldmath $u$}_{\perp i})t_{0}},
      \label{eq:noisevariance}
\end{eqnarray}
where in the third line we assume that there are no correlations between
the different visibilities 
$\Delta V^{N}_{T_{b}}(\mbox{\boldmath $u$}_{\perp i},\Delta\nu_{m})$.

Next, we introduce the baseline density
$n_{b}(\mbox{\boldmath $u$}_{\perp })$,
and the number of the antenna pairs $N_{b}(\mbox{\boldmath $u$}_{\perp })$
can be expressed as
\begin{eqnarray}
N_{b}(\mbox{\boldmath $u$}_{\perp}) = n_{b}(\mbox{\boldmath $u$}_{\perp})
\delta u \delta v,
\end{eqnarray}
where $\delta u$ and $\delta v$ are the resolutions in the u-space,
and they are given by $\delta u\delta v \approx A_{e}/\lambda^{2}$.
Here, $N_{b}(\mbox{\boldmath $u$}_{\perp})$  means
the total number of the baselines existing 
a small region of u-space,
and its ranges are from $u$ to $u+\delta u$ and from $v$ to $v + \delta v$.
By using the baseline density,
the variance-covariance matrix of the detector noise Eq.(\ref{eq:noisevariance})
is rewritten as
\cite{Tegmark:1996bz,Morales:2004ca}
\begin{eqnarray}
\left( C_{S_{T_{b}}}^{N} \right)_{ij} 
       &=&
      \frac{\lambda^{2}B}{A_{e}}
      \left\{
      \left(
      \frac{\lambda^{2}T_{{\rm sys}}}
         {A_{e}}
      \right)^{2}
      \frac{\delta_{ij}}{n_{b}(\mbox{\boldmath $u$}_{\perp i})t_{0}}
      \right\}.
      \label{eq:noisevariance2}
\end{eqnarray}
This baseline density used here 
can be calculated from the specific antenna distribution
\cite{Lidz:2007az}.
Additionally, the integration of the baseline density 
with respect to $\mbox{\boldmath $u$}_{\perp}$
becomes the total number of the antenna pairs
\begin{subequations}
\begin{eqnarray}
N_{b}^{{\rm total}} &=& \int \int n_{b}(\mbox{\boldmath $u$}_{\perp}) 
\delta u \delta v, \\
N_{b}^{{\rm total}} &=& \frac{N_{{\rm ant}}(N_{{\rm ant}}-1)}{2}.
\end{eqnarray}
\end{subequations}

\subsection{Contribution of residual foregrounds}

Here, we consider the situation of existing some residual foregrounds.
For the 21 cm line observation, 
we take account of the most dominant galactic foreground, 
namely the synchrotron radiation.
We assume that
the foreground subtraction can be done down to a given level,
and treat the contribution of the residual foreground
as a Gaussian random field.
Then, we introduce the following 
effective noise including the contribution of the residual foreground
$\Delta V^{{\rm RFg}}$,
\begin{eqnarray}
\Delta V^{N,{\rm eff}}_{T_{b}}(\mbox{\boldmath $u$}_{\perp i},\Delta\nu_{m})   
\equiv
\Delta V^{N}_{T_{b}}(\mbox{\boldmath $u$}_{\perp i},\Delta\nu_{m})   
      + \Delta V^{{\rm RFg}}
         (\mbox{\boldmath $u$}_{\perp i},\Delta\nu_{m}).
\end{eqnarray}
By using this effective noise,
we define the variance-covariance matrix of 
this effective noise as
\begin{eqnarray}
\left( C_{S_{T_{b}}}^{N,{\rm eff}}\right)_{ij} 
&\equiv& \left\langle 
      \Delta S^{N, {\rm eff}}_{T_{b}}(\mbox{\boldmath $u$}_{i})
      \Delta S^{N,{\rm eff}*}_{T_{b}}(\mbox{\boldmath $u$}_{j})
      \right\rangle \nonumber \\
= \hspace{25pt} &&
\hspace{-40pt} \sum_{m,l=1}^{B/\delta(\Delta\nu)}[\delta(\Delta\nu)]^{2}
      e^{-2\pi i ( \Delta\nu_{m} \tau_{i}-\Delta\nu_{l} \tau_{j})} \nonumber \\
\times \hspace{10pt} &&
\hspace{-25pt} \left\langle
      \left(
      \Delta V^{N}_{T_{b}}(\mbox{\boldmath $u$}_{\perp i},\Delta\nu_{m})   
      + \Delta V^{{\rm RFg}}
         (\mbox{\boldmath $u$}_{\perp i},\Delta\nu_{m})   
      \right)
\left(
\Delta V^{N*}_{T_{b}}(\mbox{\boldmath $u$}_{\perp j},\Delta\nu_{l})
+ \Delta V^{{\rm RFg}*}
(\mbox{\boldmath $u$}_{\perp j},\Delta\nu_{l})  
\right)
\right\rangle \nonumber \\
= \hspace{25pt} &&
\hspace{-40pt} \sum_{m=1}^{B/\delta(\Delta\nu)}[\delta(\Delta\nu)]^{2}
\Biggl[
(\Delta V^{N}_{T_{b}}(\mbox{\boldmath $u$}_{\perp i},\Delta\nu_{m}))^2
\delta_{ij}
\Biggl. \nonumber \\
&& \hspace{10pt} + \left.
\sum_{l=1}^{B/\delta(\Delta\nu)}
e^{-2\pi i ( \Delta\nu_{m} \tau_{i}-\Delta\nu_{l} \tau_{j})}
\left\langle 
\Delta V^{{\rm RFg}} (\mbox{\boldmath $u$}_{\perp i},\Delta\nu_{m})
\Delta V^{{\rm RFg}*}(\mbox{\boldmath $u$}_{\perp j},\Delta\nu_{l}) 
\right\rangle 
\right],
\label{eq:eff_noise_21cm}
\end{eqnarray}
where we assume that there are no correlations 
between the detector noise and the residual foreground.
%
%
When the value of $\Delta \nu_{m}$ is close to that of $\Delta \nu_{l}$ in this
frequency band, the second term becomes
\begin{eqnarray}
\left[{\rm Second \ term \ of \ Eq.(\ref{eq:eff_noise_21cm})}\right]
 &=&
\sum_{l=1}^{B/\delta(\Delta\nu)}
e^{-2\pi i ( \Delta\nu_{m} \tau_{i}-\Delta\nu_{l} \tau_{j})}
\left\langle 
\Delta V^{{\rm RFg}} (\mbox{\boldmath $u$}_{\perp i},\Delta\nu_{m})
\Delta V^{{\rm RFg}*}(\mbox{\boldmath $u$}_{\perp j},\Delta\nu_{l}) 
\right\rangle 
\nonumber \\
&\approx&
\frac{B}{\delta(\Delta\nu)}
e^{-2\pi i \Delta\nu_{m} ( \tau_{i} - \tau_{j})}
\left\langle 
\Delta V^{{\rm RFg}} (\mbox{\boldmath $u$}_{\perp i},\Delta\nu_{m})
\Delta V^{{\rm RFg}*}(\mbox{\boldmath $u$}_{\perp j},\Delta\nu_{m}) 
\right\rangle. \nonumber
\label{eq:RFg_cross-correlation}
\end{eqnarray}
%
Moreover, the correlation of the residual foreground becomes
\begin{eqnarray}
\left\langle 
\Delta V^{{\rm RFg}} (\mbox{\boldmath $u$}_{\perp i},\Delta\nu_{m})
\Delta V^{{\rm RFg}*}(\mbox{\boldmath $u$}_{\perp j},\Delta\nu_{m}) 
\right\rangle 
&=&  
\int du'^{2}_{\perp}
\left| \tilde{A}_{\nu}(\mbox{\boldmath $u$}_{\perp i}
-\mbox{\boldmath $u$}'_{\perp}) \right|^{2}
\delta_{ij}
C^{{\rm RFg}} (\mbox{\boldmath $u$}'_{\perp},\nu_{m}) \nonumber \\
&\approx &
\frac{\lambda^2}{A_{e}}
\delta_{ij}
C^{{\rm RFg}} (\mbox{\boldmath $u$}'_{\perp,i},\nu_{m}),
\label{eq:residual_F}
\end{eqnarray}
where we use the same calculation of the sample variance,
$\nu_{m}$ is the frequency corresponding to 
$\Delta \nu_{m}$ ($\equiv \nu_{m} - \nu_{*}$),
and $\nu_{*}$ is the central frequency value in this frequency band.
Therefore, the second term of the Eq.(\ref{eq:eff_noise_21cm})
can be expressed as
\begin{eqnarray}
\left[{\rm Second \ term \ of \ Eq.(\ref{eq:eff_noise_21cm})}\right]
&\approx&
\frac{B}{\delta(\Delta\nu)}
e^{-2\pi i \Delta\nu_{m}( \tau_{i} - \tau_{j})}
\left\{
\frac{\lambda^2}{A_{e}}
\delta_{ij}
C^{{\rm RFg}} (\mbox{\boldmath $u$}'_{\perp,i},\nu_{m})
\right\} \nonumber \\
&=&
\frac{\lambda^2}{A_{e}}
\frac{B}{\delta(\Delta\nu)}
C^{{\rm RFg}} (\mbox{\boldmath $u$}'_{\perp,i},\nu_{m}),
\label{eq:RFg_cross-correlation2}
\end{eqnarray}

By substituting Eq.(\ref{eq:noisevariance2}) into
the visibility noise contribution 
and Eq.(\ref{eq:RFg_cross-correlation2}) into the residual foreground contribution,
we can express the effective noise as
\begin{eqnarray}
\left( C_{S_{T_{b}}}^{N,{\rm eff}}\right)_{ij}
      &=&\frac{B}{\delta(\Delta\nu)}[\delta(\Delta\nu)]^{2}
      \left[
      \left(
      \frac{\lambda^{2}T_{{\rm sys}}}
      {A_{e}\ \sqrt[]{\delta(\Delta\nu)
      N_{b}(\mbox{\boldmath $u$}_{\perp i})t_{0}}}
      \right)^{2}
      + \frac{\lambda^2}{A_{e}}\frac{B}{\delta(\Delta\nu)}
        C^{{\rm RFg}}(\mbox{\boldmath $u$}_{\perp i},\nu_{*})
      \right]\delta_{ij} \nonumber \\
      &=&
      \left[
      \frac{B\lambda^2}{A_{e}}
      \left\{\left(
      \frac{\lambda^{2}T_{{\rm sys}}}{A_{e}}
      \right)^{2}
      \frac{\delta_{ij}}{n_{b}(\mbox{\boldmath $u$}_{\perp i})t_{0}}
      \right\}
      + \frac{B\lambda^2}{A_{e}}
      B C^{{\rm RFg}}(\mbox{\boldmath $u$}_{\perp i},\nu_{*})
      \right] \nonumber \\
      &=&
      \frac{B\lambda^2}{A_{e}}
      \left[
      \left(
      \frac{\lambda^{2}T_{\rm sys}}{A_{e}}
      \right)^{2}
      \frac{\delta_{ij}}{n_{b}(\mbox{\boldmath $u$}_{\perp i})t_{0}}   
      + 
      B C^{{\rm RFg}}(\mbox{\boldmath $u$}_{\perp i},\nu_{*})
      \right],
      \label{eq:noisevariance_FG2}
\end{eqnarray}
where we assume $\nu_{m} \sim \nu_{*}$.
%
%
When we include the contribution of the residual foreground in our analysis,
we use this effective noise as the 21cm noise power spectrum.
From now on, we introduce a foreground removal parameter 
$\sigma^{{\rm RFg}}_{{\rm 21cm}}$, which is defined as
$B C^{{\rm RFg}}(\mbox{\boldmath $u$}_{\perp i},\nu) 
=
(\sigma^{{\rm RFg}}_{{\rm 21cm}}\times 1{\rm MHz})
C^{{\rm Fg}}(\mbox{\boldmath $u$}_{\perp i},\nu)$,
where $C^{{\rm Fg}}(\mbox{\boldmath $u$}_{\perp i},\nu)$ 
represents the power of the foreground.
In our analysis, we assume 
$\sigma^{{\rm RFg}}_{{\rm 21cm}}=10^{-7}$ 
(this value corresponds to $0.03$\% at the signal).

As long as we use the flat sky approximation,
the u space variable $\mbox{\boldmath $u$}_{\perp}$
is related to the multipole $\ell$ of angular power spectrum,
$|\mbox{\boldmath $u$}_{\perp}|=\ell/2\pi$.
In this thesis, 
we use the scale dependence of synchrotron radiation 
$C_{\ell}^{S,X}(\nu)$ (Eq.(\ref{eq:synchrotron_rad}))
as
the foreground power
$C^{{\rm Fg}}(\mbox{\boldmath $u$}_{\perp i},\nu)$.

\subsection{Total variance-covariance matrix $C _{S_{T_{b}}}$}

By using the sample $C^{SV} _{S_{T_{b}}}$ 
and the noise $C^{N} _{S_{T_{b}}}$ variances,
the total variance-covariance matrix $C _{S_{T_{b}}}$ is given by
\begin{eqnarray}
\left(
C _{S_{T_{b}}}
\right)_{ij}
&=&
\left( C^{SV}_{S_{T_{b}}} \right)_{ij}
+
\left(
C^{N}_{S_{T_{b}}} 
\right)_{ij} \nonumber 
\\
       &=&
       \frac{\lambda^{2}B}{A_{e}}
       P_{T_{b}}(\mbox{\boldmath $u$}_{i}) \delta_{ij}             
      +\frac{\lambda^{2}B}{A_{e}}
      \left(
      \frac{\lambda^{2}T_{{\rm sys}}}
         {A_{e}}
      \right)^{2}  
      \frac{\delta_{ij}}{n_{b}(\mbox{\boldmath $u$}_{\perp i})t_{0}} \nonumber \\
       &=&
       \frac{\lambda^{2}B}{A_{e}}\delta_{ij}
       \left[
       P_{T_{b}}(\mbox{\boldmath $u$}_{i})      
      +
      \left(
      \frac{\lambda^{2}T_{{\rm sys}}}
         {A_{e}}
      \right)^{2}
      \frac{1}{n_{b}(\mbox{\boldmath $u$}_{\perp i})t_{0}}
      \right].
       \label{eq:samplevariance5}
\end{eqnarray}
From now on, we define and use the following noise $P_{N}$ 
and total power spectra $P_{N}$,
\begin{eqnarray}
P_{N} (\mbox{\boldmath $u$}_{\perp})
      &\equiv &     
      \left(
      \frac{\lambda^{2}T_{sys}}
         {A_{e}}
      \right)^{2}
      \frac{1}{n_{b}(\mbox{\boldmath $u$}_{\perp})t_{0}}, \\
      \nonumber \\
P^{tot}_{T_{b}}(\mbox{\boldmath $u$})
       &\equiv&
       P_{T_{b}}(\mbox{\boldmath $u$})      
       +
       P_{N}(\mbox{\boldmath $u$}_{\perp}).
       \label{eq:samplevariance6}
\end{eqnarray}
When we include the effects due to 
the residual foreground in our analysis,
the noise power becomes
\begin{eqnarray}
P_{N} (\mbox{\boldmath $u$}_{\perp})
      &=&     
      \left(
      \frac{\lambda^{2}T_{{\rm sys}}}
         {A_{e}}
      \right)^{2}
      \frac{1}{n_{b}(\mbox{\boldmath $u$}_{\perp})t_{0}}
      + (\sigma^{{\rm RFg}}_{{\rm 21cm}}\times 1 \ {\rm MHz})
      C^{{\rm Fg}}(\mbox{\boldmath $u$}_{\perp},\nu_{*}).
      \label{eq:effective_noise_power}
\end{eqnarray}
%

\subsection{Relation between
$P_{21}(\mbox{\boldmath $u$})$ and 
$P_{21}(\mbox{\boldmath $k$})$}

Here, we derive the relation between 
$P_{21}(\mbox{\boldmath $u$})$ 
and
$P_{21}(\mbox{\boldmath $k$})$,
which are defined 
by Eqs.(\ref{eq:21poweru})
and (\ref{eq:21powerspectrum}), respectively.
Below we express the former as 
$P_{21}^{{\rm u}}(\mbox{\boldmath $u$})$
and the latter as 
$P_{21}^{{\rm k}}(\mbox{\boldmath $k$})$.
%
At first, by Eq.(\ref{eq:fourierdef}),
the u-space fluctuation 
$\tilde{\delta}^{{\rm u}}_{21}(\mbox{\boldmath $u$})$ is given by
\begin{eqnarray}
\tilde{\delta}^{{\rm u}}(\mbox{\boldmath $u$}) 
      &\equiv & \int^{\infty}_{-\infty}\int^{\infty}_{-\infty}\int^{\infty}_{-\infty}
      d^{3}\Theta \delta^{{\rm u}}(\mbox{\boldmath $\Theta$}) 
      e^{-2\pi i \mbox{\boldmath $u$} \cdot \mbox{\boldmath $\Theta$}}.
\end{eqnarray}
According to the following relation of Eq.(\ref{eq:kyoudouzahyou2}),
\begin{eqnarray}
\mbox{\boldmath $x$} = (x^{1},x^{2},x^{3})
 =\left( \
   d_{A}(z_{*})\theta^{1}, d_{A}(z_{*})\theta^{2},y(z_{*})\Delta \nu \ 
   \right), \label{eq:kyoudouzahyou3}
\end{eqnarray}
we can transform the variables from $\mbox{\boldmath $\Theta $}$
to $\mbox{\boldmath $x$}$,
\begin{eqnarray}
\tilde{\delta}^{{\rm u}}(\mbox{\boldmath $u$}) 
      &= & \int^{\infty}_{-\infty}\int^{\infty}_{-\infty}\int^{\infty}_{-\infty}
      \frac{d^{3}x}{d_{A}(z_{*})^{2}y(z_{*})} 
      \delta^{{\rm u}}(\mbox{\boldmath $\Theta$}(\mbox{\boldmath $x$})) \nonumber \\
      && \hspace{100pt}
      \times 
      \exp\left[
        -2\pi i 
        \left( 
          u \frac{x^{1}}{d_{A}(z_{*})} + v\frac{x^{2}}{d_{A}(z_{*})} + \tau \frac{x^{3}}{y(z_{*})} 
        \right)
      \right] \nonumber \\
      &=&
      \frac{1}{d_{A}(z_{*})^{2}y(z_{*})} 
      \int^{\infty}_{-\infty}\int^{\infty}_{-\infty}\int^{\infty}_{-\infty}
      d^{3}x
      \delta^{{\rm u}}(\mbox{\boldmath $\Theta$}(\mbox{\boldmath $x$})) \nonumber \\
      && \hspace{100pt} 
      \times \exp\left[
        - i 
        \left( 
          \frac{2\pi u }{d_{A}(z_{*})} x^{1}+ \frac{2\pi v}{d_{A}(z_{*})}x^{2} + \frac{2\pi \tau}{y(z_{*})}  x^{3}
        \right)
      \right]. \label{eq:expyuragi}
\end{eqnarray}
%
%
Here, if we regard 
$\left(\frac{2\pi u }{d_{A}(z_{*})},  
\frac{2\pi v}{d_{A}(z_{*})},  
\frac{2\pi \tau}{y(z_{*})}\right)$
as a wave number vector $\mbox{\boldmath $k$}$, 
\begin{eqnarray}
\mbox{\boldmath $k$} = (k^{1},k^{2},k^{3}) 
\equiv 
\left(
\frac{2\pi u }{d_{A}(z_{*})} ,
\frac{2\pi v}{d_{A}(z_{*})},  
\frac{2\pi \tau}{y(z_{*})}
\right), \label{eq:kteigi}
\end{eqnarray}
the u-space fluctuation $\tilde{\delta}^{{\rm u}}(\mbox{\boldmath $u$}) $ 
reduces to
\begin{eqnarray}
\tilde{\delta}^{{\rm u}}(\mbox{\boldmath $u$}) 
&=&
\frac{1}{d_{A}(z_{*})^{2}y(z_{*})} 
\int^{\infty}_{-\infty}\int^{\infty}_{-\infty}\int^{\infty}_{-\infty}
d^{3}x
\delta^{{\rm u}}(\mbox{\boldmath $\Theta$}(\mbox{\boldmath $x$}))
\exp\left[
- i \left( k^{1} x^{1} + k^{2}x^{2} + k^{3} x^{3}
\right)
\right] \nonumber \\
&=&
\frac{1}{d_{A}(z_{*})^{2}y(z_{*})} 
\int^{\infty}_{-\infty}\int^{\infty}_{-\infty}\int^{\infty}_{-\infty}
d^{3}x
\delta^{{\rm u}}(\mbox{\boldmath $\Theta$}(\mbox{\boldmath $x$}))
e^{-i \mbox{\boldmath $k$} \cdot \mbox{\boldmath $x$} }.
\label{eq:u-space_fluc}
\end{eqnarray}
By Eq.(\ref{eq:fourier1}),
we can regard the integral part of Eq.(\ref{eq:u-space_fluc})
as the k-space fluctuation
$\tilde{\delta}^{{\rm k}}(\mbox{\boldmath $k$})$.
%
%
%
Therefore, we can obtain the following 
relation between
$\tilde{\delta}^{{\rm u}}(\mbox{\boldmath $u$})$
and
$\tilde{\delta}^{{\rm k}}(\mbox{\boldmath $k$})$,
\begin{eqnarray}
\tilde{\delta}^{{\rm u}}(\mbox{\boldmath $u$}) 
&=& 
\frac{1}{d_{A}(z_{*})^{2}y(z_{*})} 
\tilde{\delta}^{{\rm k}}(\mbox{\boldmath $k$}).
\label{eq:expyuragi2}
\end{eqnarray}

Next, by the definition of the u-space power spectrum 
Eq.(\ref{eq:21poweru}),
$P_{21}^{{\rm u}}(\mbox{\boldmath $u$})$ is given by
\begin{eqnarray}
       \langle
       \tilde{\delta}_{21}^{{\rm u}}
       (\mbox{\boldmath $u$})
       \tilde{\delta}_{21} ^{{\rm u}*}
       (\mbox{\boldmath $u$}')
       \rangle
       \equiv 
       P_{21}^{{\rm u}}(\mbox{\boldmath $u$})
       \delta^{D}(\mbox{\boldmath $u$}-\mbox{\boldmath $u$}'). 
\label{eq:21poweru2}
\end{eqnarray}
According to Eq.(\ref{eq:expyuragi2}),
the left hand side of Eq.(\ref{eq:21poweru2}) is rewritten as
\begin{eqnarray}
       \langle
       \tilde{\delta}_{21}^{{\rm u}}
       (\mbox{\boldmath $u$})
       \tilde{\delta}_{21} ^{{\rm u}*}
       (\mbox{\boldmath $u$}')
       \rangle
       &=&
       \left(
        \frac{1}{d_{A}(z_{*})^{2}y(z_{*})}
       \right)^{2}
       \langle
       \tilde{\delta}_{21}^{{\rm k}}
       (\mbox{\boldmath $k$})
       \tilde{\delta}_{21} ^{{\rm k}*}
       (\mbox{\boldmath $k$}')
       \rangle \nonumber \\
       &=& 
       \left(
        \frac{1}{d_{A}(z_{*})^{2}y(z_{*})}
       \right)^{2}
       (2\pi)^{3}P^{{\rm k}}_{21}(\mbox{\boldmath $k$})
       \delta^{D}(\mbox{\boldmath $k$}-\mbox{\boldmath $k$}'). \label{eq:powertheory1}
\end{eqnarray}
%
Besides, in consideration of Eq.(\ref{eq:kteigi}),
we find that the relation between
$\delta^{D}(\mbox{\boldmath $u$}-\mbox{\boldmath $u$}') $
and
$\delta^{D}(\mbox{\boldmath $k$}-\mbox{\boldmath $k$}') $
is given by
\begin{eqnarray}
       \delta^{D}(\mbox{\boldmath $u$}-\mbox{\boldmath $u$}') 
       =\frac{(2\pi)^{3}}{d_{A}(z_{*})^{2}y(z_{*})} \delta^{D}(\mbox{\boldmath $k$}-\mbox{\boldmath $k$}').
       \label{eq:deltafunc}
\end{eqnarray}
By using this relation, 
the right hand side of Eq.(\ref{eq:21poweru2})
can be expressed as
\begin{eqnarray}
       P_{21}^{{\rm u}}(\mbox{\boldmath $u$})
       \delta^{D}(\mbox{\boldmath $u$}-\mbox{\boldmath $u$}') 
       &=&
       P_{21}^{{\rm u}}(\mbox{\boldmath $u$})
       \frac{(2\pi)^{3}}{d_{A}(z_{*})^{2}y(z_{*})} \delta^{D}(\mbox{\boldmath $k$}-\mbox{\boldmath $k$}').
\label{eq:21poweru3}
\end{eqnarray}
According to Eqs.(\ref{eq:powertheory1}) and (\ref{eq:21poweru3}),
we can obtain the following relation between 
$P_{21}^{{\rm u}}(\mbox{\boldmath $u$})$
and
$P_{21}^{{\rm k}}(\mbox{\boldmath $k$})$,
\begin{eqnarray}
       P_{21}^{{\rm u}}(\mbox{\boldmath $u$})
       &=&
       \frac{1}{d_{A}(z_{*})^{2}y(z_{*})}P_{21}^{{\rm k}}(\mbox{\boldmath $k$}).
\label{eq:21poweru4}
\end{eqnarray}
We perform our analyses in terms of this 
u-space power spectrum $P_{21}^{{\rm u}}(\bm{u})$ 
since this quantity is directly measurable 
without any cosmological assumptions.

\subsection{Fisher matrix of 21 cm line observations}

Here, we derive the Fisher matrix of 21 cm line observations.
From Eq.(\ref{eq:likelihoodtrace3}),
the Fisher matrix for the Gaussian likelihood is given by
\begin{eqnarray}
F_{\alpha \beta} &=&
    \frac{1}{2}
    {\rm Tr} \left[\ C_{S_{T_{b}}}^{-1}C_{S_{T_{b}},\alpha}C_{S_{T_{b}}}^{-1}C_{S_{T_{b}},\beta}\right]
    +\mbox{\boldmath $\mu$}_{,\alpha}^{T} C_{S_{T_{b}}}^{-1}\mbox{\boldmath $\mu$}_{,\beta}
    \nonumber \\
    &=&
     \frac{1}{2}
    {\rm Tr} \left[\ C_{S_{T_{b}}}^{-1}C_{S_{T_{b}},\alpha}C_{S_{T_{b}}}^{-1}C_{S_{T_{b}},\beta}\right],
     \label{eq:likelihoodtrace4}
\end{eqnarray}
where $\alpha$ and $\beta$ represent indices of theoretical parameters,
and we use 
$\mbox{\boldmath $ \mu $}= \left\langle S_{T_{b}}\right\rangle = 0$.
By substituting Eq.(\ref{eq:samplevariance5}) into
Eq.(\ref{eq:likelihoodtrace4}), we can obtain
\begin{eqnarray}
F_{\alpha \beta} &=&
  \frac{1}{2}
       \sum_{i,j,k,l}
       \left( C_{S_{T_{b}}}^{-1} \right)_{ij}
       \left( C_{S_{T_{b}},\alpha} \right)_{jk}
       \left( C_{S_{T_{b}}}^{-1} \right)_{kl}
       \left( C_{S_{T_{b}},\beta} \right)_{li} \nonumber \\
   &=&
   \frac{1}{2}
       \sum_{i,j,k,l}
       \left( \frac{A_{e}}{\lambda^{2}B}
              \frac{\delta_{ij}}{P^{tot}_{T_{b}}(\mbox{\boldmath $u$}_{i})}
       \right)
       \left( \frac{\lambda^{2}B}{A_{e}}
              \delta_{jk} P^{tot}_{T_{b},\alpha}(\mbox{\boldmath $u$}_{j})
       \right)
       \left( \frac{A_{e}}{\lambda^{2}B}
              \frac{\delta_{kl}}{P^{tot}_{T_{b}}(\mbox{\boldmath $u$}_{k})}
       \right)
       \left( \frac{\lambda^{2}B}{A_{e}}
              \delta_{li} P^{tot}_{T_{b},\beta}(\mbox{\boldmath $u$}_{l})
       \right) \nonumber \\
       &=&
       \frac{1}{2}
       \sum_{i}
       \frac{1}{P^{tot}_{T_{b}}(\mbox{\boldmath $u$}_{i})^{2}}
       \frac{\partial P^{tot}_{T_{b}}(\mbox{\boldmath $u$}_{i})}{\partial \theta_{\alpha}}
       \frac{\partial P^{tot}_{T_{b}}(\mbox{\boldmath $u$}_{i})}{\partial \theta_{\beta}},
       \label{eq:21cmFisher1}
\end{eqnarray}
%
%
%
%
where $i,j,k$ and $l$ represent the raws and columns of
$S_{T_{b}i}=S_{T_{b}}(\mbox{\boldmath $u$}_{i})$,
and the numbers are determined by
independent modes of observed data 
with respect to $\mbox{\boldmath $u$}_{i}$.
The summation in the last line 
is the sum of 
$1/(P^{tot}_{T_{b}})^{2} (\partial P^{tot}_{T_{b}}/\partial \theta_{\alpha})
 (\partial P^{tot}_{T_{b}}/\partial \theta_{\beta})$
over all the independent modes in the u-space.

According to Eq.(\ref{eq:power216}),
the power spectrum of 21 cm line 
depends only on $k=|\mbox{\boldmath $k$}|$ and 
$\mu=\frac{k_{\|}}{k}=\frac{k^{3}}{k}$.
%
In other words, the power spectrum is 
determined by only $k_{\perp} \equiv \sqrt[]{k^{2}-k_{\|}^{2}}$ and $k_{\|}$.
Correspondingly, the power spectrum in the u-space
$P_{T_{b}}(\mbox{\boldmath $u$})$
also depends only on $u_{\perp}=\sqrt[]{u^{2}+v^{2}}$ 
and $u_{\|}=\tau$.
In consideration of this symmetry,
we collect up the power spectra $P^{tot}_{T_{b}}(\mbox{\boldmath $u$}_{i})$ 
which have a same value in the u-space.
According to the symmetry, 
we can see that the power spectra 
in an annular region in the u-space have same value.
The volume $dV_{A}$ 
of such annular region $A$ whose ranges are 
from $u_{\perp}$ to $u_{\perp}+\delta u_{\perp}$
and from $u_{\|}$ to $u_{\|}+\delta u_{\|}$
is given by
\begin{eqnarray}
dV_{A}=
\int_{A} d^{3}u
=
\int^{u_{\perp}+\delta u_{\perp}}_{u_{\perp}}
\int^{u_{\|}+\delta u_{\|}}_{u_{\|}}
\int^{2\pi}_{0}
u_{\perp}du_{\perp}du_{\|}d\phi
=2\pi u_{\perp}\delta u_{\perp}\delta u_{\|}.
\end{eqnarray}
Besides, we can express the resolution in the u-space as
\begin{eqnarray}
\delta^{3}u= \frac{1}{V_{\mbox{\boldmath $\Theta$}}},
\end{eqnarray}
where $V_{\Theta}$ is the survey volume in the 
$\mbox{\boldmath $\Theta $}=(\theta^{1},\theta^{2},\Delta \nu)$ 
space.
%
%
%
%
%
Therefore, the number of the independent modes 
in the annular region 
$N_{c}(u_{\perp},u_{\|})$ is given by
\begin{eqnarray}
N_{c}(u_{\perp},u_{\|}) &=& \frac{dV_{A}}
                          {\delta^{3}u} \nonumber \\
                        &=& 2\pi u_{\perp}\delta u_{\perp}\delta u_{\|}V_{\mbox{\boldmath $\Theta$}} \\
&=& 2 \pi k_\perp \delta k_\perp \delta k_\parallel 
\frac{V(z)} {(2\pi)^3},
\end{eqnarray}
where $V(z)$ is the volume of the real space.
The survey volume in the $\Theta$ space is given by
$V_{\Theta} = B \times {\rm FoV}$,
where $B$ is the bandwidth and  
FoV $\propto \lambda^{2}$ is the field of view of 
an interferometer,
and the volume of the real space is also given by
$V(z) = d_A(z)^2 y(z) V_{\Theta}$.
%
Additionally, according to 
the symmetry $\mu \longrightarrow -\mu$,
there is also a symmetry $u_{\|} \longrightarrow -u_{\|}$.
In consideration of these symmetries,
we can rewrite the Fisher matrix Eq.(\ref{eq:21cmFisher1})
as \cite{McQuinn:2005hk},
\begin{eqnarray}
F_{\alpha\beta} &= &
       \frac{1}{2}
       \sum_{i}
       \frac{1}{P^{tot}_{T_{b}}(\mbox{\boldmath $u$}_{i})^{2}}
       \frac{\partial P^{tot}_{T_{b}}(\mbox{\boldmath $u$}_{i})}{\partial \theta_{\alpha}}
       \frac{\partial P^{tot}_{T_{b}}(\mbox{\boldmath $u$}_{i})}{\partial \theta_{\beta}}
\nonumber \\
&=&       
       \frac{1}{2}
       \sum_{{\rm pixel}} 2N_{c}(u_{\perp},u_{\|})
       \frac{1}{P^{tot}_{T_{b}}(u_{\perp},u_{\|})^{2}}
       \frac{\partial P^{tot}_{T_{b}}(u_{\perp},u_{\|})}{\partial \theta_{\alpha}}
       \frac{\partial P^{tot}_{T_{b}}(u_{\perp},u_{\|})}{\partial \theta_{\beta}}
\nonumber \\
 &=&      
       \sum_{{\rm pixel}}
       \frac{1}{ \left[
                \frac{P^{tot}_{T_{b}}(u_{\perp},u_{\|})}{ \sqrt[]{N_{c}(u_{\perp},u_{\|})}}
                \right]^{2}}
       \frac{\partial P^{tot}_{T_{b}}(u_{\perp},u_{\|})}{\partial \theta_{\alpha}}
       \frac{\partial P^{tot}_{T_{b}}(u_{\perp},u_{\|})}{\partial \theta_{\beta}},
\label{eq:21cmFisher2}
\end{eqnarray}
where $\sum_{{\rm pixel}}$
means the summation of $P_{T_{b}}$ in the $u_{\perp}-u_{\|}$ plane.
Note that 
we need to sum over only the region of positive $u_{\|}$
because we have already taken account of 
the symmetry $u_{\|} \longrightarrow -u_{\|}$ in Eq.(\ref{eq:21cmFisher2}).
On the other hand,
$u_{\perp}$ is originally positive
by its definition $u_{\perp}=\sqrt[]{u^{2}+v^{2}}$.

In our analysis, to be conservative, 
when we differentiate $P_{T_{b}}({\bm u})$ with respect to 
cosmological parameters, we fix ${\cal P}_{\delta \delta}(k)$ 
in Eqs.~(\ref{eq:Pxx})~and~(\ref{eq:Pxdelta}) 
so that constraints only come from the 
${\cal P}_{\delta \delta}(k)$ terms in 
$P_{\mu^0},P_{\mu^2},P_{\mu^4}$. 
Additionally, 
we treat the parameters related to 
$\mathcal{P}_{x\delta}$ and $\mathcal{P}_{xx}$
($\bar{x}_{HI}, b_{xx}^2, b_{x\delta}^2, 
\alpha_{xx}, \alpha_{x\delta}, \gamma_{xx}, R_{xx}, R_{x\delta}$)
in same manner as the other cosmological parameters.
In other words,
they are also treated as theoretical parameters $\theta_{\alpha}$
in our analysis.

\section{Specifications of the experiments}\label{sec:21cm_spec}

Now in this section, we show the specifications 
of 21 cm line observations which are considered in this thesis.


\subsubsection{\underline{Survey range}}

In our analyses, we consider the redshift range $ z = 6.75 - 10.05$,
which we divide into 4 bins: $z = 6.75 - 7.25, 7.25 - 7.75, 7.75-8.25$
and $8.25 -  10.05$.  For the wave number, 
we set its minimum cut off $k_{{\rm min} \parallel} = 2 \pi / (y B) $ to avoid foreground
contaminations \cite{McQuinn:2005hk},
and take its maximum value $ k_{\rm max} = 2~{\rm Mpc}^{-1}$ 
in order not to be affected by nonlinear effect which
becomes important on $k \ge k_{\rm max}$. 
For methods of foreground removals, see also recent discussions about
the independent component analysis (ICA) algorithm,
FastICA~\cite{Chapman:2012yj} which will be developed in terms of the
ongoing LOFAR observation~\cite{LOFAR}.

\subsubsection{\underline{Noise power spectrum}}

From Eq.(\ref{eq:noisevariance2}), the noise power spectrum of an interferometer 
is given by
\begin{equation}
P_N (u_{\perp}) 
= \left( \frac{\lambda^{2} (z) T_{\rm sys} (z)  }{A_e (z)} \right)^2 
\frac{1}{t_0 n(u_\perp)},
\end{equation}
%
%
%
where, the system temperature $T_{\rm sys}$ is estimated as
$T_{\rm sys} = T_{{\rm sky}} + T_{{\rm rcvr}}$,
and it is dominated by the sky temperature due to synchrotron radiation.
Here, $T_{{\rm sky}} = 60 (\lambda/[m])^{2.55} $ [K] 
is the sky temperature, and
$T_{{\rm rcvr}} = 0.1T_{{\rm sky}} + 40 $[K] 
is the receiver noise \cite{SKA}.
In addition,
the  effective collecting area is proportional to the square of the 
observed wave length $A_e \propto \lambda^{2} $.
The number density of the baseline $n(u_\perp)$ 
depends on an actual realization of antenna distribution.

To obtain the future cosmological constraints from 21 cm experiments,
we consider SKA (phase1, phase2) \cite{SKA,Mellema:2012ht} 
and Omniscope \cite{Tegmark:2009kv,Tegmark:2008au},
whose specifications are shown in Table~\ref{tab:21obs}.
In the analysis of  the total neutrino mass,
the neutrino number of species and the neutrino mass
hierarchy (in the Chapter \ref{Chap:result_mass}), 
we only estimate the sensitivity of SKA.
In that of the lepton asymmetry of the Universe
(in the Chapter \ref{chap:result_lepton}),
we take account of both the experiments.
In order to calculate the number density of baseline $n(u_\perp)$,
we assume a realization of antenna distributions for these arrays as follows.
For SKA phase1, we take 95\% (866) of the total antennae (stations)
distributed with a core region of radius 3000 m.
The distribution has an antenna density profile $\rho(r)$ ($r$: a radius from center of the array) as follows \cite{Kohri:2013mxa},
\begin{equation}
\rho(r) = 
\left\{
\begin{array}{lll}
 \rho_{0}r^{-1},     &\rho_{0} \equiv \frac{13}{16\pi\left(\sqrt{10}-1\right) }  \ {\rm m}^{-2}
& \hspace{60pt} r \leq 400 \ {\rm m},\\
 \rho_{1}r^{-3/2},  &\rho_{1} \equiv \rho_{0} \times 400^{1/2}, & \ \ \ 400 \ {\rm m} < r \leq 1000 \ {\rm m}, \\
 \rho_{2}r^{-7/2},  &\rho_{2} \equiv \rho_{1} \times 1000^{2}, & \ \ 1000 \ {\rm m} < r \leq 1500 \ {\rm m}, \\
 \rho_{3}r^{-9/2},  &\rho_{3} \equiv \rho_{2} \times 1500 ,          & \ \ 1500 \ {\rm m} < r \leq 2000 \ {\rm m}, \\
 \rho_{4}r^{-17/2},&\rho_{4} \equiv \rho_{3} \times 2000^{4}, & \ \ 2000  \ {\rm m} < r \leq 3000 \ {\rm m}. \\
\end{array}
\right.
\end{equation}
This distribution agrees with the specification of the SKA phase1 
baseline design.
We ignore measurements from the sparse distribution of 
the remaining 5\% of the
total antennae that are outside this core region.
For SKA phase2, we assume that it has the 10 times larger
 total collecting area than the phase1.
Hence, we take its noise power spectrum  to be 1/100 of the phase1.
We assume that the other specifications of SKA phase2 are 
the same as values of the phase1.
For Omniscope , which is a future square-kilometer collecting area array
optimized for 21 cm tomography, we take all of antennae distributed 
with a filled nucleus in the same manner as Ref. \cite{Mao:2008ug}.
In addition, we assume an azimuthally symmetric distribution of
the antenna in both arrays.

\begin{table}[t]
  \centering 
  \begin{tabular}{cccccccc}
\hline \hline
Experiment & $N_{\rm ant}$& $A_e (z=8)$& $L_{\rm min}$
   & $L_{\rm max}$& ${\rm FOV}(z=8)$ & $t_{0}$&  $z$ \\ 
   & &$[{\rm m}^2]$&$[{\rm m}]$
   & $[{\rm km}]$ &$[{\rm deg}^2]$&[hour]&\\
\hline
SKA  phase1     & $911$ & $443$ & $35$ & $6$ & $13.12 $  
& 4000 & $6.8-10$ \\
Omniscope &  $10^6$ & $1$   & $1$    & $1$   
& $2.063\times 10^4$ & 16000 & $6.8-10$  \\ 
\hline \hline
\end{tabular}
\caption{
Specifications for 21 cm line experiments adopted 
in the analysis.
For Omniscope, we assume that the effective collecting area $A_{e}$ and 
field of view are fixed.
For SKA phase2, we assume that 
the number of antennae is 10 times as many as phase1.
Hence, we take its noise power spectrum to be 1/100 of the phase1, 
and the other specifications to be the same values.
Additionally, for SKA, we assume that 
it uses 4 multi-beaming \cite{Mellema:2012ht},
and only in the analysis of the lepton asymmetry of the Universe
(in the Chapter \ref{chap:result_lepton}),
its total observation time is the same value as that of Omniscope 
(16000 hours), but it observes 4 places in the sky 
(i.e. 4 times larger FOV and one fourth $t_{0}$.
Namely, the effective field of views are
${\rm FoV_{SKA}} = 13.21 \times 4 [{\rm deg}^2]$
in the analysis of the Chapter \ref{Chap:result_mass},
and ${\rm FoV_{SKA}} = 13.21 \times 4 \times 4 [{\rm deg}^2]$
in that of the Chapter \ref{chap:result_lepton}.
\label{tab:21obs}
}
\end{table}


\chapter{Fisher information matrix of cosmic microwave background (CMB)}
\label{chap:CMBFisher}

\section{CMB and neutrino properties}

In this thesis, 
we focus on not only the observations of the 21 cm line 
but also the CMB observations, especially CMB B-mode polarization 
produced by gravitational lensing of the matter fluctuation.
%
Although the 21 cm line observation is 
a power probe of the matter power spectrum, 
particularly, on small scales, 
observations of CMB greatly help to determine 
other cosmological parameters 
%
such as energy densities of the dark matter, 
baryons and dark energy.

Besides, CMB power spectra are sensitive to neutrino masses
through the CMB lensing.
Future precise CMB experiments are expected to 
set stringent constraints on
the sum of the neutrino masses and
the effective number of neutrino species~\cite{Wu:2014hta,Abazajian:2013oma}. 
%
%
Therefore, we propose to combine the CMB experiments with 
the 21 cm line observations.  

\section{Fisher information matrix of CMB}

We evaluate errors of cosmological parameters
by using the Fisher matrix of CMB,
which is given by~\cite{Tegmark:1996bz}.
The variance-covariance matrix of CMB is given by
\begin{eqnarray}
    C_{{\rm CMB}}= \delta_{\ell \ell'}\bm{C}_{\ell}
    = \left(
\begin{array}{ccc}
\bm{C}_{2} &            &        \\
           & \bm{C}_{3} &        \\
           &            & \ddots 
\end{array}
\right),
\hspace{10pt}
\bm{C}_{\ell}=  \left(
\begin{array}{ccc}
\hspace{5pt} \bm{C}_{\ell}^{\mathrm{TT}} \hspace{5pt} &
\hspace{5pt} \bm{C}_{\ell}^{\mathrm{TE}} \hspace{5pt} &
\hspace{5pt} \bm{C}_{\ell}^{\mathrm{Td}} \hspace{5pt} \\
\hspace{5pt} \bm{C}_{\ell}^{\mathrm{TE}} \hspace{5pt} &
\hspace{5pt} \bm{C}_{\ell}^{\mathrm{EE}} \hspace{5pt} &
0 \hspace{5pt} \\
\hspace{5pt} \bm{C}_{\ell}^{\mathrm{Td}} \hspace{5pt} &
0 \hspace{5pt} &
\hspace{5pt} \bm{C}_{\ell}^{\mathrm{dd}} \hspace{5pt}
\end{array}
\right),
\end{eqnarray}
where $\ell$ is the multipole of angular power spectra of CMB 
$\ell=2,3,\cdots$,
and $\bm{C}_{\ell}^{\mathrm{Y}}$ are the following $2\ell + 1$ 
diagonal matrices,
\begin{eqnarray}
\bm{C}_{\ell}^{\mathrm{Y}} 
= \delta_{m m'}C^{X}_{\ell}
= \left(
\begin{array}{ccc}
C^{Y}_{\ell} + N^{Y}_{\ell} &            &        \\
              &  \ddots       &                     \\
              &               & C^{Y}_{\ell} + N^{Y}_{\ell}
\end{array}
\right),
\end{eqnarray}
where $C_{\ell}^{\mathrm{Y}} 
\left(\mathrm{Y}=\mathrm{TT, EE, TE, Td, dd} \right)$ 
are the CMB power spectra 
(auto, cross correlations or deflection angle),
and $N^{Y}_{\ell}$ $\left(\mathrm{X}=\mathrm{TT, EE, dd} \right)$
are noise power spectra.
Therefore, by using the definition of the
Fisher matrix, we can obtain the following 
Fisher matrix of CMB,
\begin{equation}
F_{\alpha \beta}^{\rm (CMB)}
= \sum_{\ell}
\frac{\left( 2\ell+1\right)}{2}
\mathrm{Tr}
\left[
  C_{\ell}^{-1}
  \frac{\partial C_{\ell}}{\partial \theta_{\alpha}}
  C_{\ell}^{-1}
  \frac{\partial C_{\ell}}{\partial \theta_{\beta}}
\right],\label{eq:Fisher_CMB}
\end{equation}
\begin{align}
    C_{\ell} =  \left(
\begin{array}{ccc}
\hspace{5pt} C_{\ell}^{\mathrm{TT}} 
+ N_{\ell}^{\mathrm{TT}} \hspace{5pt} &
\hspace{5pt} C_{\ell}^{\mathrm{TE}} \hspace{5pt} &
\hspace{5pt} C_{\ell}^{\mathrm{Td}} \hspace{5pt} \\
\hspace{5pt} C_{\ell}^{\mathrm{TE}} \hspace{5pt} &
\hspace{5pt} C_{\ell}^{\mathrm{EE}} 
+N_{\ell}^{\mathrm{EE}} \hspace{5pt} &
0 \hspace{5pt} \\
\hspace{5pt} C_{\ell}^{\mathrm{Td}} \hspace{5pt} &
0 \hspace{5pt} &
\hspace{5pt} C_{\ell}^{\mathrm{dd}} 
+ N_{\ell}^{\mathrm{dd}} \hspace{5pt}
\end{array}
\right).
\end{align}
Here $C_{\ell}^{\mathrm{X}} 
\left(\mathrm{X}=\mathrm{TT, EE, TE} \right)$
are the CMB power spectra,
$C_{\ell}^{\mathrm{dd}}$ is the deflection angle spectrum,
$C_{\ell}^{\mathrm{Td}}$ is the
cross correlation between the deflection angle and the temperature,
$N_{\ell}^{\mathrm{X'}}$ 
$\left(\mathrm{X'}=\mathrm{TT, EE} \right)$
and $N_{\ell}^{\mathrm{dd}}$
are the noise power spectra, 
where $C_{\ell}^{\mathrm{dd}}$ is calculated by a lensing
potential~\cite{Okamoto:2003zw} and is related with
the lensed CMB power spectra.
The noise power spectra of CMB 
$N_{\ell}^{\mathrm{X'}}$ are expressed with a beam size
$\sigma_{\mathrm{beam}}(\nu)=$ $\theta
_{\mathrm{FWHM}}(\nu)/\sqrt{8\ln 2}$ and instrumental sensitivity
$\Delta _{\mathrm{X'}}(\nu)$ by
\begin{eqnarray}
  N_{\ell}^{\mathrm{X'}}
  = \left[ 
    \sum_{i} \frac1{n_{\ell}^{\mathrm{X'}}(\nu_{i})}
    \right]^{-1},
\end{eqnarray}
where $\nu_{i}$ is an observing frequency and
\begin{align}
n_{\ell}^{\mathrm{X'}}(\nu)=
\Delta ^2_{\mathrm{X'}}(\nu)\exp\left[\ell(\ell+1) \sigma ^2_{\mathrm{beam}}(\nu)\right].
\end{align}
%
The noise power spectrum of deflection angle 
$N^{dd}_l$ is estimated assuming lensing reconstruction with the
quadratic estimator \cite{Okamoto:2003zw},
which is computed with FUTURCMB \cite{paper:FUTURCMB}.
In this algorithm,
%
$N_{\ell}^{dd}$ is estimated from
the noise $N_{\ell}^{X'}$, 
and 
lensed and unlensed power spectra of CMB temperature, 
E-mode and B-mode polarizations.

%
Finally, the Fisher matrix in Eq.(\ref{eq:Fisher_CMB}) is modified as follows 
by taking the multipole range 
[$\ell_{min}$, $\ell_{max}$] 
and the fraction of the observed sky $f_{{\rm sky}}$ into account,
\begin{equation}
F_{\alpha \beta}^{\rm (CMB)}
= \sum_{\ell = \ell_{min}}^{\ell_{max}}
\frac{\left( 2\ell+1\right)}{2}f_{\mathrm{sky}}
\mathrm{Tr}
\left[
  C_{\ell}^{-1}
  \frac{\partial C_{\ell}}{\partial \theta_{\alpha}}
  C_{\ell}^{-1}
  \frac{\partial C_{\ell}}{\partial \theta_{\beta}}
\right].
\end{equation}
%

\section[Residual foregrounds]
{Residual foregrounds 
\normalsize{\cite{Baumann:2008aq,Verde:2005ff}}}

%
%
We consider synchrotron radiation 
and dust emission in our galaxy 
as the dominant sources of foregrounds.
%
%
These foregrounds are subtracted from each sky pixel of CMB map.
%
%
%
Here, we assume that 
foreground subtraction can be performed at 
a certain level 
($1\%$ level in the power spectra of CMB).
%
%
%
We then model the residual foregrounds in the CMB maps. 
Note that 
we only consider the residual foreground of 
CMB polarization maps as distinct from temperature.
That is because it has already been precisely measured 
by WMAP and Planck.

We model the synchrotron 
$C_{\ell}^{S, X}$ and 
dust $C_{\ell}^{D, X}$ foregrounds as
\begin{eqnarray}
C_{\ell}^{S, X}(\nu) 
&=&
A_{S}
\left(
\frac{\nu}{\nu_{S,0}}
\right)^{2\alpha_{S}}
\left(
\frac{\ell}{\ell_{S,0}}
\right)^{\beta_{S}}, \label{eq:synchrotron_rad} \\
C_{\ell}^{D, X}(\nu) 
&=& 
p^2 A_{D}
\left(
\frac{\nu}{\nu_{D,0}}
\right)^{2\alpha_{D}}
\left(
\frac{\ell}{\ell_{D,0}}
\right)^{\beta_{D}^{X}}
\left[
\frac{e^{h\nu_{D,0}/k_{B}T}-1}{e^{h\nu/k_{B}T}-1}
\right]^2,
\end{eqnarray}
where $X={\rm EE, TE, BB}$,
$\alpha_{S}=-3$, $\beta_{S}=-2.6$,
$\nu_{S,0}=30$ GHz, $\ell_{S,0}=350$,
$A_{S}=4.7\times10^{-5}\mu{\rm K}^2$,
$\alpha_{D}=2.2$, $\nu_{D,0}=94$ GHz,
$\ell_{D,0}=10$, $A_{D}=1.0 \mu {\rm K}^2$,
$\beta_{D}^{X}=-2.5$
and $p$ is the dust polarization fraction $p=5\%$.
These choices are the values used in the Refs.~\cite{Baumann:2008aq,Verde:2005ff},
and match the parameters of foregrounds observed 
by WMAP \cite{Page:2006hz}, DASI \cite{Leitch:2004gd} 
and IRAS \cite{Finkbeiner:1999aq}.
%
We then assume that residual foregrounds are modeled as follows,
\begin{eqnarray}
C^{X,{\rm RFg}}_{\ell}(\nu)=
\left[
C_{\ell}^{S,X}(\nu) + C_{\ell}^{D,X}(\nu)
\right]
\sigma^{{\rm RFg}}_{{\rm CMB}} + n_{\ell}^{{\rm RFg},X}(\nu),
\end{eqnarray}
where 
$\sigma^{{\rm RFg}}_{{\rm CMB}}$ is the foreground residual parameter
of CMB observations.
We assume $\sigma^{{\rm RFg}}_{{\rm CMB}}=0.01$ 
(this value corresponds to 10\% in the CMB maps),
and $n_{\ell}^{{\rm RFg},X}$ is the noise power spectrum
of the foreground template maps,
which is created by taking map differences and 
thus are somewhat affected by the instrumental noise.
We assume that
this noise power spectrum of the template maps is given by
\begin{eqnarray}
n_{\ell}^{{\rm RFg},X}(\nu) = 
\frac{n_{\ell}^{X}(\nu)}{N_{{\rm chan}}(N_{{\rm chan}}-1)/4}
\left\{
\left(
\frac{\nu}{\nu_{S,{\rm ref}}}
\right)^{2\alpha_{S}}
+
\left(
\frac{\nu}{\nu_{D,{\rm ref}}}
\right)^{2\alpha_{D}}
\right\},
\end{eqnarray}
where $N_{{\rm chan}}$ is the total frequency channels
which are used for the foregrounds removal,
$\nu_{D,{\rm ref}}$ and $\nu_{S,{\rm ref}}$ are
the highest and lowest frequency channel included 
in the foregrounds removal, respectively.
Furthermore, we introduce the following 
effective noise power spectrum 
including the residual foregrounds,
\begin{eqnarray}
N^{{\rm eff},X}_{\ell}
=
\left[
\sum_{i}
\frac{1}{n^{X}_{\ell}(\nu_{i}) + C^{X,{\rm RFg}}_{\ell}(\nu_{i})}
\right]^{-1},
\end{eqnarray}
%
where $i$ means the frequency band.
When we include the effects due to 
the residual foregrounds in our analysis,
we use this effective noise as the CMB noise power spectrum.

By making the modifications given above
to FUTURCMB \cite{paper:FUTURCMB},
we calculate the estimated errors of the deflection
angle of CMB and use them in our Fisher matrix analysis.

\section{Specifications of the experiments}

Now in this section, we show the specifications 
of the observations of CMB
which are considered in this thesis.

\subsection{Analysis of the neutrino mass 
and the mass hierarchy}

In the analysis of the total neutrino mass,
the number of the neutrino species,
and the mass hierarchy (in the Chapter \ref{Chap:result_mass}),
in order to obtain the future constraints, 
we consider Planck~\cite{Planck:2006aa},
\textsc{Polarbear}-2 and Simons Array,
whose experimental specifications 
are summarized in Table~\ref{tab:cmb_obs_mass}.
The latter two experiments are 
ground-based precise CMB polarization observations.

For the analysis about Planck and \textsc{Polarbear}-2 or Simons Array, 
we combine both the experiments, 
%
and assume that a part of the whole sky $(f_{{\rm sky}}\times 100\%)$ 
is observed by both the experiments, 
and the remaining observed region $(65\%-f_{{\rm sky}}\times 100\%)$ 
is observed only by Planck. 
Therefore, we evaluate a total Fisher matrix of CMB
$F^{({\rm CMB})}$ by summing the two Fisher matrices,
\begin{align}
F^{({\rm CMB})} = F^{({\rm Planck})}(65\%-f_{{\rm sky}}\times 100\%)
+F^{({\rm Planck + PB-2 \ or \ SA})}(f_{{\rm sky}}\times 100\%),
\end{align}
%
%
where 
$F^{({\rm Planck + PB-2 \ or \ SA})}$
is the Fisher matrix of the 
region observed by both Planck and \textsc{Polarbear}-2 (PB-2)
or Simons Array (SA), 
and $F^{({\rm Planck})}$ is that by Planck only.

In addition, we calculate noise power spectra 
$N_{\ell}^{\mathrm{X,Planck+PB-2 \ or \ SA}}$ of 
the CMB polarization ($X = {\rm EE}$ or ${\rm BB}$)
in $F^{\mathrm{Planck+PB-2 \ or \ SA}}$ with 
the following operation. 
\begin{description}
\item[(1)] $2\leq \ell < 25$
\begin{align}
N_{\ell}^{\mathrm{X,Planck + PB-2 \ or \ SA}} 
= N_{\ell}^{\mathrm{X,Planck}}
\end{align}
\item[(2)] $25\leq \ell \leq 3000$
\begin{align}
N_{\ell}^{\mathrm{X,Planck + PB-2 \ or \ SA}} 
= [1/N_{\ell}^{\mathrm{X,Planck}} 
+ 1/N_{\ell}^{\mathrm{X,PB-2 \ or \ SA}}]^{-1}
\end{align}
\end{description}
Since we assume that 
the CMB temperature fluctuation observed by
\textsc{Polarbear}-2 or Simons Array 
is not used for constraints on the cosmological parameters,
the temperature noise power spectrum 
$N_{\ell}^{\mathrm{TT,\,Planck+PB-2 \ or \ SA}}$ is equal to 
{\bf $N_{\ell}^{\mathrm{TT,\,Planck}}$}.
This reason is that the CMB temperature fluctuation 
observed by Planck reaches almost cosmic variance limit.
Therefore, the constraints are not strongly improved
if we include the CMB temperature fluctuation
observed by \textsc{Polarbear}-2 or Simons Array.

\begin{table}[ht]
\begin{center}
\begin{tabular}{c|ccccccc}
\hline
\hline
\shortstack{Experiment\\ \,}&
\shortstack{$\nu$ \\ $[\mathrm{GHz}]$}&
\shortstack{$\Delta _{\mathrm{TT}}$\\ $[\mathrm{\mu K-'}]$}&
\shortstack{$\Delta _{\mathrm{PP}}$\\ $[\mathrm{\mu K-'}]$}& 
\shortstack{$\theta_{\mathrm{FWHM}}$\\ $[\mathrm{-'}]$}&
\shortstack{$f_{\mathrm{sky}}$\\ $ $} &
\shortstack{$\ell_{{\rm min}}$\\ $ $} &
\shortstack{$\ell_{{\rm max}}$\\ $ $} 
\\
\hline
\hline
Planck & 30 & 145 & 205  & 33  &      &   &      \\ 
       & 44 & 150 & 212  & 23  &      &   &      \\ 
       & 70 & 137 & 195  & 14  &      &   &      \\
       & 100& 64.6& 104  & 9.5 & 0.65 & 2 & 3000 \\
       & 143& 42.6& 80.9 & 7.1 &      &   &      \\
       & 217& 65.5& 134  & 5   &      &   &      \\ 
       & 353& 406 & 406  & 5   &      &   &      \\ 
\hline
\textsc{Polarbear}-2  & 95  & - & 3.09 & 5.2 & 0.016 & 25 & 3000 \\ 
$f_{{\rm sky}}=0.016$ & 150 & - & 3.09 & 3.5 &       &    &      \\
\hline
\textsc{Polarbear}-2  & 95  & - & 10.9 & 5.2 & 0.2 & 25 & 3000 \\ 
$f_{{\rm sky}}=0.2$   & 150 & - & 10.9 & 3.5 &     &    &      \\
\hline
Simons Array          & 95  & - & 2.18& 5.2 &       &    &      \\ 
$f_{{\rm sky}}=0.016$ & 150 & - & 1.78& 3.5 & 0.016 & 25 & 3000 \\
                      & 220 & - & 4.72& 2.7 &       &    &      \\
\hline
Simons Array          & 95  & - & 7.72& 5.2 &       &    &      \\ 
$f_{{\rm sky}}=0.2$   & 150 & - & 6.30& 3.5 & 0.2   & 25 & 3000 \\
                      & 220 & - & 16.7& 2.7 &       &    &      \\
                      
\hline
\hline
\end{tabular}
\caption{
Experimental specifications of Planck, \textsc{Polarbear}-2
and Simons Array assumed in the analysis. 
Here $\nu$ is the observation frequency, 
$\Delta_{\rm TT}$ is the temperature sensitivity per $1'\times1'$
pixel, $\Delta_{\rm PP}=\Delta_{\rm EE}=\Delta_{\rm BB}$ 
is the polarization (E-mode and B-mode) sensitivity 
per $1'\times1'$ pixel,
$\theta_{\rm FWHM}$ is the angular resolution defined as the full width at
half-maximum, and $f_{\rm sky}$ is the observed fraction of the sky.
For Planck experiment, we assume that 
the three frequency bands ($70, 100, 143$ GHz)
are only used for the observation of CMB,
and the other bands ($30,44,217,353$ GHz) for foregrounds removal.
For Simons Array, we consider two situations:
One situation is that 220 GHz band is used for the observation of CMB,
and the other is that the band is used for the foreground removal.
}
\label{tab:cmb_obs_mass}
\end{center}
\end{table}

\subsection{Analysis of the lepton asymmetry}

In the analysis of the lepton asymmetry of the
Universe (in the Chapter \ref{chap:result_lepton}),
to obtain the future constraints, 
we consider Planck~\cite{Planck:2006aa}
and CMBPol~\cite{Baumann:2008aq},
whose experimental specifications  
are summarized in Table~\ref{tab:cmb_obs_lepton}.


\begin{table}[tbp]
  \centering
\begin{tabular}{c|ccccccc}
\hline
\hline
\shortstack{Experiment\\ \,}&
\shortstack{$\nu$ \\ $[\mathrm{GHz}]$}&
\shortstack{$\Delta _{\mathrm{TT}}$\\ $[\mathrm{\mu K-'}]$}&
\shortstack{$\Delta _{\mathrm{PP}}$\\ $[\mathrm{\mu K-'}]$}& 
\shortstack{$\theta_{\mathrm{FWHM}}$\\ $[\mathrm{-'}]$}&
\shortstack{$f_{\mathrm{sky}}$\\ $ $} &
\shortstack{$\ell_{{\rm min}}$\\ $ $} &
\shortstack{$\ell_{{\rm max}}$\\ $ $} 
\\
\hline
\hline
Planck
& 100 & 64.6 & 104  & 9.5 &  & \\
& 143 & 42.6 & 80.9 & 7.1 & 0.65 & 2 & 3000 \\ 
& 217 & 65.5 & 134  & 5   &  & \\ 
\hline
CMBpol 
& 45   & 5.85   & 8.27 & 17  &   & \\
& 70   & 2.96   & 4.19 & 11  &   & \\
& 100  & 2.29   & 3.24 & 8   & 0.65 & 2 &3000\\
& 150  & 2.21   & 3.13 & 5   &   & \\
& 220  & 3.39   & 4.79 & 3.5 &   & \\
\hline \hline
\end{tabular} 
\caption{
Specifications for Planck and CMBpol adopted in the analysis. 
For CMBpol, we assumed the mid-cost (EPIC-2m)  mission and 
only used five frequency bands 
for a realistic foreground removal.
}
\label{tab:cmb_obs_lepton}
\end{table}



\chapter{Fisher information matrix of baryon acoustic oscillation (BAO) observations}
\label{chap:BAOFisher}

In this chapter, we briefly review
analysis methods about 
the baryon acoustic oscillation (BAO).
In the early Universe,
baryons and photons are strongly coupled
and their fluctuations (Fourier components) of 
the mixed fluid oscillate by the pressure of radiation.
At the time of the decouple between them,
a characteristic peak feature remains at
the sound horizon.
The scale can be used for a standard ruler of distance. 
Therefore, we can get the information of 
the distance and the Hubble expansion rate
by measurements of the BAO scale for matter fluctuations.
In this thesis, we especially consider galaxy survey 
for the BAO observation.

\section[Fisher matrix of BAO]
{Fisher matrix of BAO \normalsize{\cite{Albrecht:2006um}}}

In this section, we introduce the Fisher matrix of BAO experiments.
The observables of BAO are the comoving angular diameter
distance $d_{A}(z)$ and the Hubble parameter $H(z)$
(and more specifically, $\ln(d_{A}(z))$ and $\ln(H(z))$
are the observables).
For the observables, the Fisher matrix is given by 
\begin{eqnarray}
F^{({\rm BAO}) \ d,H}_{\alpha \beta} &=&
\sum_{i} \frac{1}{\sigma_{d,H}^2(z_{i})+(\sigma_{s}^i)^{2}} 
\frac{\partial f_{i}^{d,H}}{\partial \theta_{\alpha}}
\frac{\partial f_{i}^{d,H}}{\partial \theta_{\beta}}, \\
f_{i}^{d} &=& \ln(d_{A}(z_{i})), \\
f_{i}^{H} &=& \ln(H(z_{i})),
\end{eqnarray}
where $\sigma_{d}(z_{i})$ and $\sigma_{H}(z_{i})$
are the variances of $\ln(d_{A}(z_{i}))$
and $\ln(H(z_{i}))$ in the BAO observation respectively,
$\sigma_{s}^{i}$ is the error of the systematics,
%
and we assume that 
the observed redshift range is divided into bins,
whose width and central redshift values
are $\Delta z_{i}$ and $z_{i}$, respectively.
Here, $i$ is the index of the redshift bins.

The variances of  $\ln(d_{A}(z_{i}))$
and $\ln(H(z_{i}))$ are determined by the fitting formulae
of BAO presented by \cite{Blake:2005jd}, and they are given by
\begin{eqnarray}
\sigma_{d}(z_{i}) 
&=&
x^{d}_{0} \frac{4}{3}
\sqrt{\frac{V_{0}}{V_{i}}}
f_{nl}(z_{i}), \\
\sigma_{d}(z_{i})
&=&
x^{H}_{0} \frac{4}{3}
\sqrt{\frac{V_{0}}{V_{i}}}
f_{nl}(z_{i}),
\end{eqnarray}
Here, $V_{i}$ is the comoving survey volume
and expressed as
\begin{eqnarray}
V_{i} =
\frac{(d_{A}(z_{i}))^{2}}{H(z_{i})}\Omega_{{\rm sky}}\Delta z_{i},
\end{eqnarray}
where $\Omega_{{\rm sky}}$ is the survey solid angle.
$f_{nl}(z_{i})$ is the non-linear evolution factor,
which represents the erasure of baryon oscillation features by the non-linear evolution of density fluctuations.
In our analysis,
we use the following function as $f_{nl}(z_{i})$,
\begin{eqnarray}
f_{nl}(z_{i}) = \left\{
\begin{array}{c}
1 \hspace{40pt} z>z_{m},\\
\left(
\frac{z_{m}}{z_{i}}
\right)^{\gamma} \hspace{10pt} z<z_{m}.
\end{array}
\right.
\end{eqnarray}
where $z_{m}$ is the redshift at which the improvement in the baryon oscillation 
accuracy saturates (for fixed survey volume and number density).
Additionally, in the analysis of the BAO observation,
we use the following parameters,
\begin{subequations}
\begin{eqnarray}
x_{0}^{d}&=&0.0085, \\
x_{0}^{H}&=&0.0148, \\
V_{0}    &=& \frac{2.16}{h^3} {\rm Gpc}^3, \\
\gamma &=& \frac{1}{2}, \\
z_{m}  &=& 1.4,
\end{eqnarray}
\end{subequations}
where $h\equiv H_{0}/(100{\rm km}/s/{\rm Mpc})$ 
is the dimensionless Hubble parameter.
According to \cite{Albrecht:2006um},
we assume the following systematic error,
\begin{eqnarray}
\sigma_{s}^{i} = 0.01 \times \sqrt{\frac{0.5}{\Delta z_{i}}}.
\end{eqnarray}

The set of cosmological parameters 
related to the BAO observation
are only $(\Omega_{m}h^2, \Omega_{\Lambda})$
or $(h, \Omega_{\Lambda})$
when we assume that the Universe is flat and
the dark energy is the cosmological constant.

\section{Specification of the BAO observation}

We estimate the sensitivity of BAO observation
only in the analysis of the neutrino mass, 
the number of neutrino species and the mass hierarchy.
In the analysis, we focus on the 
Dark Energy Spectroscopic Instrument (DESI) \cite{DESI:web,Font-Ribera:2013rwa},
which is a future large volume galaxy survey. 
The survey redshift range is 
$0.1<z<1.9$ 
(we do not include 
the Ly-$\alpha$ forest at $1.9<z$ for simplicity)
and the sold angle is
$\Omega_{{\rm sky}}=14000 [{\rm deg}^2]$.
In our analysis, we divide the redshift range
into 18 bins, in other words $\Delta z_{i}=0.1$ \cite{Wu:2014hta}.

Additionally,
in the same manner as \cite{Wu:2014hta},
when we combine BAO with the other observations,
we put 1\% prior on the present Hubble parameter $H_{0}$,
which is achievable in the next decade.
The Fisher matrix of the Hubble prior is given by 
\begin{eqnarray}
F^{(H_{0} \ {\rm prior})}_{\theta_{\alpha}\theta_{\beta}} 
= \left\{
\begin{array}{c}
 \hspace{-40pt} \frac{1}{(1\%\times H_{0,{\rm fid}})^2}, 
 \hspace{30pt} \theta_{\alpha}=\theta_{\beta}=H_{0},\\
 \hspace{20pt}0,
 \hspace{60pt} {\rm the \ other \ components},
\end{array}
\right.
\end{eqnarray}
where $H_{0,{\rm fid}}$ is the fiducial value of $H_{0}$.
%
If we choose the Hubble parameter as 
a dependent parameter,
it is necessary to translate the Fisher matrix
into that of the chosen parameter space.
Under the transformation of a parameter space
$\theta \longrightarrow \tilde{\theta}$,
the translated Fisher matrix is give by \cite{Albrecht:2006um}
\begin{eqnarray}
\tilde{F}_{l,m} = 
\frac{\partial \theta_{j}}{\partial \tilde{\theta}_{l}}
\frac{\partial \theta_{k}}{\partial \tilde{\theta}_{m}}
F_{jk}.
\end{eqnarray}
By using this formula,
under the translation of 
$(h, \Omega_{\Lambda})$ $\longrightarrow$
$(\Omega_{m}h^2, \Omega_{\Lambda})$,
the Fisher matrix in the new parameter space is written as
\begin{align}
\tilde{F}^{H_{0} \ {\rm prior}} = 
\left(
\begin{array}{cc}
\tilde{F}_{\Omega_{m}h^2 \Omega_{m}h^2} &
\tilde{F}_{\Omega_{m}h^2 \Omega_{\Lambda}} \\
\tilde{F}_{\Omega_{m}h^2 \Omega_{\Lambda}} & 
\tilde{F}_{\Omega_{\Lambda} \Omega_{\Lambda}}
\end{array}
\right)
=
\frac{1}{(1\%\times H_{0,{\rm fid}})^2}
\left(\frac{1}{2\Omega_{m}h^2}\right)^2
\left(
\begin{array}{cc}
h^2 & h^4 \\
h^4 & h^6
\end{array}
\right).
\end{align}

\chapter{Forecasts for the neutrino mass}
\label{Chap:result_mass}

\section{Future constraints}

In this chapter, we present our results for projected constraints 
by the 21cm, CMB and BAO observations on cosmological parameters, 
paying particular attention to parameters related to neutrino
(the total neutrino mass, the effective number of neutrino species  
and the neutrino mass hierarchy).
When we calculate the Fisher matrices, we choose the following basic set
of cosmological parameters:
the energy density of matter $\Omega_{m}h^{2}$, 
baryons $\Omega_{b}h^{2}$ and dark energy $\Omega_{\Lambda}$, 
the scalar spectral index $n_{s}$, the scalar fluctuation amplitude $A_{s}$ 
(the pivot scale is taken to be $k_{{\rm pivot}}=$ $0.05 \ {\rm Mpc}^{-1}$), 
the reionization optical depth $\tau$, the primordial helium 4 mass fraction $Y_{{\rm p}}$ 
and the total neutrino mass $\Sigma m_{\nu} = m_{1}+m_{2}+m_{3}$.
Fiducial values of these parameters (except for $\Sigma m_{\nu}$) are adopted 
to be $(\Omega_{m}h^{2},\Omega_{b}h^{2},\Omega_{\Lambda},n_{s},A_{s},\tau,Y_{\rm p})$
$=( 0.1417, 0.02216, 0.6914, 0.9611, 2.214\times 10^{-9}, 0.0952, 0.25)$,
which are the best fit values of the Planck result \cite{Ade:2013zuv}.

Here, we numerically  evaluate how we can determine 
the effective number of neutrino species (in section \ref{subsec:const_hie}), 
and the neutrino mass hierarchy (in section \ref{subsec:const_hie}),  
by combining the  21 cm line observations (SKA phase1 or phase2)
with the CMB experiments (Planck + \textsc{Polarbear}-2 or Simons Array) and the BAO observation (DESI).
In the former analysis, we fix the neutrino mass hierarchy to be the normal one,
and set the fiducial value of the total neutrino mass $\Sigma m_{\nu}$
and the effective number of neutrino species $N_{\nu}$ 
to be $\Sigma m_{\nu} = 0.1$ or $0.06 \ {\rm eV}$ and $N_{\nu} = 3.046$.
%
Next, in the latter analysis, we fix $N_{\nu}$ to be $3.046$,
and set the fiducial values of the $\Sigma m_{\nu}$ and the mass hierarchy parameter $r_{\nu}$ to be 
$(\Sigma m_{\nu},r_{\nu})=(0.06 \ {\rm eV}, 0.82)$ (normal hierarchy) or
$(\Sigma m_{\nu},r_{\nu})=(0.1 \ {\rm eV}, -0.46)$ (inverted hierarchy).

To obtain Fisher matrices we use CAMB \cite{Lewis:1999bs,CAMB}~\footnote{
In this analysis, we use non-linear power spectra
for the calculations by performing a public code HALOFIT~\cite{Lewis:1999bs,CAMB}.
}
for calculations of CMB anisotropies $C_{l}$ and matter power spectra $P_{\delta \delta}(k)$.
In order to combine the CMB experiments with the 21 cm line and BAO observations, 
we calculate the combined Fisher matrix to be
\begin{equation}
F_{\alpha\beta} 
= F^{\rm (21cm)}_{\alpha\beta} + F^{\rm (CMB)}_{\alpha\beta} 
+ F^{\rm (BAO)}_{\alpha\beta},
\end{equation}
In this thesis, we do not use information for a possible correlation 
between fluctuations of the 21 cm and the CMB.

\section{Constraints on $\Sigma m_{\nu}$ and $N_{\nu}$}
\label{subsec:const_N_nu}

\begin{figure*}[htbp]
 \begin{center}
   \includegraphics[bb= 26 152 571 690, width=1\linewidth]{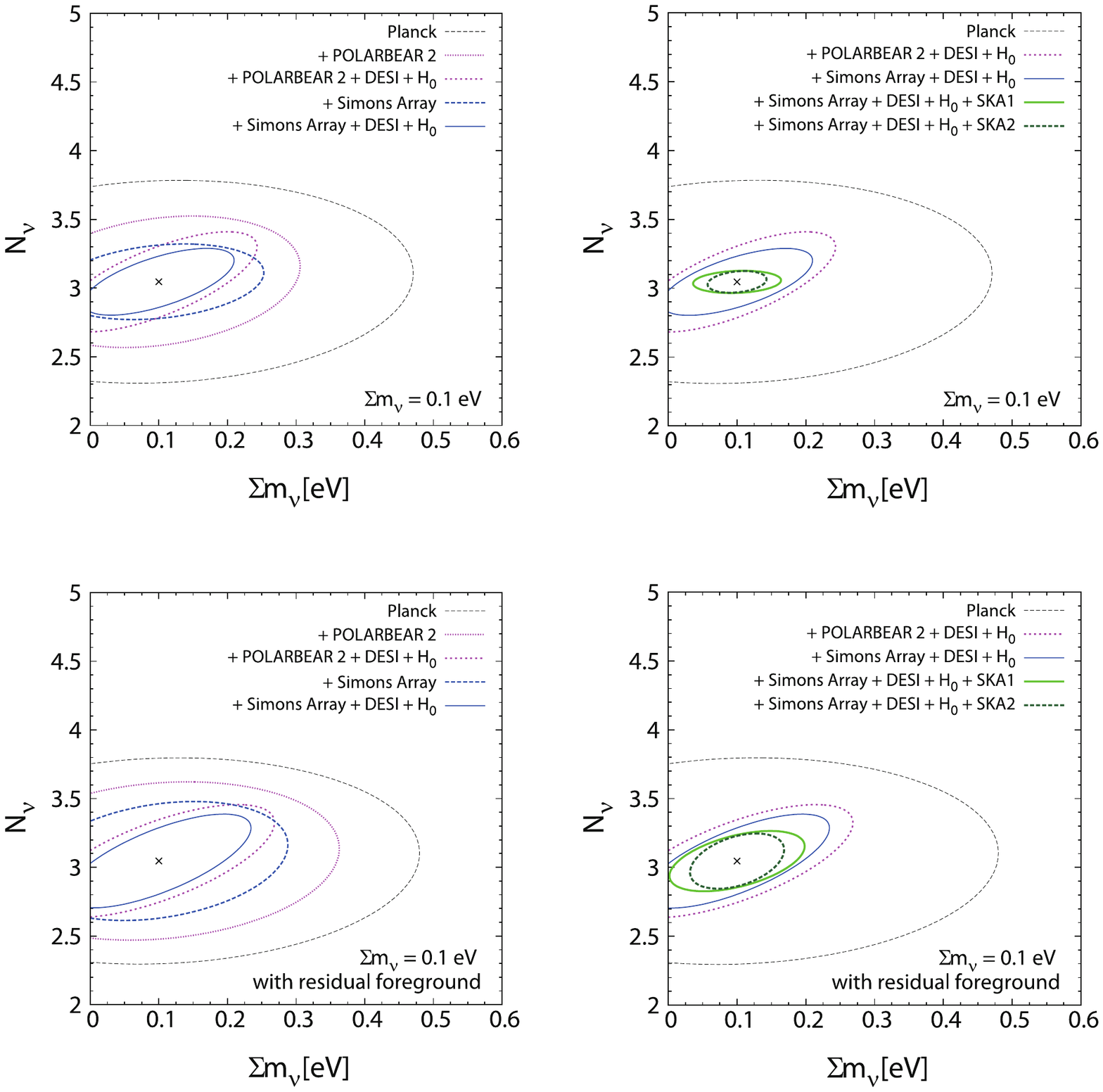}
   \caption{ 
   Contours are 95\% C.L. forecasts in $\Sigma m_{\nu}$-$N_{\nu}$ plane.
   Fiducial values of neutrino parameters, $N_{\nu}$ and $\Sigma m_{\nu}$, are taken to be
   $N_{\nu} = 3.046$ and $\Sigma m_{\nu} = 0.1$~eV.
   In the left two panels, the contours are constraints 
   by adopting Planck (outer dashed line), 
   Planck combined with \textsc{Polarbear}-2 ($f_{{\rm sky}}=0.2$) (outer dotted line) or 
   Simons Array (inner thick dashed line),
   Planck + BAO(DESI) + Hubble prior + \textsc{Polarbear}-2 ($f_{{\rm sky}}=0.2$) (inner thick dotted line) 
   or Simons Array (thin solid line), respectively.
   In the right two panels, they are constraints by adopting Planck (outer dashed line), 
   Planck + BAO(DESI) + Hubble prior combined with \textsc{Polarbear}-2 ($f_{{\rm sky}}=0.2$) (dotted line) or 
   Simons Array (outer thin solid line),
   Planck + BAO(DESI) + Hubble prior + Simons Array 
   combined with SKA phase1 (inner thick solid line) or phase2 (inner thick dashed line), respectively.
}
   \label{fig:Nnu01_fsky02}
 \end{center}
\end{figure*}

\begin{figure*}[htbp]
 \begin{center}
   \includegraphics[bb= 26 152 571 690, width=1\linewidth]{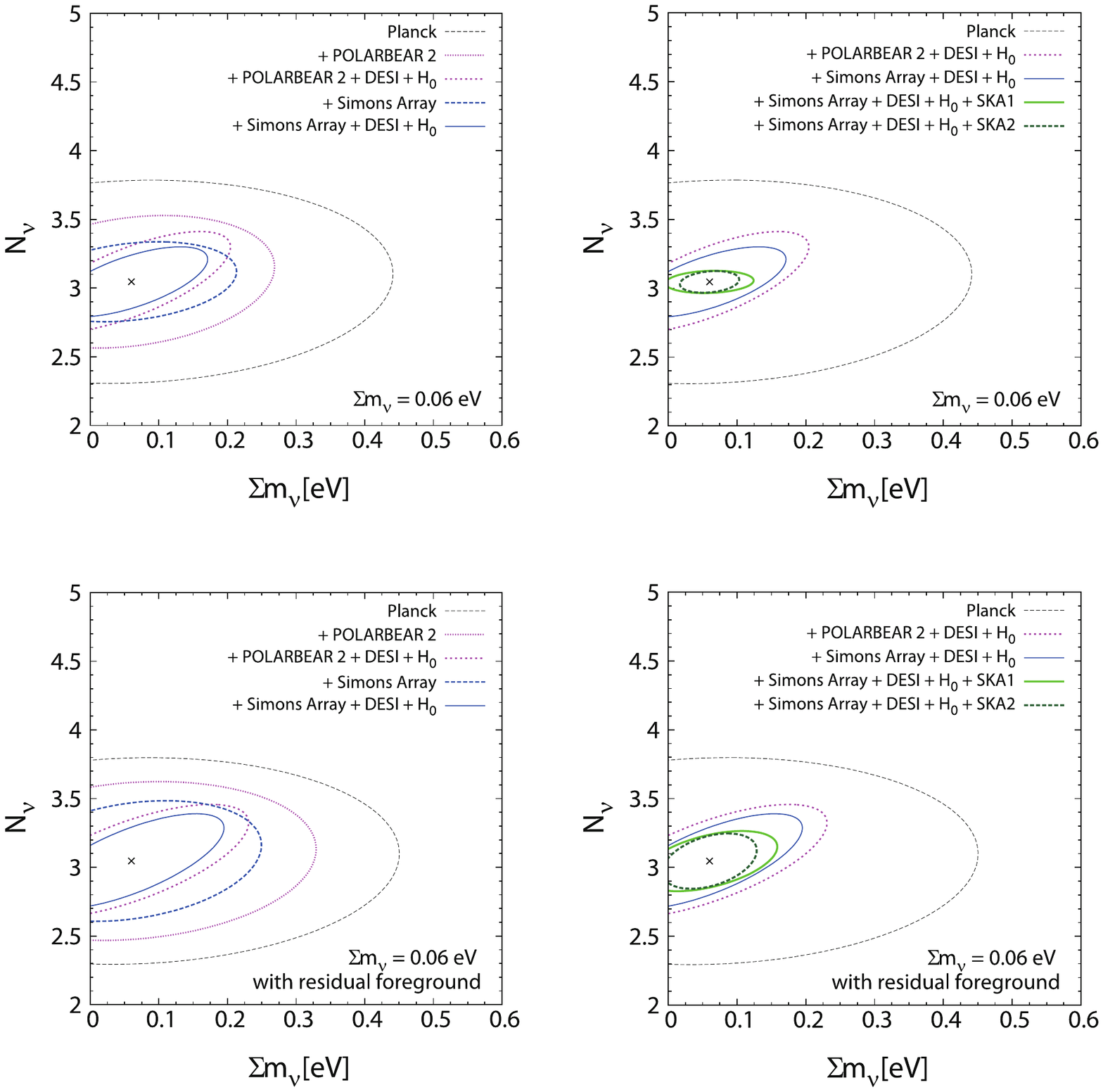}
   \caption{
   Same as Fig.\ref{fig:Nnu01_fsky02}, but the
   fiducial values of neutrino parameters, $N_{\nu}$ and $\Sigma m_{\nu}$, are taken to be
   $N_{\nu} = 3.046$ and $\Sigma m_{\nu} = 0.06$~eV.
}
   \label{fig:Nnu006_fsky02}
 \end{center}
\end{figure*}

\begin{figure*}[htbp]
 \begin{center}
   \includegraphics[bb= 26 152 571 690, width=1\linewidth]{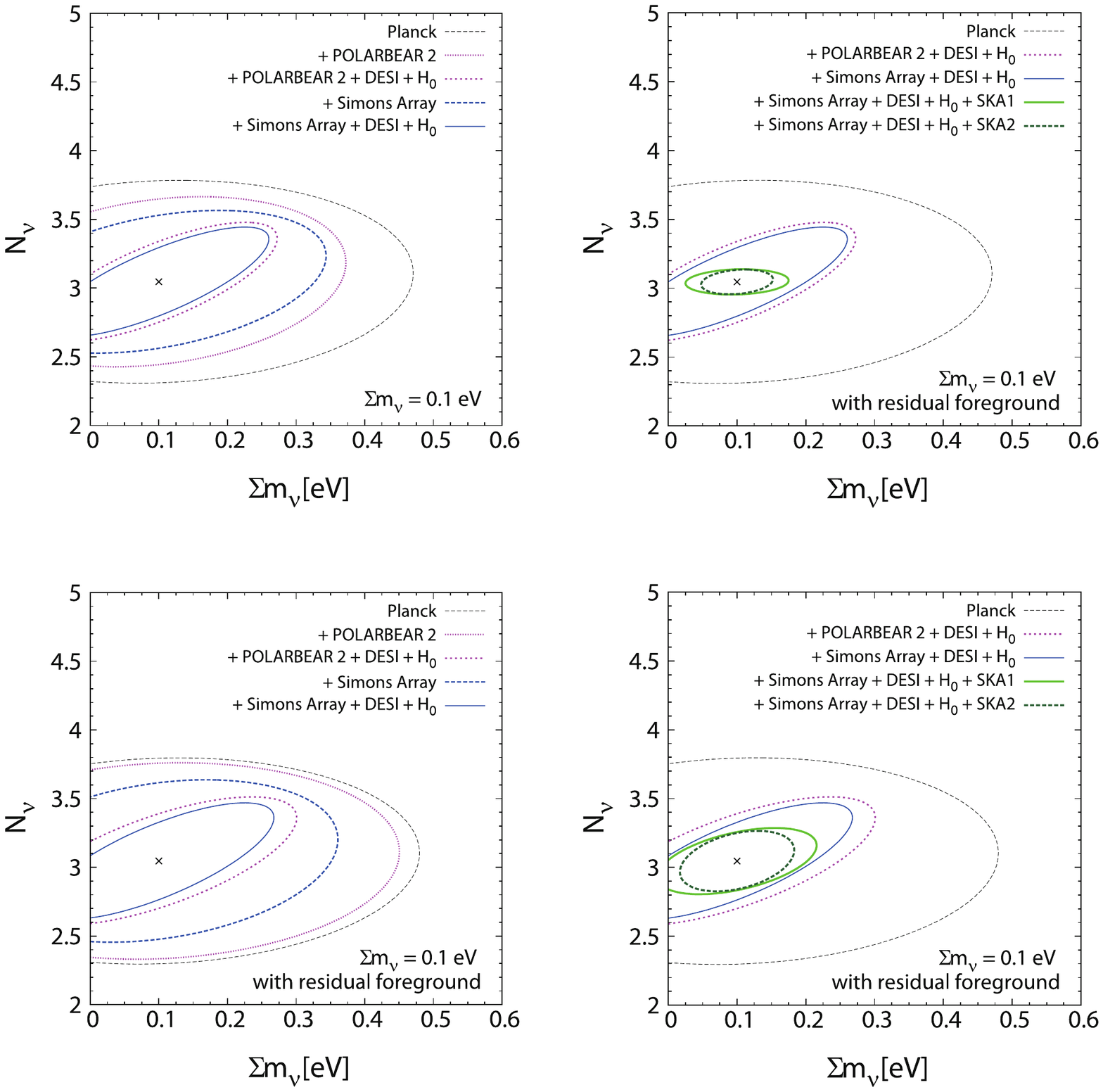}
   \caption{ 
   Same as Fig.\ref{fig:Nnu01_fsky02}, 
   but the sky coverages of \textsc{Polarbear}-2 and Simons Array
   are $f_{{\rm sky}}=0.016$
}
   \label{fig:Nnu01_fsky0016}
 \end{center}
\end{figure*}

\begin{figure*}[htbp]
 \begin{center}
   \includegraphics[bb= 26 152 571 690, width=1\linewidth]{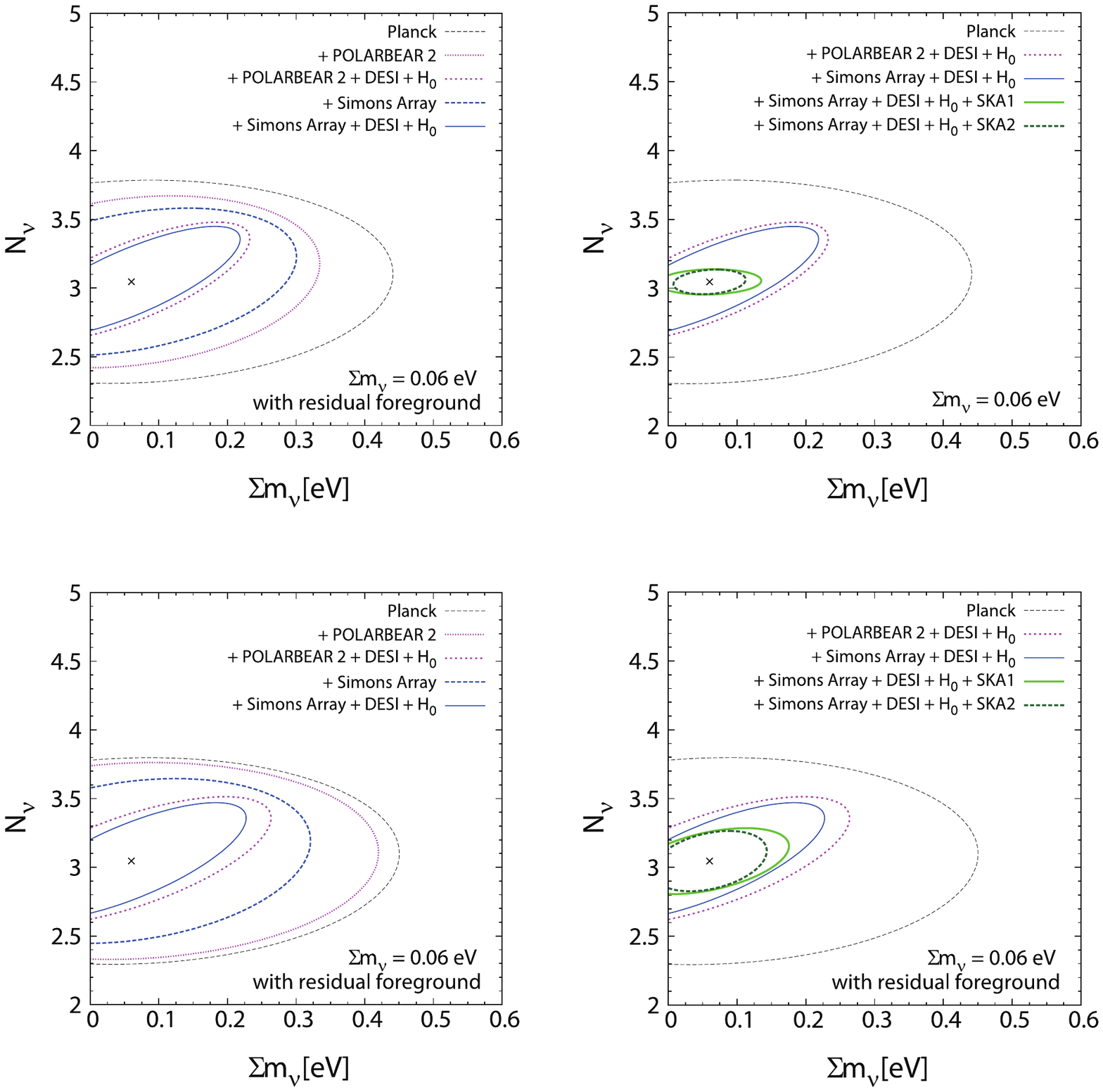}
   \caption{ 
   Same as Fig.\ref{fig:Nnu01_fsky0016}, but 
   the fiducial values of neutrino parameters, $N_{\nu}$ and $\Sigma m_{\nu}$, are taken to be
   $N_{\nu} = 3.046$ and $\Sigma m_{\nu} = 0.06$~eV.}
   \label{fig:Nnu006_fsky0016}
 \end{center}
\end{figure*}


\begin{figure*}[htbp]
 \begin{center}
   \includegraphics[bb= 170 295 427 545, width=0.5\linewidth]{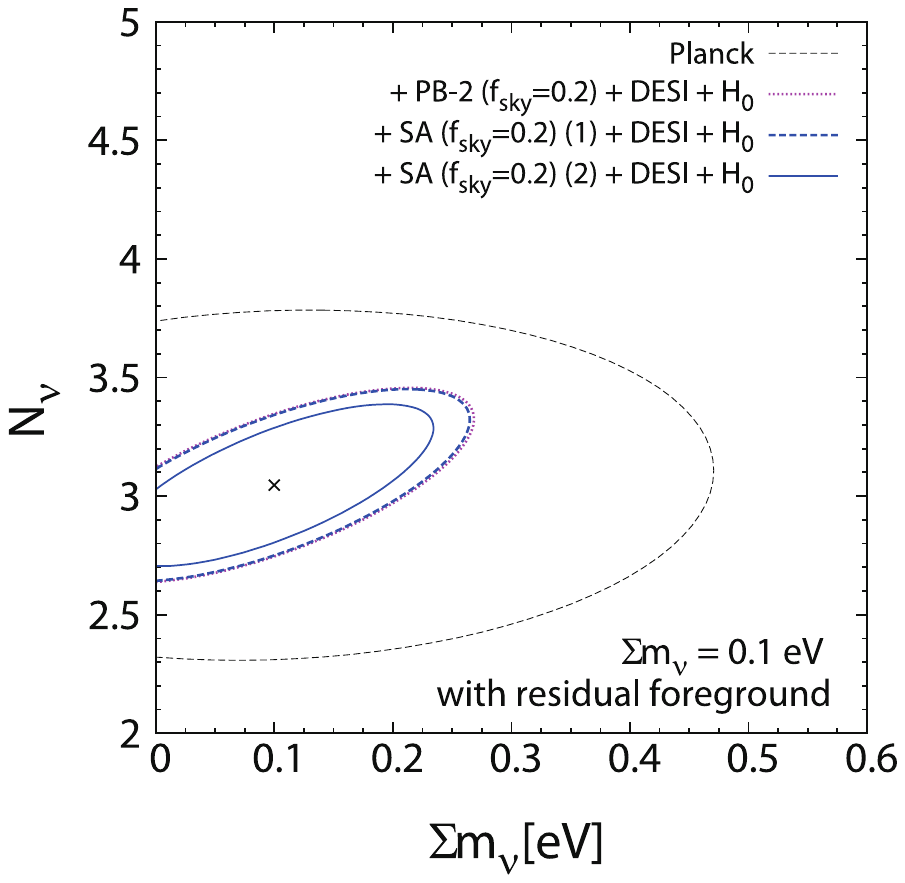}
   \caption{
   Contours are 95\% C.L. forecasts in $\Sigma m_{\nu}$-$N_{\nu}$ plane.
   The fiducial values of neutrino parameters, $N_{\nu}$ and $\Sigma m_{\nu}$, are taken to be
   $N_{\nu} = 3.046$ and $\Sigma m_{\nu} = 0.1$~eV.
   The contours are the constraints by adopting Planck (outer dashed line),
   Planck + BAO(DESI) + Hubble prior + \textsc{Polarbear}-2 ($f_{{\rm sky}}=0.2$) 
   (outer dotted line).
   For Simons Array ($f_{{\rm sky}}=0.2$), we plot results of two different situations.
   At first, we assume that the 220~GHz band of Simons Array 
   is used for only observation of CMB,
   and not used for the foreground removal (outer thick dashed line, 
   we call this situation Simons Array (1)).
   Secondary, we consider that the 220~GHz band is used 
   for the foreground removal
   (inner solid line: we call this situation Simons Array (2)).
   The constraint of Simons Array (1) almost laps over that of \textsc{Polarbear}-2.
}
   \label{fig:Nnu01_fsky02_Simons_fg}
 \end{center}
\end{figure*}


In Figs.\ref{fig:Nnu01_fsky02}-\ref{fig:Nnu01_fsky02_Simons_fg}, 
we plot contours of 95\% confidence levels (C.L.) forecasts in $\Sigma m_{\nu}$-$N_{\nu}$ plane. 
The fiducial values of the total neutrino mass is $\Sigma m_{\nu} = 0.1$~eV 
(Figs.\ref{fig:Nnu01_fsky02}, \ref{fig:Nnu01_fsky0016} and \ref{fig:Nnu01_fsky02_Simons_fg}) 
or $\Sigma m_{\nu} = 0.06$~eV (Figs.\ref{fig:Nnu006_fsky02} and \ref{fig:Nnu006_fsky0016}).
%
Additionally,
sky coverages of \textsc{Polarbear}-2 and Simons Array are $f_{{\rm sky}}=0.2$
(Figs.\ref{fig:Nnu01_fsky02}, \ref{fig:Nnu006_fsky02} and \ref{fig:Nnu01_fsky02_Simons_fg})
or $f_{{\rm sky}}=0.016$ (Figs.\ref{fig:Nnu01_fsky0016} and \ref{fig:Nnu006_fsky0016}).

In the left two panels of Figs.\ref{fig:Nnu01_fsky02}-\ref{fig:Nnu006_fsky0016},
each contour represents a constraint by CMB only or CMB + BAO~(DESI) + Hubble prior,
in the right panels, by Planck only or CMB + BAO~(DESI) + Hubble prior + 21cm~(SKA).
In these four figures,
the upper two panels are results when we assume that residual foregrounds are completely removed.
In contrast, the lower panels are results 
when the residual foregrounds are remaining
and we assume that 
the 220 GHz band of Simons Array are used for the foreground removal.

From these figures, adding the BAO experiments to the CMB ones,
we see that there is a strong improvement for the sensitivities to $\Sigma m_{\nu}$ and $N_{\nu}$
because several parameter degeneracies are broken by those combinations.
Besides, we find that larger sky coverage is more effective than smaller one.
However, it is difficult to detect the non-zero neutrino mass at 2$\sigma$ level
even by using the combination of Simons Array and DESI.
On the other hand, adding the 21 cm experiment (SKA phase1) to the CMB experiment, 
we see that there is a substantial improvement,
and the combination has enough sensitivity to detect the non-zero neutrino mass 
in the case of $\Sigma m_{\nu}=0.1$ eV to be fiducial 
when there are not residual foregrounds. 
Of course, CMB + SKA phase2 can obviously do the same job.
In the case with the residual foregrounds,
Planck + Simons Array ($f_{{\rm sky}}=0.2$) + BAO + SKA phase1 can
detect the non-zero neutrino mass, however 
Planck + Simons Array ($f_{{\rm sky}}=0.016$) + BAO + SKA phase1 
does not have enough sensitivity, and
only combination with SKA phase2 can do it.

In the case of $\Sigma m_{\nu}=0.06$ eV to be fiducial
(which corresponds to the lowest value for the normal hierarchy)
without  the residual foregrounds, 
only CMB + BAO + SKA phase2 can detect the non-zero neutrino mass.
However, in the case with the residual foregrounds,
even combination with SKA phase2 can not do it.
Hence, we see that stronger foreground removal in the 21 cm line observation is necessary.

In Fig.\ref{fig:Nnu01_fsky02_Simons_fg}, 
we show the two different situations about the observation of Simons Array.
At first, we assume that 
the 220 GHz band of Simons Array are used for the foreground removal (solid inner line),
and use this assumption in the analysis
of the under panels of Figs.\ref{fig:Nnu01_fsky02}-\ref{fig:Nnu006_fsky0016}.
Secondary, we consider that
the band is used only observation of CMB and not used for the foreground removal,
which are done by only Planck.
we plot the both results in the Fig.\ref{fig:Nnu01_fsky02_Simons_fg}.
From this figure,
we find that the constraint in the latter situation 
is almost the same level as that of \textsc{Polarbear}-2
because the strength of the residual foregrounds 
depends only on the Planck sensitivity.
Therefore, we find that it is better to use 220 GHz band of Simon Array 
for the foreground removal.

\section{Constraints on the neutrino mass hierarchy}
\label{subsec:const_hie}


\begin{figure*}[tbp]
 \begin{center}
   \resizebox{160mm}{!}{
   \includegraphics[bb= 84 261 514 578, width=1\linewidth]{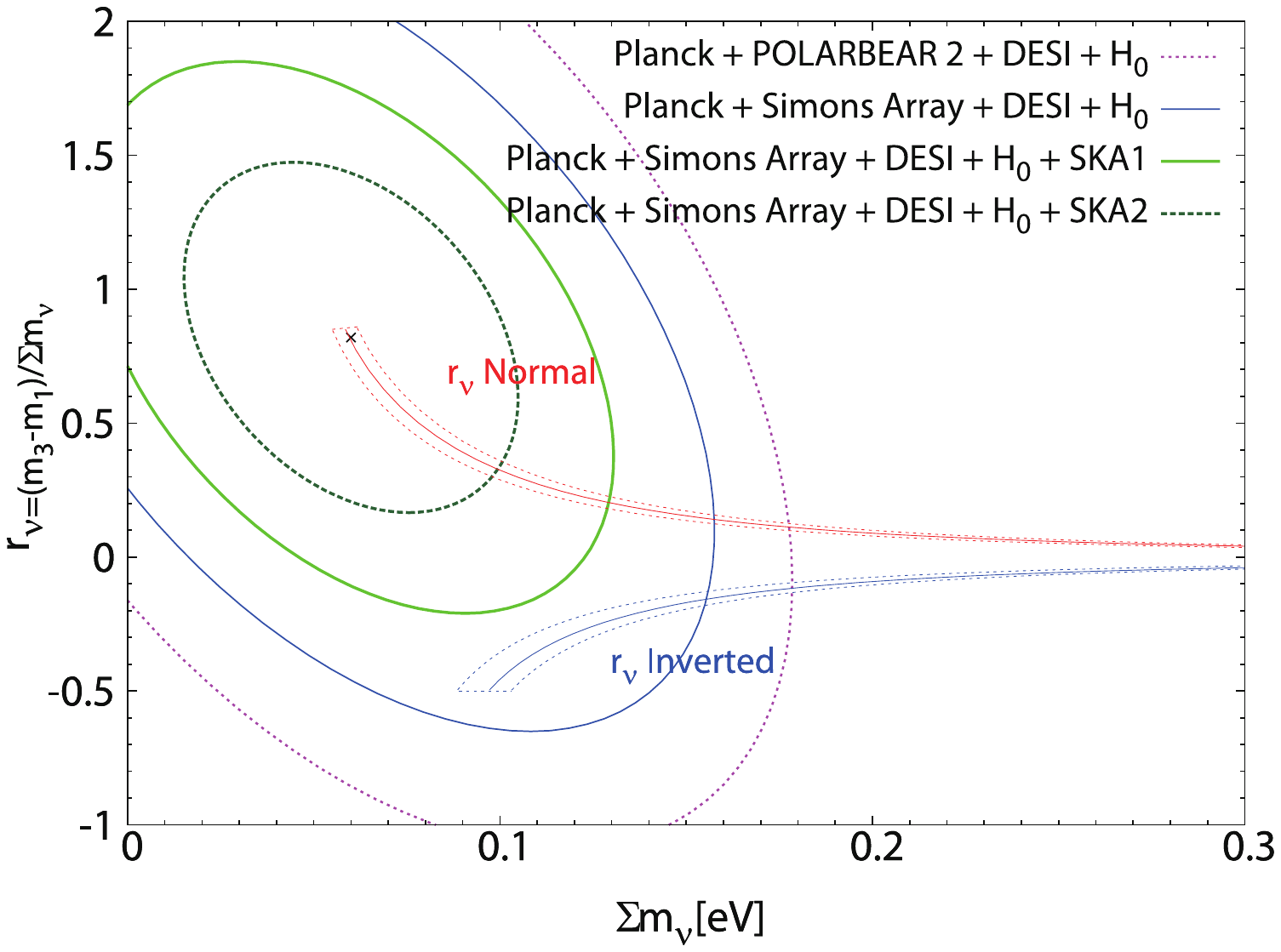} 
   \includegraphics[bb= 84 261 514 578, width=1\linewidth]{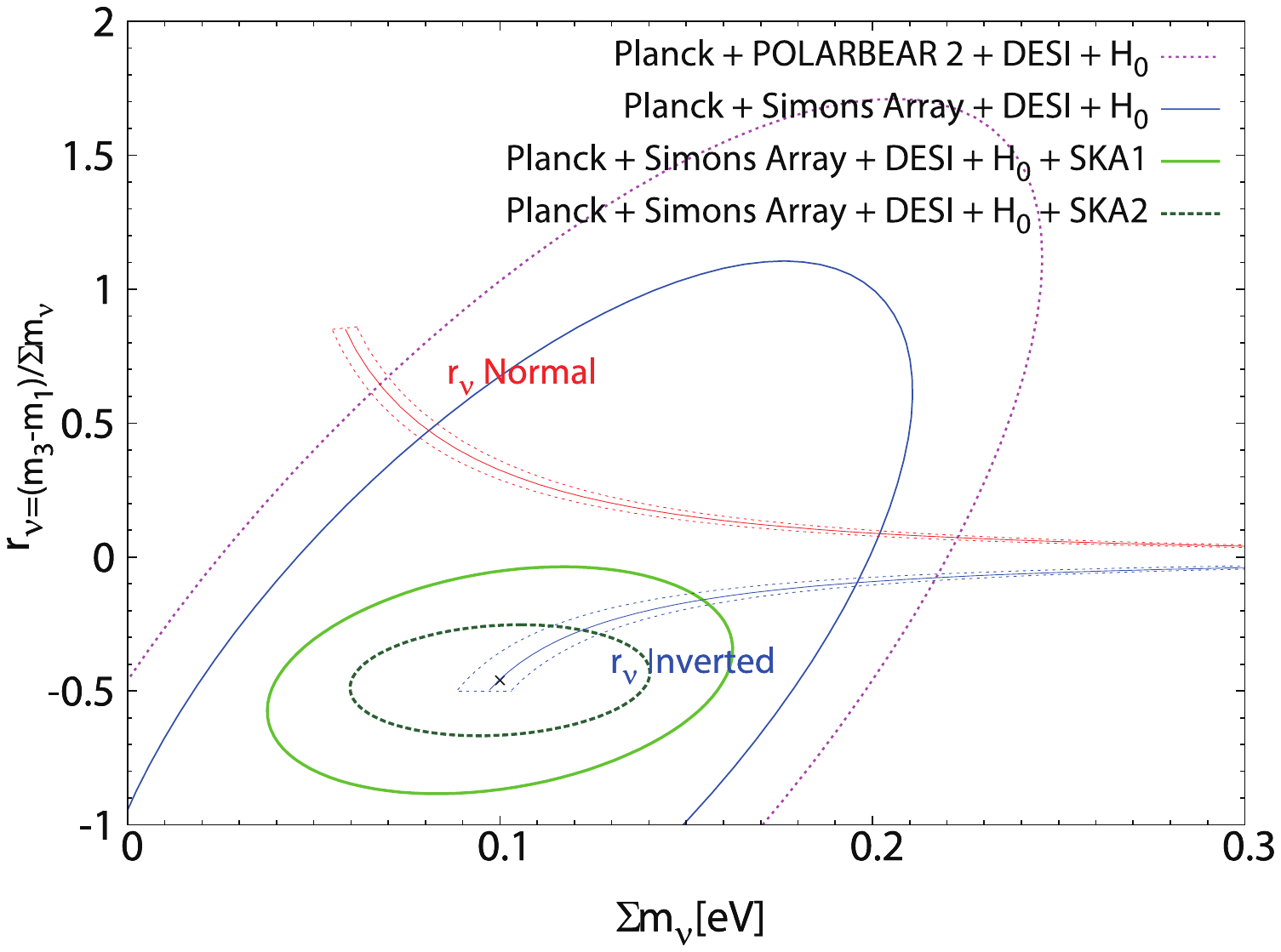} 
   }
   \resizebox{160mm}{!}{
   \includegraphics[bb= 84 261 514 578, width=1\linewidth]{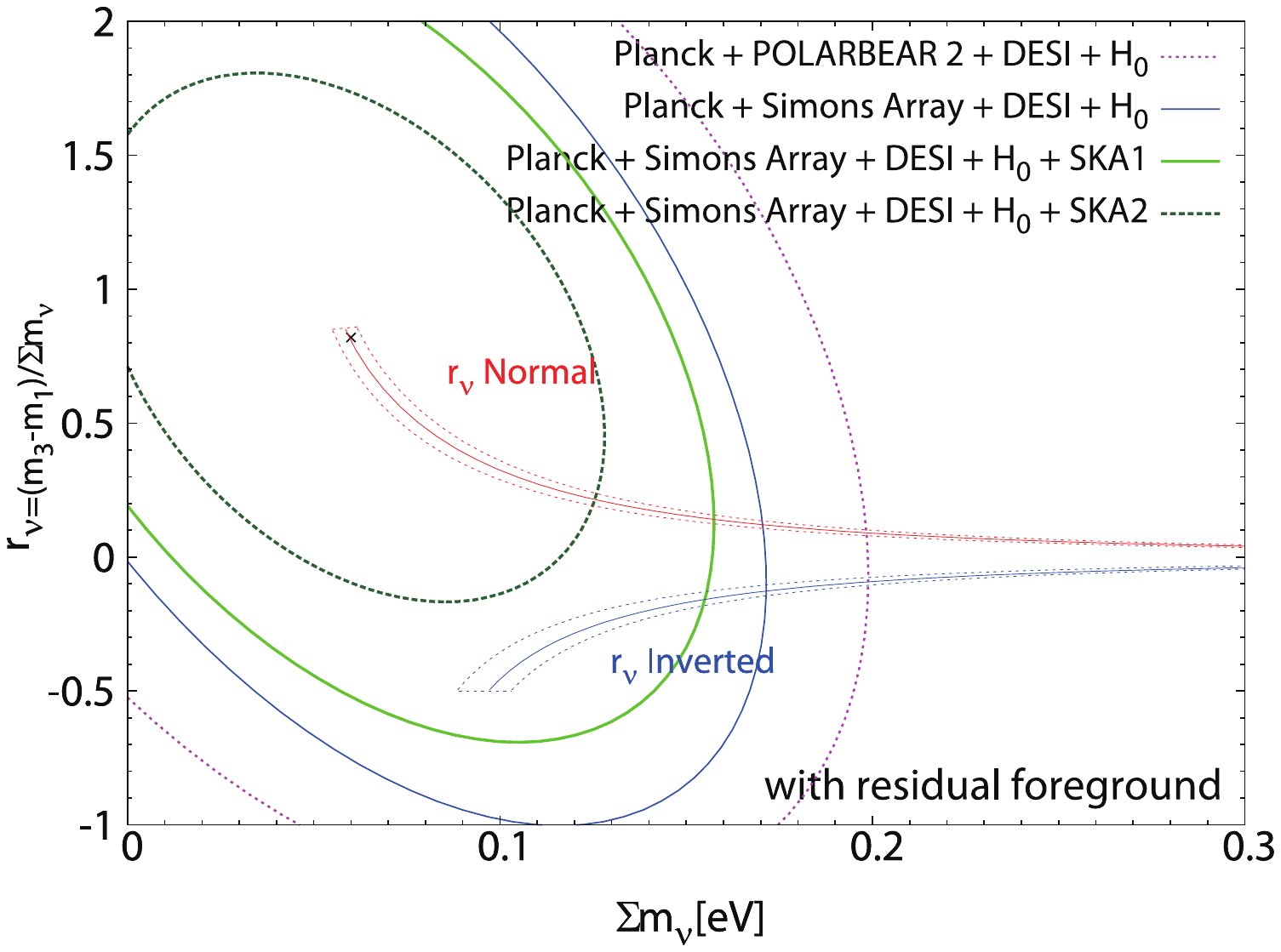} 
   \includegraphics[bb= 84 261 514 578, width=1\linewidth]{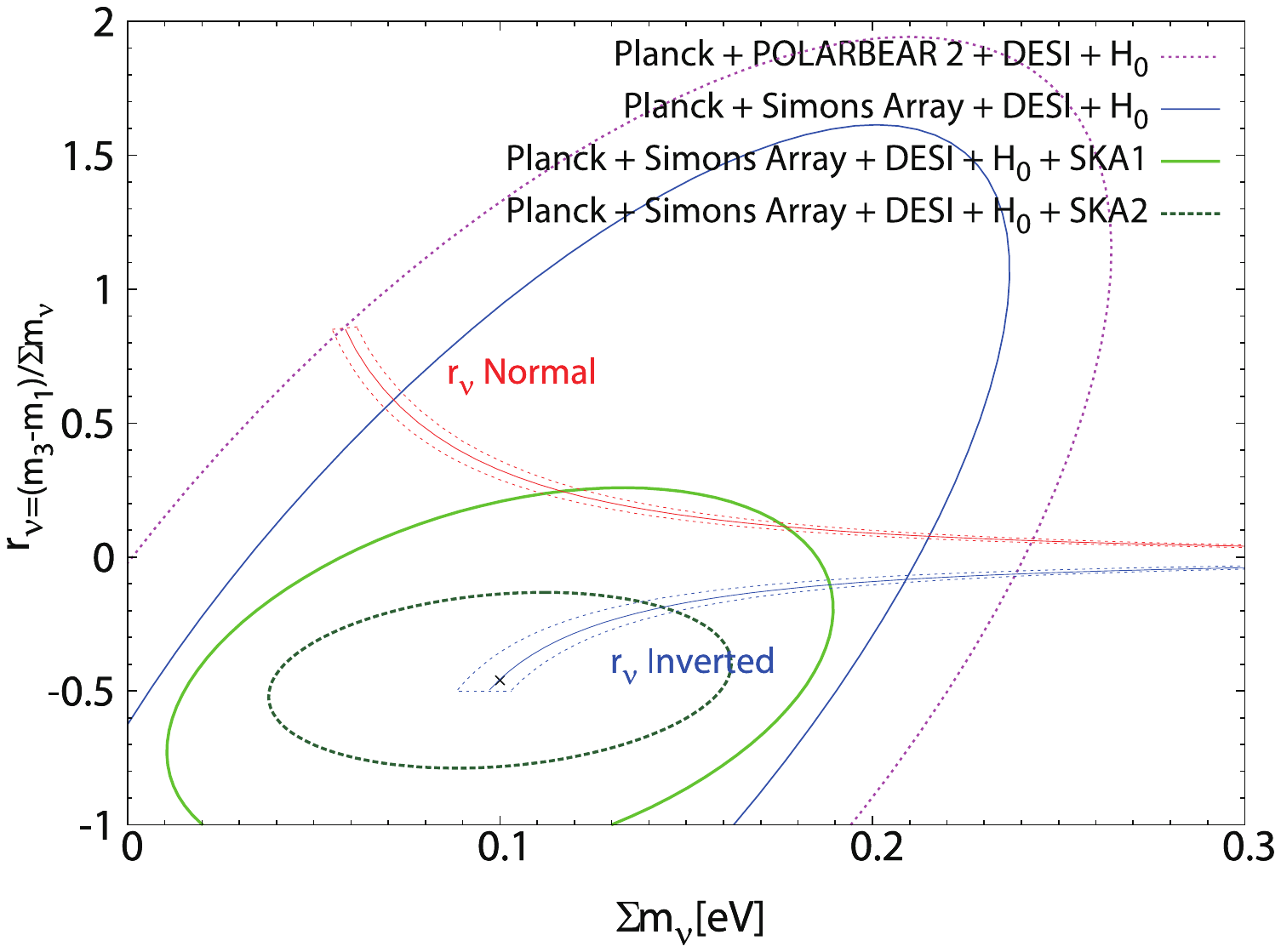} 
   }
   \caption{
   Contours are 95\% C.L. forecasts in $\Sigma m_{\nu}$-$r_{\nu}$ plane, 
   by adopting Planck (outer dashed line), 
   Planck + BAO(DESI) + Hubble prior combined with \textsc{Polarbear}-2 ($f_{sky}=0.2$) (dotted line) or 
   Simons Array (outer thin solid line),
   Planck + BAO(DESI) + Hubble prior + Simons Array 
   combined with SKA phase1 (inner thick solid line) or phase2 (inner thick dashed line), respectively.
   Allowed parameters on $r_{\nu}$ by neutrino oscillation experiments 
   are plotted as two bands for the inverted and the normal hierarchies,
   respectively (the name of each hierarchy is written in the close vicinity of the line). 
   The solid lines inside the bands are the central values of $r_{\nu}$ by oscillation experiments,
   and the fiducial points are denoted by cross-marks.
}
   \label{fig:hie_ellipse}
 \end{center}
\end{figure*}


Next, we discuss whether we will be able to determine the neutrino mass
hierarchies by using the future 21 cm line and CMB observations. 
%
In Fig.~\ref{fig:hie_ellipse}, we plot 2${\bf \sigma}$ errors of the parameter 
$r_{\nu}\equiv (m_{3} - m_{1} )/\Sigma m_{\nu}$ constrained by both
the 21 cm line and the CMB observations in case of the inverted hierarchy
to be fiducial (the left two panels), and the normal hierarchy to be fiducial (the right two panels). 
In this figure, the upper two panels are results 
when we assume that the residual foregrounds are completely removed.
On the other hand, the lower panels show the results
when the residual foregrounds are remaining
and the 220 GHz band of Simons Array are used for the foreground removal.

It is notable that the difference between $r_{\nu}$'s of these two hierarchies 
becomes larger as the total mass $\Sigma m_{\nu}$ becomes smaller. 
Therefore, $r_{\nu}$ is quite useful to distinguish a true mass hierarchy from the other.  
Allowed parameters on $r_{\nu}$ by neutrino oscillation experiments are plotted as two
bands for the inverted and the normal hierarchies, respectively.  
The thin solid lines inside the bands are the experimental mean values by
oscillations. 

%
As is clearly shown in Fig.~\ref{fig:hie_ellipse}, 
actually those combinations of the observations will be able to determine 
the neutrino mass hierarchy to be inverted  or normal for 
$\Sigma m_{\nu} \sim 0.06$ eV or 
$\Sigma m_{\nu} \sim 0.1 \ {\rm eV} $ at 95 \% C.L.,
%
respectively.  Although the determination is possible only at around
$\Sigma m_{\nu}\lesssim {\cal O}(0.1) \ {\rm eV}$, those results should be
reasonable. That is  because a precise discrimination of the mass
hierarchy itself may have no meaning if the masses are highly
degenerate, i.e., if $0.1$ eV $\ll \Sigma m_{\nu} $.

Once a clear signature $\Sigma m_{\nu} \ll 0.1 \ {\rm eV}$ were
determined by observations or experiments, it should be obvious that
the hierarchy must be normal without any ambiguities. On the other
hand, if the hierarchy were inverted, we could not determine it only by
using $\Sigma m_{\nu}$.  However, it is remarkable that our method is
quite useful because we can discriminate the hierarchy from the other
even if the fiducial values were $\Sigma m_{\nu} \gtrsim 0.1$~eV for
both the normal and inverted cases.
This is clearly shown in Fig~\ref{fig:hie_ellipse}.  
In case that a fiducial value of $\Sigma m_{\nu}$ is taken to be the lowest values in
neutrino oscillation experiments, 
this figure indicates that even Simons Array + SKA can discriminate the inverted (normal)
mass hierarchy from the normal (inverted) one.

\chapter{Forecasts for the lepton asymmetry}
\label{chap:result_lepton}

In this chapter, we discuss future prospects of  
the determination of the lepton asymmetry, or the chemical 
potentials for neutrino \cite{Kohri:2014hea}. 

\begin{figure}[htbp]
  \begin{center}
    \resizebox{120mm}{!}{
     \includegraphics[bb= 39 23 576 769]{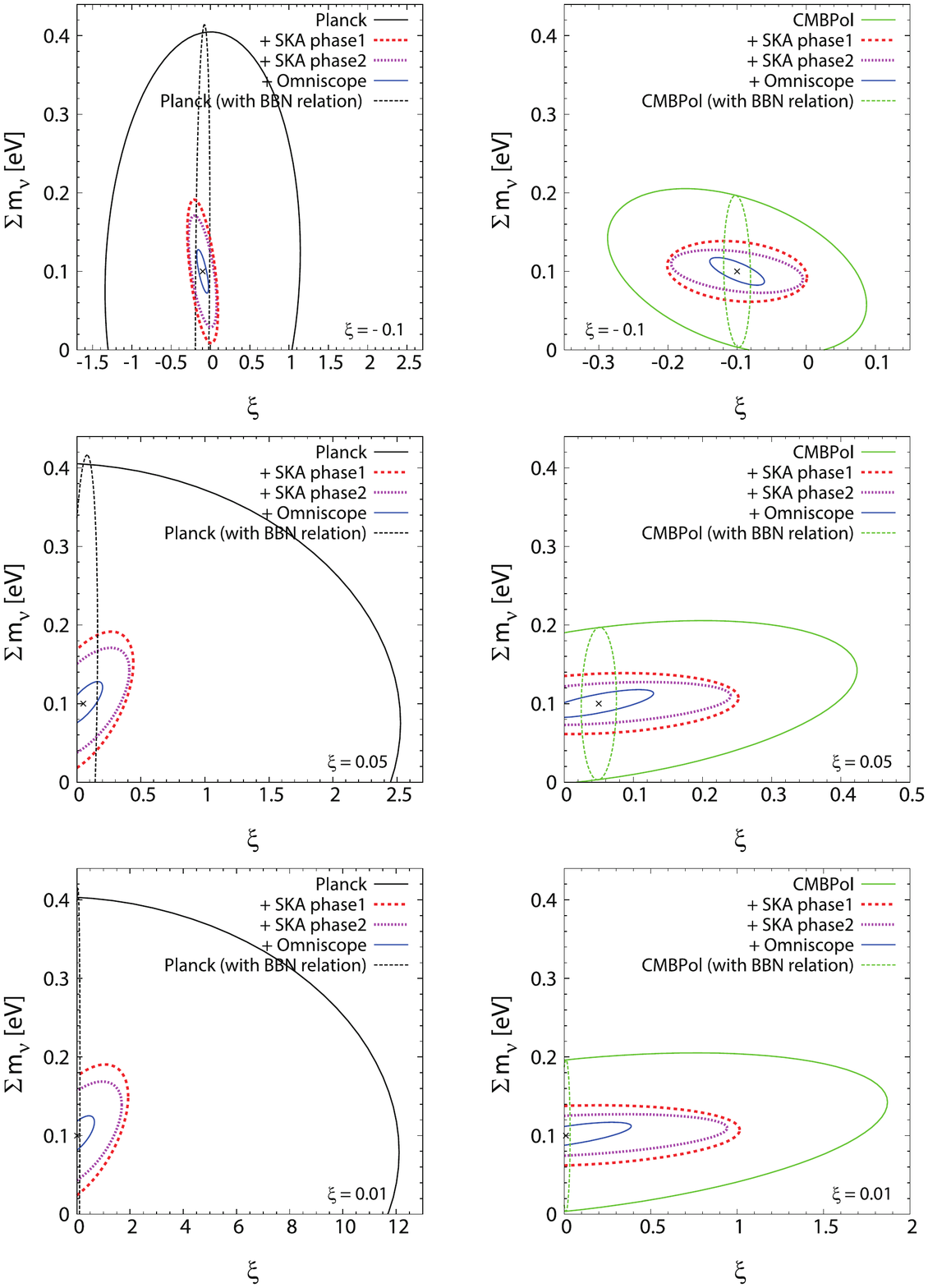}
}
\end{center}
\caption{Expected 2$\sigma$ constraints on the $\sum m_\nu$--$\xi$ plane. 
  As CMB data, the Planck and CMBPol surveys are adopted in the left and right panels, respectively.
  In order from top to bottom, the fiducial values of $\xi$ are set to $-0.1$, 0.05 and 0.01. 
  Here, we mainly present constraints for fixed $Y_p=0.25$. 
  Shown are the constraints from CMB alone (solid black/green line) as well as the ones from CMB data combined with 21~cm data from 
  SKA phase1 (red line), SKA phase2 (magenta line) and Omniscope (blue line). 
  As a reference, the constraints from CMB data alone with the BBN relation
  are also shown (dotted black/green line).
  Note that scales in $x$-axis differ among different panels.
  }
  \label{fig:xi_mnu_noBBN}
\end{figure}

\begin{figure}[htbp]
  \begin{center}
    \resizebox{120mm}{!}{
     \includegraphics[bb= 43 23 571 769]{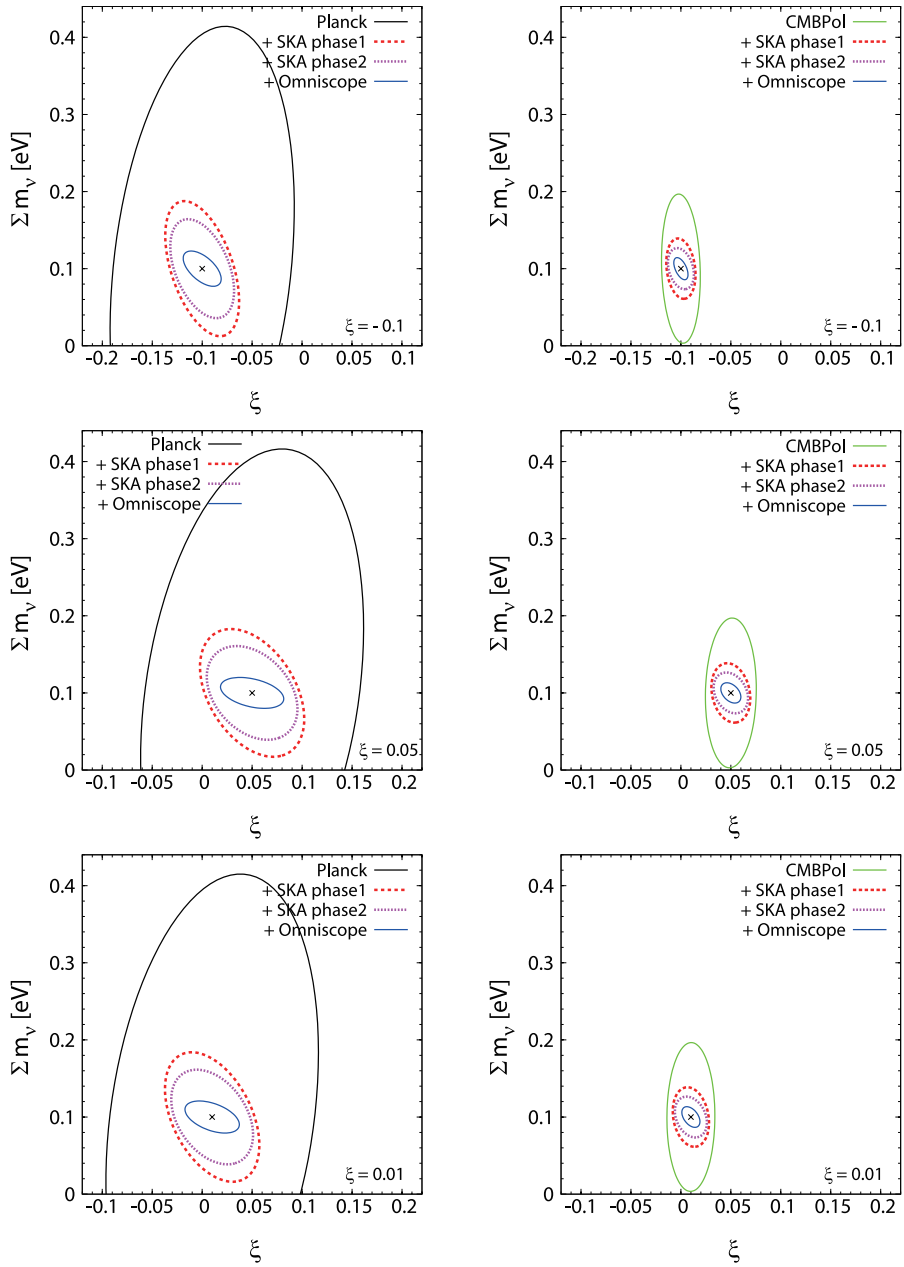}
}
  \end{center}
  \caption{Expected 2$\sigma$ constraints on the $\sum m_\nu$--$\xi$ plane. In this figure, the BBN relation is assumed.}
  \label{fig:xi_mnu_wBBN}
\end{figure}

In the following analysis, we explore the parameter space which includes 
the degeneracy parameter 
$\xi=\xi_{\nu_{e}}=\xi_{\nu_{\mu}}=\xi_{\nu_{\tau}}$ assuming the universal 
lepton asymmetry~\footnote{
Regardless of the initial value of $\xi_{\nu i}$ (with $ i= e,\mu,\tau$) at the decoupling, 
the lepton asymmetry would be universal, due to the large mixing in neutrino mass matrix~\cite{Dolgov:2002ab}.}
and neutrino mass $m_\nu$ as well as 
the six standard cosmological parameters
($\Omega_\Lambda,~\Omega_b h^2,~\Omega_mh^2,~\tau,~A_s,~n_s$).
In addition to these parameters, in some cases, we also include the primordial $^4$He mass fraction 
$Y_p$ and the effective number of neutrino species for extra (dark) radiation $\Delta N_\nu$
which gives its energy density in units of a single massless neutrino species as 
\begin{equation}
\bar\rho_{\rm ext}=\Delta N_\nu \frac{7\pi^2}{120}{T_\nu}^4.
\end{equation}
Although the chemical potential $\xi$ contributes the changes to 
$N_\nu$,
that is, the effective number of neutrino species for total dark radiation (neutrinos and extra radiation) as seen from 
Eqs.~\eqref{eq:rho_r2} and \eqref{eq:rho_nr}, $\Delta N_\nu$ counts for possible other contribution to $N_\nu$.
Furthermore, in BBN theory, $Y_p$ is related to
$\Omega_bh^2, \xi$  and 
$\Delta N_\nu$.
Therefore, we make the analysis with/without assuming so-called BBN relation among these parameters in some analysis.  
When the BBN relation is not adopted, we vary $Y_p$ freely or fix it to $Y_p=0.25$. 

Regarding fiducial parameters, we often present constraints for several fiducial values of $\xi$ and $\Delta N_\nu$.
On the other hand, fiducial values of $\sum m_\nu$ is fixed  to be 0.1~eV and 
those of other cosmological parameters are fixed to be
($\Omega_\Lambda,~\Omega_b h^2,~\Omega_mh^2,~\tau,~A_s,~n_s$)
$=( 0.6914, 0.02216, 0.1417, 0.0952, 2.214\times10^{-9}, 0.9611 )$,
which are the best fit values from the Planck result \cite{Ade:2013zuv}.
%
To obtain Fisher matrices we use CAMB \cite{Lewis:1999bs,CAMB}~\footnote{
In this analysis, we use linear power spectra.
By performing a public code HALOFIT~\cite{Lewis:1999bs,CAMB},
we have checked that modifications by including
nonlinear  effects for evolutions of the matter power spectrum are
much smaller than typical errors in our analyses and negligible for
parameter fittings
}
for calculations of CMB anisotropies $C_{l}$ and matter power spectra $P_{\delta \delta}(k)$.
In order to combine the CMB experiments with the 21 cm line
experiments, we calculate the combined fisher matrix to be
\begin{equation}
F_{\alpha\beta} 
= F^{\rm (21cm)}_{\alpha\beta} + F^{\rm (CMB)}_{\alpha\beta}.
\end{equation}
%

\section{Cases without extra radiation}

Let us first see the cases without extra radiation.
Fig.~\ref{fig:xi_mnu_noBBN} shows constraints on the $\xi$--$\sum
m_\nu$ plane for mainly the cases where we fixed $Y_p$ to 0.25 without
assuming the BBN relation.  On the other hand, constraints only from
CMB observations with the BBN relation $Y_p(\Omega_bh^2,~\xi,~\Delta
N_\nu)$ are also shown as well, for the purpose of comparison.
Regarding fiducial values of $\xi$, we adopted $\xi=0.01$, 0.05 and
$-0.1$ here. Note that $\xi=0.05$ and $-0.1$ roughly correspond to the
upper and lower bounds at 2$\sigma$ from primordial abundance of the
light elements (See Fig.~\ref{fig:etaXi2014} in
Appendix~\ref{sec:BBNrelation} and
Ref. \cite{Steigman:2012ve}).
From the figure, we can immediately see that 21~cm line observations 
can be a powerful probes of the lepton asymmetry of the Universe.
Compared with the constraints on $\xi$ from Planck alone, 
the error is improved by a factor around 5 (10) by combining SKA (Omniscope).
Even though CMBPol can by itself give much tighter constraints than Planck, combinations with 21~cm observations are still 
able to improve the constraints further by a factor around 2 (SKA) and 4 (Omniscope).
We also note that constraints on the neutrino masses from CMB observations can 
be also improved by combining 21~cm line observations.
As an illustrative example, constraints on cosmological parameters for the cases with fiducial $\xi=0.05$
are summarized in Table \ref{tab:fixedYp}. 

In Fig.~\ref{fig:xi_mnu_noBBN}, one may notice that the uncertainties in $\xi$, which we denote as $\sigma_\xi$, 
is dependent on the fiducial value of $\xi$.
This is because, in the absence of  the BBN relation, there is no difference between neutrinos and anti-neutrinos 
in their effects both on the CMB and 21~cm power spectra. 
Therefore, these power spectra are even functions of $\xi$, as can be also read from Eqs.\eqref{eq:rho2}-\eqref{eq:p2}
and \eqref{eq:dlnfdlny}, which respectively govern effects on the background and perturbation evolutions.
In particular for small $\xi\ll1$, these power spectra should respond linearly to $\xi^2$.
This leads that  $\sigma_\xi$ is proportional to the inverse of the fiducial $\xi$, 
while the error $\sigma_{\xi^2}\propto \xi\,\sigma_\xi$ is almost independent of the fiducial $\xi$, which 
is confirmed from Table \ref{tab:xi}, where we summarized constraints on $\xi$ for various setups  
(e.g. without the BBN relation)
and fiducial values of $\xi$ for cases of $\Delta N_\nu=0$. 

Although $\sigma_\xi$ is dependent on fiducial $\xi$, we can still see that $\xi=-0.1$, which is roughly the current lower bound from 
the primordial light elements, can be detected marginally by CMBPol+SKA and significantly by CMBPol+Omniscope.
This is remarkable as this indicates that even without assuming the BBN relation, 
we may be able to obtain a constraint on $\xi$ better than one from the primordial light elements.

On the other hand, from the above figure, one may think 21~cm line alone is powerful enough to give similar constraints on $\xi$ 
as those from CMB+21~cm line. However, this is not true.
This can be understood by seeing that  provided a very precise observation of 21~cm line, e.g., Omniscope, 
its combinations with Planck and CMBPol still differ non-negligibly.
This is due to that some cosmological parameters which degenerate with $\xi$ when only a 21~cm line observation 
is adopted can be determined well by CMB. 

Let us next see the cases with the BBN relation $Y_p(\Omega_bh^2,~\xi,~\Delta N_\nu)$, though 
we here still assume $\Delta N_\nu$ to vanish. 
In this case, $\xi$ affects CMB and 21~cm observations also through $Y_p$ 
in addition to the effects we have taken into account in the case of fixed $Y_p$.
Regarding effects of $\xi$ on the CMB power spectrum, 
this indirect effect through the BBN relation is more significant than direct ones.
This can be noticed in Fig.~\ref{fig:xi_mnu_noBBN}, 
where the contours of constraints from CMB alone can be squeezed in the direction of $\xi$ 
by an order of magnitude with the BBN relation.

Fig.~\ref{fig:xi_mnu_wBBN} shows the same constraints 
as in Fig.~\ref{fig:xi_mnu_noBBN} except that the BBN relation is now taken into account in any combinations 
of observations. Compared with the previous figure, improvements brought about 
by the combination of 21~cm line observations are not as dramatic as in the cases without the BBN relation.
This indirectly suggests that 21cm line observations are not as sensitive to $Y_p$ as CMB.
However, the combination with SKA can reduce the size of error in $\xi$ by a few times 
from Planck alone and a similar level of improvement can be brought about by Omniscope compared to CMBPol alone.
We note that with the BBN relation being assumed, a combination of CMB and 21~cm line observations
can constrain the lepton asymmetry substantially better than the primordial abundances of light elements.

Different from the cases without the BBN relation, 
one can notice that the sizes of errors in $\xi$ little depend on fiducial $\xi$ with the BBN relation. 
This is because prediction of BBN is sensitive to the sign of $\xi$. 
Therefore, $Y_p$ responses linearly to $\xi$ at the lowest order. 
In particular, the most significant effect of $\xi$ on $Y_p$ is that $\xi_e$ changes the ratio of 
neutron number density to proton one when BBN starts.
Positive (negative) $\xi$ effectively boosts (suppresses) $n\to p$ 
conversion and reduces (increases) $Y_p$. 
Such an effect can break the degeneracy between $\xi$ and $-\xi$ existing without the BBN relation.

Constraints on cosmological parameters are summarized in Tables~\ref{tab:fixedYp}, \ref{tab:withBBN} and \ref{tab:variedYp}, 
where we fixed $Y_p$ to 0.25, assumed the BBN relation and varied $Y_p$ as a free parameter, respectively.
In these tables, we present constraints only for the fiducial $\xi=0.05$, as we found that 
dependencies of errors on the fiducial $\xi$ is not significant except for $\sigma_\xi$; 
as long as one considers a fiducial $\xi\le0.1$, 
errors of cosmological parameters differ by no more than 25\%. 
The only exception is $\sigma_ \xi$ which has been shown to depend on fiducial $\xi$ in the absence of the BBN relation.
In Table~\ref{tab:xi}, we summarize the dependence of $\sigma_ \xi$ on fiducial values of $\xi$.
Except for the cases with the BBN relation, we see that $\sigma_ \xi$ scales almost 
proportionally to the inverse of fiducial $\xi$.

\begin{table}
\begin{center}
\begin{tabular}{l|cccc}
\hline \hline \\
& $\Omega_{m}h^{2}$ &  $\Omega_{b}h^{2}$ &  $\Omega_{\Lambda}$ & $n_s $ \\ 
\hline
Planck & $ 2.86 \times 10^{  -3} $ & $ 1.95 \times 10^{  -4} $ & $ 2.01 \times 10^{  -2} $ & $ 6.06 \times 10^{  -3} $ \\
\quad + SKA phase1 & $ 3.40 \times 10^{  -4} $ & $ 7.63 \times 10^{  -5} $ & $ 2.33 \times 10^{  -3} $ & $ 2.03 \times 10^{  -3} $ \\
\quad + SKA phase2 & $ 2.52 \times 10^{  -4} $ & $ 7.40 \times 10^{  -5} $ & $ 9.26 \times 10^{  -4} $ & $ 1.42 \times 10^{  -3} $ \\
\quad + Omniscope & $ 8.16 \times 10^{  -5} $ & $ 2.42 \times 10^{  -5} $ & $ 4.18 \times 10^{  -4} $ & $ 4.81 \times 10^{  -4} $\\ 
\hline
CMBPol & $ 1.16 \times 10^{  -3} $ & $ 3.78 \times 10^{  -5} $ & $ 7.48 \times 10^{  -3} $ & $ 1.75 \times 10^{  -3} $ \\
\quad + SKA phase1 & $ 3.11 \times 10^{  -4} $ & $ 2.91 \times 10^{  -5} $ & $ 2.14 \times 10^{  -3} $ & $ 1.20 \times 10^{  -3} $ \\
\quad + SKA phase2 & $ 2.12 \times 10^{  -4} $ & $ 2.74 \times 10^{  -5} $ & $ 9.06 \times 10^{  -4} $ & $ 9.16 \times 10^{  -4} $ \\
\quad + Omniscope & $ 5.13 \times 10^{  -5} $ & $ 1.31 \times 10^{  -5} $ & $ 4.09 \times 10^{  -4} $ & $ 3.68 \times 10^{  -4} $ \\ 
\hline \\
&  $A_{s} \times 10^{10} $ & $\tau_{\rm reion}$  & $\Sigma m_{\nu}$ & $\xi$  \\ \hline
Planck & $ 2.31 \times 10^{  -1} $ & $ 4.58 \times 10^{  -3} $ & $ 1.23 \times 10^{  -1} $ & $ 9.99 \times 10^{  -1} $ \\
\quad + SKA phase1 & $ 1.88 \times 10^{  -1} $ & $ 4.36 \times 10^{  -3} $ & $ 3.69 \times 10^{  -2} $ & $ 1.58 \times 10^{  -1} $ \\
\quad + SKA phase2 & $ 1.87 \times 10^{  -1} $ & $ 4.28 \times 10^{  -3} $ & $ 2.86 \times 10^{  -2} $ & $ 1.45 \times 10^{  -1} $ \\
\quad + Omniscope & $ 1.84 \times 10^{  -1} $ & $ 4.15 \times 10^{  -3} $ & $ 1.13 \times 10^{  -2} $ & $ 6.09 \times 10^{  -2} $ \\ 
\hline
CMBPol & $ 1.10 \times 10^{  -1} $ & $ 2.46 \times 10^{  -3} $ & $ 4.26 \times 10^{  -2} $ & $ 1.51 \times 10^{  -1} $ \\
\quad + SKA phase1 & $ 1.01 \times 10^{  -1} $ & $ 2.41 \times 10^{  -3} $ & $ 1.56 \times 10^{  -2} $ & $ 8.15 \times 10^{  -2} $ \\
\quad + SKA phase2 & $ 9.95 \times 10^{  -2} $ & $ 2.37 \times 10^{  -3} $ & $ 1.10 \times 10^{  -2} $ & $ 7.69 \times 10^{  -2} $ \\
\quad + Omniscope & $ 7.81 \times 10^{  -2} $ & $ 1.78 \times 10^{  -3} $ & $ 7.15 \times 10^{  -3} $ & $ 3.19 \times 10^{  -2} $ \\ 
\hline\hline
\end{tabular} 
\caption{1$\sigma$ errors on cosmological parameters for fiducial $\xi=0.05$ for the cases with fixed $Y_p=0.25$.}
\label{tab:fixedYp}
\end{center}
\end{table}

\begin{table}
\begin{center}
\begin{tabular}{l|cccc}
\hline \hline \\
&  $\Omega_{m}h^{2}$ &  $\Omega_{b}h^{2}$ &  $\Omega_{\Lambda}$ & $ n_s $ \\ 
\hline
Planck & $ 2.41 \times 10^{  -3} $ & $ 2.13 \times 10^{  -4} $ & $ 2.09 \times 10^{  -2} $ & $ 7.06 \times 10^{  -3} $ \\
\quad + SKA phase1 & $ 3.04 \times 10^{  -4} $ & $ 9.35 \times 10^{  -5} $ & $ 2.30 \times 10^{  -3} $ & $ 2.22 \times 10^{  -3} $ \\
\quad + SKA phase2 & $ 2.02 \times 10^{  -4} $ & $ 8.64 \times 10^{  -5} $ & $ 9.21 \times 10^{  -4} $ & $ 1.44 \times 10^{  -3} $ \\
\quad + Omniscope & $ 7.94 \times 10^{  -5} $ & $ 1.54 \times 10^{  -5} $ & $ 4.15 \times 10^{  -4} $ & $ 3.54 \times 10^{  -4} $ \\ 
\hline
CMBPol & $ 9.27 \times 10^{  -4} $ & $ 4.83 \times 10^{  -5} $ & $ 7.16 \times 10^{  -3} $ & $ 2.54 \times 10^{  -3} $ \\
\quad + SKA phase1 & $ 2.75 \times 10^{  -4} $ & $ 4.16 \times 10^{  -5} $ & $ 2.11 \times 10^{  -3} $ & $ 1.46 \times 10^{  -3} $ \\
\quad + SKA phase2 & $ 1.43 \times 10^{  -4} $ & $ 4.05 \times 10^{  -5} $ & $ 9.00 \times 10^{  -4} $ & $ 1.04 \times 10^{  -3} $ \\
\quad + Omniscope & $ 4.81 \times 10^{  -5} $ & $ 1.24 \times 10^{  -5} $ & $ 4.08 \times 10^{  -4} $ & $ 3.17 \times 10^{  -4} $ \\ 
\hline \\
&  $A_{s} \times 10^{10} $ & $\tau_{\rm reion}$  & $\Sigma m_{\nu}$ & $\xi$ \\ 
\hline
Planck & $ 2.07 \times 10^{  -1} $ & $ 4.64 \times 10^{  -3} $ & $ 1.28 \times 10^{  -1} $ & $ 4.50 \times 10^{  -2} $ \\
\quad + SKA phase1 & $ 1.92 \times 10^{  -1} $ & $ 4.31 \times 10^{  -3} $ & $ 3.34 \times 10^{  -2} $ & $ 2.10 \times 10^{  -2} $ \\
\quad + SKA phase2 & $ 1.89 \times 10^{  -1} $ & $ 4.25 \times 10^{  -3} $ & $ 2.45 \times 10^{  -2} $ & $ 1.83 \times 10^{  -2} $ \\
\quad + Omniscope & $ 1.85 \times 10^{  -1} $ & $ 4.14 \times 10^{  -3} $ & $ 8.08 \times 10^{  -3} $ & $ 1.28 \times 10^{  -2} $ \\ 
\hline
CMBPol & $ 1.07 \times 10^{  -1} $ & $ 2.48 \times 10^{  -3} $ & $ 3.92 \times 10^{  -2} $ & $ 1.03 \times 10^{  -2} $ \\
\quad + SKA phase1 & $ 1.01 \times 10^{  -1} $ & $ 2.39 \times 10^{  -3} $ & $ 1.55 \times 10^{  -2} $ & $ 7.85 \times 10^{  -3} $ \\
\quad + SKA phase2 & $ 9.78 \times 10^{  -2} $ & $ 2.33 \times 10^{  -3} $ & $ 1.07 \times 10^{  -2} $ & $ 6.95 \times 10^{  -3} $ \\
\quad + Omniscope & $ 6.86 \times 10^{  -2} $ & $ 1.56 \times 10^{  -3} $ & $ 5.30 \times 10^{  -3} $ & $ 4.04 \times 10^{  -3} $ \\ 
\hline\hline
\end{tabular} 
\caption{Same as in Table~\ref{tab:fixedYp} but for the cases with the BBN relation.}
\label{tab:withBBN}
\end{center}
\end{table}

\begin{table}
\begin{center}
\begin{tabular}{l|ccccc}
\hline \hline \\
&  $\Omega_{m}h^{2}$ &  $\Omega_{b}h^{2}$ &  $\Omega_{\Lambda}$ & $ n_s $ \\ 
\hline
Planck & $ 3.31 \times 10^{  -3} $ & $ 2.27 \times 10^{  -4} $ & $ 2.11 \times 10^{  -2} $ & $ 7.56 \times 10^{  -3} $ \\
\quad + SKA phase1 & $ 3.46 \times 10^{  -4} $ & $ 1.09 \times 10^{  -4} $ & $ 2.34 \times 10^{  -3} $ & $ 2.25 \times 10^{  -3} $\\
\quad + SKA phase2 & $ 2.66 \times 10^{  -4} $ & $ 1.05 \times 10^{  -4} $ & $ 9.26 \times 10^{  -4} $ & $ 1.46 \times 10^{  -3} $ \\
\quad + Omniscope & $ 8.31 \times 10^{  -5} $ & $ 3.88 \times 10^{  -5} $ & $ 4.18 \times 10^{  -4} $ & $ 4.87 \times 10^{  -4} $ \\ 
\hline
CMBPol & $ 1.29 \times 10^{  -3} $ & $ 4.90 \times 10^{  -5} $ & $ 8.03 \times 10^{  -3} $ & $ 2.72 \times 10^{  -3} $ \\
\quad + SKA phase1 & $ 3.17 \times 10^{  -4} $ & $ 4.29 \times 10^{  -5} $ & $ 2.14 \times 10^{  -3} $ & $ 1.49 \times 10^{  -3} $  \\
\quad + SKA phase2 & $ 2.23 \times 10^{  -4} $ & $ 4.20 \times 10^{  -5} $ & $ 9.06 \times 10^{  -4} $ & $ 1.05 \times 10^{  -3} $ \\
\quad + Omniscope & $ 5.27 \times 10^{  -5} $ & $ 2.28 \times 10^{  -5} $ & $ 4.10 \times 10^{  -4} $ & $ 3.69 \times 10^{  -4} $  \\ 
\hline \\
&  $A_{s} \times 10^{10} $ & $\tau_{\rm reion}$  & $\Sigma m_{\nu}$ & $\xi$ & $Y_{p}$ \\ \hline
Planck & $ 2.32 \times 10^{  -1} $ & $ 4.66 \times 10^{  -3} $ & $ 1.28 \times 10^{  -1} $ & $ 1.12 $ & $ 1.13 \times 10^{  -2} $ \\
\quad + SKA phase1 & $ 1.92 \times 10^{  -1} $ & $ 4.36 \times 10^{  -3} $ & $ 3.70 \times 10^{  -2} $ & $ 2.10 \times 10^{  -1} $ & $ 5.90 \times 10^{  -3} $ \\
\quad + SKA phase2 & $ 1.89 \times 10^{  -1} $ & $ 4.29 \times 10^{  -3} $ & $ 2.88 \times 10^{  -2} $ & $ 2.05 \times 10^{  -1} $ & $ 5.41 \times 10^{  -3} $ \\
\quad + Omniscope & $ 1.85 \times 10^{  -1} $ & $ 4.17 \times 10^{  -3} $ & $ 1.16 \times 10^{  -2} $ & $ 8.99 \times 10^{  -2} $ & $ 3.83 \times 10^{  -3} $ \\ 
\hline
CMBPol & $ 1.10 \times 10^{  -1} $ & $ 2.49 \times 10^{  -3} $ & $ 4.47 \times 10^{  -2} $ & $ 1.85 \times 10^{  -1} $ & $ 2.83 \times 10^{  -3} $ \\
\quad + SKA phase1 & $ 1.02 \times 10^{  -1} $ & $ 2.42 \times 10^{  -3} $ & $ 1.57 \times 10^{  -2} $ & $ 1.01 \times 10^{  -1} $ & $ 2.15 \times 10^{  -3} $ \\
\quad + SKA phase2 & $ 1.00 \times 10^{  -1} $ & $ 2.37 \times 10^{  -3} $ & $ 1.11 \times 10^{  -2} $ & $ 9.89 \times 10^{  -2} $ & $ 1.96 \times 10^{  -3} $ \\
\quad + Omniscope & $ 7.94 \times 10^{  -2} $ & $ 1.91 \times 10^{  -3} $ & $ 7.47 \times 10^{  -3} $ & $ 4.93 \times 10^{  -2} $ & $ 1.31 \times 10^{  -3} $  \\ 
\hline\hline
\end{tabular} 
\caption{Same as in Table~\ref{tab:fixedYp} but for the cases with freely varying $Y_p$.}
\label{tab:variedYp}
\end{center}
\end{table}

\begin{table}
\begin{itemize}
\item Fixing $Y_p=0.25$ 
\begin{center}
\begin{tabular}{l|ccc}
\hline \hline 
& $\xi=-0.1$ & $\xi=0.05$ & $\xi=0.01$ \\
\hline
Planck & $5.01\times 10^{-1}$ & $ 9.99 \times 10^{  -1} $ & $4.88$ \\
\quad + SKA phase1 & $ 7.85 \times 10^{  -2} $ & $ 1.58 \times 10^{  -1} $ & $ 7.73 \times 10^{  -1} $ \\
\quad + SKA phase1  & $ 7.23 \times 10^{  -2} $ & $ 1.45 \times 10^{  -1} $ & $ 6.76 \times 10^{  -1} $ \\
\quad + Omniscope  & $ 3.02 \times 10^{  -2} $  & $ 6.09 \times 10^{  -2} $ & $ 2.62 \times 10^{  -1} $ \\
\hline
CMBPol & $ 7.55 \times 10^{  -2} $ & $ 1.51 \times 10^{  -1} $ & $ 7.50 \times 10^{  -1} $ \\
\quad + SKA phase1 & $ 4.07 \times 10^{  -2} $ & $ 8.15 \times 10^{  -2} $ & $ 4.05 \times 10^{  -1} $ \\
\quad + SKA phase2 & $ 3.84 \times 10^{  -2} $ & $ 7.69 \times 10^{  -2} $ & $ 3.76 \times 10^{  -1} $ \\
\quad + Omniscope & $ 1.59 \times 10^{  -2} $ & $ 3.19 \times 10^{  -2} $ & $ 1.52 \times 10^{  -1} $ \\
\hline\hline
\end{tabular}
\end{center}
\item With the BBN relation 
\begin{center}
\begin{tabular}{l|ccc}
\hline\hline
& $\xi=-0.1$ & $\xi=0.05$ & $\xi=0.01$ \\
\hline
Planck & $3.72 \times 10^{-2}$ & $ 4.50 \times 10^{  -2} $ & $ 4.29\times 10^{-2}$ \\
\quad + SKA phase1 & $ 1.49 \times 10^{  -2} $ & $ 2.10 \times 10^{  -2} $ & $ 1.90 \times 10^{  -2} $ \\
\quad + SKA phase2 & $ 1.29 \times 10^{  -2} $ & $ 1.83 \times 10^{  -2} $ & $ 1.65 \times 10^{  -2} $ \\
\quad + Omniscope & $ 7.66 \times 10^{  -3} $ & $ 1.28 \times 10^{  -2} $ & $ 1.10 \times 10^{  -2} $ \\
\hline
CMBPol & $ 7.82 \times 10^{  -3} $ & $ 1.03 \times 10^{  -2} $ & $ 9.68 \times 10^{  -3} $ \\
\quad + SKA phase1 & $ 5.89 \times 10^{  -3} $ & $ 7.85 \times 10^{  -3} $ & $ 7.31 \times 10^{  -3} $ \\
\quad + SKA phase2 & $ 5.25 \times 10^{  -3} $ & $ 6.95 \times 10^{  -3} $ & $ 6.47 \times 10^{  -3} $ \\
\quad + Omniscope & $ 2.86 \times 10^{  -3} $ & $ 4.04 \times 10^{  -3} $ & $ 3.65 \times 10^{  -3} $ \\
\hline\hline
\end{tabular}
\end{center}
\item Freely varying $Y_p$
\begin{center}
\begin{tabular}{l|ccc}
\hline\hline
& $\xi=-0.1$ & $\xi=0.05$ & $\xi=0.01$ \\
\hline
Planck & $ 5.61 \times 10^{  -1} $ & $ 1.12 $ & $ 5.42 $ \\
\quad + SKA phase1 & $ 1.05 \times 10^{  -1} $ & $ 2.10 \times 10^{  -1} $ & $ 1.02 $ \\
\quad + SKA phase2 & $ 1.02 \times 10^{  -1} $ & $ 2.05 \times 10^{  -1} $ & $ 9.06 \times 10^{  -1} $ \\
\quad + Omniscope & $ 4.48 \times 10^{  -2} $ & $ 8.99 \times 10^{  -2} $ & $ 3.39 \times 10^{  -1} $ \\
\hline
CMBPol & $ 9.24 \times 10^{  -2} $ & $ 1.85 \times 10^{  -1} $ & $ 9.17 \times 10^{  -1} $\\
\quad + SKA phase1 & $ 5.07 \times 10^{  -2} $ & $ 1.01 \times 10^{  -1} $ & $ 5.03 \times 10^{  -1} $ \\
\quad + SKA phase2 & $ 4.95 \times 10^{  -2} $ & $ 9.89 \times 10^{  -2} $ & $ 4.79 \times 10^{  -1} $ \\
\quad + Omniscope & $ 2.46 \times 10^{  -2} $ & $ 4.93 \times 10^{  -2} $ & $ 2.24 \times 10^{  -1} $ \\
\hline\hline
\end{tabular}
\end{center}
\end{itemize}
\caption{Dependence of $\sigma_\xi$ on the fiducial value of $\xi$.}
\label{tab:xi}
\end{table}

%

\section{Cases with extra radiation}

So far we have been investigating constraints on $\xi$ in combination with CMB and 21~cm  line observations.
Having observed that the combination of observations can improve constraints on $\xi$ from only CMB ones, 
we extend our analysis to consider cosmological models with not only non-zero $\xi$ but also 
extra (dark) radiation other than active neutrinos. 
Throughout this section, we assume that the extra radiation is massless. 
In addition, we assume the BBN relation 
 $Y_p(\Omega_bh^2,~\xi,~\Delta N_\nu)$, 
which allows us to distinguish $\xi$ and $\Delta N_\nu$ even if the active neutrinos are almost massless.

\begin{figure}[htbp]
  \begin{center}
    \resizebox{135mm}{!}{
     \includegraphics[bb=0 30 595 832]{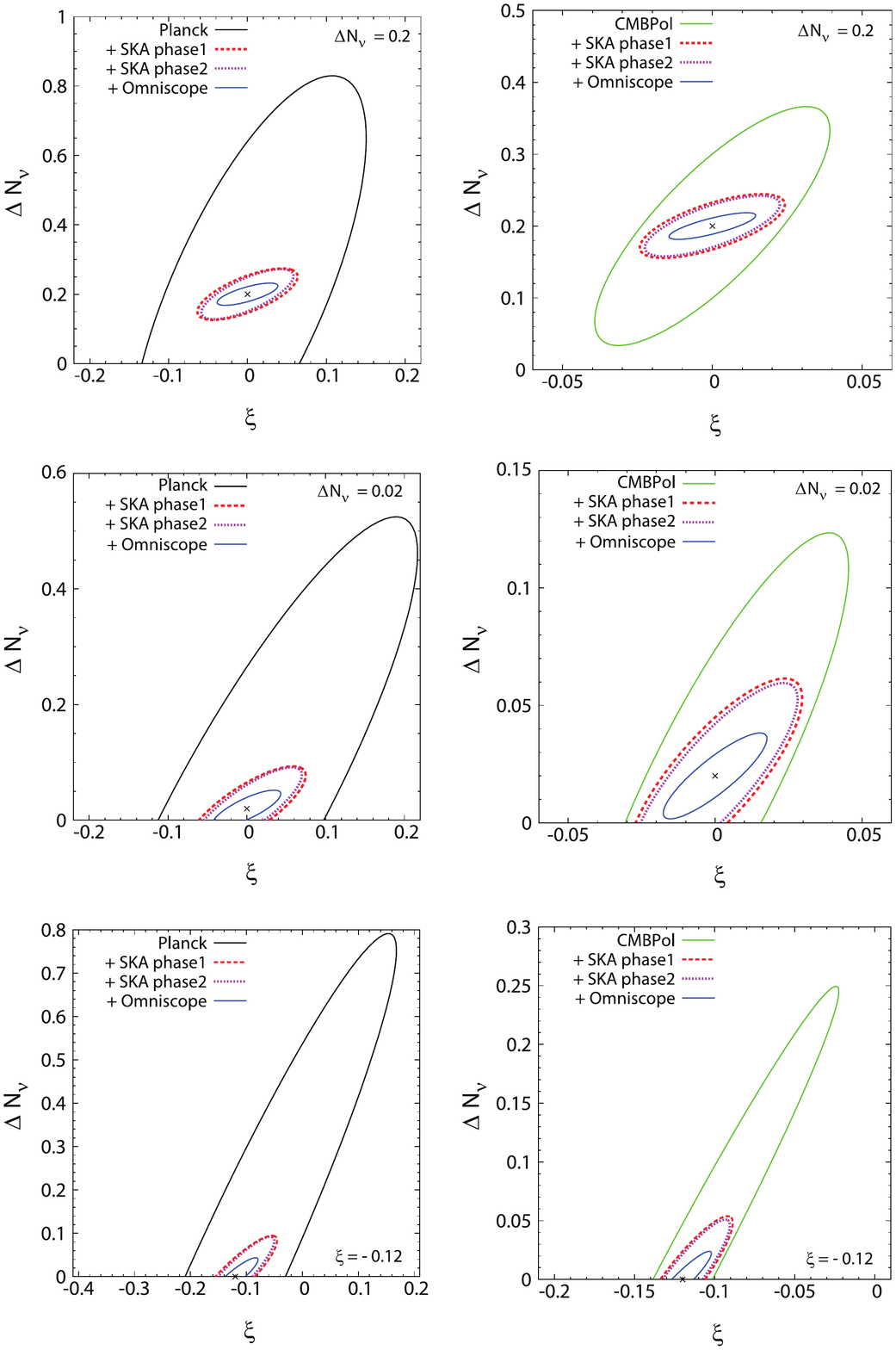}}
  \end{center}
  \caption{Expected 2$\sigma$ constraints on the $\xi$--$\Delta N_\nu$ plane. In this figure, the BBN relation is assumed.
  As fiducial values of ($\xi,~\Delta N_\nu$), we here adopt (0.2, 0), (0.02, 0) and (0, $-0.12$) in the top, middle and bottom panels, respectively.
  Note that scales differ among different panels.}
  \label{fig:xi_dNnu_wBBN}
\end{figure}

In Fig.~\ref{fig:xi_dNnu_wBBN}, we plot 2$\sigma$ constraints in the $\xi$--$\Delta N_\nu$ plane from
CMB alone as well as combinations of CMB and 21~cm line.  Three different fiducial models 
$(\xi, \Delta N_\nu)=(0,~0.2)$, (0,~0.02) and ($-0.12$,~0) are adopted here.
We note that the latter two fiducial models give the similar effective numbers of neutrino species
when neutrinos are relativistic.
We can see that CMB alone cannot constrain $\Delta N_\nu$ tightly. 
Moreover, the sizes of 2$\sigma$ contours in the $\Delta N_\nu$ direction are dependent on fiducial 
parameters $\xi$ and $\Delta N_\nu$.
This dependency should be suggesting that 
observations are not enough constraining and the likelihood surface in the $\xi$-$\Delta N_\nu$ plane
deviates from Gaussian cases to some extent.
This may lead that when one explores constraints in a full parameter space using the Markov chain Monte Carlo, e.g., CosmoMC~\cite{Lewis:2002ah}, 
resulting constraints would be somewhat less stringent than forecasts based on the Fisher matrix analysis.
However, once we combine 21~cm observations, the constraints on $\Delta N_\nu$ greatly improve.
Moreover, the size of errors become almost independent of the fiducial values of $\xi$ and $\Delta N_\nu$ 
by an order of magnitude.
This shows that combinations of CMB and 21~cm line observations will be promising 
to disentangle degenerating $\xi$ and $\Delta N_\nu$.
In Table~\ref{tab:xi_dNnu_wBBN}, we present the 1$\sigma$ constraints only for $\xi$ and $\Delta N_\nu$.
We note that regarding the constraints on other cosmological parameters, 
the inclusion of $\Delta N_\nu$ does not degrade most of them significantly, or, by at most 50 \%.
Only exceptions are the constants on $\Omega_mh^2$ from Planck alone and
$\Omega_bh^2$ from Planck+Omniscope and CMBPol+Omniscope, which are degraded by 2-3 times.

\begin{table}
\begin{itemize}
\item fiducial $(\xi,~\Delta N_\nu)=(0,~0.2)$
\begin{center}
\begin{tabular}{l|cc}
\hline \hline 
\\
& $\xi$ & $\Delta N_\nu$ \\
\hline 
Planck & $ 6.07 \times 10^{-  2} $ & $ 2.54 \times 10^{  -1} $ \\
\quad + SKA phase1 & $ 2.56 \times 10^{  -2} $ & $ 2.99 \times 10^{  -2} $ \\
\quad + SKA phase2  & $ 2.36 \times 10^{  -2} $ & $ 2.91 \times 10^{  -2} $ \\
\quad + Omniscope  & $ 1.55 \times 10^{  -2} $  & $ 1.29 \times 10^{  -2} $ \\
\hline
CMBPol & $ 1.58 \times 10^{-  2} $ & $ 6.71 \times 10^{  -2} $ \\
\quad + SKA phase1 & $ 9.77 \times 10^{  -3} $ & $ 1.79 \times 10^{  -2} $ \\
\quad + SKA phase2  & $ 9.09 \times 10^{  -3} $ & $ 1.70 \times 10^{  -2} $ \\
\quad + Omniscope  & $ 5.83 \times 10^{  -3} $  & $ 7.47 \times 10^{  -3} $ \\
\hline\hline
\end{tabular}
\end{center}
\item fiducial $(\xi,~\Delta N_\nu)=(0,~0.02)$
\begin{center}
\begin{tabular}{l|cc}
\hline \hline 
\\
& $\xi$ & $\Delta N_\nu$ \\
\hline 
Planck & $ 8.74 \times 10^{-  2} $ & $ 2.04 \times 10^{  -1} $ \\
\quad + SKA phase1 & $ 3.01 \times 10^{  -2} $ & $ 2.94 \times 10^{  -2} $ \\
\quad + SKA phase2  & $ 2.82 \times 10^{  -2} $ & $ 2.88 \times 10^{  -2} $ \\
\quad + Omniscope  & $ 1.74 \times 10^{  -2} $  & $ 1.28 \times 10^{  -2} $ \\
\hline
CMBPol & $ 1.83 \times 10^{-  2} $ & $ 4.17 \times 10^{  -2} $ \\
\quad + SKA phase1 & $ 1.20 \times 10^{  -2} $ & $ 1.67 \times 10^{  -2} $ \\
\quad + SKA phase2  & $ 1.13 \times 10^{  -2} $ & $ 1.59 \times 10^{  -2} $ \\
\quad + Omniscope  & $ 7.11 \times 10^{  -3} $  & $ 7.37 \times 10^{  -3} $ \\
\hline\hline
\end{tabular}
\end{center}
\item fiducial $(\xi,~\Delta N_\nu)=(-0.12,~0)$
\begin{center}
\begin{tabular}{l|cc}
\hline \hline 
\\
& $\xi$ & $\Delta N_\nu$ \\
\hline 
Planck & $ 1.16 \times 10^{-  1} $ & $ 3.19 \times 10^{  -1} $ \\
\quad + SKA phase1 & $ 3.02 \times 10^{  -2} $ & $ 3.81 \times 10^{  -2} $ \\
\quad + SKA phase2  & $ 2.82 \times 10^{  -2} $ & $ 3.71 \times 10^{  -2} $ \\
\quad + Omniscope  & $ 1.64 \times 10^{  -2} $  & $ 1.75 \times 10^{  -2} $ \\
\hline
CMBPol & $ 3.93 \times 10^{-  2} $ & $ 1.01 \times 10^{  -1} $ \\
\quad + SKA phase1 & $ 1.26 \times 10^{  -2} $ & $ 2.17 \times 10^{  -2} $ \\
\quad + SKA phase2  & $ 1.19 \times 10^{  -2} $ & $ 2.06 \times 10^{  -2} $ \\
\quad + Omniscope  & $ 7.22 \times 10^{  -3} $  & $ 9.65 \times 10^{  -3} $ \\
\hline\hline
\end{tabular}
\end{center}
\end{itemize}
\caption{1 $\sigma$ errors on $\xi$ and $\Delta N_\nu$ for the case with the BBN relation 
and their dependence on fiducial $(\xi,~\Delta N_\nu)$}
\label{tab:xi_dNnu_wBBN}
\end{table}

\chapter{Summary}
\label{sec:conclusion}

In this thesis,
we have studied how we can constrain the total neutrino mass $\Sigma
m_{\nu}$, the effective number of neutrino species $N_{\nu}$, 
the neutrino mass hierarchy, and the lepton asymmetry $\xi$ in the Universe
by using 21 cm line (SKA or Omniscope) and CMB (Planck,
\textsc{Polarbear}-2, Simons Array or CMBPol) observations.
It is essential to combine the 21 cm line observation 
with the precise CMB polarization observation
to break various degeneracies in cosmological parameters
when we perform multiple-parameter fittings.

About the constraints on the $\Sigma m_{\nu}$--$N_{\nu}$ plane, we
have found that there is a significant improvement in the
sensitivities to $\Sigma m_{\nu}$ and $N_{\nu}$ by adding the BAO
experiments to the experiments of CMB. However, for a fiducial value $\Sigma
m_{\nu}=0.1$ eV, it is impossible to detect the non-zero neutrino mass
at 2$\sigma$ level even by using the combination of Simons Array and
DESI.
On the other hand, by adding the 21 cm experiments (SKA phase1) to the
CMB experiment, we find that there is a substantial improvement. By
using Planck + Simons Array + BAO(DESI) + SKA phase1, we can detect
the non-zero neutrino mass (but it is necessary to remove the foregrounds
with high degree of accuracy).
For a fiducial value $\Sigma m_{\nu}=0.06$ eV, which corresponds to
the lowest value in the normal hierarchy of the neutrino mass, we need
the sensitivity of SKA phase2 in order to detect the non-zero
neutrino mass.

Next, as for the determination of the neutrino mass hierarchy, we have
introduced the parameter $r_{\nu}= (m_{3} - m_{1} )/\Sigma m_{\nu}$,
and studied how to discriminate a true hierarchy from the other by
constraining $r_{\nu}$.
As was clearly shown in
Fig.~\ref{fig:hie_ellipse}, by adopting the combinations of the Planck
+ Simons Array + BAO(DESI) + SKA phase2,
we will be able to determine the hierarchy to be inverted or normal at 2$\sigma$
unless the mass structure is  degenerated.

Finally, for the constraints on the lepton asymmetry, when we consider
constraints on $\xi$ in the absence of extra radiation, we have found
that, even without assuming the BBN relation, by combining the 21~cm line
observations with the CMB observations, we can constrain $\xi$ with a
better accuracy than the primordial abundances of light elements,
which cannot be achieved by the CMB observation alone.
Next, once the BBN relation has been taken into account,
even the sensitivity of the CMB observations alone to $\xi$
substantially improves. However the 21~cm line observations can still
improve the constraints and be useful in constraining the lepton
asymmetry.
In addition, we have also investigated constraints on $\xi$ in the
presence of some extra radiation. We have shown that the 21~cm line
observations can substantially improve the constraints on $\Delta
N_\nu$ compared with the case of the CMB observations alone, and allow
us to distinguish between the lepton asymmetry and extra radiation.

Our results indicate that the 21~cm line 
and CMB polarization observations
can become a powerful probe of the neutrino properties 
and the origin of matter in the Universe.
%

\chapter*{Acknowledgments}

First of all, I would like to show my greatest appreciation 
to my supervisor, Dr.~Kazunori Kohri 
at the Graduate University for Advanced Studies (SOKENDAI)
and at High Energy Accelerator Research Organization (KEK),
whose enormous support and insightful comments 
were invaluable during the course of my study,
and discussions with him have been illuminating for me.
I have learned and study a lot of things from him during these five years.

Also my deepest appreciation goes to Prof.~Hideo Kodama 
at the Graduate University for Advanced Studies (SOKENDAI)
and at High Energy Accelerator Research Organization (KEK),
whose comments were innumerably valuable at all times,
and I received generous support from him.
I am also indebted to Dr.~Kunihito Ioka 
at the Graduate University for Advanced Studies (SOKENDAI)
and at High Energy Accelerator Research Organization (KEK),
who provided very useful discussion and sincere encouragement. 

I would also like to thank Prof.~Masashi Hazumi 
at the Graduate University for Advanced Studies (SOKENDAI)
and at High Energy Accelerator Research Organization (KEK).
For the analysis of CMB, he gives insightful comments and
technical help about CMB observation and 
treatment of its foregrounds.
Besides, I also grateful to Dr.~Tomo Takahashi
at Saga University 
and Dr.~Toyokazu Sekiguchi.
they give me a enormous help and provided
significant contributions for our work.

Additionally, I am deeply grateful to 
the members of the KEK Theory center Cosmophysics Group:
Dr.~Seiju Ohashi, Dr.~Shota Kisaka
Dr.~Hajime Takami, Dr.~Kentaro Tanabe,
Dr.~Hirotaka Yoshino.
They gave me constructive comments and warm encouragement.

My junior fellow Eunseong So and other students in Department of Particle 
and Nuclear Physics at SOKENDAI provided advice on life as well as science.
I am thankful to all the colleagues at SOKENDAI and KEK,
and secretaries of the KEK Theory center Tamao Shishido and Kieko Iioka.

Financially, I appreciate the support of 
the Grant-in-Aid for Scientific research from 
the Ministry of Education, Science, Sports, and Culture, Japan, 
Nos. 25.4260, which made it possible to complete my study.

Finally, I would also like to express my gratitude to my family
for their moral support and warm encouragements.

\appendix

\chapter
[Hyperfine splitting of neutral hydrogen atom]
{Hyperfine splitting of neutral hydrogen atom
\label{ap:bisai} \normalsize{\cite{igikawai} }}

Here, we show the energy splitting due to the hyperfine structure
of neutral hydrogen atom.
This splitting is caused by an interaction 
between the magnetic moment of nucleus and that of electron.
This splitting is much smaller than 
that of the fine structure,
which is caused by the interaction between the spin and 
orbital angular momentum.
Below we calculate the energy splitting
by considering the spin-spin interaction.

Since a nucleus can be regarded as a magnetic dipole,
the magnetic moment $\mbox{\boldmath $M$}_{p}$ is given by
\begin{eqnarray}
\mbox{\boldmath $M$}{p} &=& 
     \frac{|e|g_{p}}{2M_{p}c}\mbox{\boldmath $\hat{I}$}
     =
     g_{p}\mu_{p}
     \frac{\mbox{\boldmath $\hat{I}$}}{\hbar}, \\
     \mu_{p} &\equiv & \frac{|e|\hbar}{2M_{p}c},
\end{eqnarray}
where $\mbox{\boldmath $\hat{I}$}$,
$M_{p}$, $e$ and $g_{p}$ are 
the spin, the mass, the electric charge and the $g$ factor
of the nucleus, respectively.
The vector potential due to the magnetic moment
is expressed as
\begin{eqnarray}
\mbox{\boldmath $A$}(\mbox{\boldmath $r$}) =
     - \left(
             \mbox{\boldmath $M$}{p} \times 	\nabla
       \right)
       \left(
             \frac{1}{r}
       \right),
\end{eqnarray}
and the magnetic field due to the potential can
be written as
\begin{eqnarray}
\mbox{\boldmath $B$}(\mbox{\boldmath $r$}) &=&
     \nabla \times \mbox{\boldmath $A$} \nonumber \\
     &=&
     -g_{p}\mu_{p}\nabla \times
     \left(
           \frac{
                 \mbox{\boldmath $\hat{I}$}
                 }
                 {\hbar}
           \times \nabla
     \right)
     \left( 
          \frac{1}{r}
     \right).
\end{eqnarray}
Therefore,
the potential $V_{hfs}$ which is caused by
the interaction between the magnetic field
and the spin of the electron
\mbox{\boldmath $\hat{S}$} is given by
\begin{eqnarray}
V_{hfs}
     &=&
     \frac{|e|}{m_{e}c}
     \mbox{\boldmath $\hat{S}$}
     \cdot
     \mbox{\boldmath $B$}(\mbox{\boldmath $r$})
     =2 \mu_{B}\frac{\mbox{\boldmath $\hat{S}$}}{\hbar}\cdot 
\mbox{\boldmath $B$}(\mbox{\boldmath $r$}) \nonumber\\
     &=&
     -2g_{p}\mu_{B}\mu_{p}
     \frac{\mbox{\boldmath $\hat{S}$}}{\hbar} 
     \cdot
     \left[
     \nabla \times
     \left(
           \frac{
                 \mbox{\boldmath $\hat{I}$}
                 }
                 {\hbar}
           \times \nabla
     \right)
     \left( 
          \frac{1}{r}
     \right)
     \right] \nonumber \\
     &=&
      -2g_{p}\mu_{B}\mu_{p}
     \left[
     \frac{\mbox{\boldmath $\hat{S}$}}{\hbar} \cdot
     \left\{
     \frac{
           \mbox{\boldmath $\hat{I}$}
           }
           {\hbar} \cdot
     \nabla^{2}
     -
     \nabla
     \left(
          \nabla \cdot
          \frac{
           \mbox{\boldmath $\hat{I}$}
           }
           {\hbar} 
     \right)
     \right\}
     \right] \frac{1}{r},
\end{eqnarray}
where $\mu_{B}=|e|\hbar/(2m_{e}c)$
is the Bohr magnet and 
$m_{e}$ is the mass of the electron.
In the S state, the first order perturbation of the potential 
is written as
\begin{eqnarray}
\hspace{-20pt}
\left\langle V_{hfs} \right\rangle
     =
      -2g_{p}\mu_{B}\mu_{p}
     \int dr^{3} |\phi_{100}(r)|^{2}
     \left[\left\langle
     \frac{\mbox{\boldmath $\hat{S}$}\cdot\mbox{\boldmath $\hat{I}$}}{\hbar^{2}} 
     \right\rangle
     \nabla^{2} 
     -
     \left\langle  \left(
     \frac{\mbox{\boldmath $\hat{S}$}}{\hbar}  \cdot \nabla     
     \right)
     \left(
          \frac{ \mbox{\boldmath $\hat{I}$}}{\hbar} 
           \cdot \nabla
     \right)\right\rangle
     \right] \frac{1}{r},
\end{eqnarray}
where $\phi_{100}$ is the wave function of 
the the S state, and it is expressed as
\begin{eqnarray}
\phi_{100}&=&
 \frac{1}{\sqrt[]{4\pi}}
 \left( \frac{1}{a_{0}} \right)^{\frac{3}{2}} 2 \exp
 \left(
    -\frac{r}{a_{0}}
 \right), \\ 
 a_{0} &= &\frac{\hbar^{2}}{m_{e}e^{2}}.
\end{eqnarray}
According to 
the following spherical symmetric property
of the S state, 
\begin{eqnarray}
\left\langle  \left(
     \mbox{\boldmath $\hat{S}$} \cdot \nabla     
     \right)
     \left(
          \mbox{\boldmath $\hat{I}$}
          \cdot \nabla
\right)\right\rangle
\longrightarrow
       \frac{1}{3} 
       \left\langle \mbox{\boldmath $\hat{S}$} \cdot \mbox{\boldmath $\hat{I}$} \right\rangle
       \nabla^{2},
\end{eqnarray}
the potential can be rewritten as
\begin{eqnarray}
\left\langle V_{hfs} \right\rangle
     &=&
      -\frac{4}{3}
     g_{p}\mu_{B}\mu_{p}
     \int dr^{3} |\phi_{100}(r)|^{2}
     \left\langle
     \frac{\mbox{\boldmath $\hat{S}$}\cdot\mbox{\boldmath $\hat{I}$}}{\hbar^{2}} 
     \right\rangle
     \nabla^{2} 
     \frac{1}{r} \nonumber \\ 
     &=&
     -\frac{4}{3}
     g_{p}\mu_{B}\mu_{p}
     \int dr^{3} |\phi_{100}(r)|^{2}
     \left\langle
     \frac{\mbox{\boldmath $\hat{S}$}\cdot\mbox{\boldmath $\hat{I}$}}{\hbar^{2}} 
     \right\rangle
     (-4\pi \delta^{D}(\mbox{\boldmath $r$}) )    
\nonumber \\ 
     &=&
     \frac{16\pi}{3}
     g_{p}\mu_{B}\mu_{p}
     |\phi_{100}(0)|^{2}
     \left\langle
     \frac{\mbox{\boldmath $\hat{S}$}\cdot\mbox{\boldmath $\hat{I}$}}{\hbar^{2}} 
     \right\rangle.
\end{eqnarray}
Here, 
the square of the  absolute value of 
the wave function $\phi_{100}$
at $r=0$ is give by
\begin{eqnarray}
|\phi_{100}(0)|^{2}=\frac{1}{\pi a_{0}^{3}},
\end{eqnarray}
Therefore, we can obtain 
\begin{eqnarray}
\left\langle V_{hfs} \right\rangle
     &=&
     \frac{16\pi}{3}
     g_{p}\mu_{B}\mu_{p}
     \frac{1}{\pi a_{0}^{3}}
     \left\langle
     \frac{\mbox{\boldmath $\hat{S}$}\cdot\mbox{\boldmath $\hat{I}$}}{\hbar^{2}} 
     \right\rangle \nonumber \\
     &=&
     \frac{8}{3}
     \left( \frac{e^{2}}{2a_{0}}\right) g_{p}
     \frac{m_{e}}{M_{p}}\alpha_{EM}^{2}
      \left\langle
     \frac{\mbox{\boldmath $\hat{S}$}\cdot\mbox{\boldmath $\hat{I}$}}{\hbar^{2}} 
     \right\rangle, 
\end{eqnarray}
where $\alpha_{EM} = e^{2}/(\hbar c)$
is the fine structure constant.
By using the total spin of the nucleus
$\mbox{\boldmath $\hat{F}$}$
(=$\mbox{\boldmath $\hat{S}$}+\mbox{\boldmath $\hat{I}$}$),
we can express this potential as
\begin{eqnarray}
 \frac{\left\langle \mbox{\boldmath $\hat{S}$}
      \cdot\mbox{\boldmath $\hat{I}$}\right\rangle }{\hbar^{2}} 
&=&\frac{\left\langle
\mbox{\boldmath $\hat{F}$}^{2}-\mbox{\boldmath $\hat{S}$}^{2}-\mbox{\boldmath $\hat{I}$}^{2}
\right\rangle}
      {2\hbar^{2}}=
 \frac{F(F+1)-3/4-I(I+1)}{2}\nonumber \\
&=&
\frac{1}{2}
\left\{ \begin{array}{ll}
I & \left(F=I+\frac{1}{2}\right), \\
-I-1 & \left(F=I-\frac{1}{2}\right). \\
\end{array} \right.
\end{eqnarray}
In the case of $I=1/2$,
the difference between the upper and lower states is 1.
Therefore, the energy splitting of the hyperfine structure 
$\Delta E_{hfs}$ is  given by
\begin{eqnarray}
\Delta E_{hfs}
     &=&
     \frac{8}{3}
     \left( \frac{e^{2}}{2a_{0}}\right) g_{p}
     \frac{m_{e}}{M_{p}}\alpha_{EM}^{2}.
\end{eqnarray}
By substituting the $g$ factor of the proton $g_{p}=5.56$
into this equation, we obtain the following value of the energy,
\begin{eqnarray}
\Delta E_{hfs}
     &=&
     \frac{8}{3}
     \left( 13.6 \ {\rm eV} \right) (5.56)
     \frac{1}{1840}\left( \frac{1}{137} \right)^{2}\nonumber \\
     &\simeq&
     5.8 \times 10^{-6} {\rm eV}.
\end{eqnarray}
In this case, the transition frequency is 
\begin{eqnarray}
\nu \simeq 1.4 \ {\rm GHz},
\end{eqnarray}
and the wave length is
\begin{eqnarray}
\lambda \simeq 21 \ {\rm cm}.
\end{eqnarray}
This is the 21 cm line due to the 
neutral hydrogen atom.

\chapter
[Einstein coefficients]
{Einstein coefficients
 \label{ap:Ein} \normalsize{\cite{text:rybicki,text:Nakai}}
}

Here, we show the derivation of the 
relation between the Einstein A and B coefficient
(Eqs.(\ref{eq:Einsteinrelation1}) and (\ref{eq:Einsteinrelation2})),
by considering the equilibrium between the upper and lower states.
By using the definition of the Einstein coefficients,
the time derivatives of the number densities of 
the upper $n_{u}$ and lower $n_{l}$ states are give by
\begin{eqnarray}
\frac{d n_{u}}{dt}&=&n_{l}B_{lu}I_{\nu_{ul}} - n_{u}(A_{ul} + B_{ul}I_{\nu_{ul}}), \label{eq:appenB1}\\
\frac{d n_{l}}{dt}&=&-n_{l}B_{lu}I_{\nu_{ul}} + n_{u}(A_{ul} + B_{ul}I_{\nu_{ul}}), \label{eq:appenB2}
\end{eqnarray}
where $I_{\nu}$ is the specific intensity of 
the incident photon and 
$\nu_{ul}$ is the transition frequency.
Since the number of particles does not vary
in the equilibrium state,
the derivatives of the number densities are zero, 
i.e. $dn_{u}/dt=0$, $dn_{l}/dt=0$.
In this case, by Eqs.(\ref{eq:appenB1}) or (\ref{eq:appenB2}), 
we  can find
\begin{eqnarray}
n_{l}B_{lu}I_{\nu_{ul}} - n_{u}(A_{ul} + B_{ul}I_{\nu_{ul}})=0, 
\nonumber \\ \longrightarrow
I_{\nu_{ul}} = \frac{A_{ul}}{B_{ul}}
\frac{1}{\frac{B_{lu}n_{l}}{B_{ul}n_{u}} -1}. \label{eq:appenB3}
\end{eqnarray}
Furthermore, the following Boltzmann distribution is valid
in thermal equilibrium,
\begin{eqnarray}
\frac{n_{u}}{n_{l}}
     = \frac{g_{u}}{g_{l}}\exp\left(-\frac{h_{P}\nu_{ul}}{k_{B}T} \right).
\end{eqnarray}
By substituting this equation into Eq.(\ref{eq:appenB3}),
$I_{\nu_{ul}}$ is expressed as
\begin{eqnarray}
I_{\nu_{ul}} = \frac{A_{ul}}{B_{ul}}
               \frac{1}{\frac{B_{lu}}{B_{ul}} \frac{g_{l}}{g_{u}}\exp\left(\frac{h_{P}\nu_{ul}}{k_{B}T} \right)
               -1} \label{eq:appenB4}.
\end{eqnarray}  
Here, the specific intensity of the black body is given by
\begin{eqnarray}
I_{\nu_{ul}}=I_{\nu_{ul}}^{BB}
            =\frac{2h_{P}\nu_{ul}^{3}}{c^{2}}
             \frac{1}{\exp \left(\frac{h_{P}\nu_{ul}}{k_{B}T}\right)-1}, \label{eq:appenB5}
\end{eqnarray}
In comparison between Eqs.(\ref{eq:appenB4}) and (\ref{eq:appenB5}),
we obtain the following relations between the Einstein coefficients,
\begin{eqnarray}
&&\frac{A_{ul}}{B_{ul}}=\frac{2h_{P}\nu_{ul}^{3}}{c^{2}} 
           \ \ \longrightarrow \ \ 
           A_{ul}= \frac{2h_{P}\nu_{ul}^{3}}{c^{2}}B_{ul}, \\
&&\frac{B_{lu}}{B_{ul}}\frac{g_{l}}{g_{u}}=1
           \ \ \ \ \  \longrightarrow  \ \
           B_{ul}=\frac{g_{l}}{g_{u}}B_{lu}.
\end{eqnarray}

\chapter[Non-relativistic limit of $\rho_{\nu}+\rho_{\bar{\nu}}$]
{Non-relativistic limit of $\rho_{\nu}+\rho_{\bar{\nu}}$
\normalsize{\cite{Kohri:2014hea}} }
\label{sec:app_lept}

\section{Expressions for the coefficients $C_i$}
\label{sec:app_lept1}

Below we give explicit expressions for the coefficients $C_i$, which are necessary to 
obtain Eqs.~\eqref{eq:rho_nr} and \eqref{eq:p_nr}. 
\begin{eqnarray}
C_0 & = & \frac{2}{e^y+1}, \\
C_2 & = & \frac{e^y \left(e^y-1\right)}{\left(e^y+1\right)^3}, \\
C_4 & = & \frac{e^y \left(11 e^y-11 e^{2 y}+e^{3 y}-1\right)}{12 \left(e^y+1\right)^5}, \\
C_6 & = &  \frac{e^y \left(57 e^y-302 e^{2 y}+302 e^{3 y}-57 e^{4 y}+e^{5 y}-1\right)}{360
   \left(e^y+1\right)^7}, \\ 
C_8 & = &  \frac{e^y \left(247 e^y-4293 e^{2 y}+15619 e^{3 y}-15619 e^{4 y}+4293 e^{5
   y}-247 e^{6 y}+e^{7 y}-1\right)}{20160 \left(e^y+1\right)^9}, \nonumber \\
   \\
C_{10} & = &   \frac{e^y \! \left(1013 e^y\!-\!47840 e^{2 y}\!+\!455192 e^{3 y}\!-\!1310354 e^{4 y}\!+\!1310354
   e^{5 y}\!-\!455192 e^{6 y}\right)}
   {1814400\left(e^y+1\right)^{11}} \nonumber \\
   && + \frac{e^y \! \left(47840 e^{7 y}\!-\!1013 e^{8 y}\!+\!e^{9 y}\!-\!1\right)}
   {1814400\left(e^y+1\right)^{11}}.
\end{eqnarray}

\section{Non-relativistic limit of $\rho_{\nu}+\rho_{\bar{\nu}}$ and $p_{\nu}+p_{\bar{\nu}}$ \\ for any $\xi$}\label{sec:app_lept2}

Below we show the exact solutions for the $\rho_{\nu}+\rho_{\bar{\nu}}$ and $p_{\nu}+p_{\bar{\nu}}$
for any $\xi$ in non-relativistic limit by using polylogarithm ${\rm Li}_{s}(z)$, which is one of special functions.
\begin{eqnarray}
\!\!\!\!\!\!\!\!\!\!
\rho_\nu \!+\! \rho_{\bar{\nu}} \!
&\simeq &\!
\frac{T_{\nu }^4 a \tilde{m}}{2\pi^2 } \int_0^{\infty} y^2 dy  
\left[ 1 + \frac12 \left( \frac{y}{a \tilde{m}} \right)^2  \right]
\left( \frac{1}{e^{y + \xi} +1}  +  \frac{1}{e^{y - \xi} +1} \right) \notag \\
&=&\!
\frac{T_{\nu }^4 a \tilde{m}}{2\pi^2 }\!
\left[
\!-2\!\left\{\!{\rm Li}_{3}(\!-e^{-\xi})+{\rm Li}_{3}(\!-e^{\xi})\!\right\}\!
\right] 
\! +\! \frac{T_{\nu }^4}{4\pi^2  a \tilde{m}}\! \!
\left[
\!-24\!\left\{\!{\rm Li}_{5}(\!-e^{-\xi})+{\rm Li}_{5}(\!-e^{\xi})\!\right\}\!
\right]\!, \\
\!\!\!\!\!\!\!\!\!\!
p_{\nu}\!+\!p_{\bar{\nu}}\!
&\simeq&\!
\frac{T_{\nu }^4 }{6\pi^2 a \tilde{m}} \int_0^{\infty} y^4 dy  
\left[ 1 - \frac12 \left( \frac{y}{a \tilde{m}} \right)^2  \right]
\left( \frac{1}{e^{y + \xi} +1}  +  \frac{1}{e^{y - \xi} +1} \right) \notag \\
&=&\!
\frac{T_{\nu }^4 }{6\pi^2 a \tilde{m}} \!\!
\left[
\!-24\!\left\{\!{\rm Li}_{5}(\!-e^{-\xi})\!+\!{\rm Li}_{5}(\!-e^{\xi})\!\right\}\!
\right]
\! - \! \frac{T_{\nu }^4}{12\pi^2  (a \tilde{m})^{3}}\! \!
\left[
\!-720\!\left\{\!{\rm Li}_{7}(\!-e^{-\xi})\!+\!{\rm Li}_{7}(\!-e^{\xi})\!\right\}\!
\right]. \nonumber \\
\end{eqnarray} 
If we expand these formulas around  $\xi = 0$,
they reduce to Eqs.~\eqref{eq:rho_nr} and \eqref{eq:p_nr}.

\chapter[BBN relation]
{BBN relation \normalsize{\cite{Kohri:2014hea}} }
\label{sec:BBNrelation}

In the early universe with a higher temperature than ${\mathcal O}(1)$~MeV,
the inter-converting reactions between
neutron and proton through the weak interaction ($n+ e^+
\leftrightarrow p + \nu_e$, $n+ \bar{\nu}_e \leftrightarrow p + e^-$,
and $n \leftrightarrow p + e^- + \nu_e$) were sufficiently rapid. 
In this case, the neutron to proton ratio obeys its thermal equilibrium
value,
\begin{eqnarray}
  \label{eq:n2pth}
 \frac{n}{p} = \exp\left[ - \frac{\Delta m_{np} + \mu_{\nu_e}}{T}\right] = \exp\left[ -\frac{\Delta m_{np}}{T} - \xi_{\nu_e}\right],
\end{eqnarray}
with the mass difference $\Delta m_{np} = 1.3$~MeV. 
Here, we explicitly
wrote the degeneracy parameter of $\nu_e$ to be $\xi_{\nu_e} =
\mu_{\nu_e}/T_{\nu}$ with $\mu_{\nu_e}$ being the chemical potential of
$\nu_e$. It is remarkable that the electron's chemical potential $\mu_{e^{-}}$ must
be much smaller than that of $\nu_e$ because of the neutrality of the
Universe $\xi_{e} = \mu_{e^{-}}/T \sim {\mathcal O}(\eta) \ll \xi_{\nu_e}$ 
with  $T$ and $\eta$ being the photon temperature and the baryon-to-photon ratio, respectively.
Accordingly $\xi_{\nu_e}$ affects the freezeout value of $n/p$, which can change
the light element abundances. 
In particular, $Y_p$ depends on
$\xi_{\nu_e}$ in addition to $\eta$ (or $\Omega_bh^2$) and $N_\nu$.
Thus, $Y_p$ is related to those three parameters,
i.e. $Y_p$=$Y_p(\Omega_bh^2,~\xi_{\nu_e},~\Delta N_\nu)$,
which is called "the BBN relation".

Since we need quite a precise value of $Y_p$ in the current studies,
we numerically compute $Y_p$ as functions of those three parameters
without adopting known fitting formula (e.g., given in
Ref.~\cite{Steigman:2007xt}). In this computation, we have used the
most recent data for nuclear reaction
rates~\cite{Smith:1992yy,Angulo:1999zz,Cyburt:2001pp,Serpico:2004gx, Cyburt:2008up}.

\begin{figure}[htbp]
\begin{center}
\resizebox{120mm}{!}{
\includegraphics[bb= 10 142 584 699]{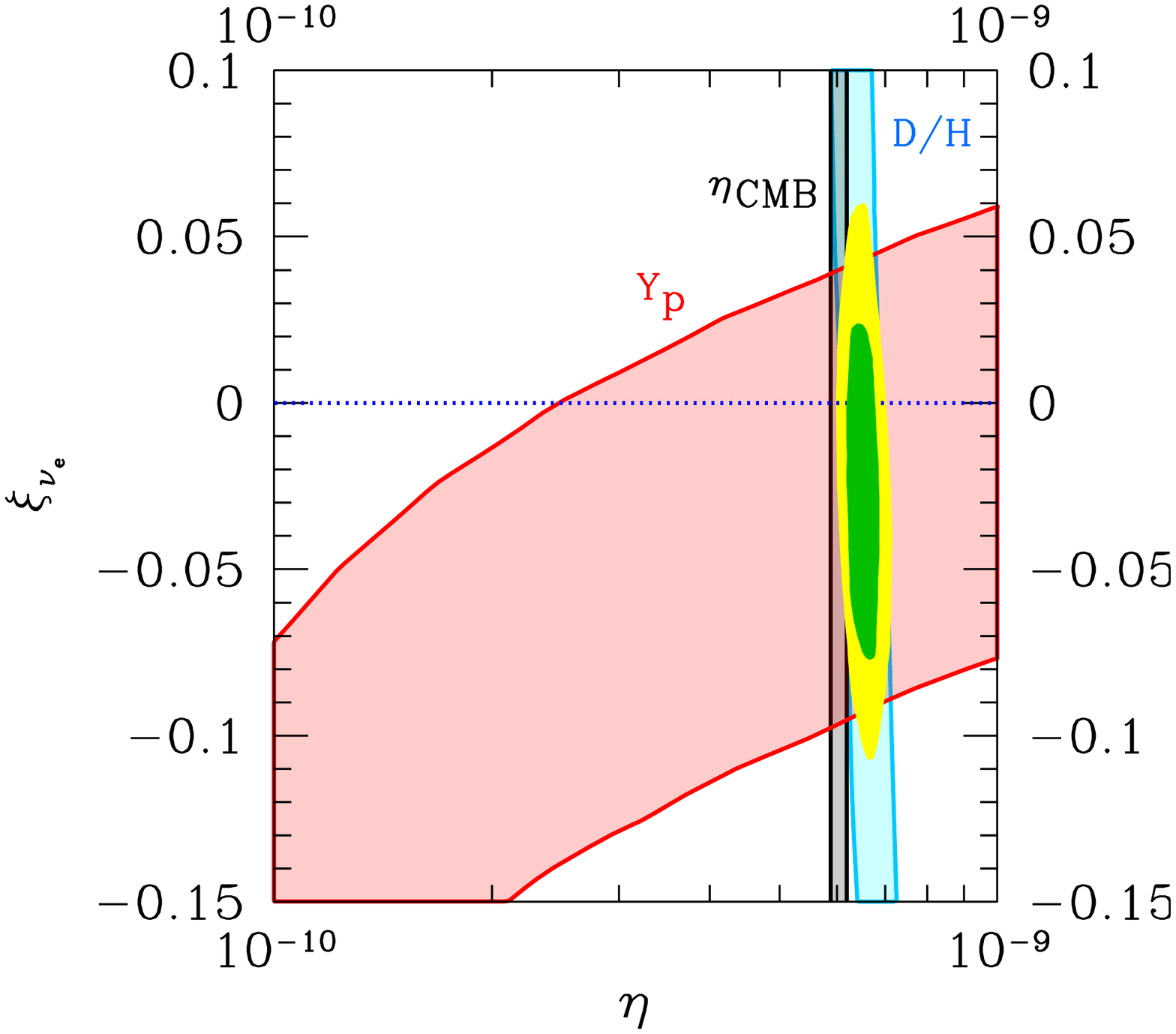}
}
\end{center}
\caption{Regions allowed by the BBN alone in the
  $\eta-\xi_{\nu_e}$ plane. The 68$\%$ and the 95$\%$
  C.L. contours are plotted, respectively.  Here, we set $\Delta
  N_\nu=0$. 
The vertical band represents the baryon to photon ratio reported by Planck
$\eta = (6.04 \pm 0.15)\times10^{-10}$ at 95$\%$C.L..
The line of each light element corresponds to the individual constraint at 95$\%$ C.L..
}
  \label{fig:etaXi2014}
\end{figure}

In Fig.~\ref{fig:etaXi2014}, as a reference, we plotted allowed
regions in the $\eta-\xi_{\nu_e}$ plane at the 68$\%$ and the
95$\%$ C.L, respectively. Here we set $\Delta N_\nu=0$. We have
adopted the following observational light element abundances, $Y_p =
0.2534 \pm 0.0083$ (68$\%$) \cite{Aver:2011bw} and D/H=
$n_{\rm  D}/n_{\rm H} = (2.535 \pm 0.050) \times 10^{-5}$
(68$\%$)~\cite{Pettini:2012ph}.

\end{document}